\documentclass[twocolumn]{aastex631}

\shorttitle{Kinematics in NGC 6153}
\shortauthors{Richer et al.}
\graphicspath{{./}{figures/}}

\begin{document}

\title{NGC 6153:  Reality is complicated\footnote{Based on observations collected at the European Southern Observatory under ESO programme(s) 69.D-0174(A).}}

\correspondingauthor{Michael G. Richer}
\email{richer@astro.unam.mx}

\author[0000-0003-4757-1153]{Michael G. Richer}
\affiliation{Instituto de Astronom\'ia, Universidad Nacional Aut\'onoma de M\'exico, Apartado Postal 106, CP 22800 Ensenada, Baja California, M\'exico; richer@astro.unam.mx, jal@astro.unam.mx}

\author[0000-0000-0000-0000]{Anabel Arrieta}
\affiliation{Departamento de F\'isica y Matem\'aticas, Universidad Iberoamericana, Prolongaci\'on Paseo de la Reforma 880, Lomas de Santa Fe, CP 01210, Ciudad de M\'exico, M\'exico; anabel.arrieta@ibero.mx, lorena.arias@ibero.mx, eadolfogalindo@gmail}

\author[0000-0000-0000-0000]{Lorena Arias}
\affiliation{Departamento de F\'isica y Matem\'aticas, Universidad Iberoamericana, Prolongaci\'on Paseo de la Reforma 880, Lomas de Santa Fe, CP 01210, Ciudad de M\'exico, M\'exico; anabel.arrieta@ibero.mx, lorena.arias@ibero.mx, eadolfogalindo@gmail}

\author[0000-0000-0000-0000]{Lesly Casta\~neda-Carlos}
\affiliation{Instituto de Astronom\'ia, Universidad Nacional Aut\'onoma de M\'exico, Apartado Postal 70-264, CP 04510 Ciudad de M\'exico, M\'exico; silvia@astro.unam.mx, lcastaneda@astro.unam.mx}

\author[0000-0000-0000-0000]{Silvia Torres-Peimbert}
\affiliation{Instituto de Astronom\'ia, Universidad Nacional Aut\'onoma de M\'exico, Apartado Postal 70-264, CP 04510 Ciudad de M\'exico, M\'exico; silvia@astro.unam.mx, lcastaneda@astro.unam.mx}

\author[0000-0000-0000-0000]{Jos\'e Alberto L\'opez}
\affiliation{Instituto de Astronom\'ia, Universidad Nacional Aut\'onoma de M\'exico, Apartado Postal 106, CP 22800 Ensenada, Baja California, M\'exico; richer@astro.unam.mx, jal@astro.unam.mx}

\author[0000-0000-0000-0000]{Adolfo Galindo}
\affiliation{Departamento de F\'isica y Matem\'aticas, Universidad Iberoamericana, Prolongaci\'on Paseo de la Reforma 880, Lomas de Santa Fe, CP 01210, Ciudad de M\'exico, M\'exico; anabel.arrieta@ibero.mx, lorena.arias@ibero.mx, eadolfogalindo@gmail.com}



\begin{abstract}

We study the kinematics of emission lines that arise from many physical processes in NGC 6153 based upon deep, spatially-resolved, high resolution spectra acquired with the UVES spectrograph at the ESO VLT.  Our most basic finding is that the plasma in NGC 6153 is complex, especially its temperature structure.  The kinematics of most emission lines defines a classic expansion law, with the outer part expanding fastest (normal nebular plasma).  However, the permitted lines of \ion{O}{1}, \ion{C}{2}, \ion{N}{2}, \ion{O}{2}, and \ion{Ne}{2} present a constant expansion velocity that defines a second kinematic component (additional plasma component).  The physical conditions imply two plasma components, with the additional plasma component having lower temperature and higher density.  The [\ion{O}{2}] density and the [\ion{N}{2}] temperature are anomalous, but may be understood considering the contribution of recombination to these forbidden lines.  The two plasma components have very different temperatures.  The normal nebular plasma appears to be have temperature fluctuations in part of its volume (main shell), but only small fluctuations elsewhere.  The additional plasma component contains about half of the mass of the N$^{2+}$ and O$^{2+}$ ions, but only $3-5$\% of the mass of H$^+$ ions, so the two plasma components have very different chemical abundances.  We estimate abundances of $12+\log(\mathrm O^{2+}/\mathrm H^+)\sim 9.2$\,dex and $\mathrm{He}/\mathrm H\sim 0.13$.  Although they are all complications, multiple plasma components, temperature fluctuations, and the contributions of multiple physical processes to a given emission line are all part of the reality in NGC 6153, and should generally be taken into account.  

\end{abstract}



\section{Introduction} \label{sec:intro}

NGC 6153 is a bright, southern planetary nebula that has played an important part in the abundance discrepancy problem.  The abundance discrepancy was first noted by \citet{wyse1942}, who observed permitted lines of \ion{O}{2} and found that they indicated a much higher oxygen abundance than the forbidden [\ion{O}{3}] lines that originate from the same O$^{2+}$ ions.  Over the decades, study after study has found that the permitted lines yield systematically larger abundances of C, N, O, and Ne than do the forbidden lines, which is known as the abundance discrepancy problem.  The abundance discrepancy occurs in both \ion{H}{2} regions and planetary nebulae.  Whether it also occurs in active galactic nuclei is unknown since their kinematics impede investigating it.  

Until the 1990's, the magnitude of the abundance discrepancy was typically a factor of $2-3$ (the permitted lines indicated abundances $2-3$ times higher than the forbidden lines).  However, as CCD detectors came into common use, it became clear that many planetary nebulae had abundance discrepancies that were much higher, e.g., a factor of 5 in NGC 7009 \citep{liuetal1995}, 10 in NGC 6153 \citep{liuetal2000}, and values in excess of 100 for A46 \citep{corradietal2015}.  In their study of NGC 6153, \citet{liuetal2000} proposed their model of a chemically-inhomogeneous plasma containing hydrogen-deficient clumps as a possible explanation of the abundance discrepancy.  This model and the model of temperature fluctuations that \citet{peimbert1967} originally suggested to explain the factor of 2 abundance discrepancies in \ion{H}{2} regions have remained the main contenders to explain the abundance discrepancy.  (\citet{torrespeimbertetal1990} had shown how a chemically-inhomogeneous model of NGC 4361 could explain the discrepant C abundances found from permitted and forbidden lines.)  

Much of the debate regarding the abundance discrepancy, at least in planetary nebulae, has focussed upon which of the abundances, derived from permitted or forbidden lines, are the correct abundance ratios to characterize the chemical composition of the plasma in planetary nebulae.  This has happened even though evidence for a difference in the spatial distributions of the emission from the permitted and forbidden lines has long existed, with the permitted emission being more centrally-concentrated \cite[e.g.,][]{barker1982, barker1991, garnettdinerstein2001, tsamisetal2008, garciarojasetal2016}.  Likewise, chemically-inhomogeneous models demonstrating that the two abundances may not be contradictory, but instead provide information concerning multiple plasmas in planetary nebulae have existed almost since the \citet{liuetal2000} study of NGC 6153 \citep[e.g.,][]{pequignotetal2002, ercolanoetal2003, tylenda2003, tsamispequignot2005, yuanetal2011}.  More recently, the kinematics of the permitted and forbidden lines are often found to be distinct, with the permitted lines apparently arising from more highly ionized plasma, i.e., in agreement with the spatial distribution \citep[][]{sharpeeetal2004, barlowetal2006, otsukaetal2010, richeretal2013, richeretal2017, penaetal2017}.  These results as well as the work of \citet{gomezllanosmorisset2020} strongly influence our view that the structure of the plasma in planetary nebulae is more complex than hitherto considered in the analyses of their chemical abundances.  

Given the length of this paper, we provide a general roadmap here, and more detailed versions at the beginning of each section. In \S\ref{sec_observations}, we describe the observations, their reduction, the construction of position-velocity (PV) diagrams, and our estimate of the interstellar reddening for NGC 6153.  In \S\ref{sec_results}, we present the results related to the structure of the plasma, with various lines of evidence indicating the presence of two plasma components.  In \S\ref{sec_discussion}, we focus upon the complex temperature structure of the nebular plasma and its consequences concerning the chemical composition of NGC 6153's nebular shell.  In \S\ref{sec_conclusions}, we present our conclusions.  In summary, we analyze spectroscopy of NGC 6153 at high spectral resolution and conclude that the kinematics and physical conditions of the nebular plasma are consistent with the presence of two plasma components of different composition, density, and temperature.  Ignoring this complexity inevitably leads to the conclusion that there is a large abundance discrepancy in NGC 6153.

\section{The observations, their reduction, PV diagrams, and reddening} \label{sec_observations}

We begin this section presenting the data used and their reduction.  We then continue with a description of the construction of the PV diagrams we use to study the kinematics, physical conditions, and chemical abundances in NGC 6153.  We conclude this section with our determination of the reddening for NGC 6153, which also allows us to set a common flux scale for all of the wavelength intervals.  Throughout, we identify the wavelengths of emission lines rounded to integer values, except when that would allow confusion, in which cases we provide more precise wavelengths. Generally, we adopt the wavelengths from \citet[Atomic Line List version v3.00b4;][]{vanhoof2018}\footnote{https://www.pa.uky.edu/peter/newpage/} and \citet[National Institute of Standards and Technology][]{kramidaetal2021}\footnote{https://physics.nist.gov/asd}, but \citet{bowen1960} for forbidden lines and \citet{cleggetal1999} for the fine structure components of \ion{H}{1}.  We use a lot of atomic data, which we cite in Table \ref{tab_atomic_data} and refer to as needed.

\begin{deluxetable}{p{0.25truein}p{2.75truein}}
\tablecaption{Atomic data used\label{tab_atomic_data}}
\tablewidth{0pt}
\tablehead{
\colhead{Ion} & \colhead{Atomic data} 
}
\startdata
\ion{H}{1} & \citet{storeyhummer1995, ercolanostorey2006} \\
\ion{He}{1} & \citet{ercolanostorey2006, porteretal2013} \\
\ion{He}{2} & \citet{storeyhummer1995, ercolanostorey2006} \\
\ion{N}{2} & \citet{nussbaumerstorey1984, pequignotetal1991, fangetal2011, fangetal2013, tayal2011} \\
\ion{O}{2} & \citet{zeippen1982, nussbaumerstorey1984, wenaker1990, pequignotetal1991, wieseetal1996, tayal2007, kisieliusetal2009, storeyetal2017} \\
\ion{O}{3} & \citet{dalgarnoetal1981, nussbaumerstorey1984, roueffdalgarno1988, pequignotetal1991, dalgarnosternberg1989, liudanziger1993, storeyzeippen2000, tachievfischer2001, froesefischertachiev2004, storeyetal2014, kramidaetal2021} \\
\ion{S}{2} & \citet{rynkunetal2019, tayalzatsarinny2010} \\
\ion{Cl}{3} & \citet{rynkunetal2019, butlerzeippen1989} \\
\ion{Ar}{3} & \citet{munosburgosetal2009} \\
\ion{Ar}{4} & \citet{rynkunetal2019, ramsbottombell1997} \\
\enddata
\end{deluxetable}

\subsection{The observations and their reduction}

The data used here were acquired via programme 69.D-0174A (PI Danziger) on 8 June 2002 using the Ultraviolet and Visual Echelle Spectrograph \citep[UVES;][]{dekkeretal2000} on the Very Large Telescope (VLT) Kueyen (UT2) of the European Southern Observatory (ESO).  We retrieved the raw data from the ESO data archive.  \citet{mcnabbetal2016} previously analyzed this same data.  

UVES is a two-arm, cross-dispersed echelle spectrograph with 
common pre-slit optics, but independent slit and post-slit optics.  The blue and red entrance slits are 10\arcsec\ and 13\arcsec\ long.  The detector for the blue arm was an EEV 44-82 CCD with $2048\times 4096$ 15\,$\mu$m pixels.  The detector for the red arm was a mosaic of an EEV 44-82 CCD and a MIT-LL CCID-20 CCD, both of which had $2048\times 4096$ 15\,$\mu$m pixels.  For these observations, all detectors were used in $2\times 2$ binning.  Four cross dispersers (CD\#1, CD\#2, CD\#3, and CD\#4) were used to cover the wavelength interval 3043--10655\AA\ (except for three gaps, see below).  The effective spectral resolution ($\lambda/\delta\lambda$) in all cases was in the neighborhood of 30,000.  UVES' atmospheric dispersion corrector was not used.

\begin{figure}
\plotone{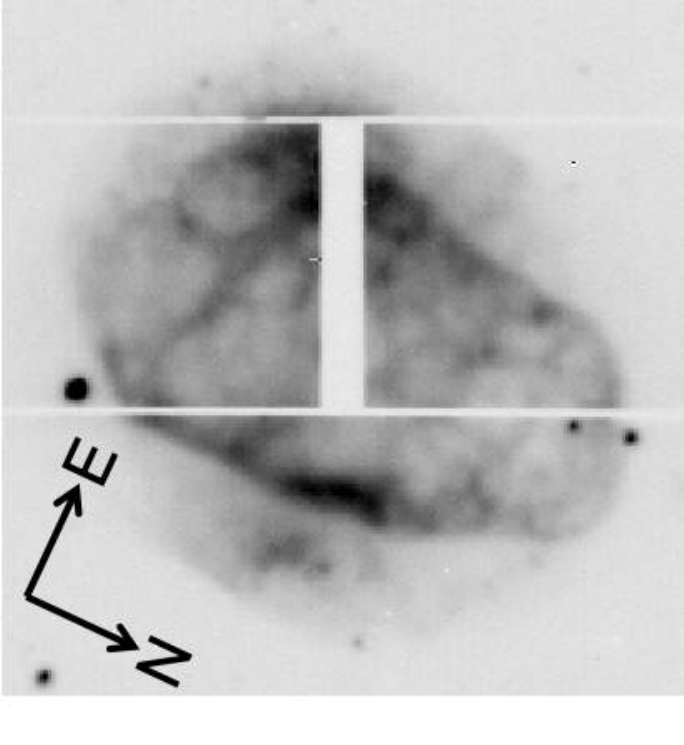}
\caption{This image from the (red) slit camera shows the slit superposed upon NGC 6153.  The arrows indicate the image orientation.  The slit size is $15.75\arcsec\times 2\arcsec$ and its position angle is 118$^{\circ}$ (N through E).  The decker limits the slit length to 10\arcsec (CD1 and CD2) or 13\arcsec (CD3 and CD4).  
}
\label{slit_on_image}
\end{figure}

\begin{figure*}
\begin{center}\includegraphics[width=0.76\linewidth]{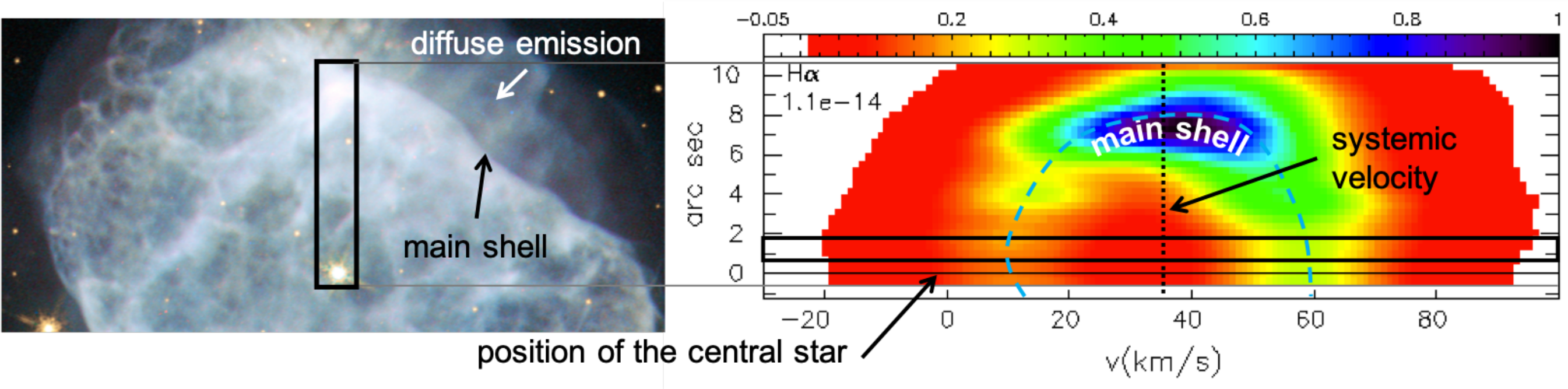}\end{center}
\caption{We compare an image (left; Hubble Space Telescope proposal ID 8594) and PV diagram (right; CD\#2) of NGC 6153 in the light of H$\alpha$.  The spectrograph slit is shown on the left.  The position of the central star, whose continuum is subtracted from the PV diagram, would appear at the position of the horizontal line (right panel; y-coordinate of zero).  The main structure of NGC 6153 is its main shell, which is the origin of the ``$\bigcap$"-shaped structure in the PV diagram (dashed line):  The Doppler effect separates the emission from the approaching and receding sides, except at the top where the expansion is perpendicular to the line of sight.  The rectangle (right panel) indicates the three rows used to compute the velocity separation between the approaching and receding sides of the nebula (\S\ref{sec_ionization_structure}).  The systemic velocity is the average radial velocity for the object.  
}
\label{fig_pv_diagram}
\end{figure*}

Figure \ref{slit_on_image} shows the location of the red spectrograph slit superposed upon the object.  The observations of both NGC 6153 and the standard stars CD\,$-32^{\circ}$\,9927, LTT 3864, and Feige 56 occurred on 8 June 2002, all at low airmass ($1.0-1.14$ for NGC 6153; $1.01-1.25$ for the standards).  The observations consist of two standard UVES instrument configurations with simultaneous coverage of CD\#1 and CD\#3 and then CD\#2 and CD\#4.  For NGC 6153, three 1200\,s exposures were obtained for all wavelength intervals.  In addition, single short exposures of 60\,s (CD\#1/CD\#3) and 120\,s (CD\#2/CD\#4) were obtained to avoid saturating the brightest lines. 
While NGC 6153 was east of the meridian, the long exposures and then the short for the CD\#2/CD\#4 configuration were acquired, followed by the long exposures and then the short exposure for the CD\#1/CD\#3 configuration as it crossed the meridian.  For the standard stars, all exposures were of 120\,s duration.  The slit width was 2\arcsec\ for NGC 6153 and 10\arcsec\ for the standard stars.  The observing conditions were not photometric, as the fluxes for NGC 6153 and the standard stars varied from exposure to exposure and between the CD\#1/CD\#3 and CD\#2/CD\#4 instrument configurations.  

We reduced the data using the Image Reduction and Analysis Facility \citep[IRAF\footnote{IRAF is distributed by the National Optical Astronomical Observatory, which is operated by the Association of Universities for Research in Astronomy, Inc., under cooperative agreement with the National Science Foundation.};][]{tody1986, tody1993}.  We reduced the images from each of the three individual CCDs separately, so we work with six wavelength intervals: CD1 (3043--3875\AA), CD2 (3759--4987\AA), CD3b (4980--5965\AA), CD3r (6034--7009\AA), CD4b (7101--8915\AA), and CD4r (9050--10655\AA; ``CD3" and ``CD4" are shorthand for CD3b+CD3r and CD4b+CD4r, respectively).  The reduction steps that follow were applied on a CCD-by-CCD basis.  Hence, our data reduction establishes a proper relative flux scale only \emph{within} each of the six wavelength intervals.  Given the lack of overlapping spectral coverage between the wavelength intervals (except CD1 and CD2) and the non-photometric observing conditions, we use the interstellar reddening to establish a common relative flux scale among all of the wavelength intervals (\S\ref{sec_reddening}).

The bias images were combined and subtracted from all other images.  The flat field images were processed to remove the scattered light between and within the spectral orders.  All of the spectra were divided by their flat field images to remove pixel-to-pixel variations, the spectrograph's blaze function, and the fringing that occurs in the far-red.  The exposure times for the flat field images were very short (0.44\,s for CD1, 1.8\,s for CD2, 0.33\,s for CD3, i.e., CD3b and CD3r, and 0.7\,s CD4), so we created shutter pattern images using other flat field images obtained on the same night and used them to correct the spectra of NGC 6153 and the standard stars.  The three long exposures for each wavelength interval for NGC 6153 were combined and then the object spectra were extracted.  For the short exposures of NGC 6153 and the standard stars, the object spectra were extracted from the reduced images.  When extracting the spectra, the position of the central star of NGC 6153 was traced, as was the standard star's position.  Since NGC 6153 filled the slit, we did not subtract the sky background.  Exposures of the ThAr arc lamp were extracted in the same way as the spectra of NGC 6153 and the standard stars and used to determine the wavelength solution.  For the wavelength calibration of the CD4 wavelength intervals, we also used the sky lines in the deep spectra of NGC 6153 
\citep[wavelengths from][]{osterbrocketal1996, osterbrocketal1997}.  We combined the order-separated echelle spectra of the standard stars into a single spectrum for each wavelength interval.  To derive the system sensitivity function, we used one spectrum of each of the standard stars in the CD1/CD3 wavelength intervals (the ones with the most signal) and one spectrum of LTT 3864 (most signal) and two of CD\,$-32^{\circ}$\,9927 (all available) for the CD2/CD4 wavelength intervals \citep{hamuyetal1992, hamuyetal1994}.  We then applied the system sensitivity function to flux-calibrate the spectra of NGC 6153.  

\subsection{The construction of PV diagrams}

We constructed position-velocity (PV) diagrams (or maps) for the emission lines of interest in NGC 6153 from the two-dimensional spectra, of which Figure \ref{fig_pv_diagram} is an example.  PV diagrams present the spatial structure within the slit as a function of radial velocity along the line of sight.  At each spatial coordinate in the PV diagram (vertical position in the slit), the emission from the plasma appears at its velocity with respect to the observer due to the Doppler effect.  In Figure \ref{fig_pv_diagram}, NGC 6153's main structure is a shell.  There is a more diffuse structure outside it (left panel).  Since the shell is expanding, the approaching side is shifted to bluer wavelengths (more negative velocities) while the receding side is shifted to redder wavelengths (more positive velocities).  The expansion of the main shell produces the ``$\bigcap$"-shaped structure in the PV diagram.  The approaching and receding sides merge to the same velocity at the edge of the shell since the expansion there is perpendicular to the line of sight.  

To construct the PV diagrams, we adopt the position of the central star as the spatial zero point and slice the two-dimensional spectra into a collection of one-dimensional spectra one pixel high.  These slices correspond to spatial slices of 0.492\arcsec\ and 0.362\arcsec\ for the blue and red arms, respectively.  The number of spatial slices varied for each cross disperser setting, but covered the full extent of the slits in all cases.  These spatial slices were then interpolated in a Python script to construct PV diagrams for each line with a uniform spatial sampling of 0.36\arcsec\ per row in all of the PV diagrams and spanning a common spatial extent for all wavelengths (i.e., we discard the extra extension of the red slit).  When converting the spectra to PV diagrams, it is necessary to correct the intensities for the change in units from wavelength to velocity, equivalent to multiplying the intensity by a factor of $\lambda/c$, where $\lambda$ is the wavelength of the emission line.  The PV diagrams span a velocity range of 130\,km/s about the systemic velocity of NGC 6153, which we adopted as 35.0\,km/s.  

By adopting the position of the central star as the spatial zero point, we eliminate the effect of atmospheric refraction in the spatial direction within the slit.  However, we cannot compensate for its effect in the direction perpendicular to the slit.  Based upon the tabulated atmospheric refraction for Cerro Paranal\footnote{https://www.eso.org/gen-fac/pubs/astclim/lasilla/diffrefr.html}, the  image displacements perpendicular to slit amount to a maximum of 0.14\arcsec, 0.07\arcsec, -0.26\arcsec, and -0.08\arcsec\ for the cross dispersers CD\#1, CD\#2, CD\#3, and CD\#4, respectively.  Within each wavelength interval, the maximum image offsets perpendicular to the slit is smaller, 0.10\arcsec, 0.02\arcsec, 0.10\arcsec, and 0.01\arcsec for the cross dispersers CD\#1, CD\#2, CD\#3, and CD\#4, respectively.  (CD\#2/CD\#4 are affected less since the refraction vector was more parallel to the slit.)  To the extent possible, we compare emission lines within a given wavelength interval to minimize the effects of spatial mis-matches due to atmospheric refraction.

We shall often need to compare two PV diagrams.  
In these comparisons, small errors in wavelength calibration or uncertainties in laboratory wavelengths could introduce spurious structures.  To avoid this, we first align the PV diagrams for the individual emission lines to a common velocity scale.  The simplest means of doing so is to use the measurements of the velocities for the approaching and receding sides of the nebula, used to determine the velocity splittings in Table \ref{tab_line_splitting}, setting the mean value for each emission line to a common value.  This process should be reasonable since we will usually be comparing PV diagrams for emission lines from ions that arise from the same plasma.

\subsection{Interstellar reddening}\label{sec_reddening}

We use the interstellar reddening to establish a consistent flux scale across our six wavelength intervals.  A common flux scale is required in order to study the temperature structure or to investigate the contributions of distinct physical processes to the PV diagrams of a particular transition, e.g., to understand why the forbidden nebular and auroral lines have different velocity splitting.  For this reason, we include the reddening as part of the data reductions.  

We determine both the interstellar reddening and the scale factors for the six wavelength intervals using two one-dimensional spectra (spatially-integrated).  The first of these is obtained by summing the one-pixel spatial extractions of the two-dimensional spectra to simulate a traditional one-dimensional spectrum.  The second spectrum is a ``standard" one-dimensional spectrum from the long exposure spectra of NGC 6153 that was reduced independently.  In both cases, we consider only the spatial extent of the blue slit.  The results from both spectra are equivalent, so we present those for the first spectrum.  We include this spectrum as online data.

We compute the reddening using the \ion{H}{1} and \ion{He}{1} lines and the \citet{fitzpatrick1999} reddening law scaled for a total-to-selective extinction ratio of $R_V=3.07$ \citep{mccallarmour}.  We refer the intensity of the \ion{H}{1} lines to H$\beta$ and that of the \ion{He}{1} lines to \ion{He}{1} $\lambda$4922 or \ion{He}{1} $\lambda$5876 using the intrinsic intensity ratios supposing an electron temperature of $10,000$\,K and the mean of electron densities of 1,000 and 10,000\,cm$^{-3}$ (atomic data: Table \ref{tab_atomic_data}).  We do not use the H8 line or \ion{He}{1} $\lambda$3889 because the two are blended.  We cannot use the \ion{He}{1} $\lambda$5048 line since it is contaminated by charge overflow from [\ion{O}{3}] $\lambda$5007 in the adjacent order.  Only for the CD2 wavelength interval is the wavelength/reddening baseline long enough to establish a secure, independent value for the interstellar reddening.  Hence, our relative flux calibration among the six wavelength intervals is relative to the long exposure spectrum for the CD2 wavelength interval.  Only the CD1 and CD2 wavelength intervals have emission lines in common that allow checking their relative flux calibration.  

\begin{deluxetable}{lll}[t]
\tablecaption{The \ion{H}{1} and \ion{He}{1} lines used for reddening\label{tab_interstellar_reddening}}
\tablewidth{0pt}
\tablehead{
\colhead{Interval} & \colhead{\ion{H}{1} lines} & \colhead{\ion{He}{1} lines} 
}
\decimalcolnumbers
\startdata
CD1 & H9-H25 & 3187, 3614 \\
CD2 & {\bf H$\beta$}, H$\gamma$, H$\delta$, & 3965, 4026, 4388, 4438, \\
    & H$\epsilon$, H9-H11 & 4471, 4713, {\bf 4922} \\
CD3b & & 5015, {\bf 5876} \\
CD3r & H$\alpha$ & 6678 \\
CD4b & P11-P25 & 7281 \\
CD4r & P7-P9 & \\
\enddata
\end{deluxetable}

Table \ref{tab_interstellar_reddening} presents the \ion{H}{1} and \ion{He}{1} lines available in each wavelength interval for which theoretical line intensities exist.    The lines shown in boldface are the reference lines used to calculate the reddening.  Table \ref{tab_HI_int_reddening} presents the wavelength interval/order (col. 1; the nomenclature $n$/$x$ implies wavelength interval CD$n$ and order $x$), line wavelength (col. 2), the observed (col. 4) and intrinsic (col. 5) relative line intensities for \ion{H}{1} lines, the reddening law from \citet{fitzpatrick1999} (col. 3; $A_1(\lambda)=A(\lambda)/E(B-V)$ for $E(B-V)=1.0$\,mag), and the reddening we find (col. 6).  Table \ref{tab_HeI_int_reddening} presents the analogous information for the \ion{He}{1} lines, with \ion{He}{1} $\lambda$4922 used as the reference line in the top half and \ion{He}{1} $\lambda$5876 as the reference line in the bottom half.  All of the line intensities in Tables \ref{tab_HI_int_reddening} and \ref{tab_HeI_int_reddening} are from the long exposure spectra, except H$\alpha$, which is from the short exposure spectrum of interval CD3r (3rs in Table \ref{tab_HI_int_reddening}).  We compute the reddening according to 
$$ E(B-V) = \frac{\log{(F(\lambda)/F(\lambda_{ref}))}-\log{(I(\lambda)/I(\lambda_{ref}))}}{-0.4 (A_1(\lambda)-A_1(\lambda_{ref}))} $$
\noindent where $F(\lambda)/F(\lambda_{ref})$ are the observed flux ratios, after scaling if they are not from the CD2 wavelength interval, $I(\lambda)/I(\lambda_{ref})$ are the intrinsic flux ratios, and $A_1(\lambda)$ is the reddening law \citep[][see above]{fitzpatrick1999}.  

\startlongtable
\begin{deluxetable}{lccccc}
\tablecaption{\ion{H}{1} line intensities and reddenings\label{tab_HI_int_reddening}}
\tablewidth{0pt}
\tablehead{
\colhead{CD/\#} & \colhead{line} & \colhead{$A_1(\lambda)$} & \colhead{$\frac{\displaystyle F(\lambda)}{\displaystyle F(\mathrm H\beta)}$} & \colhead{$\frac{\displaystyle I(\lambda)}{\displaystyle I(\mathrm H\beta)}$} & $E(B-V)$ \\
 & \colhead{(\AA)} & \colhead{(mag)} & (obs) & (int) & \colhead{(mag)} 
}
\decimalcolnumbers
\startdata
1/127       &         3669.45   &         4.67      &         0.0027    &         0.0041    &         0.40      \\                  
1/127       &         3671.32   &         4.67      &         0.0031    &         0.0045    &         0.35      \\                  
1/127       &         3673.81   &         4.67      &         0.0030    &         0.0050    &         0.51      \\                  
1/127       &         3676.38   &         4.66      &         0.0036    &         0.0056    &         0.43      \\                  
1/127       &         3679.37   &         4.66      &         0.0038    &         0.0063    &         0.49      \\                  
1/127       &         3682.82   &         4.66      &         0.0042    &         0.0072    &         0.52      \\                  
1/126       &         3682.82   &         4.66      &         0.0046    &         0.0072    &         0.44      \\                  
1/127       &         3686.83   &         4.65      &         0.0053    &         0.0082    &         0.43      \\                  
1/126       &         3686.83   &         4.65      &         0.0051    &         0.0082    &         0.48      \\                  
1/127       &         3691.55   &         4.65      &         0.0064    &         0.0095    &         0.40      \\                  
1/126       &         3691.55   &         4.65      &         0.0058    &         0.0095    &         0.49      \\                  
1/127       &         3697.16   &         4.64      &         0.0071    &         0.011     &         0.45      \\                  
1/126       &         3697.16   &         4.64      &         0.0066    &         0.011     &         0.53      \\                  
1/126       &         3703.86   &         4.64      &         0.0078    &         0.013     &         0.54      \\                  
1/126       &         3711.98   &         4.63      &         0.0094    &         0.016     &         0.53      \\                  
1/125       &         3711.98   &         4.63      &         0.0091    &         0.016     &         0.56      \\                  
1/126       &         3721.95   &         4.62      &         0.018     &         0.020     &         0.07      \\                  
1/125       &         3721.95   &         4.62      &         0.017     &         0.020     &         0.17      \\                  
1/125       &         3734.37   &         4.61      &         0.014     &         0.024     &         0.57      \\                  
1/125       &         3750.15   &         4.59      &         0.020     &         0.031     &         0.46      \\                  
1/124       &         3750.15   &         4.59      &         0.020     &         0.031     &         0.46      \\                  
1/124       &         3770.63   &         4.57      &         0.024     &         0.040     &         0.57      \\                  
1/123       &         3770.63   &         4.57      &         0.023     &         0.040     &         0.59      \\                  
1/123       &         3797.91   &         4.54      &         0.032     &         0.053     &         0.56      \\                  
1/122       &         3797.91   &         4.54      &         0.033     &         0.053     &         0.51      \\                  
1/122       &         3835.40   &         4.51      &         0.045     &         0.073     &         0.56      \\                  
1/121       &         3835.40   &         4.51      &         0.044     &         0.073     &         0.58      \\                  
2/124       &         3770.63   &         4.57      &         0.025     &         0.040     &         0.50      \\                  
2/123       &         3770.63   &         4.57      &         0.023     &         0.040     &         0.60      \\
2/123       &         3797.91   &         4.54      &         0.033     &         0.053     &         0.52      \\
2/122       &         3797.91   &         4.54      &         0.031     &         0.053     &         0.60      \\
2/122       &         3835.40   &         4.51      &         0.048     &         0.073     &         0.49      \\
2/121       &         3835.40   &         4.51      &         0.045     &         0.073     &         0.56      \\
2/118       &         3970.08   &         4.38      &         0.11      &         0.16      &         0.48      \\
2/117       &         3970.08   &         4.38      &         0.11      &         0.16      &         0.52      \\
2/114       &         4101.73   &         4.26      &         0.18      &         0.26      &         0.59      \\
2/114       &         4101.73   &         4.26      &         0.18      &         0.26      &         0.59      \\
2/108       &         4340.47   &         4.05      &         0.38      &         0.47      &         0.45      \\
2/107       &         4340.47   &         4.05      &         0.37      &         0.47      &         0.51      \\
2/96        &         4861.35   &         3.56      &         1         &         1         &                   \\
3rs/93       &         6563.00   &         2.30      &         5.00      &         2.86      &         0.49      \\
4b/74        &         8323.42   &         1.57      &         0.0039    &         0.0013    &         0.60      \\
4b/74        &         8333.78   &         1.57      &         0.0045    &         0.0015    &         0.61      \\
4b/74        &         8345.54   &         1.56      &         0.0046    &         0.0016    &         0.56      \\
4b/74        &         8359.00   &         1.56      &         0.0059    &         0.0019    &         0.63      \\
4b/74        &         8374.48   &         1.55      &         0.0059    &         0.0021    &         0.56      \\
4b/74        &         8392.40   &         1.55      &         0.0066    &         0.0024    &         0.55      \\
4b/74        &         8413.32   &         1.54      &         0.0066    &         0.0028    &         0.47      \\
4b/73        &         8413.32   &         1.54      &         0.0067    &         0.0028    &         0.48      \\
4b/73        &         8437.95   &         1.53      &         0.0091    &         0.0032    &         0.56      \\
4b/73        &         8467.26   &         1.52      &         0.011     &         0.0038    &         0.55      \\
4b/73        &         8502.48   &         1.51      &         0.012     &         0.0045    &         0.51      \\
4b/72        &         8545.38   &         1.50      &         0.016     &         0.0055    &         0.56      \\
4b/72        &         8598.39   &         1.48      &         0.019     &         0.0067    &         0.54      \\
4b/71        &         8665.02   &         1.47      &         0.025     &         0.0084    &         0.57      \\
4b/71        &         8750.46   &         1.44      &         0.030     &         0.011     &         0.53      \\
4b/70        &         8862.89   &         1.41      &         0.040     &         0.014     &         0.54      \\
4r/235       &         9229.01   &         1.32      &         0.079     &         0.025     &         0.55      \\
4r/233       &         9545.97   &         1.24      &         0.100     &         0.037     &         0.47      \\
4r/230       &         10049.37  &         1.13      &         0.21      &         0.055     &         0.59      \\
\enddata
\end{deluxetable}

\begin{deluxetable}{lccccc}
\tablecaption{\ion{He}{1} line intensities and reddenings\label{tab_HeI_int_reddening}}
\tablewidth{0pt}
\tablehead{
\colhead{CD/\#} & \colhead{line} & \colhead{$A_1(\lambda)$} & \colhead{$\frac{\displaystyle F(\lambda)}{\displaystyle F(\mathrm{ref})}$} & \colhead{$\frac{\displaystyle I(\lambda)}{\displaystyle I(\mathrm{ref})}$} & $E(B-V)$ \\
 & \colhead{(\AA)} & \colhead{(mag)} & (obs) & (int) & \colhead{(mag)} 
}
\decimalcolnumbers
\startdata
1/147       &         3187.75   &         5.24      &         0.77    &         3.44      &         0.93      \\
1/146       &         3187.75   &         5.24      &         0.80    &         3.44      &         0.91      \\
1/129       &         3613.64   &         4.73      &         0.11    &         0.42      &         1.16      \\
2/118       &         3964.73   &         4.39      &         0.39    &         0.86      &         0.96      \\
2/117       &         3964.73   &         4.39      &         0.37    &         0.86      &         1.02      \\
2/116       &         4026.19   &         4.33      &         1.22    &         1.73      &         0.45      \\
2/107       &         4387.93   &         4.01      &         0.40    &         0.46      &         0.30      \\
2/106       &         4387.93   &         4.01      &         0.36    &         0.46      &         0.53      \\
2/105       &         4437.55   &         3.96      &         0.05    &         0.06      &         0.48      \\
2/105       &         4471.48   &         3.93      &         2.94    &         3.67      &         0.55      \\
2/104       &         4471.48   &         3.93      &         2.85    &         3.67      &         0.63      \\
2/99        &         4713.15   &         3.70      &         0.37    &         0.44      &         0.86      \\
2/99        &         4713.15   &         3.70      &         0.37    &         0.44      &         0.86      \\
2/95        &         4921.93   &         3.50      &         1       &         1         &                   \\
3b/122       &         5015.68   &         3.40      &         1.53    &         2.19      &         -4.17     \\
3b/104       &         5875.62   &         2.71      &         15.80   &         10.71     &         0.54      \\
3r/92        &         6678.15   &         2.24      &         5.98    &         3.02      &         0.59      \\
3r/91        &         6678.15   &         2.24      &         5.94    &         3.02      &         0.59      \\
4b/85        &         7281.35   &         1.96      &         0.83    &         0.64      &         0.18      \\
4b/85        &         7281.35   &         1.96      &         0.83    &         0.64      &         0.18      \\
\\
1/147       &         3187.75   &         5.24      &         0.048   &         0.32      &         0.81      \\
1/146       &         3187.75   &         5.24      &         0.050   &         0.32      &         0.79      \\
1/129       &         3613.64   &         4.73      &         0.0071  &         0.039     &         0.92      \\
2/118       &         3964.73   &         4.39      &         0.025   &         0.080     &         0.76      \\
2/117       &         3964.73   &         4.39      &         0.024   &         0.080     &         0.79      \\
2/116       &         4026.19   &         4.33      &         0.077   &         0.16      &         0.49      \\
2/107       &         4387.93   &         4.01      &         0.025   &         0.043     &         0.44      \\
2/106       &         4387.93   &         4.01      &         0.023   &         0.043     &         0.54      \\
2/105       &         4437.55   &         3.96      &         0.0033  &         0.0060    &         0.52      \\
2/105       &         4471.48   &         3.93      &         0.19    &         0.34      &         0.54      \\
2/104       &         4471.48   &         3.93      &         0.18    &         0.34      &         0.57      \\
2/99        &         4713.15   &         3.70      &         0.024   &         0.041     &         0.60      \\
2/99        &         4713.15   &         3.70      &         0.024   &         0.041     &         0.60      \\
2/95        &         4921.93   &         3.50      &         0.063   &         0.093     &         0.54      \\
3b/122       &         5015.68   &         3.40      &         0.097   &         0.20      &         1.16      \\
3b/104       &         5875.62   &         2.71      &         15.80   &         1         &                   \\
3r/92        &         6678.15   &         2.24      &         0.38    &         0.28      &         0.69      \\
3r/91        &         6678.15   &         2.24      &         0.38    &         0.28      &         0.67      \\
4b/85        &         7281.35   &         1.96      &         0.053   &         0.060     &         -0.18     \\
4b/85        &         7281.35   &         1.96      &         0.053   &         0.060     &         -0.18     \\
\enddata
\end{deluxetable}

\begin{figure*}
\center{
\includegraphics[width=0.7\linewidth]{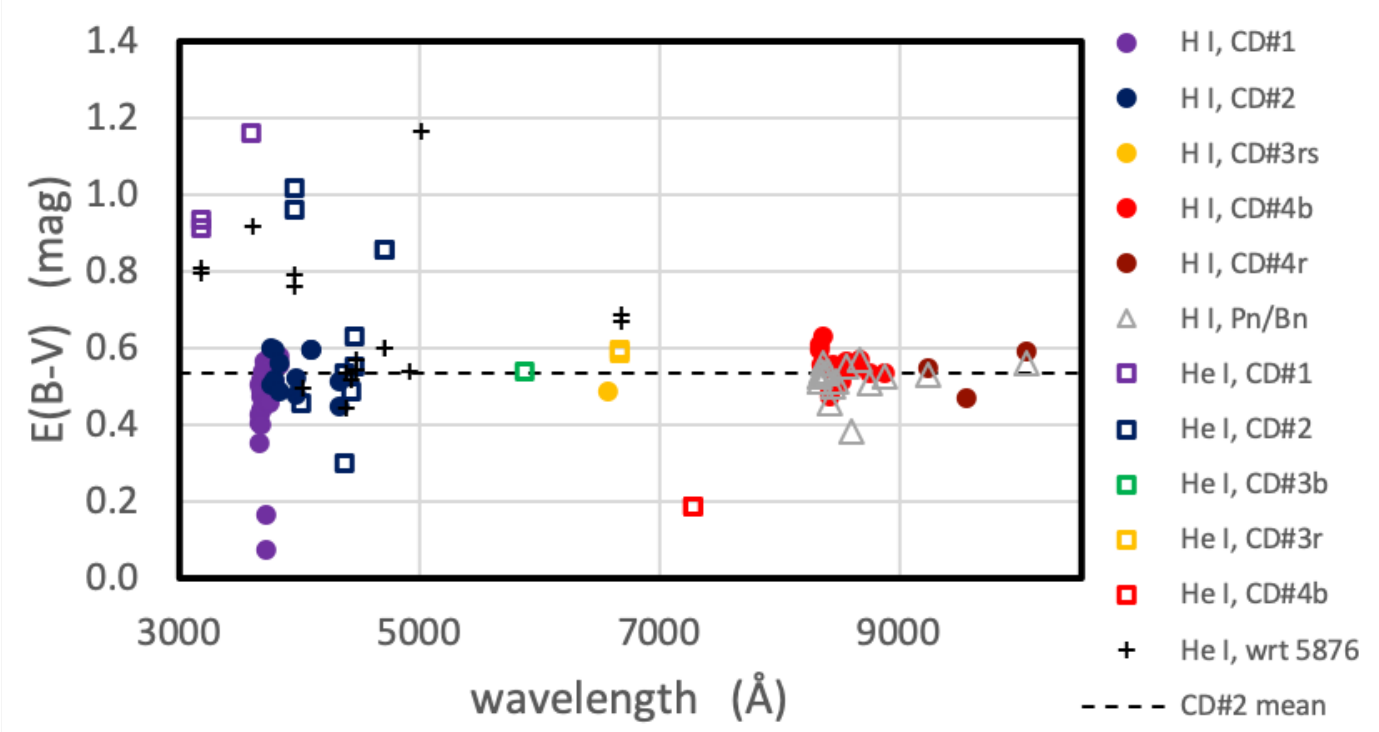}
}
\caption{We plot the reddening, $E(B-V)$, from Tables \ref{tab_HI_int_reddening} and \ref{tab_HeI_int_reddening} as a function of wavelength for \ion{H}{1} and \ion{He}{1} lines.  For the lines indicated as ``\ion{H}{1}, CD\#$X$", the line ratio is $\mathrm Hn/\mathrm H\beta$ or $\mathrm Pn/\mathrm H\beta$ and plotted at the wavelength of $\mathrm Hn$ or $\mathrm Pn$, respectively, the Balmer and Paschen lines originating at upper level $n$.  For the reddening computed from the ratio of Paschen to Balmer lines, $\mathrm Pn/\mathrm Bn$, the reddening is plotted at the wavelength of the Paschen line.  The horizontal dashed line represents the mean value of $E(B-V)$ for the \ion{H}{1} lines in the CD2 wavelength interval (reference value).  For the \ion{He}{1} lines, the squares represent ratios using \ion{He}{1} $\lambda$4922 as the reference line while the crosses use \ion{He}{1} $\lambda$5876 as the reference line.  The reddening computed from the \ion{He}{1} $\lambda\lambda$5015/4922 and \ion{He}{1} $\lambda\lambda$7281/5876 line ratios fall at negative values (and are not shown).  Both lines are apparently much too weak.  
}
\label{fig_reddening_wavelength}
\end{figure*}

\begin{deluxetable}{lllc}
\tablecaption{Flux scale factors\label{tab_flux_scale_factors}}
\tablewidth{0pt}
\tablehead{
\colhead{Interval} & \colhead{Long exposure} & \colhead{Short exposure } & \colhead{short/long}
}
\colnumbers
\startdata
CD1  &0.95 (0.915-0.974)& $0.778\pm 0.028$ & $1.221\pm 0.023$ \\
CD2  & 1.0 (reference)  & $0.991\pm 0.047$ & $1.009\pm 0.047$ \\
CD3b & 2.0 (1.9-2.1)    & $1.525\pm 0.091$ & $1.312\pm 0.042$ \\
CD3r & 0.6 (0.57-0.63)  & $0.430\pm 0.047$ & $1.40\pm 0.13$ \\
CD4b & 0.5 (0.48-0.51)  & $0.515\pm 0.019$ & $0.971\pm 0.021$ \\
CD4r & 0.4 (0.38-0.42)  & $0.402\pm 0.020$ & $0.995\pm 0.005$ \\
\enddata
\end{deluxetable}

We cross-check the CD2-referenced reddenings in several ways.  We compute the reddening using the \ion{H}{1} $\mathrm Pn/\mathrm Bn$ ratios, where $\mathrm Pn$ and $\mathrm Bn$ are the Paschen and Balmer lines, respectively, with upper level $n$.  This uses CD1/CD4b and CD2/CD4r line ratios, but not including H$\beta$.  There are no \ion{H}{1} lines in the CD3b interval.  However, we tie its calibration more tightly to other intervals by computing the reddening from the \ion{He}{1} lines using \ion{He}{1} $\lambda$5876 as the reference line.  

When a line is well-detected in adjacent orders, we use both measurements.  Note that the relative line intensities in Tables \ref{tab_HI_int_reddening} and \ref{tab_HeI_int_reddening} already include the scale factors required to place all six wavelength intervals on a common flux scale.  These flux scale factors are given in column 2 of Table \ref{tab_flux_scale_factors} with the acceptable range for the scale factor in parentheses.  These flux scale factors apply to the PV diagrams.  The scale factors between the long and short exposures (column 4) are determined by comparing line fluxes measured in both spectra and so are independent of the flux scale factors.  The scale factors for the short exposure spectra (column 3) are the ratio of the two previous scale factors.  Usually, the uncertainty is dominated by the range allowed for the scale factors of the long exposure spectra.  

Figure \ref{fig_reddening_wavelength} presents the results, showing the reddening, $E(B-V)$, as a function of wavelength for the \ion{H}{1} and \ion{He}{1} lines.  For the \ion{H}{1} lines, the filled circles show the line ratio with respect to H$\beta$ with the reddening plotted at the wavelength of the \ion{H}{1} lines (numerator).  The open gray triangles show the reddening computed using the ratios of Paschen and Balmer lines arising from the same upper level, plotted at the wavelength of the Paschen lines (the plot is less confusing that way).  The horizontal dashed line is the mean value of $E(B-V)$ for the CD2 wavelength interval.  For the \ion{He}{1} lines, the open squares show the reddening computed using \ion{He}{1} $\lambda$4922 as the reference line and plotted at the position of each He I line (numerator).  The black crosses present the reddening computed from the \ion{He}{1} lines using \ion{He}{1} $\lambda$5876 as the reference wavelength, again plotted at the wavelength of the \ion{He}{1} line in the numerator.  For the \ion{H}{1} and \ion{He}{1} lines plotted as filled circles and open squares, the colour indicates the wavelength interval.  

For the \ion{H}{1} lines, the most anomalous is \ion{H}{1} $\lambda$3722, which is blended with [\ion{S}{3}] $\lambda$3722.  For the CD1 interval, the reddening decreases for the bluest lines.  In contrast, the bluest Paschen lines imply slightly higher extinction.  This trend is not apparent when the reddening is computed using the ratio of Paschen and Balmer lines.  This could well be the same effect noted by \citet{mesadelgadoetal2009} due to $l$-changing collisions causing departures from Case B theory since its effect will largely disappear when comparing Paschen and Balmer lines from the same upper level.  The \ion{He}{1} lines show considerable scatter.  The scatter is more pronounced when using \ion{He}{1} $\lambda$4922 as the reference wavelength, in part because the wavelength baseline for the blue lines is shorter and also because it is the weaker reference line.  \ion{He}{1} $\lambda\lambda$3187, 3614, 3964, 5015 are all apparently too weak, regardless of the reference line.  Since they all have the metastable 2s levels as their lower levels (\ion{He}{1} $\lambda$3187 is a triplet state, the others singlet states), this may be due to radiative transfer effects.  \ion{He}{1} $\lambda$7281 is too weak, independent of the reference line we choose, but it is not clear why.  \ion{He}{1} $\lambda$6678 has the same lower level and its intensity is, if anything, slightly too high.  Most of the Paschen lines from the CD4b wavelength interval are not anomalous.

\begin{deluxetable}{lcc}[t]
\tablecaption{Average $E(B-V)$\label{tab_EBV_values}}
\tablewidth{0pt}
\tablehead{
\colhead{Interval} & \colhead{\ion{H}{1} lines} & \colhead{\ion{He}{1} lines} 
}
\startdata
CD1                           &         $0.496     \pm       0.064\ ^{\tablenotemark{a}}$     &                                       \\
CD1, overlap                  &         $0.560     \pm       0.026$     &                                       \\
CD2                           &         $0.535     \pm       0.053$     &         $0.528     \pm       0.078\ ^{\tablenotemark{b}}$     \\
CD3b                          &                                         &         $0.536\ ^{\tablenotemark{b}}$                         \\
CD3r                          &         $0.486$                         &         $0.589     \pm       0.005\ ^{\tablenotemark{c}}$     \\
CD4b                          &         $0.561     \pm       0.032$     &                                       \\
CD4b, $\mathrm Pn/\mathrm Hn$ &         $0.532     \pm       0.021$     &                                       \\
CD4r                          &         $0.536     \pm       0.061$     &                                       \\
CD4r, $\mathrm Pn/\mathrm Hn$ &         $0.545     \pm       0.023$     &                                       \\
\enddata
\tablenotetext{a}{Excludes \ion{H}{1} $\lambda$3722}
\tablenotetext{b}{$\lambda_{ref}=$\,\ion{He}{1} $\lambda$5876}
\tablenotetext{c}{$\lambda_{ref}=$\,\ion{He}{1} $\lambda$4922}
\end{deluxetable}

Table \ref{tab_EBV_values} presents the mean reddening values for lines in each wavelength interval by the different means described above.  The mean reddening for the \ion{H}{1} lines from the CD2 interval is compatible with that measured for all other wavelength intervals, and we adopt its value, $E(B-V)=0.535\pm 0.053$\,mag, as our reddening for NGC 6153.  Based upon the \citet{fitzpatrick1999} reddening law, this is equivalent to $c(\mathrm H\beta)=1.424 E(B-V) = 0.76$\,dex.  

The reddening we find is lower than that reported by others.  
\citet{kingsburghbarlow1994} find $c(\mathrm H\beta)= 0.96$\,dex from their optical spectrum, \citet{liuetal2000} report $c(\mathrm H\beta)= 1.27\pm 0.06$ and $1.30\pm 0.01$\,dex based upon their scanned (entire object) and minor axis optical spectra, respectively, \citet{pottaschetal2003} deduce $c(\mathrm H\beta)= 1.19$\,dex comparing the observed H$\beta$ flux with various 6 cm radio fluxes, and \citet{mcnabbetal2016} obtain $c(\mathrm H\beta)= 1.32$\,dex based upon the \ion{H}{1} Balmer lines.  It is not obvious why our reddening is different since we checked that our flux calibrations are similar to the historical calibrations in the ESO archives\footnote{https://www.eso.org/observing/dfo/quality/UVES/qc/SysEffic\_qc1.html}.  Also, the \citet{fitzpatrick1999} extinction law is similar to the \citet{cardellietal1989} extinction law used by \citet{mcnabbetal2016}.

\begin{deluxetable}{lccccc}
\tablecaption{\ion{He}{2} line intensities and abundances\label{tab_HeII_int_abun}}
\tablewidth{0pt}
\tablehead{
\colhead{CD/\#} & \colhead{line} & \colhead{$A_1(\lambda)$} & \colhead{$\frac{\displaystyle F(\lambda)}{\displaystyle F(\mathrm{ref})}$} & \colhead{$\frac{\displaystyle I(\lambda)}{\displaystyle I(\mathrm{ref})}$} & \colhead{$\frac{\displaystyle \mathrm{He}^{+i}}{\displaystyle \mathrm H^+}$} \\
 & \colhead{(\AA)} & \colhead{(mag)} & (obs) & (int) &  
}
\decimalcolnumbers
\startdata
\ion{He}{1} \\
2/104       &         4471.50   &         3.93      &         5.00    &         6.05      &         0.115      \\
3b/104      &         5875.60   &         2.71      &         26.9    &         17.5      &         0.115      \\
3r/92       &         6678.15   &         2.24      &         9.74    &         5.03      &         0.117      \\
3r/91       &         6678.15   &         2.24      &         9.66    &         4.99      &         0.116      \\
\ion{He}{2} \\
1/146       &         3203.17   &         5.22      &         4.49    &         10.4      &         0.021      \\
1/145       &         3203.17   &         5.22      &         4.01    &         9.26      &         0.019      \\
2/100       &         4685.68   &         3.73      &         15.1    &         16.5      &         0.014      \\
3b/113      &         5411.52   &         3.04      &         1.43    &         1.10      &         0.012      \\
3r/101      &         6074.11   &         2.58      &         0.052   &         0.032     &         0.015      \\
3r/100      &         6074.11   &         2.58      &         0.048   &         0.030     &         0.014      \\
3r/100      &         6118.26   &         2.55      &         0.056   &         0.034     &         0.014      \\
3r/99       &         6170.60   &         2.52      &         0.077   &         0.046     &         0.016      \\
3r/96       &         6406.38   &         2.39      &         0.14    &         0.079     &         0.016      \\
3r/92       &         6683.20   &         2.24      &         0.19    &         0.099     &         0.013      \\
3r/91       &         6683.20   &         2.24      &         0.19    &         0.099     &         0.013      \\
3r/89       &         6890.90   &         2.14      &         0.26    &         0.127     &         0.013      \\
3r/88       &         6890.90   &         2.14      &         0.29    &         0.135     &         0.013      \\
\enddata
\end{deluxetable}

Whether our reddening is correct is less important than whether our intrinsic, reddening-corrected line intensity ratios are correct.  For wavelength intervals where the density of \ion{H}{1} or \ion{He}{1} lines is high (CD2, the red end of CD4b, the blue half of the CD4r), our line ratios corrected for our reddening must necessarily give the correct intrinsic ratios, because that is how we determine the reddening.  For the same reason, the flux scale in the vicinity of the \ion{He}{1} $\lambda\lambda$5876,6678 and H$\alpha$ lines will also be correct.  To check that there are no problems across the CD3b or CD3r wavelength intervals where there are few \ion{H}{1} or \ion{He}{1} lines, Table \ref{tab_HeII_int_abun} presents the $\mathrm{He}^{+}/\mathrm H^+$ and $\mathrm{He}^{2+}/\mathrm H^+$ abundances for lines that fall in those wavelength intervals and for \ion{He}{2} $\lambda\lambda$3203,4686 from the CD1 and CD2 intervals.  The format of Table \ref{tab_HeII_int_abun} is identical to that of Tables \ref{tab_HI_int_reddening} and \ref{tab_HeI_int_reddening}, except that the last column presents the He ionic abundance ratios and the intensity ratios are given on the scale where $F(\mathrm H\beta)=I(\mathrm H\beta)=100$.  We assume a density of $10^4$\,cm$^{-3}$ and temperatures of 8,000\,K and 10,000\,K when computing the He$^+$ and He$^{2+}$ abundances, respectively (atomic data: Table \ref{tab_atomic_data}).  After correcting for reddening (col. 5), our \ion{He}{1} line intensities are very similar to those found by \citet{liuetal2000}, as expected given that they are used to derive the reddening and flux scale factors.   Given the constancy of the $\mathrm{He}^{2+}/\mathrm H^+$ abundance ratio for the lines in the CD3b and CD3r wavelength intervals, there is no problem with flux scale.  Our only test of the flux scale across the blue half of the CD4b wavelength interval is the [\ion{Ar}{3}] $\lambda\lambda$7135,7751 lines.  Since they indicate very similar [\ion{Ar}{3}] temperatures (Figure \ref{fig_forb_TeAr3}), it would appear that the shape of the flux scale in this part of the CD4b wavelength interval is secure, but we have no objective measure of whether its normalization is correct.  (The \ion{He}{2} $\lambda\lambda$7177,7592 lines are badly affected by an instrumental artifact and atmospheric absorption, respectively.)  

So, while we are unable to explain why the reddening we determine for NGC 6153 is different from others, we are confident that our relative line intensities, once corrected for our reddening, yield the correct intrinsic relative intensities.  Note that the flux calibration does not affect any result based upon kinematics.

\section{Results:  The plasma structure} \label{sec_results}

The multi-component structure of the plasma in NGC 6153's nebular shell is the unifying theme of this section.  We first analyze the ionization structure of NGC 6153 in detail.  We consider both the information available in the morphology of the PV diagrams as well as the more common Wilson diagram, which presents the velocity splitting as a function of the ionization energy.  Following this, we investigate the physical conditions based upon kinematics, forbidden lines, and permitted lines.  Finally, we show that, by assuming two plasma components in NGC 6153, we can explain the anomalous [\ion{N}{2}] temperature and [\ion{O}{2}] density as the result of excitation of the [\ion{N}{2}] $\lambda$5755 and [\ion{O}{2}] lines due to recombination, mostly in the additional plasma component.  

Both the kinematics and the physical conditions imply the presence of two distinct plasma components.  One component has the kinematics and physical conditions typically-associated with planetary nebulae \citep[e.g., as described in textbooks][]{aller1984, osterbrockferland2006}, so we shall refer to it as the {\bf ``normal nebular plasma"}.  The other component has unusual kinematics and is colder and denser than the former, so we shall call it the {\bf ``additional plasma component"}.  

\subsection{The ionization structure of NGC 6153}\label{sec_ionization_structure}

\begin{figure}
\includegraphics[width=\linewidth]{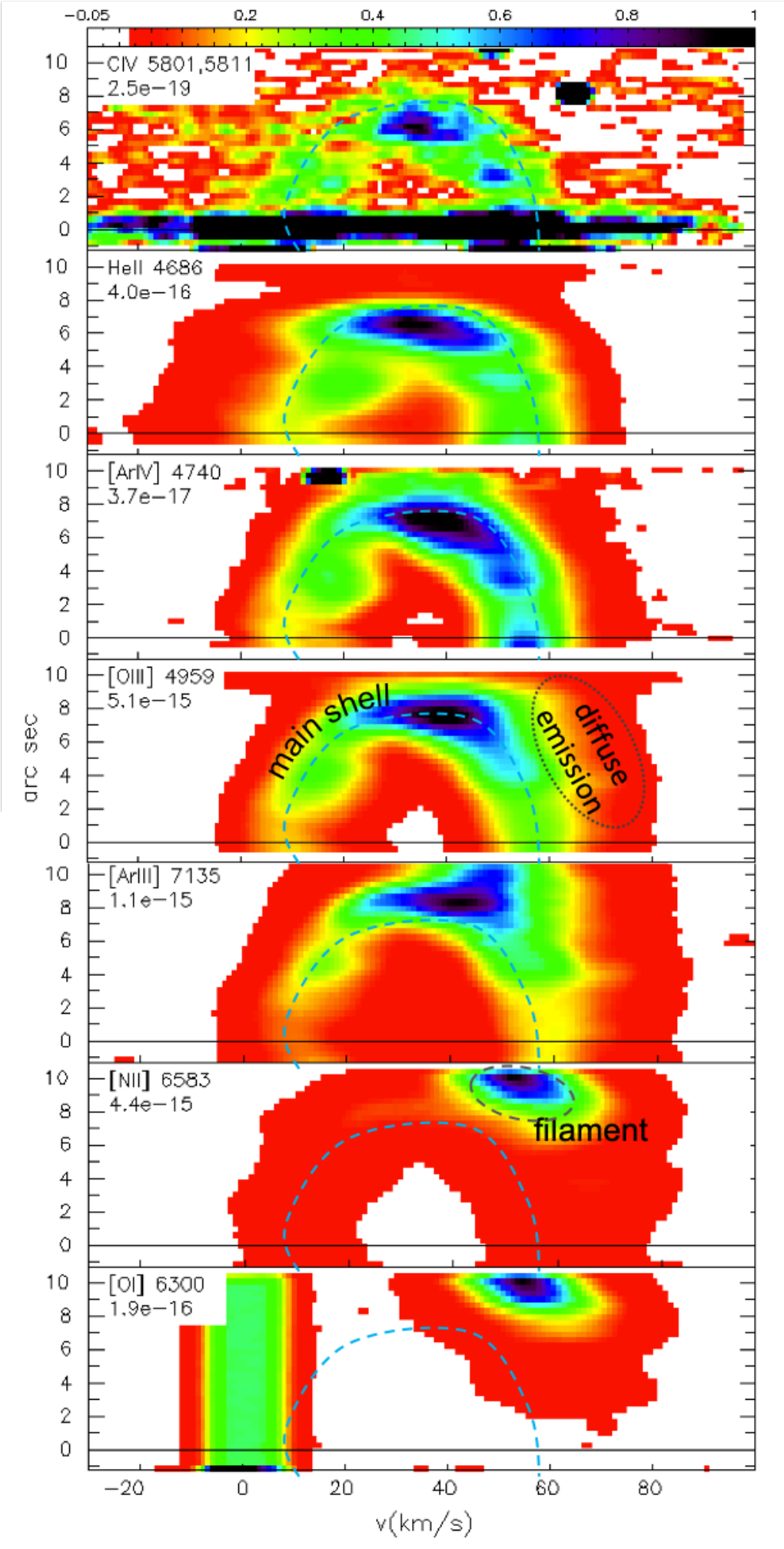}
\caption{We present the PV diagrams for lines spanning the full range of ionization energies in NGC 6153:  \ion{C}{4} $\lambda\lambda$5801,5811, \ion{He}{2} $\lambda$4686, [\ion{Ar}{4}] $\lambda$4740, [\ion{O}{3}] $\lambda$4959, [\ion{Ar}{3}] $\lambda$7135, [\ion{N}{2}] $\lambda$6583, and [\ion{O}{1}] $\lambda$6300 (top to bottom).  The intensity in each panel is normalized to its maximum value, indicated at upper left in each panel, below the line identification.  The color bar at top is common to all panels (-5\% to 100\% of the maximum value; linear scale).  We identify several morphological features that we refer to in the text.  There are clear, continuous, and systematic changes in the shape of the emission line profile as a function of the ionization energy.  Likewise, the difference in velocity between the approaching and receding sides varies with ionization energy.
}
\label{fig_PV_ionization}
\end{figure}

A PV diagram (e.g., Figure \ref{fig_pv_diagram}) is the result of the spatial distribution, the kinematics, and the physical conditions of the of plasma that emits a given line within the nebular volume intercepted by the spectrograph's slit convolved with the fine structure of that line and the mass of the emitting ion.  
Figure \ref{fig_PV_ionization} presents a gallery of PV diagrams that span the full range of ionization energies we could usefully study in NGC 6153, from \ion{C}{4} $\lambda\lambda$5801,5811 (64.5\,eV) to [\ion{O}{1}] $\lambda$6300 (0\,eV).  In Figure \ref{fig_PV_ionization}, we indicate features that we shall refer to later, such as the emission from the nebula's main shell, the filament, and the diffuse emission, both of which are beyond the receding side of the main shell.  The line that schematically indicates the velocity ellipse for the main shell is drawn for the [\ion{O}{3}] $\lambda$4959 line and repeated in each panel, making it easier to appreciate the systematic changes in the extent of the main shell (away from the central star) or its velocity splitting as a function of ionization energy.  In Appendix \ref{app_ionization_structure} (Figures \ref{fig_app_PVC4}-\ref{fig_app_PVSi2}), we present a large collection of PV diagrams of all stages of ionization.

There are clear and systematic changes in the shape of the line emission profiles as a function of ionization energy in Figure \ref{fig_PV_ionization}.  In the most highly ionized gas (\ion{C}{4} $\lambda\lambda$5801,5811), there is little structure apart from a round velocity ellipse (the low S/N is an impediment; the central star also emits in this line; \cite{liuetal2000}).  In the next most highly ionized line, \ion{He}{2} $\lambda$4686, much more structure is apparent, and the main shell's spatial extent increases, creating a profile that is more square in shape.  Then, in [\ion{Ar}{4}] $\lambda$4740, the emission from the main shell changes, especially on the approaching side, to a more rounded shape, mostly as a result of a greater velocity difference between the approaching and receding sides of the main shell, but with little increase in spatial extent.  With [\ion{O}{3}] $\lambda$4959, the increasing velocity difference between the two sides of the main shell continues, but there is also less emission on the receding side of the main shell along the direction to the central star (the horizontal line; ordinate value of zero).  Continuing to lower ionization energies, in [\ion{Ar}{3}] $\lambda$7135, an extension beyond the main shell on its receding side begins to appear and there is a dimming of both sides of the main shell along the line of sight to the central star.  Proceeding to [\ion{N}{2}] $\lambda$6583, the emission from the main shell decreases substantially and the filament is by far the brightest feature.  Finally, in the lowest ionization stage, [\ion{O}{1}] $\lambda$6300, the main shell almost disappears while the filament remains very bright (the other emission is telluric).  

\begin{figure}
\includegraphics[width=\linewidth, viewport=0in 0.3in 5in 3in]{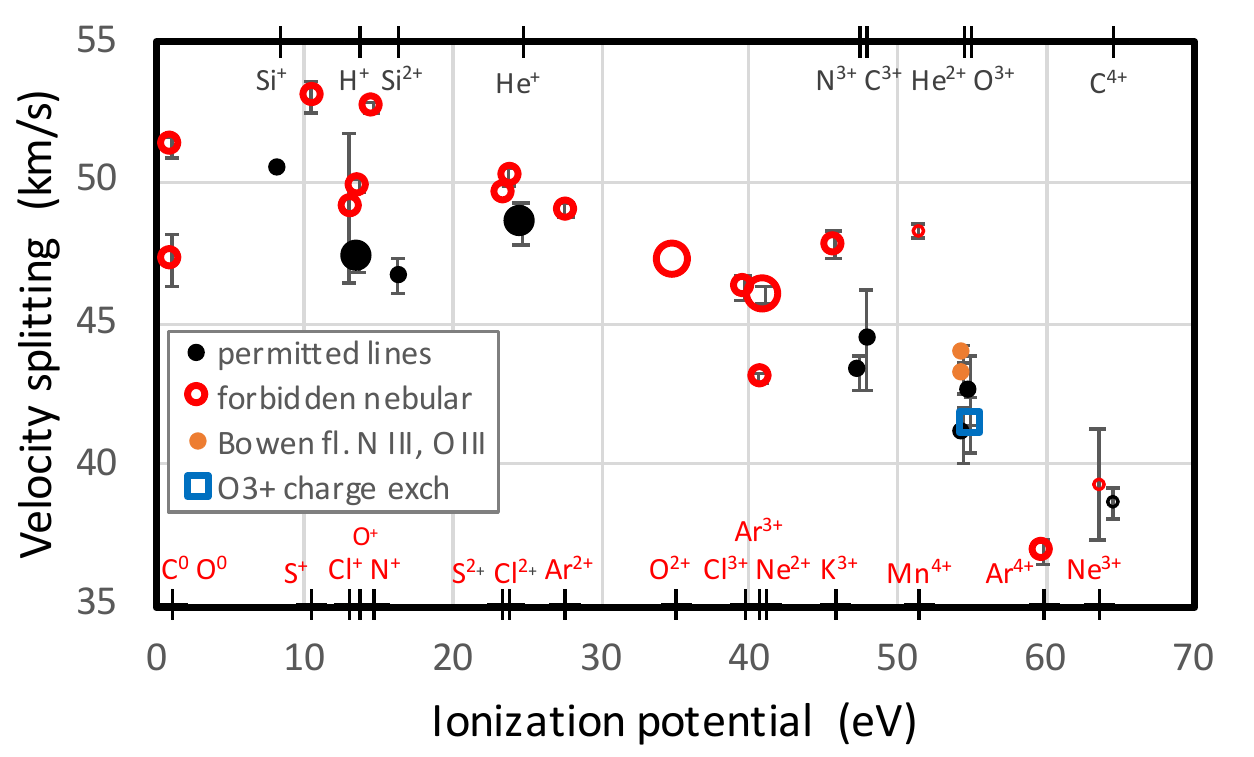}
\includegraphics[width=\linewidth, viewport=0in 0.3in 5in 3in]{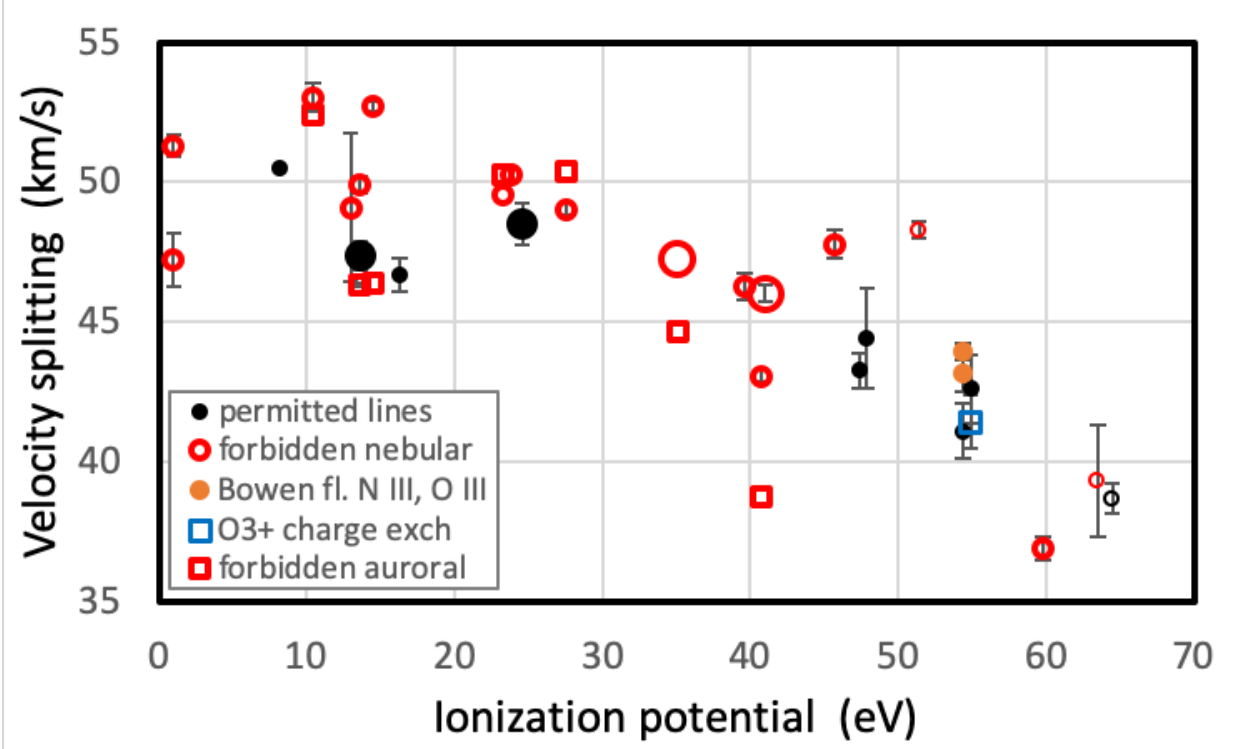}
\includegraphics[width=\linewidth]{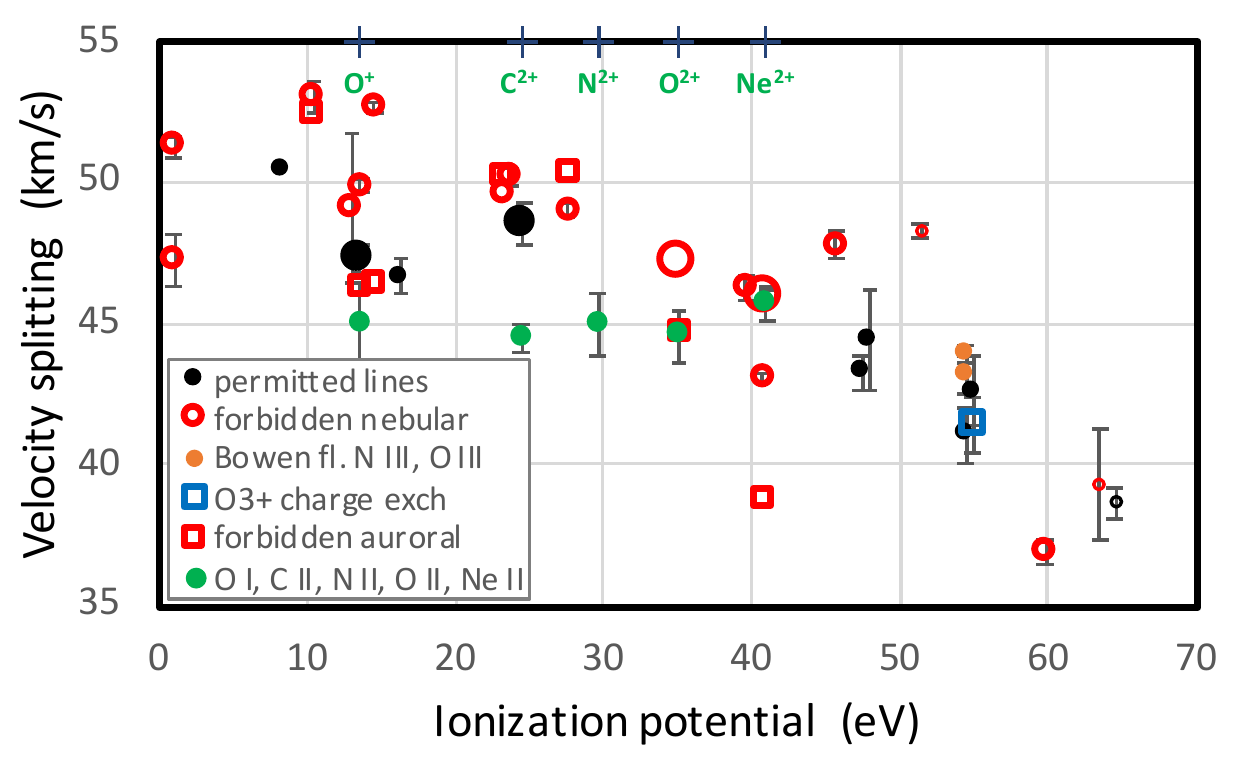}
\caption{These panels present the Wilson diagram, adding details from top to bottom.  The top panel presents the velocity splitting as a function of ionization energy from all forbidden nebular lines and all permitted lines, 
except those of O$^0$, C$^+$, N$^+$, O$^+$, and Ne$^+$.  These lines establish a well-defined relation between velocity splitting and ionization potential.  
In the second panel, we add the auroral forbidden lines of N$^+$, O$^+$, O$^{2+}$, S$^+$, S$^{2+}$, Ar$^{2+}$, and Ar$^{3+}$ (open squares).  For N$^+$, O$^+$, O$^{2+}$, and Ar$^{3+}$, these transitions have smaller velocity splitting than the nebular transitions.  In the bottom panel, we add the permitted lines of O$^0$, C$^+$, N$^+$, O$^+$, and Ne$^+$.  Although these lines span a wide range of ionization potentials, they share a common velocity splitting.  The permitted lines of H$^0$ and He$^0$ and the forbidden lines of O$^{2+}$ and Ne$^{2+}$ are shown in larger symbols since these lines sample the majority of the nebular volume.  The small symbols for Mn$^{4+}$, Ne$^{3+}$ (forbidden), and C$^{3+}$ (permitted) indicate that they are difficult measurements.  The tick marks and labels indicate the ionization energies.  
}
\label{fig_Wilson_diagram}
\end{figure}

Figure \ref{fig_Wilson_diagram} presents three \citet{wilson1950} diagrams for NGC 6153 in which we quantify the foregoing, adding details from top to bottom.  For emission lines arising from different ions and different excitation processes, we compute the line splitting, measured as the velocity difference between the approaching and receding sides of the nebula.  In Figure \ref{fig_Wilson_diagram}, we plot the line splitting as a function of the ionization potential of the ``parent" ion, i.e., the ion that initiates the emission process.  For example, O$^{2+}$ ions are responsible for initiating the emission of the permitted lines of \ion{O}{2}, and the emission of forbidden [\ion{O}{3}] lines.  
The Bowen fluorescence lines of \ion{O}{3} and \ion{N}{3} are a special case.  These lines arise due to fluorescence from \ion{He}{2} Ly$\alpha$, so we plot them at the ionization energy of He$^{+}$ and not the ionization energies of O$^{2+}$ and N$^{2+}$.  

We measure the line splitting near the position of central star, within the three rows of the PV diagram in the rectangle in Figure \ref{fig_pv_diagram}, over the spatial extent 0.72\arcsec--1.80\arcsec\ NE of the central star.  Due to the symmetry with respect to the line of sight towards the central star, measuring the line splitting in this region minimizes projection effects.  All of the measurements of the line splittings used to construct Figure \ref{fig_Wilson_diagram} are given in Table \ref{tab_line_splitting}.  When several lines from a given ion and process are available, we plot the average line splitting and its standard deviation (error bars) in Figure \ref{fig_Wilson_diagram}.  We adopt ionization energies from \citet{kramidaetal2021}.   

\begin{figure*}
\begin{center}\includegraphics[width=0.86\linewidth]{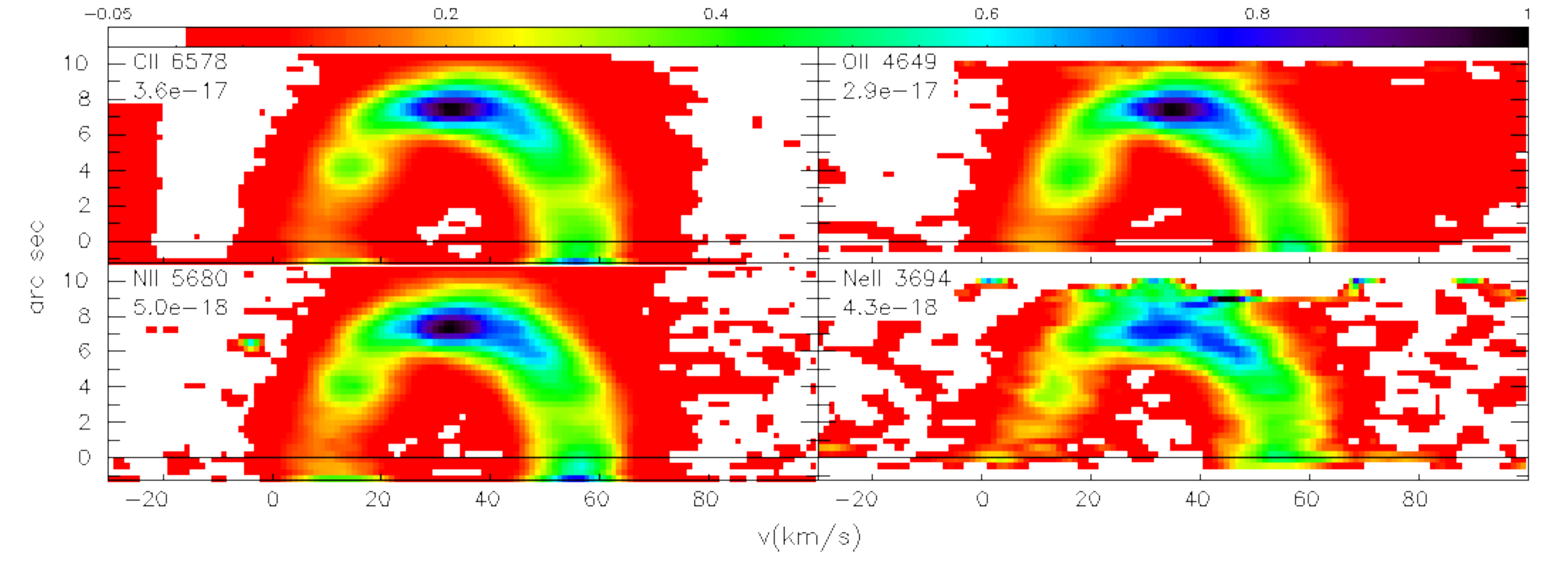}\end{center}
\caption{These panels present the PV diagrams in the lines of \ion{C}{2} $\lambda$6578, \ion{O}{2} $\lambda$4649, \ion{N}{2} $\lambda$5680 , and \ion{Ne}{2}$\lambda$3694.  Although the PV diagram for \ion{Ne}{2} $\lambda$3694 is of lower S/N, all of the lines share the same morphology, indicating that they arise from the same volume of plasma.  Note that the morphology of these PV diagrams is quite different from the morphology of the PV diagrams for [\ion{O}{3}] $\lambda\lambda$4959,5007 and [\ion{Ne}{3}] $\lambda$3869 (Figure \ref{fig_PV_ionization}, Figure \ref{fig_app_PVHbeta}, Figure \ref{fig_app_PVAr3}).  They are more similar to the PV diagrams in Figure \ref{fig_app_PVC3}, but their overall shape is even more circular because of their larger velocity splitting.  
}
\label{fig_PVC2}
\end{figure*}

In the top panel of Figure \ref{fig_Wilson_diagram}, the velocity splitting of the emission lines in NGC 6153 decreases as the ionization energy for the parent ion increases \citep{wilson1950}.  This well-known result arises since the most highly ionized ions are in the innermost plasma that expands most slowly, i.e., the inner plasma cannot overrun the plasma outside it.  This panel includes all lines due to forbidden nebular transitions (transitions between the first excited state and the ground state or ground term), Bowen fluorescence lines of \ion{O}{3} and \ion{N}{3}, charge exchange lines of \ion{O}{3}, and all permitted lines, except those of \ion{O}{1}, \ion{C}{2}, \ion{N}{2}, \ion{O}{2}, and \ion{Ne}{2}.  The sequence, 
from [\ion{Ar}{5}], [\ion{Ne}{4}], \ion{O}{3}, and \ion{He}{2} at the smallest velocity splitting to [\ion{S}{2}], [\ion{O}{2}], and [\ion{N}{2}] at the largest, is therefore expected.  With the exception of the \ion{H}{1} lines, 
the permitted and forbidden lines follow a single, continuous ionization structure.  Even lines such as \ion{O}{3} $\lambda\lambda$3757,5592 that arise from charge exchange and the \ion{O}{3} and \ion{N}{3} Bowen fluorescence emission fall, as they should, at velocities similar to those for the emission from \ion{He}{2} and \ion{O}{3} recombination (higher ionization energies).  

In the second panel, we add the auroral forbidden lines (transitions from the second excited state to the first excited state).  For the lines of [\ion{O}{2}], [\ion{N}{2}], [\ion{O}{3}], and [\ion{Ar}{4}] (Table \ref{tab_line_splitting}), the auroral lines have smaller velocity splitting than their nebular counterparts, e.g., [\ion{O}{2}] $\lambda\lambda$7320,7330 have a smaller velocity splitting than [\ion{O}{2}] $\lambda\lambda$3726,3729.  
The auroral and nebular lines of [\ion{S}{2}], [\ion{S}{3}], and [\ion{Ar}{3}] differ by of order 1 km/s, which is similar to the typical dispersion (standard deviation) of the line splitting in Table \ref{tab_line_splitting} for many ions.  As we shall see, the cases of [\ion{N}{2}] and [\ion{O}{2}] arise due to contamination of the auroral lines from recombination (\S\ref{sec_contamination}), while the difference for [\ion{O}{3}] appears to be real and due to the ionization structure (\S\ref{sec_contamination}).  For the [\ion{S}{3}] lines (similar velocity splitting) and the [\ion{Ar}{4}] lines, 
atmospheric absorption confuses the issue (details in Appendix \ref{app_ionization_structure}).  

In the bottom panel of Figure \ref{fig_Wilson_diagram}, we add the permitted lines of \ion{O}{1}, \ion{C}{2}, \ion{N}{2}, \ion{O}{2}, and \ion{Ne}{2}.  Though the lines from these ions span a large range in ionization energy, they all have similar velocity splitting.  In particular, the \ion{O}{2} and \ion{Ne}{2} lines arise from O$^{2+}$ and Ne$^{2+}$ ions, respectively, as do the [\ion{O}{3}] and [\ion{Ne}{3}] lines.  Though the \ion{Ne}{2} and [\ion{Ne}{3}] lines have similar velocity splitting, the \ion{O}{2} and [\ion{O}{3}] lines do not.  Since we expect all lines emitted by a given ion to share the same volume of the nebula, the different velocity splitting for the \ion{O}{2} and [\ion{O}{3}] lines is striking.  Similarly, the \ion{C}{2} and \ion{N}{2} lines would normally be expected to present velocity splittings similar to those of \ion{He}{1}, [\ion{S}{3}], [\ion{Cl}{3}], and [\ion{Ar}{3}], but the velocity splitting for these lines is systematically too low given their ionization potentials, by $\sim 3-5$\,km/s.  

The velocity splitting for these lines, $44-45$\,km/s, is very similar to that of the \ion{C}{3}, \ion{N}{3}, and [\ion{Ar}{4}] lines at $43-44$\,km/s.  Figure \ref{fig_PVC2} presents examples of the PV diagrams of the \ion{C}{2}, \ion{N}{2}, \ion{O}{2}, and \ion{Ne}{2} lines.  It is notable that the morphology of these PV diagrams differs from the morphologies of the PV diagrams for [\ion{Ar}{4}] $\lambda$4740 (Figure \ref{fig_PV_ionization}) or those in Figure \ref{fig_app_PVC3} that have very similar velocity splitting, 
indicating that they arise from a different volume of plasma.  The PV diagrams illustrate more clearly than the Wilson diagram (Figure \ref{fig_Wilson_diagram}) that the kinematics of the \ion{C}{2}, \ion{N}{2}, \ion{O}{2}, and \ion{Ne}{2} lines is different from the kinematics of other lines of either the same ionization potential or velocity splitting.  (See Figures \ref{fig_PV_ionization} and \ref{fig_app_PVHbeta}, respectively, for the PV diagrams of the [\ion{O}{3}] $\lambda$4959 and [\ion{Ne}{3}] $\lambda$3869 lines.)

Finally, there is a cluster of permitted lines from \ion{H}{1}, \ion{O}{1}, and \ion{Si}{2}, at approximately 14 eV and 46 km/s (the auroral lines of [O II] and [N II] also fall here; middle and bottom panels of Figure \ref{fig_Wilson_diagram}).  The case of \ion{H}{1} is the simplest to understand.  Given that it is emitted throughout the whole nebular volume, its velocity splitting should be similar to that of other lines that sample the majority of this volume, such as the lines of \ion{He}{1}, [\ion{O}{3}], and [\ion{Ne}{3}].  Indeed, this is the case, so its deviation from the general trend is not unexpected.  

\begin{figure}
\includegraphics[width=\linewidth]{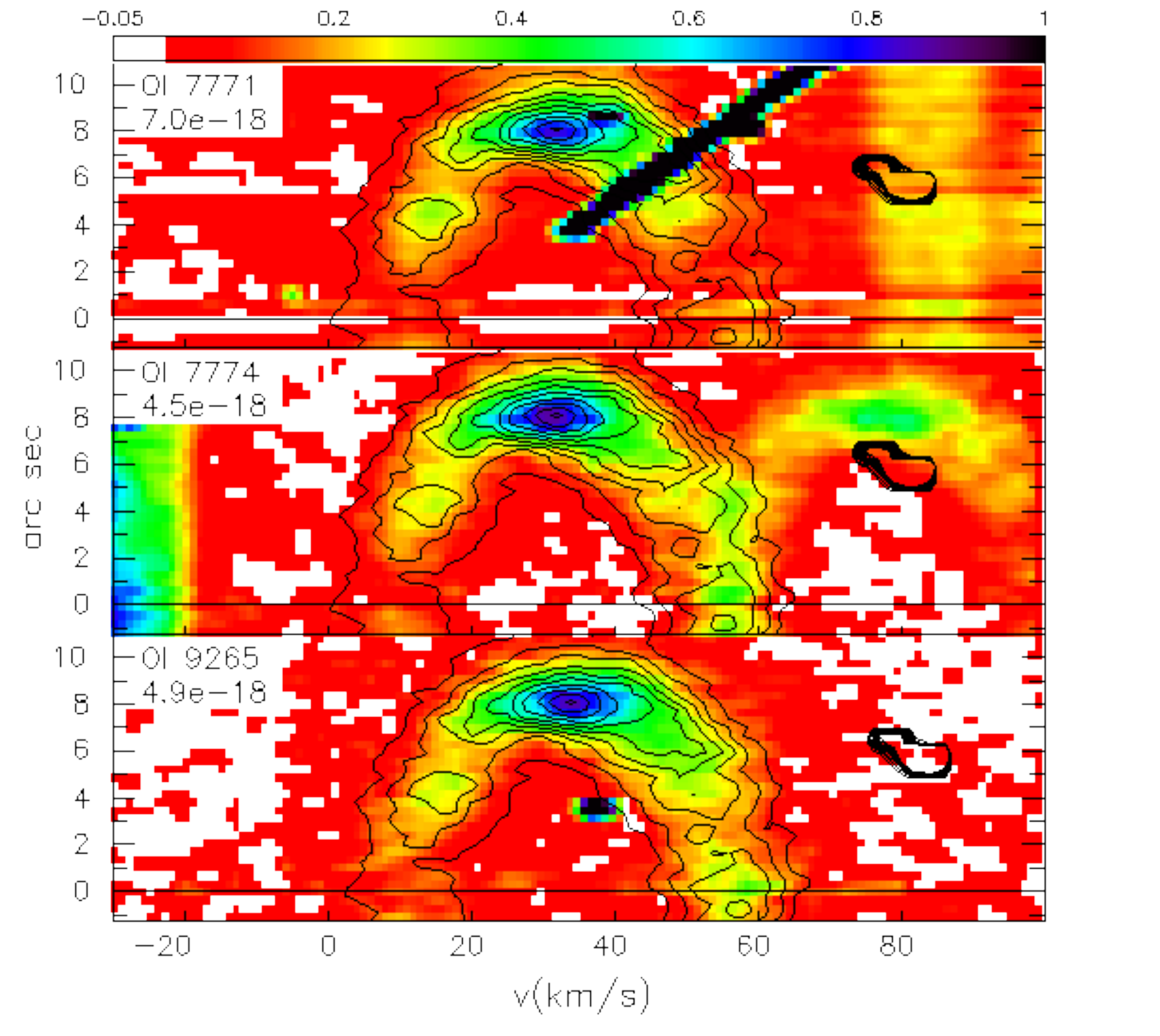}
\caption{These panels present the PV diagrams of the \ion{O}{1} $\lambda\lambda$ 7771, 7774 (7775 to the red), and 9265 lines.  In all cases, some sky lines have been subtracted, which is what produces the faint vertical bands.  The black streak is a cosmic ray.  The contours are of the intensity of the emission of the \ion{O}{2} $\lambda$9982 line.  In this and following figures, we plot the contours at 10\%, 20\%, ..., 90\% of the maximum intensity.  The kinematics of these \ion{O}{1} quintuplet lines are compatible with those of the \ion{O}{2} $\lambda$9982 line.  
}
\label{fig_PV_O1_perm}
\end{figure}

Second, the \ion{O}{1} $\lambda\lambda$7771,9265 lines in question are quintuplet lines, whose lower level is the meta-stable 3s\,$^5\mathrm S^o$ state that can decay to the O$^0$ ground state only through a forbidden transition.  Hence, they 
cannot be excited by fluorescence from the ground state and so are likely the result of recombination.  
Figure \ref{fig_PV_O1_perm} compares the PV diagrams of these lines with the emission from \ion{O}{2} $\lambda$9982, 
obtained simultaneously.  The morphology and velocity splitting of the PV diagrams for these \ion{O}{1} lines, especially \ion{O}{1} $\lambda$9265 (least affected by sky lines), is very similar to that of \ion{O}{2} $\lambda$9982.  So, it appears that the \ion{O}{1} $\lambda\lambda$7771,9265 lines arise from the same plasma that emits the \ion{O}{2} lines.  \citet{garciarojasetal2022} find the same result.  

Third, the \ion{Si}{2} lines involved are \ion{Si}{2} $\lambda\lambda$5041, 6347, 6374 from multiplets V5 (5041\AA) and V2, respectively, and we detect them with good S/N (see Figure \ref{fig_app_PVSi2}).  We suspect that these lines in NGC 6153 are partly excited by recombination and partly by fluorescence as a minority ion in zones of higher ionization.  So, the velocity splitting from these \ion{Si}{2} lines in NGC 6153 may not reflect the kinematics of the volume where Si$^{2+}$ is the dominant ionization state, which is how this data point is plotted in Figure \ref{fig_Wilson_diagram}.  We provide more details in Appendix \ref{app_ionization_structure}.

The general trend of a velocity splitting that decreases as the ionization energy of the parent ion increases in Figure \ref{fig_Wilson_diagram} is the expected one.  This trend defines what we refer to as the \emph{normal nebular plasma}.  There are two novelties though.  First, the auroral forbidden lines of [\ion{O}{2}], [\ion{N}{2}], and [\ion{O}{3}] clearly have lower velocity splitting than their nebular counterparts, the first two due to contamination from recombination and the last due to real temperature structure (\S\ref{sec_contamination}).  This also occurs for [\ion{Ar}{4}], but may be due to atmospheric absorption.  Second, the lines of \ion{O}{1}, \ion{C}{2}, \ion{N}{2}, \ion{O}{2}, and \ion{Ne}{2} share a common velocity despite spanning a large range in the ionization energy of the parent ions.  These lines appear to define a second kinematic component that does not participate in the general trend and which we call the \emph{additional plasma component}.  

\begin{deluxetable*}{lccl}
\tablecaption{Line splitting\label{tab_line_splitting}}
\tablewidth{0pt}
\tablehead{
\colhead{Parent} & \colhead{$E_{ion}$} & \colhead{Velocity splitting} & \colhead{Lines} \\
 & \colhead{(eV)} & \colhead{(km/s)} 
}
\startdata
H$^+$     & 13.6 & $47.37\pm 0.48$ & \ion{H}{1} $\lambda\lambda$3770, 3797, 3835, 3970, 4101, 4340, 4861, 6563, 9229, 9545, 10049 \\
He$^+$    & 24.6 & $48.51\pm 0.75$ & \ion{He}{1} $\lambda\lambda$3187, 3614, 3965, 4026, 4387, 4471, 4713, 4922, 5015, 5876, 6678, 7281 \\
He$^{2+}$ & 54.4 & $41.09\pm 1.00$ & \ion{He}{2} $\lambda\lambda$3203, 4541, 4686, 4859, 5411, 6560, 10123 \\
C$^0$     & 0.0  & $47.23\pm 0.95$ & {[}\ion{C}{1}{]} $\lambda\lambda$9850 (nebular)\\
C$^{2+}$  & 24.4 & $44.48\pm 0.46$ & \ion{C}{2} $\lambda\lambda$4267, 5342, 6151, 6461, 6578, 7231, 9903 \\
C$^{3+}$  & 47.9 & $44.4\pm 1.8$   & \ion{C}{3} $\lambda$4647 \\
C$^{4+}$  & 64.5 & $38.67\pm 0.54$ & \ion{C}{4} $\lambda\lambda$5801, 5811 \\
N$^+$     & 14.5 & $52.72\pm 0.18$ & {[}\ion{N}{2}{]} $\lambda\lambda$6548, 6583 (nebular)\\
N$^+$     & 14.5 & $46.4$          & {[}\ion{N}{2}{]} $\lambda$5755 (auroral)\\
N$^{2+}$  & 29.6 & $45.0\pm 1.1$   & \ion{N}{2} $\lambda\lambda$4035, 4041, 5666, 5676, 5680, 5686, 5710 \\
N$^{2+}$  & 54.4 & $43.92\pm 0.29$ & \ion{N}{3} $\lambda\lambda$4097, 4103, 4634, 4641 (Bowen fluorescence) \\
N$^{3+}$  & 47.4 & $43.26\pm 0.63$ & \ion{N}{3} $\lambda\lambda$4379, 5147 \\
O$^0$     & 0.0  & $51.30\pm 0.38$ & {[}\ion{O}{1}{]} $\lambda\lambda$6300, 6364 (nebular)\\
O$^+$     & 13.6 & $49.90\pm 0.30$ & {[}\ion{O}{2}{]} $\lambda\lambda$3726, 3729 (nebular)\\
O$^+$     & 13.6 & $46.32\pm 0.04$ & {[}\ion{O}{2}{]} $\lambda\lambda$7319, 7320, 7330, 7331 (auroral)\\
O$^+$     & 13.6 & $45.1\pm 1.4$   & \ion{O}{1} $\lambda\lambda$7771, 7774, 9265 \\
O$^{2+}$  & 35.1 & $47.2$          & {[}\ion{O}{3}{]} $\lambda$4959 (nebular) \\
O$^{2+}$  & 35.1 & $44.6$          & {[}\ion{O}{3}{]} $\lambda$4363 (auroral) \\
O$^{2+}$  & 35.1 & $44.58\pm 0.96$ & \ion{O}{2} $\lambda\lambda$4069.6, 4069.9, 4072, 4076, 4084, 4089, 4317, 4349, 4366, 4591, 4596, 4639, \\ 
          &      &                 & 4649, 4662, 4676, 6500.8, 6501.4, 9982 \\
O$^{2+}$  & 54.4 & $43.18\pm 0.69$ & \ion{O}{3} $\lambda\lambda$3133, 3299, 3312, 3340, 3444, 3754, 3759 (Bowen fluorescence)\\
O$^{3+}$  & 54.9 & $42.6\pm 1.2$   & \ion{O}{3} $\lambda$3260, 3707 \\
O$^{3+}$  & 54.9 & $41.43\pm 0.98$ & \ion{O}{3} $\lambda\lambda$3757, 3774, 5592 (charge exchange)\\
Ne$^{2+}$ & 41.0 & $46.02\pm 0.32$ & {[}\ion{Ne}{3}{]} $\lambda\lambda$3869, 3967 (nebular)\\
Ne$^{2+}$ & 41.0 & $45.68\pm 0.55$ & \ion{Ne}{2} $\lambda\lambda$3335, 3568, 3694, 5130, 9096 \\
Ne$^{3+}$ & 63.5 & $39.3\pm 2.0$   & {[}\ion{Ne}{4}{]} $\lambda$4724 (nebular) \\
Si$^+$    &  8.2 & $50.5$          & \ion{S}{1} $\lambda$6486 \\
Si$^{2+}$ & 16.3 & $46.67\pm 0.60$ & \ion{Si}{2} $\lambda\lambda$5041, 6347, 6371 \\
S$^+$     & 10.4 & $53.03\pm 0.50$ & {[}\ion{S}{2}{]} $\lambda\lambda$6716, 6731 (nebular)\\
S$^+$     & 10.4 & $52.4$          & {[}\ion{S}{2}{]} $\lambda$4069 (auroral)\\
S$^{2+}$  & 23.3 & $49.6$          & {[}\ion{S}{3}{]} $\lambda\lambda$9531 (nebular)\\
S$^{2+}$  & 23.3 & $50.27\pm 0.37$ & {[}\ion{S}{3}{]} $\lambda$6312 (auroral)\\
Cl$^+$    & 13.0 & $49.1\pm 2.7$   & {[}\ion{Cl}{2}{]} $\lambda\lambda$8578, 9123 (nebular)\\
Cl$^{2+}$ & 23.8 & $50.26\pm 0.30$ & {[}\ion{Cl}{3}{]} $\lambda\lambda$5517, 5537 (nebular)\\
Cl$^{3+}$ & 39.6 & $46.27\pm 0.47$ & {[}\ion{Cl}{4}{]} $\lambda\lambda$7530, 8045 (nebular)\\
Ar$^{2+}$ & 27.6 & $49.02\pm 0.22$ & {[}\ion{Ar}{3}{]} $\lambda$7135, 7751 (nebular)\\
Ar$^{2+}$ & 27.6 & 50.4            & {[}\ion{Ar}{3}{]} $\lambda$5191 (auroral)\\
Ar$^{3+}$ & 40.7 & $43.05\pm 0.21$ & {[}\ion{Ar}{4}{]} $\lambda\lambda$4711, 4740 (nebular)\\
Ar$^{3+}$ & 40.7 & 38.7            & {[}\ion{Ar}{4}{]} $\lambda$7262 (auroral)\\
Ar$^{4+}$ & 59.8 & $36.87\pm 0.43$ & {[}\ion{Ar}{5}{]} $\lambda$6435, 7006 (nebular)\\
K$^{3+}$  & 45.7 & $47.78\pm 0.52$ & {[}\ion{K}{4}{]} $\lambda\lambda$6101, 6795 (nebular)\\
Mn$^{4+}$ & 51.4 & $48.29\pm 0.29$ & {[}\ion{Mn}{5}{]} $\lambda\lambda$5701, 5861, 6083, 6393 \\
\enddata
\end{deluxetable*}

\subsection{Physical conditions}\label{sec_physical_conditions}

In the following subsections, we study the temperature and density in the nebular shell of NGC 6153 based upon a variety of methods.  Table \ref{tab_main_results} summarizes these results.  The temperatures derived from forbidden lines and the \ion{He}{1} lines are substantially higher than the temperatures derived using the permitted lines of \ion{N}{2} and \ion{O}{2}.  Likewise, a dichotomy exists concerning the density, with the \ion{N}{2} and \ion{O}{2} lines indicating higher densities than other methods.  Thus, the physical conditions reflect the results of the kinematics, i.e., that two plasma components are present.

\subsubsection{Thermal broadening of line profiles}\label{sec_kinematic_temperature}

Ions belonging to the same plasma component should share common spatial, velocity, density, and temperature structures provided that they occupy the same (or similar) volume within the nebula.  Even so, emission lines from two ions that share the same volume of the nebula will have different PV diagrams if they have different intrinsic fine structure or different thermal line widths.  Therefore, we can exploit differences in atomic mass to infer the plasma temperature for ions that share the same nebular volume if we can account for the differences in the intrinsic line structure of the two lines \citep[e.g.,][]{courtesetal1968, dysonmeaburn1971, garciadiazetal2008}.  Determining the temperature in this way circumvents issues related to atomic data, but requires comparing only ions that share the same nebular volume.  

We will consider lines arising from ions of H$^+$, He$^+$, and O$^{++}$.  The latter two are expected to occupy very similar volumes within the nebula, especially since NGC 6153 is highly ionized and optically thin.  The volume of the nebula occupied by H$^+$ will differ from that occupied by He$^+$ and O$^{++}$, but the effect of this difference is predictable and clear:  the PV diagrams for lines from H$^+$ will present an excess at low expansion velocities and near the central star (the He$^{2+}$ zone) compared to PV diagrams for lines arising from He$^+$ and O$^{++}$.  
The sensitivity of the process depends upon the difference in atomic mass, so it is most instructive to compare H and He with heavy elements.  
The case of O$^{++}$ allows us to extend our results to the permitted lines of \ion{C}{2}, \ion{O}{2}, \ion{N}{2}, and \ion{Ne}{2}.  In order to minimize instrumental and observational effects, we consider only lines from the CD2 wavelength interval, for which 
the slit's position angle was closest to the parallactic angle, so they suffer least from differential atmospheric refraction.

The width observed for an emission line will depend upon the velocity structure of the nebula, the instrumental resolution, the thermal broadening, and the intrinsic line structure.  The thermal and instrumental broadening are Gaussian or closely so in profile, so we approximate the line width as $\sigma^2=\sigma_{neb}^2 + \sigma_{ins}^2 + \sigma_{th}^2 +\sigma_{fs}^2$, where $\sigma_{neb}$ is the contribution due to the nebula's kinematic structure, $\sigma_{ins}$ the instrumental broadening, $\sigma_{th}$ the thermal broadening, and $\sigma_{fs}$ the fine structure of the emission line itself.  

\begin{figure*}[]
\includegraphics[width=0.49\linewidth]{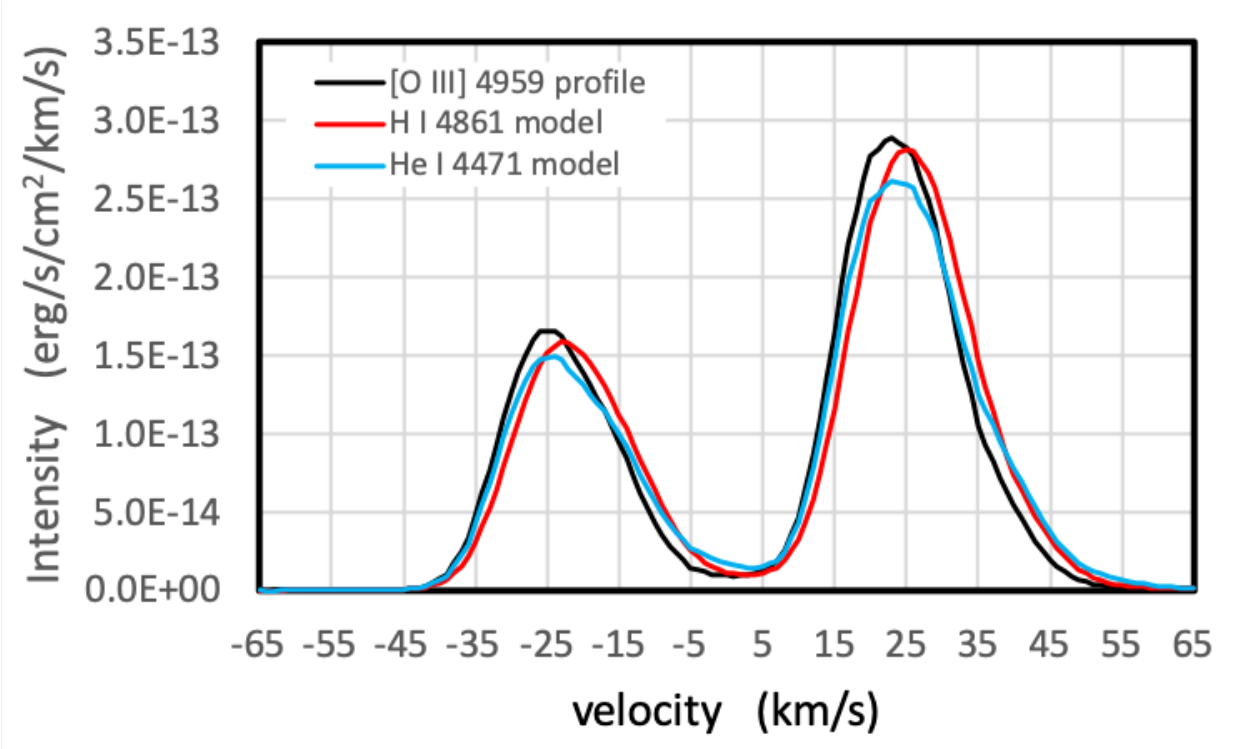}
\includegraphics[width=0.49\linewidth]{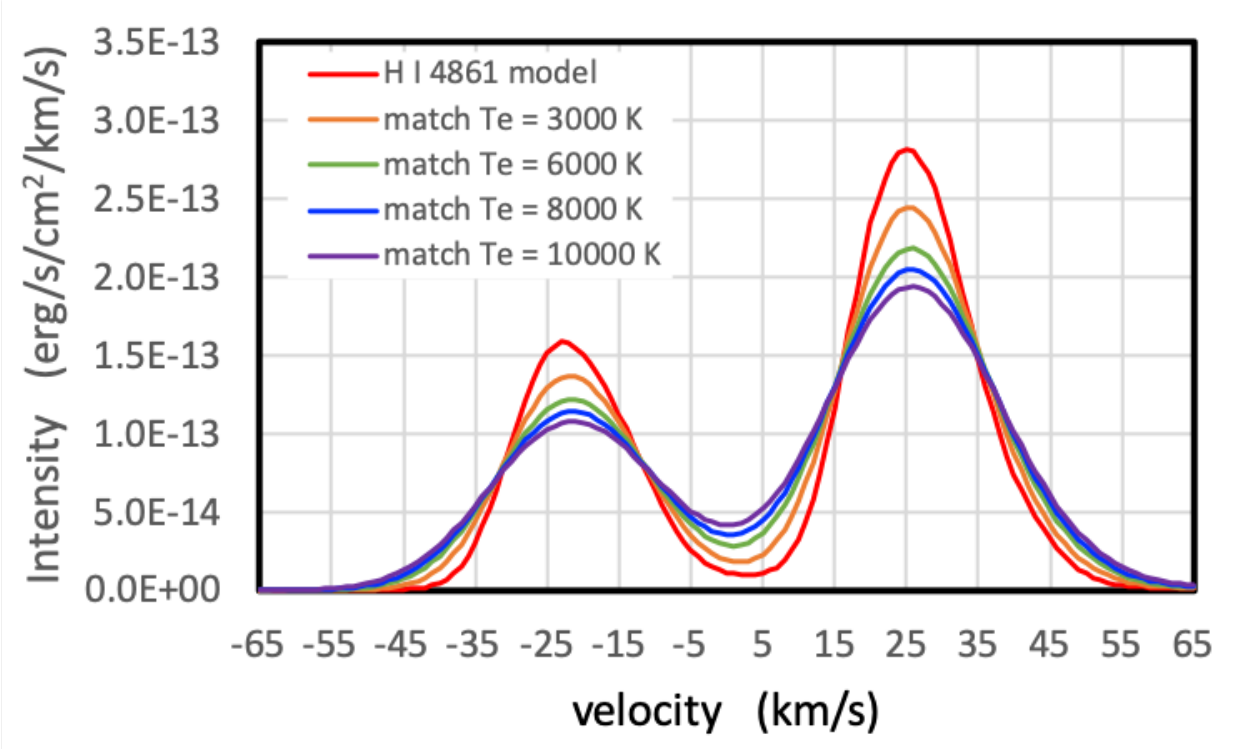}
\caption{left panel: This plot presents the line profile for the spatial interval used to compute the velocity splitting (see Figure \ref{fig_pv_diagram}) for the [\ion{O}{3}] $\lambda$4959 line as well as the models based upon this line for the H$\beta$ and \ion{He}{1} $\lambda$4471 lines.  
Right panel: This plot presents the model of the H$\beta$ line based upon the [\ion{O}{3}] $\lambda$4959 line from the left panel as well as the profiles of this model after broadening with a Gaussian to match the thermal line width of the H$\beta$ line for temperatures of 3,000 K, 6,000 K, 8,000 K, and 10,000 K.  The latter is the largest modification made to the [O III] line profile.  
}
\label{fig_OIII4959models}
\end{figure*}

\begin{figure*}
\begin{center}\includegraphics[width=0.86\linewidth]{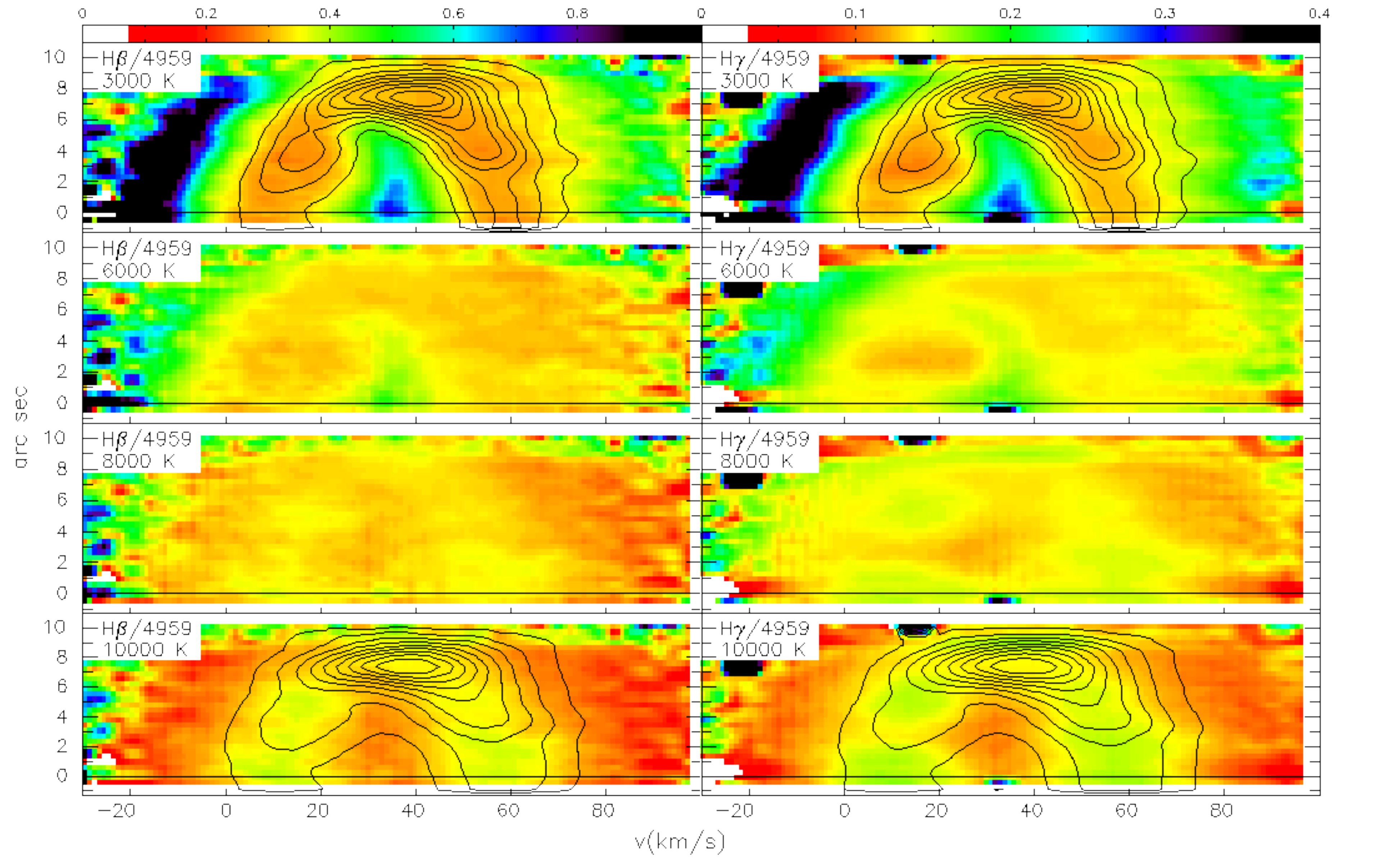}\end{center}
\caption{These panels present the ratio of H$\beta$ (left column) and H$\gamma$ (right column) with respect to [\ion{O}{3}] $\lambda$4959.  The ratio images assume temperatures of 3,000 K, 6,000 K, 8,000 K, and 10,000 K (top to bottom) when broadening the PV diagram of [\ion{O}{3}] $\lambda$4959.  In the top row, the contours are those of the PV diagram for [\ion{O}{3}] $\lambda$4959.  In the bottom row, the contours are those of H$\beta$ (left column) and H$\gamma$ (right column).  The ratios 
best approximate a constant value for a temperature of 8,000 K.  
}
\label{fig_tkin_HbHgO3}
\end{figure*}

\begin{figure*}
\begin{center}\includegraphics[width=0.86\linewidth]{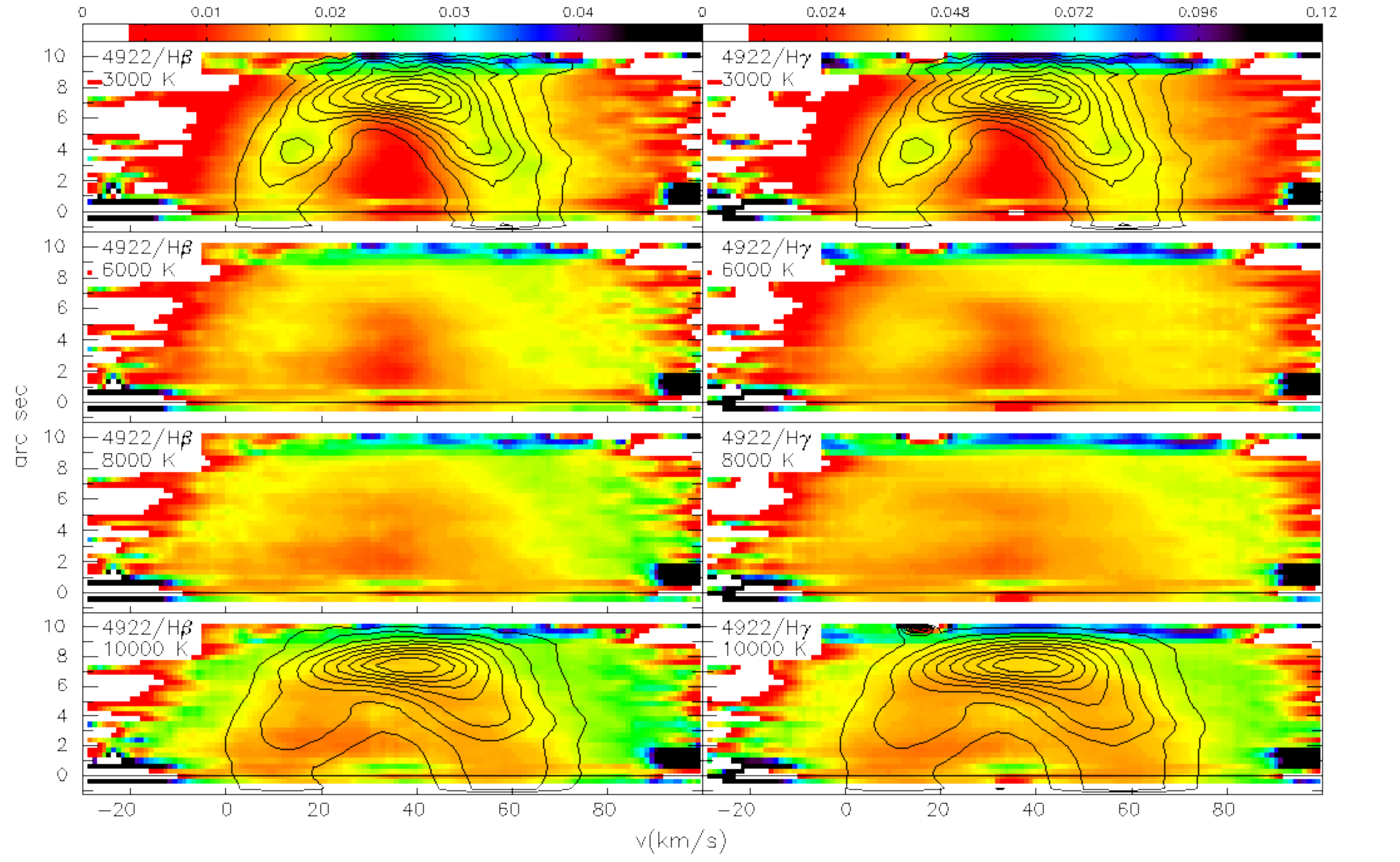}\end{center}
\caption{These panels present the ratio of the \ion{He}{1} $\lambda$4922 line with respect to H$\beta$ (left column) and H$\gamma$ (right column), assuming temperatures of 3,000 K, 6,000 K, 8,000 K, and 10,000 K (top to bottom).  The contours are those of the PV diagram for \ion{He}{1} $\lambda$4922 in the top row, but of H$\beta$ (left) and H$\gamma$ (right) in the bottom row.  The ratios are most constant when the assumed temperature to broaden the He I lines is 8,000 K.  
}
\label{fig_tkin_HeIHbHg}
\end{figure*}

The effect of atomic mass appears through the thermal broadening of the lines emitted by a given ion.  The thermal width is given by $\sigma_{th} = \sqrt{82.5((10^{-4}T_e)/m_{ion})}$, where 
$T_e$ is the temperature of the plasma (K) and $m_{ion}$ is the mass of the ion \citep[atomic mass units;][]{garciadiazetal2008}.  
Thus, the difference in thermal line width can be accounted for by convolving the PV diagram of the emission line from the more massive ion with a Gaussian to account for the difference in atomic mass between it and the ion with lower atomic mass.  

The differences in the line structure can always be accommodated if one of the lines is a single, spectrally-resolved emission component.  In that case, we can modify the PV diagram of the single line to match the line structure of the other, reflected in its PV diagram:  Multiple copies of the single line are shifted and scaled appropriately to mimic the structure of the line with multiple components.  

The \ion{H}{1} and \ion{He}{1} triplet lines have substructure that must be taken into account.  The \ion{H}{1} Balmer lines have 7 components, spread over a small range in velocity \citep{cleggetal1999}.  The seven components divide into two groups whose spread in velocity is significantly less than the difference in velocity between the two groups (blue and red).  
Within the blue group, about 80\% of the total intensity is split between two components while, within the red group, a single component accounts for about 80\% of the total intensity \citep[e.g., Fig. 1 of][]{cleggetal1999}.  We adopt the fraction of the emission from each group from \citet{cleggetal1999}.  The He I lines are either triplet or singlet states.  
The triplets are formed by three ($n\,^3\mathrm S-n^{\prime}\,^3\mathrm P$) or six ($n\,^3D-n^{\prime}\,^3\mathrm P$) individual transitions; 
all but one conform a blue group spread over 1-3 km/s while the last transition falls 10-20 km/s to the red \citep{vanhoof2018}.   We follow \citet{axneretal2004} to determine the relative intensities of the individual transitions in \ion{He}{1}, finding that the blue group contains 89\% of the total intensity.  Within the blue group, two transitions separated by about half of the total spread account for about 72\% of its total intensity.  

So, we approximate the \ion{H}{1} and \ion{He}{1} triplet lines as two components, one representing the blue group and the other the red line/group.  Two copies of the PV diagram of a single line are created and added together after appropriately shifting them in wavelength and scaling them in flux according to the details of the \ion{H}{1} or \ion{He}{1} line involved.  
Then, this model PV diagram is broadened by convolving it with a Gaussian so as to match the thermal width of the \ion{H}{1} or \ion{He}{1} line.  
Matching the thermal width is a process of trial and error in which we consider electron temperatures between 3,000\,K and 15,000\,K.  
We then compare this model to the observed PV diagram, 
determining the best thermal broadening by searching for the PV diagram with the most constant ratio.  

This kinematic temperature is an emission-weighted mean temperature, measured by the motions of the ions in the plasma.  It does not imply that there is a single, uniform temperature within the plasma emitting in the \ion{H}{1}, \ion{He}{1}, \ion{O}{2}, and [\ion{O}{3}] lines.  
The plasma may well contain temperature gradients or small-scale structure, but the PV diagrams inherently reflect the result of that structure.  (Given the thermal width of the \ion{H}{1} and \ion{He}{1} lines, the kinetic temperature will not be very sensitive to small-scale structure.)  One worry is that the temperature sensitivity varies among the lines considered.  The intensity of the [\ion{O}{3}] lines increases with increasing temperature, but the intensities of the other lines decreases, an effect that cannot be compensated exactly.  To mitigate the issue, we consider only [\ion{O}{3}] $\lambda$4959.   

\begin{figure*}
\begin{center}\includegraphics[width=0.86\linewidth]{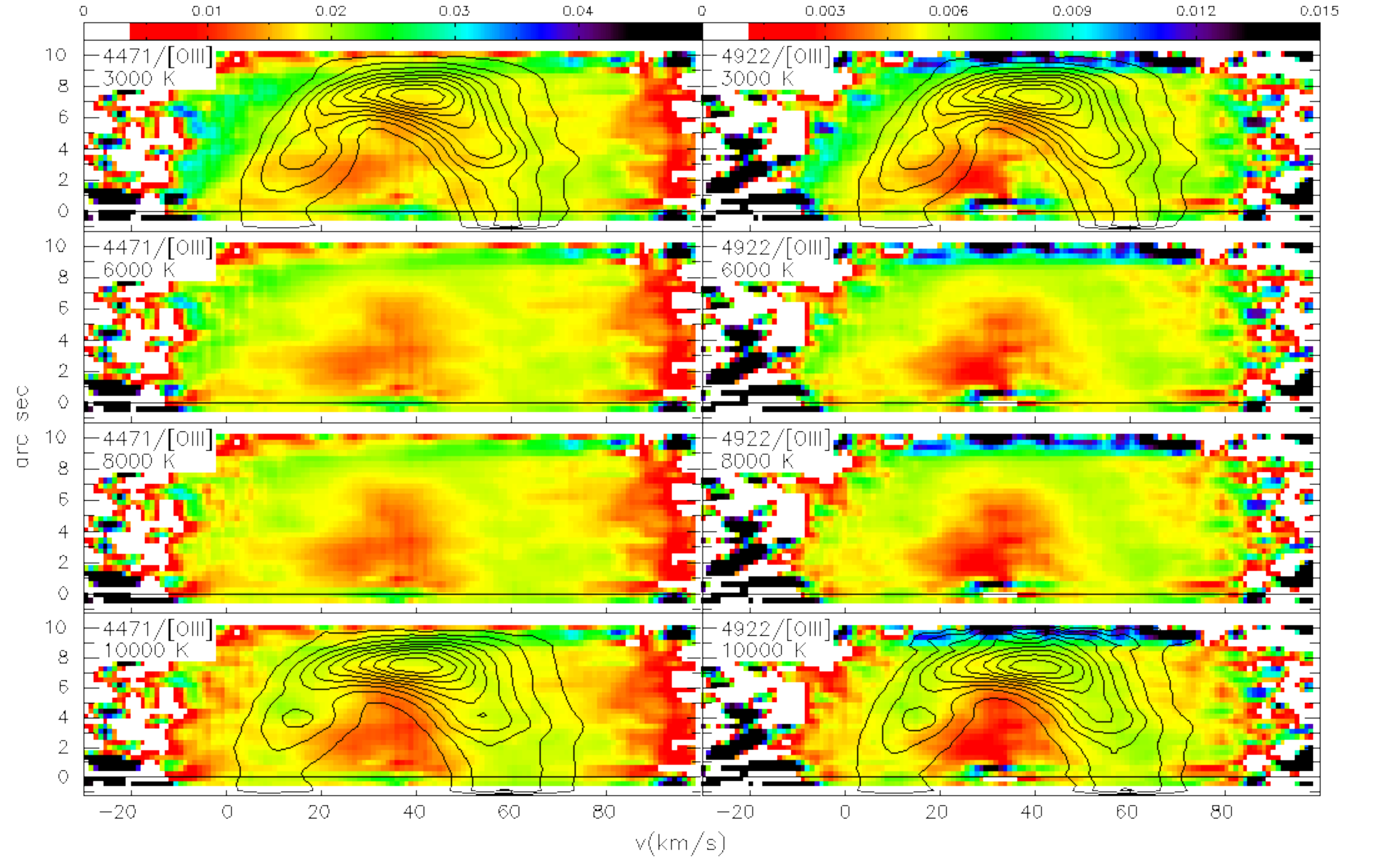}\end{center}
\caption{These panels present the ratio of the PV diagram of \ion{He}{1} $\lambda$4471 (left column) and \ion{He}{1} $\lambda$4922 (right column) with respect to [\ion{O}{3}] $\lambda$4959.  The ratios assume temperatures of 3,000 K, 6,000 K, 8,000 K, and 10,000 K (top to bottom).  In the top row, the contours are those of the PV diagram for [\ion{O}{3}] $\lambda$4959.  In the bottom row, the contours are those of \ion{He}{1} $\lambda$4471 (left) and \ion{He}{1} $\lambda$4922 (right).  Again, the ratios are most constant for a temperature of 8,000\,K.
}
\label{fig_tkin_HeIO3}
\end{figure*}

\begin{figure}
\center{
\includegraphics[width=\linewidth]{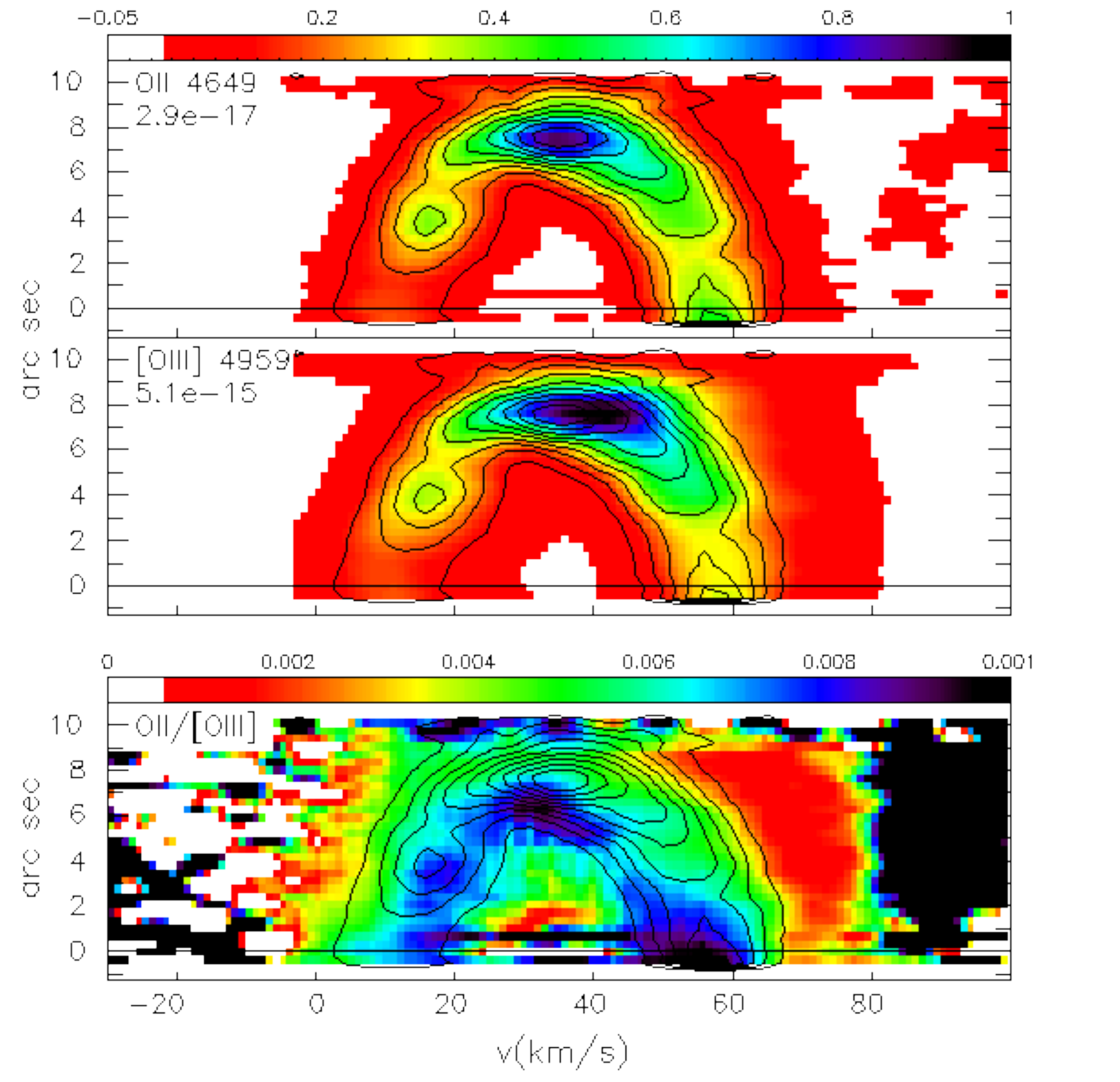}
}
\caption{These panels present the PV diagrams for the \ion{O}{2} $\lambda$4649 and [\ion{O}{3}] $\lambda$4959 lines 
and the ratio \ion{O}{2} $\lambda$4649/[\ion{O}{3}] $\lambda$4959.  The contours represent the intensity of the \ion{O}{2} $\lambda$4649 line.  Both lines are single lines from the CD2 wavelength interval from the same ion, so there are no corrections for the intrinsic line structure or thermal broadening (supposing they arise from the same plasma component).  The solid black color at right is due to contamination of the PV diagram of \ion{O}{2} $\lambda$4649 by \ion{C}{3} $\lambda$4650.
}
\label{fig_tkin_O2O3}
\end{figure}

\begin{figure*}
\begin{center}\includegraphics[width=0.86\linewidth]{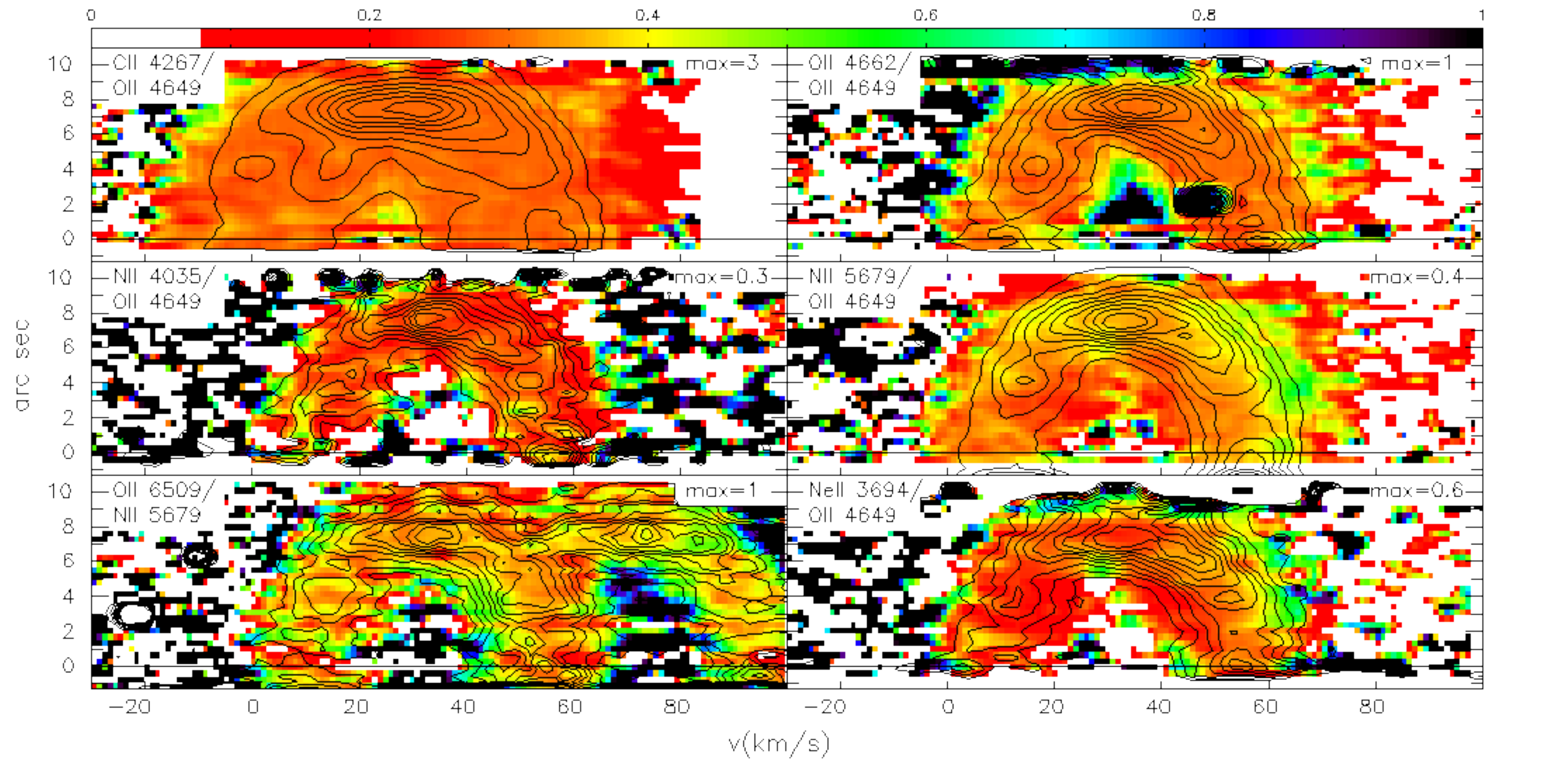}\end{center}
\caption{These panels present the ratios of PV diagrams \ion{C}{2}, \ion{N}{2}, \ion{O}{2}, and \ion{Ne}{2} lines, as indicated in each panel.  
In all panels, the contours are of the intensity of the numerator.  No correction for thermal broadening was applied to any of the PV diagrams.  The \ion{C}{2} $\lambda$4267/\ion{O}{2} $\lambda$4649 and \ion{O}{2} $\lambda\lambda$6509,6510/\ion{N}{2} $\lambda$5680 ratios required modeling multiple \ion{C}{2} and \ion{O}{2} lines.  The horizontal bands in the \ion{O}{2} $\lambda$6509/\ion{N}{2} $\lambda$5680 ratio are due to the poor background subtraction for \ion{O}{2} $\lambda$6509.  The ratio of PV diagrams all approximate a constant value, indicating that these permitted lines of \ion{C}{2}, \ion{N}{2}, \ion{O}{2}, and \ion{Ne}{2} arise from the same plasma component.  
}
\label{fig_tkin_C2N2Ne2O2}
\end{figure*}

Most of the structure in the line profiles for the lines of the heavy elements is due to the velocity structure of the nebula, not the thermal width of the line (nor the instrumental resolution).  For example, the observed line width of [\ion{O}{3}] $\lambda$4959 is slightly more than 18\,km/s FWHM.  Of this, the thermal broadening contributes 5.4\,km/s (supposing a temperature of 10,000\,K, less if the temperature is lower), while the instrumental broadening of order 10\,km/s.  Since these add in quadrature, the broadening due to the velocity structure dominates, amounting to 14\,km/s.  

Even so, the substantial modifications required to match the [\ion{O}{3}] $\lambda$4959 line to the \ion{H}{1} Balmer lines gives the method good sensitivity.  The left panel of Figure \ref{fig_OIII4959models} demonstrates that accounting for the line structure of either H$\beta$ or \ion{He}{1} $\lambda$4471 requires only minor modifications.  However, the broadening required to match the thermal line width of 
a hydrogen line can be substantial, as shown in the right panel of Figure \ref{fig_OIII4959models}, amounting to as much as 20.7\,km/s for a temperature of 10,000\,K.  The modifications required to match the thermal width of a He line to a H line are less important because of the larger thermal widths of the He lines.  Likewise, matching the thermal width of an O line to a He line also requires less broadening than illustrated in Figure \ref{fig_OIII4959models}.  

Figure \ref{fig_tkin_HbHgO3} compares the PV diagrams of the H$\beta$, H$\gamma$, and [\ion{O}{3}] $\lambda$4959 lines.  In the left column, we present the ratio of H$\beta$ with respect to [\ion{O}{3}] $\lambda$4959.  (By ``[\ion{O}{3}] $\lambda$4959", we mean the model of the H$\beta$/H$\gamma$ line constructed with the [\ion{O}{3}] $\lambda$4959 PV diagram.)  The four panels consider temperatures of 3,000\,K, 6,000\,K, 8,000\,K, and 10,000\,K when broadening the PV diagram for [\ion{O}{3}] $\lambda$4959.  Since the volumes occupied by H$^+$ and O$^{2+}$ largely coincide, the ratio should be approximately constant when the appropriate temperature is used to broaden the PV diagram for [\ion{O}{3}] $\lambda$4959.  If the assumed temperature is too low, the PV diagram for [\ion{O}{3}] $\lambda$4959 will not be sufficiently broadened, and we expect a minimum in the H$\beta$/[\ion{O}{3}] $\lambda$4959 ratio where the emission from [\ion{O}{3}] $\lambda$4959 is most intense.  Conversely, if the assumed temperature is too high, the PV diagram for [\ion{O}{3}] $\lambda$4959 will be broadened too much, diffusing its emission too far, to velocities too far from and too close to the systemic velocity.  In this case, we expect that the H$\beta$/[\ion{O}{3}] $\lambda$4959 ratio will present a maximum where the H$\beta$ emission is most intense.  The main difference in the volumes occupied by H$^+$ and O$^{2+}$ is that O$^{2+}$ does not sample as completely as H$^+$ the innermost volume of the nebular shell occupied by He$^{2+}$ (closest to the central star and with velocities closest to the systemic velocity).  

The four panels in the left column of Figure \ref{fig_tkin_HbHgO3} present a clear trend.  In the first panel, with [\ion{O}{3}] $\lambda$4959 broadened assuming a temperature of 3,000\,K, the ratio varies the most.  The H$\beta$/[\ion{O}{3}] $\lambda$4959 ratio falls strongly at the velocities and spatial positions where the emission from [\ion{O}{3}] $\lambda$4959 is most concentrated (the contour lines) because the PV diagram for H$\beta$ is broader than that of  [\ion{O}{3}] $\lambda$4959.  As a result, the ratio is too high at the velocities both closest to and farthest from the systemic velocity.  As the assumed temperature increases to 6,000\,K (second row) and 8,000\,K (third row), the ratio better approximates a constant value.  For a temperature of 8,000\,K, the ratio is very constant.  For an assumed temperature of 10,000\,K, we see an increase in the ratio at the velocities and positions where the H$\beta$ emission is most intense, implying that the broadening of the [\ion{O}{3}] $\lambda$4959 PV diagram has now gone too far.  For a temperature of 10,000\,K, the ratio of the H$\beta$/[\ion{O}{3}] $\lambda$4959 decreases at the velocities that most differ from the systemic velocity, also a result of the [\ion{O}{3}] $\lambda$4959 PV diagram being now too diffuse.   Hence, we conclude that a temperature between 6,000\,K and 8,000\,K is that which best matches the kinematics of the lines of H$\beta$ and [\ion{O}{3}] $\lambda$4959.  Comparing the last two panels in the left column of Figure \ref{fig_tkin_HbHgO3}, the pattern of the emission from H$\beta$ is also becoming visible in the third panel, for an assumed temperature of 8,000\,K.  So, the temperature that will best characterize the kinematics of both the H$\beta$ and [\ion{O}{3}] $\lambda$4959 lines will be somewhat below 8,000\,K (we return to this in \S\ref{sec_H_mass_additional}).  

The right column of Figure \ref{fig_tkin_HbHgO3} presents the analogous results for the ratio H$\gamma$/[\ion{O}{3}] $\lambda$4959.  Again, the most constant value of this ratio occurs for an assumed temperature of 8,000 K, though, again, the best temperature is likely to be somewhat lower than this.  

In Figure \ref{fig_tkin_HeIHbHg}, we consider the ratio of the \ion{He}{1} $\lambda$4922 singlet line with respect to H$\beta$ and H$\gamma$ for assumed temperatures of 3,000\,K, 6,000\,K, 8,000\,K, and 10,000\,K when broadening the \ion{He}{1} $\lambda$4922 line.  For 3,000\,K, the \ion{He}{1} $\lambda$4922/H$\beta$ and \ion{He}{1} $\lambda$4922/H$\gamma$ ratios indicate that the \ion{He}{1} $\lambda$4922 line is narrower than the H I lines, since the ratio is high where there is \ion{He}{1} $\lambda$4922 emission.  To some extent, this persists at 6,000\,K, especially considering the \ion{He}{1} $\lambda$4922/H$\gamma$ ratio.  At 8,000\,K the  two ratios are approximately constant.  At 10,000\,K, there is a bright rim in the PV diagram of the \ion{He}{1} $\lambda$4922/H$\beta$ and \ion{He}{1} $\lambda$4922/H$\gamma$ ratios and the ratios are depressed where the emission from H$\beta$ and H$\gamma$ is strong, all of which indicate that the \ion{He}{1} $\lambda$4922 line is too broad in these cases.  Hence, it appears that 8,000\,K is the temperature that best permits matching the \ion{He}{1} $\lambda$4922, H$\beta$, and H$\gamma$ PV diagrams.  In the first three rows of Figure \ref{fig_tkin_HeIHbHg}, there is a depression in the \ion{He}{1} $\lambda$4922/H$\beta$ and \ion{He}{1} $\lambda$4922/H$\gamma$ ratios for the velocities closest to the systemic velocity and spatial positions closest to the central star 
because He$^+$ is supplanted by He$^{2+}$ as the dominant ionization stage of helium in the innermost part of the nebular shell.  

We present the ratios of the PV diagrams of \ion{He}{1} $\lambda$4471 (left column) and \ion{He}{1} $\lambda$4922 (right column) with respect to that of [\ion{O}{3}] $\lambda$4959 in Figure \ref{fig_tkin_HeIO3}.  
These ratios are computed after broadening the [\ion{O}{3}] $\lambda$4959 line assuming temperatures of 3,000\,K, 6,000\,K, 8,000\,K, and 10,000\,K.  Regardless of the temperature assumed, there is a depression in the \ion{He}{1} $\lambda$4471/[\ion{O}{3}] $\lambda$4959 and \ion{He}{1} $\lambda$4922/[\ion{O}{3}] $\lambda$4959 ratios at velocities near the systemic velocity and positions near the central star.  This arises because (1) O$^{2+}$ is the dominant ionization stage of oxygen in the central part of the nebula while He$^+$ is not and (2) the smaller thermal widths better resolve this central region.  
Outside this central zone, the \ion{He}{1} $\lambda$4471/[\ion{O}{3}] $\lambda$4959 and \ion{He}{1} $\lambda$4922/[\ion{O}{3}] $\lambda$4959 ratios vary by approximately 10\% for temperatures of 6,000\,K and 8,000\,K.  

Figure \ref{fig_tkin_O2O3} presents the PV diagrams for the lines of \ion{O}{2} $\lambda$4649 and [\ion{O}{3}] $\lambda$4959 as well as the ratio of these two lines.  Both lines arise from the O$^{2+}$ ion and are single lines, so there is no correction for either line structure or thermal broadening, assuming that they arise from the same plasma component.  Hence, the ratio \ion{O}{2} $\lambda$4649/[\ion{O}{3}] $\lambda$4959 is of no use in determining the temperature of the plasma from which these lines arise.  However, since this ratio is very non-uniform, it clearly indicates that the two lines do not arise from the same plasma component.  

\begin{figure}
\includegraphics[width=\linewidth]{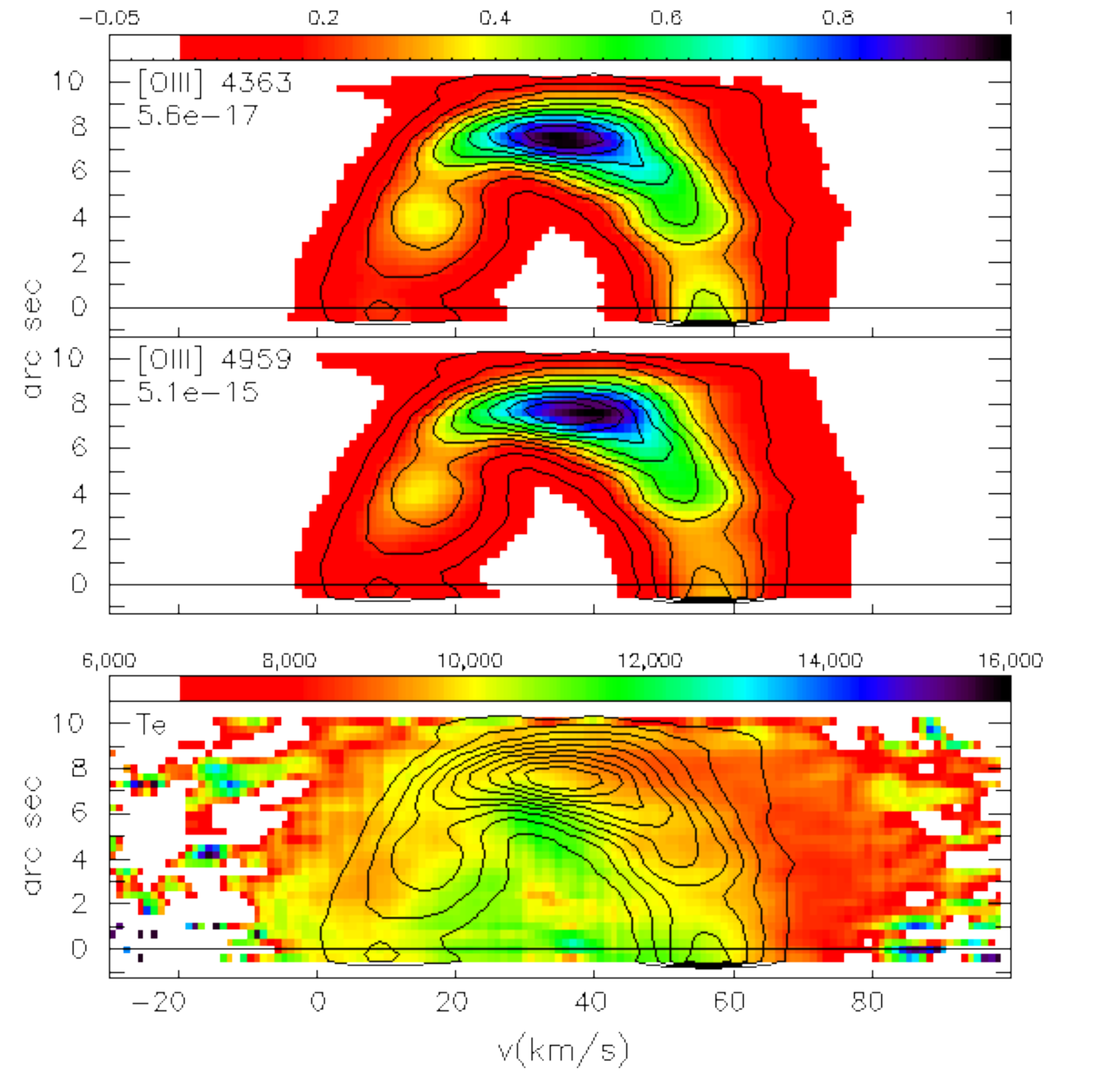}
\caption{These panels present the PV diagrams of the [\ion{O}{3}] $\lambda$4363,4959 
lines and the [\ion{O}{3}] temperature (bottom).  The contours in all panels are those of the intensity of [\ion{O}{3}] $\lambda$4363.  
The white colour in the temperature map includes positions with zero values or unphysical values.  There is a mild temperature gradient from the zone close to the central star (near the systemic velocity and in the middle of the slit) to zones farther away (the smallest and largest velocities and towards the top of the slit).  
}
\label{fig_forb_TeO3}
\end{figure}

Given that all of these PV diagrams are very similar to \ion{O}{2} $\lambda$4649, the \ion{C}{2}, \ion{N}{2}, \ion{O}{2}, and \ion{Ne}{2} lines presumably all arise in the same plasma component.  
In Figure \ref{fig_tkin_C2N2Ne2O2}, we test this hypothesis.  All of the panels in Figure \ref{fig_tkin_C2N2Ne2O2} present ratios of different \ion{C}{2}, \ion{N}{2}, \ion{O}{2}, and \ion{Ne}{2} lines, all of which yield approximately constant values.  In calculating these ratios, we do not correct for thermal broadening since the broadening involved varies from 0.7 to 1.3\,km/s over the $3,000-10,000$\,K temperature range.  The \ion{O}{2} $\lambda$4662/\ion{O}{2} $\lambda$4649 ratio, is a sanity check to test the method.  The two lines are from the same ion and the same spectral order.  In four of the six ratios, the lines are from the same wavelength interval (CD2 in three cases, CD3 in the other), so differential atmospheric refraction should not be a problem.  In two cases, \ion{N}{2} $\lambda$5680/\ion{O}{2} $\lambda$4649 and \ion{Ne}{2} $\lambda$3694/\ion{O}{2} $\lambda$4649, the ratios compare lines from wavelength intervals CD3 and CD1 with respect to CD2.  These are the cases where the ratio varies most. 
The basic result of Figure \ref{fig_tkin_C2N2Ne2O2} is that ratios of the lines of \ion{C}{2}, \ion{N}{2}, \ion{O}{2}, and \ion{Ne}{2} produce constant values, as expected if these lines arise in the same plasma component.  

Figures \ref{fig_tkin_HbHgO3}-\ref{fig_tkin_HeIO3} argue that the kinematics of \ion{H}{1}, \ion{He}{1}, and [\ion{O}{3}] lines are compatible with these emission lines arising from the same plasma component whose temperature is $\sim 8,000$\,K.  
On the other hand, Figure \ref{fig_tkin_O2O3} indicates that the \ion{O}{2} $\lambda$4649 does not arise from the same plasma component that gives rise to the \ion{H}{1}, \ion{He}{1}, and [\ion{O}{3}] lines.  Figure \ref{fig_tkin_C2N2Ne2O2} indicates that this line, as well as other lines of \ion{C}{2}, \ion{N}{2}, \ion{O}{2}, and \ion{Ne}{2}, have kinematics that are compatible with an origin in a common plasma component.  So, by extension, all of these lines arise from a different plasma component from that producing the \ion{H}{1}, \ion{He}{1}, and [\ion{O}{3}] lines.  

\subsubsection{Forbidden lines}\label{sec_physcond_forb}

\begin{figure*}
\begin{center}\includegraphics[width=0.86\linewidth]{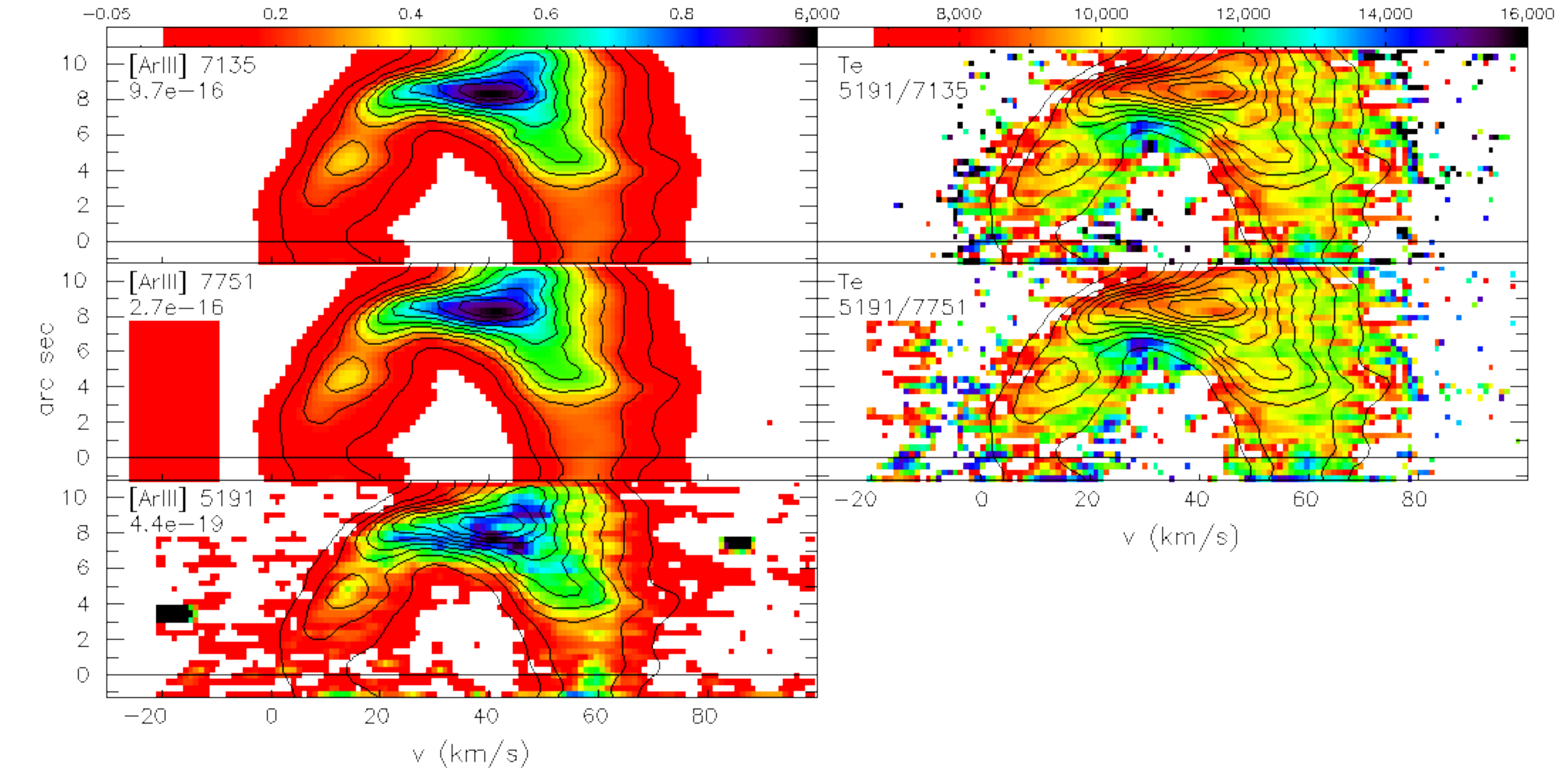}\end{center}
\caption{These panels present the PV diagrams of the lines of [\ion{Ar}{3}] $\lambda$7135,7751,5191 
the [Ar III] temperature from the [\ion{Ar}{3}] $\lambda\lambda$5191/7135 and 
5191/7751 line ratios in the right column.  The contours are of the intensity of the [\ion{Ar}{3}] $\lambda$7135 line.  The color scale in the right column is the same as used for the [\ion{O}{3}] temperature in Figure \ref{fig_forb_TeO3}.  
}
\label{fig_forb_TeAr3}
\end{figure*}

\begin{figure}
\includegraphics[width=\linewidth]{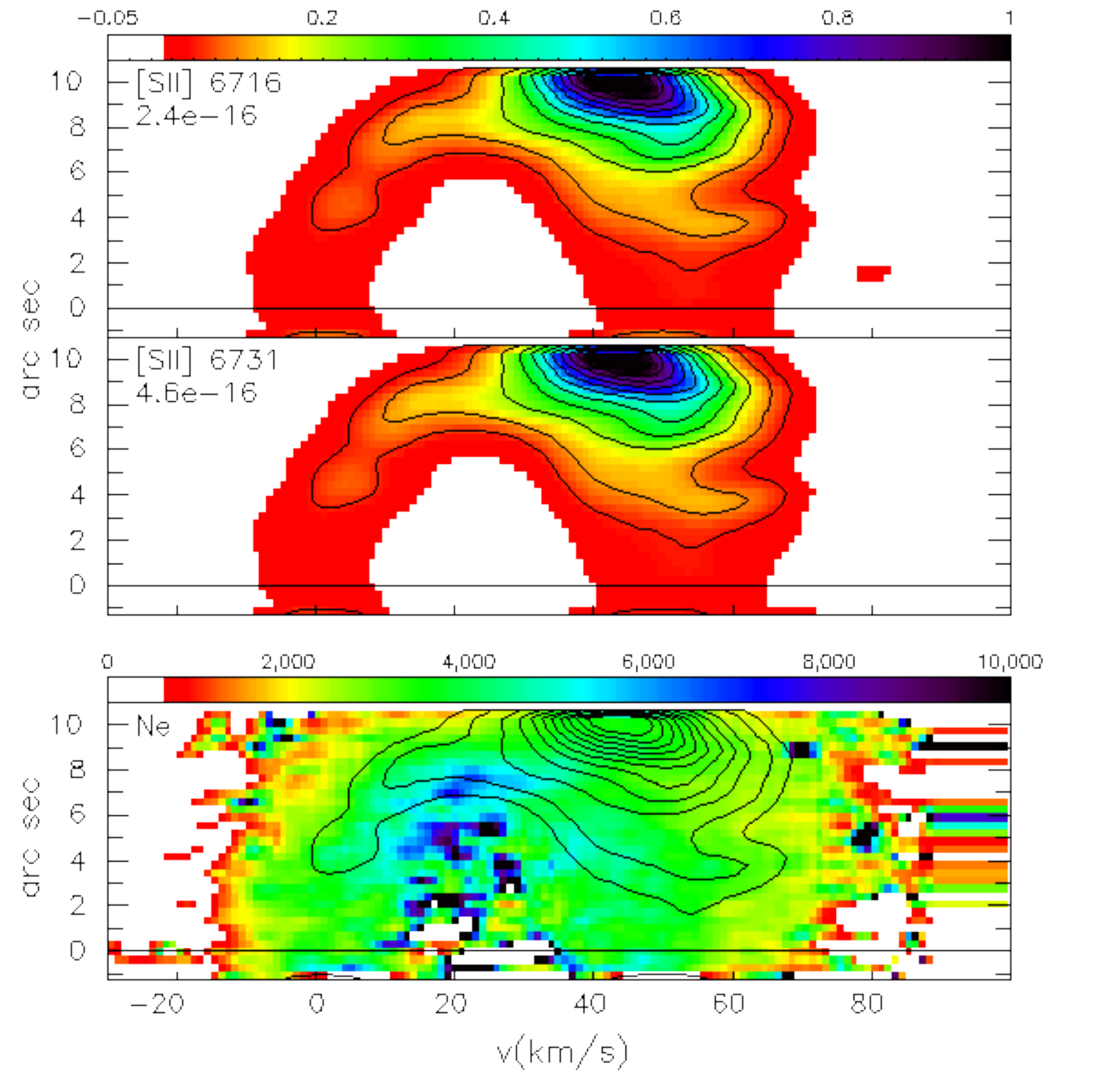}
\caption{These panels present the PV diagrams of the [\ion{S}{2}] $\lambda\lambda$6716,6731 lines and of the [\ion{S}{2}] density.  In all panels the contours are of the [\ion{S}{2}] $\lambda$6716 line intensity.  The white color in the PV diagram of the [\ion{S}{2}] density represent non-physical values or where there is no [\ion{S}{2}] emission.  The color scale for the density spans the range $0-10,000$\,cm$^{-3}$.  
}
\label{fig_forb_NeS2_PV}
\end{figure}

\begin{figure}
\includegraphics[width=\linewidth]{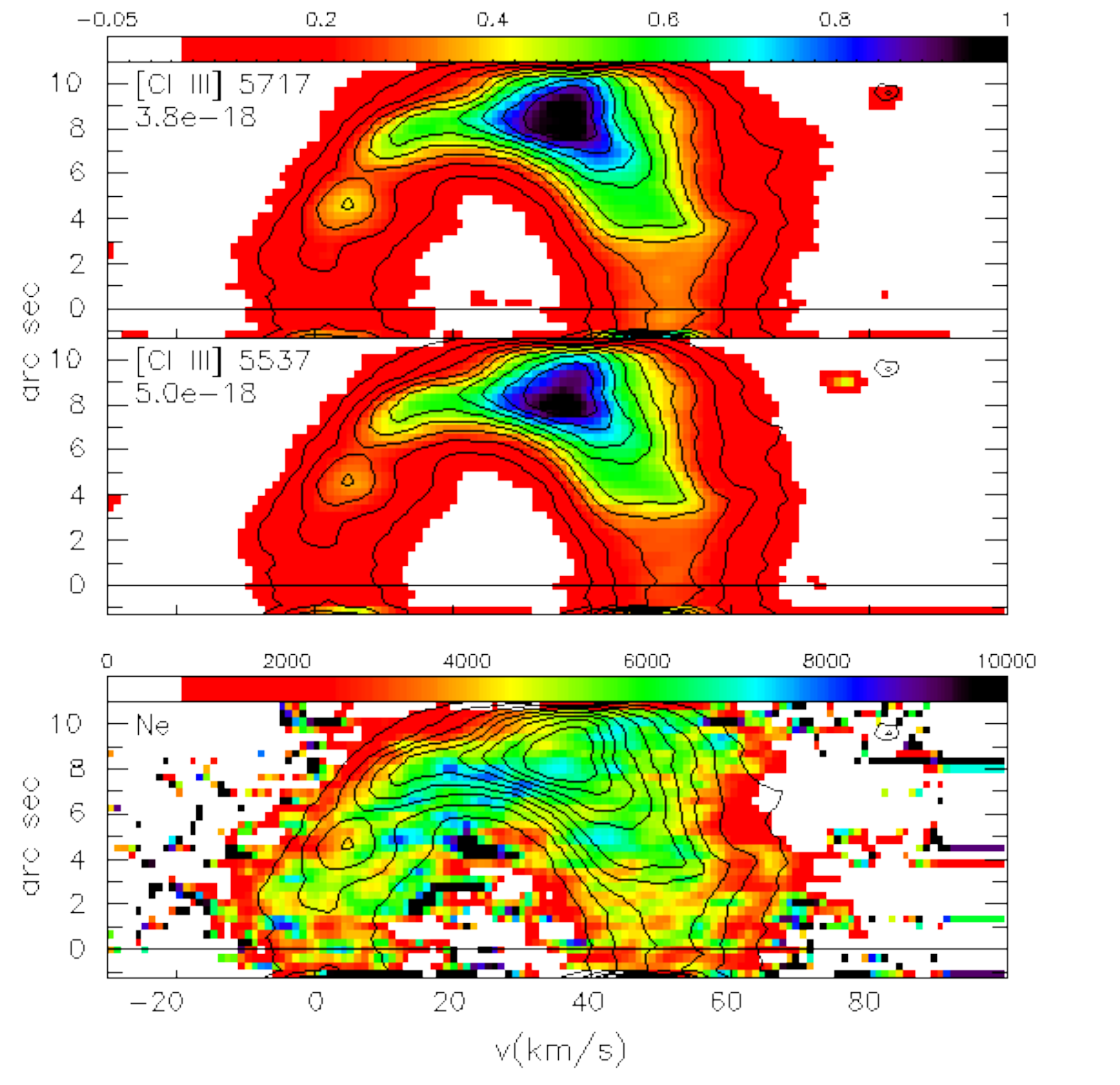}
\caption{These panels present the PV diagrams of the [\ion{Cl}{3}] $\lambda\lambda$5717,5537 lines and of the [\ion{Cl}{3}] density.  In all panels the contours are of the [\ion{Cl}{3}] $\lambda$5717 line intensity.  The white color in the PV diagram of the [\ion{Cl}{3}] density represent non-physical values or where there is no [\ion{Cl}{3}] emission.  The color scale for the density spans the range $0-10,000$\,cm$^{-3}$.  
}
\label{fig_forb_NeCl3_PV}
\end{figure}

\begin{figure}
\includegraphics[width=\linewidth]{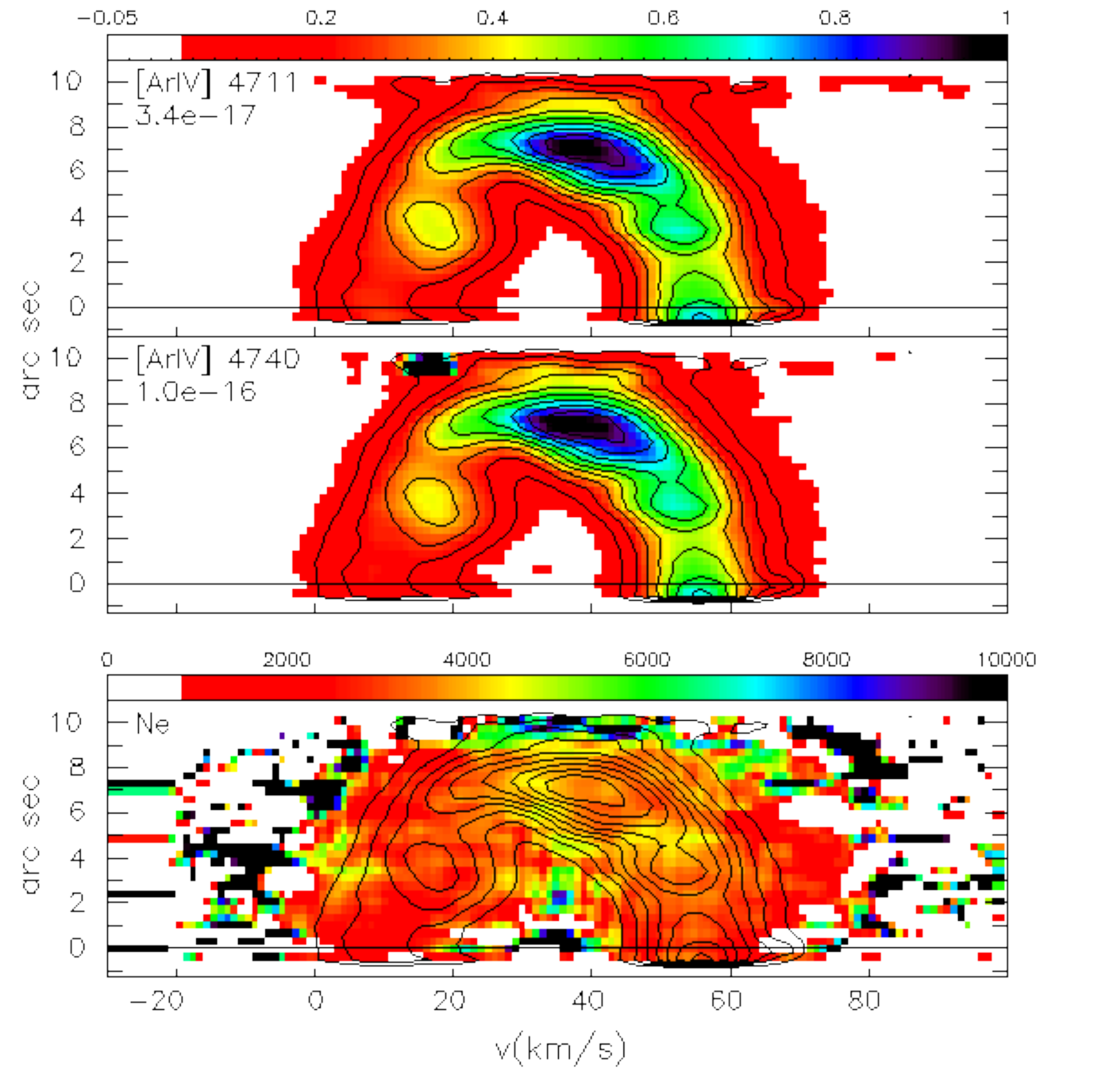}
\caption{These panels present the PV diagrams for the lines of [\ion{Ar}{4}] $\lambda\lambda$4711,4740 and of the [\ion{Ar}{4}] density.  In all panels, the contours are of the intensity of the [\ion{Ar}{4}] $\lambda$4711 line.  The white color in the PV diagram of the [\ion{Ar}{4}] density represent non-physical values or where there is no [\ion{Ar}{4}] emission.  The color scale for the density spans the range $0-10,000$\,cm$^{-3}$.  
}
\label{fig_forb_NeAr4_PV}
\end{figure}

We construct PV diagrams 
for the electron temperature and density diagnostics from collisionally-excited lines.  
Since we resolve both the spatial structure along the slit and the velocity structure along the line of sight, regions expected to have higher temperatures, such as those close to the central star, have different PV coordinates from other volumes of the nebula.  We compute the physical conditions using PyNeb \citep{luridianaetal2015} to convert an intensity ratio into an electron temperature or density.  

We consider the electron temperatures derived from the [\ion{O}{3}] and [\ion{Ar}{3}] lines as well as the electron densities derived from the [\ion{S}{2}], [\ion{Cl}{3}] and [\ion{Ar}{4}] lines.  As we shall see (\S\ref{sec_contamination}), the [\ion{N}{2}] and [\ion{O}{2}] forbidden lines have an important excitation component due to recombination, in addition to the usual collisional excitation process, so we defer their discussion for later.  
Finally, due to the radial velocity of NGC 6153 at the time of the observations, the [\ion{S}{3}] $\lambda\lambda$9069,9531 lines coincided with strong telluric absorption lines, so they cannot be used to determine a reliable electron temperature \citep{stevenson1994}.  

Figure \ref{fig_forb_TeO3} presents the PV diagrams of the [\ion{O}{3}] $\lambda\lambda$4363,4959 lines as well as the PV diagrams of the [\ion{O}{3}] temperature.  
In computing the [\ion{O}{3}] temperature, we adopt 
an electron density of 4,000\,cm$^{-3}$ (atomic data: Table \ref{tab_atomic_data}).  The [\ion{O}{3}] temperature map presents three regimes.  In the main shell, there is a slight gradient from the outermost parts of the nebula (largest distance from the central star, largest velocities w.r.t. the systemic velocity) to the innermost part (spatial positions closer to the central star, velocities near the systemic velocity), that varies only slightly $9,000-10,000$\,K.  Inside the main shell, the plasma is hotter, $10,000-11,000$\,K.  In the diffuse emission beyond the main shell at the largest recession velocities, the temperature is lowest $8,000-9,000$\,K.  

Figure \ref{fig_forb_TeAr3} presents the PV diagrams of the [\ion{Ar}{3}] $\lambda\lambda$5191, 7135, 7751 lines as well as the PV diagrams of the temperature based upon the ratios of [\ion{Ar}{3}] $\lambda\lambda$5191/7135 and [\ion{Ar}{3}] $\lambda\lambda$5191/7751.  In this case, lines from different wavelength intervals must be used (CD3b and CD4b).  Again, 
we adopt an electron density of 4,000\,cm$^{-3}$ (atomic data: Table \ref{tab_atomic_data}).  The [\ion{Ar}{3}] temperature map is limited to the main shell and the filament on its receding side.  The [\ion{Ar}{3}] temperature map is more uniform than the [\ion{O}{3}] temperature, though with larger uncertainties, and slightly lower, by $\sim 500$\,K.  

\begin{figure*}
\includegraphics[width=0.49\linewidth]{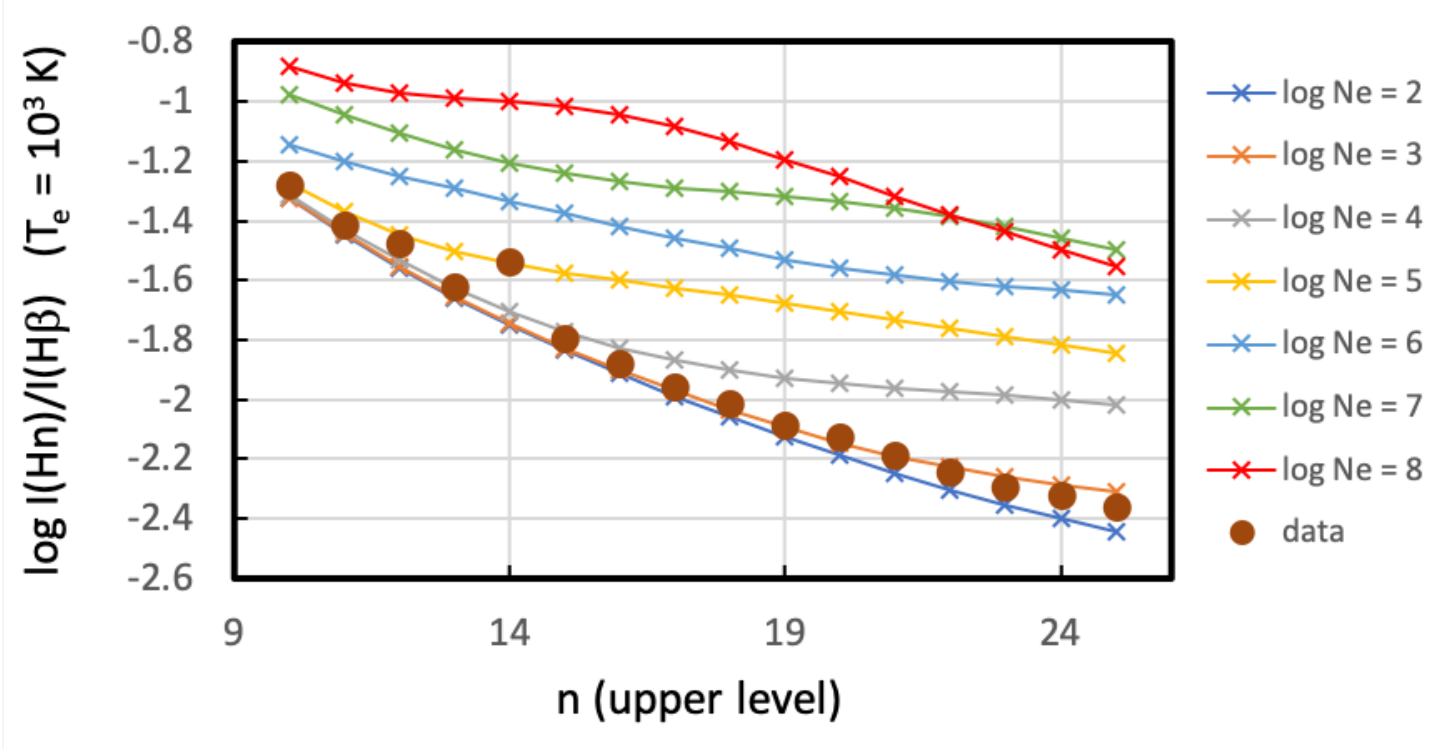}
\includegraphics[width=0.49\linewidth]{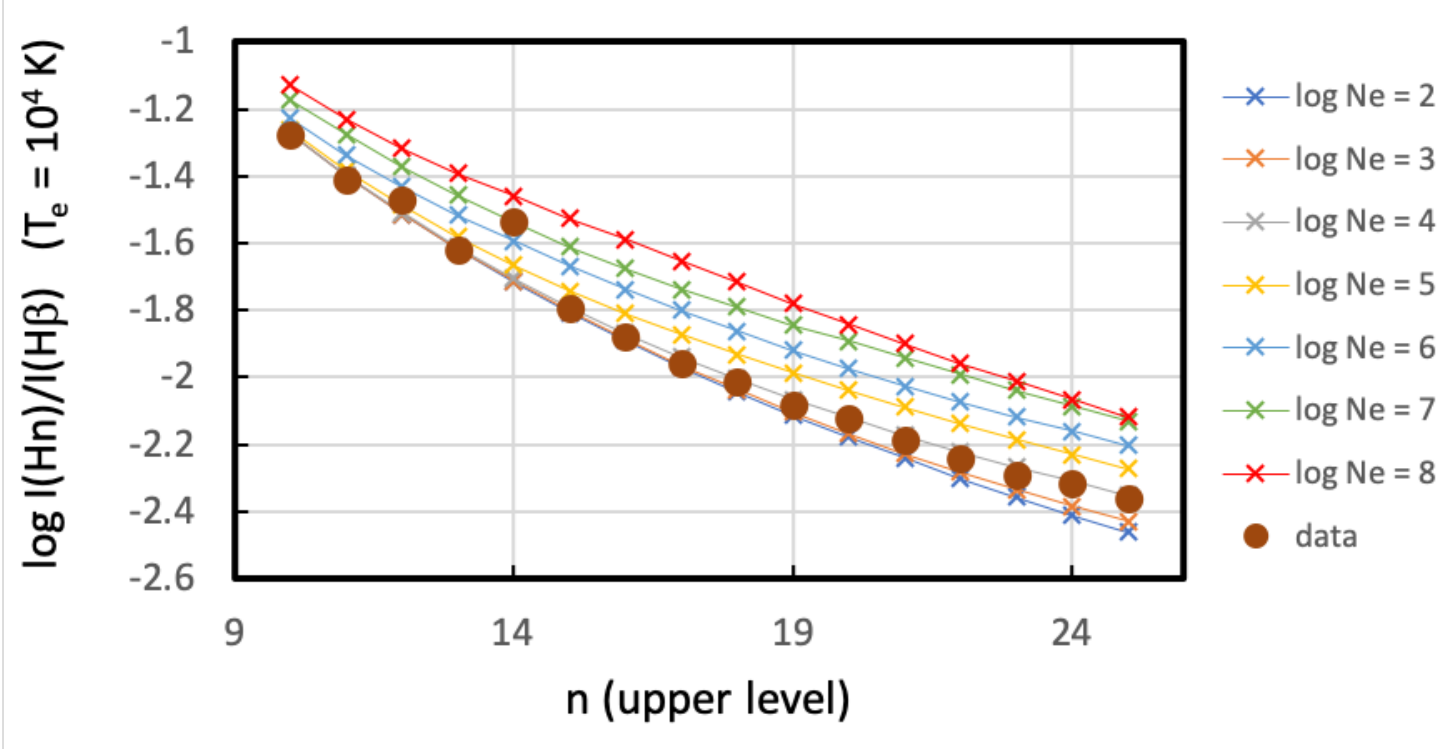}
\caption{These panels present the intensities of reddening-corrected, high-$n$ Balmer lines with respect to the intensity of H$\beta$ as a function of the upper level for assumed electron temperatures of 1,000\,K (left) and 10,000\,K (right).  The implied densities are between $<1,000$\,cm$^{-3}$ and $\sim 10,000$\,cm$^{-3}$.  
}
\label{fig_perm_NeH1_Balmer}
\end{figure*}

\begin{figure*}
\includegraphics[width=0.49\linewidth]{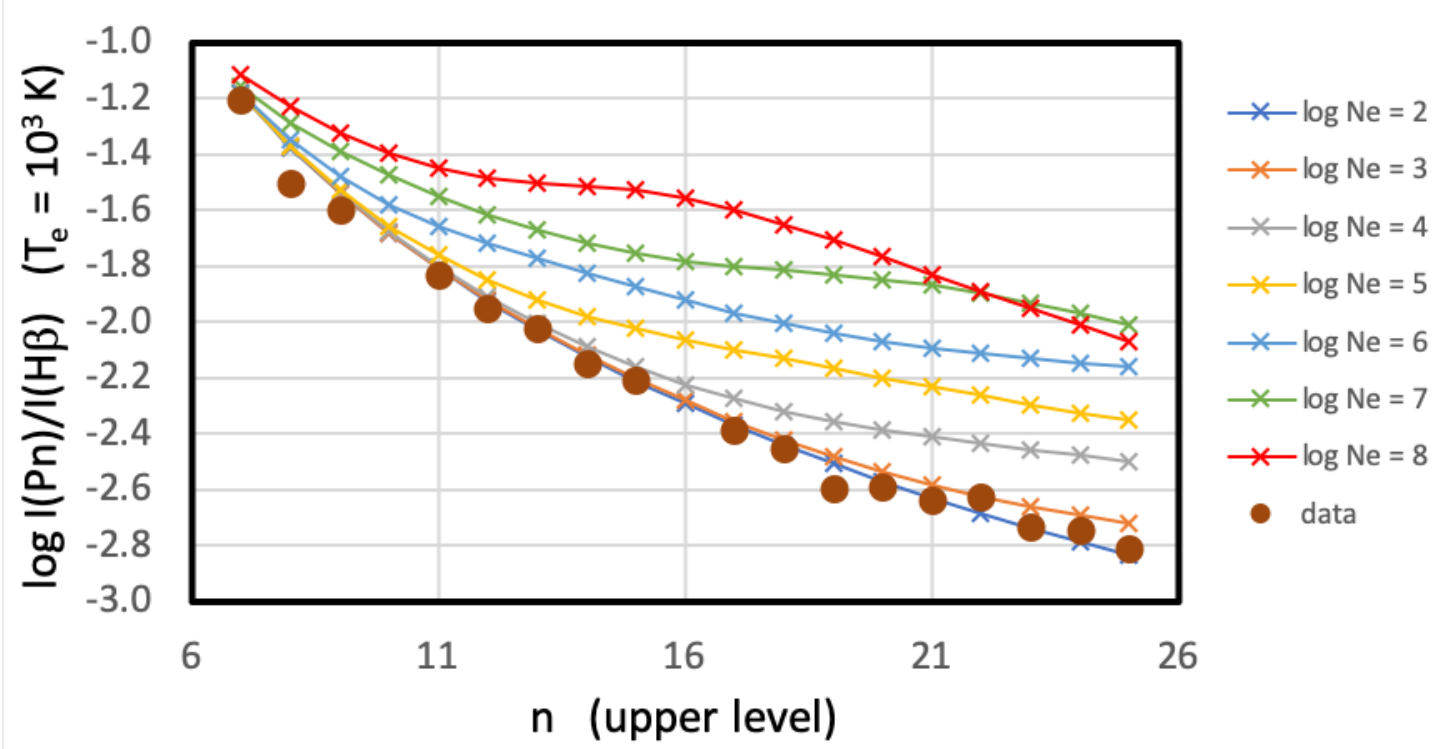}
\includegraphics[width=0.49\linewidth]{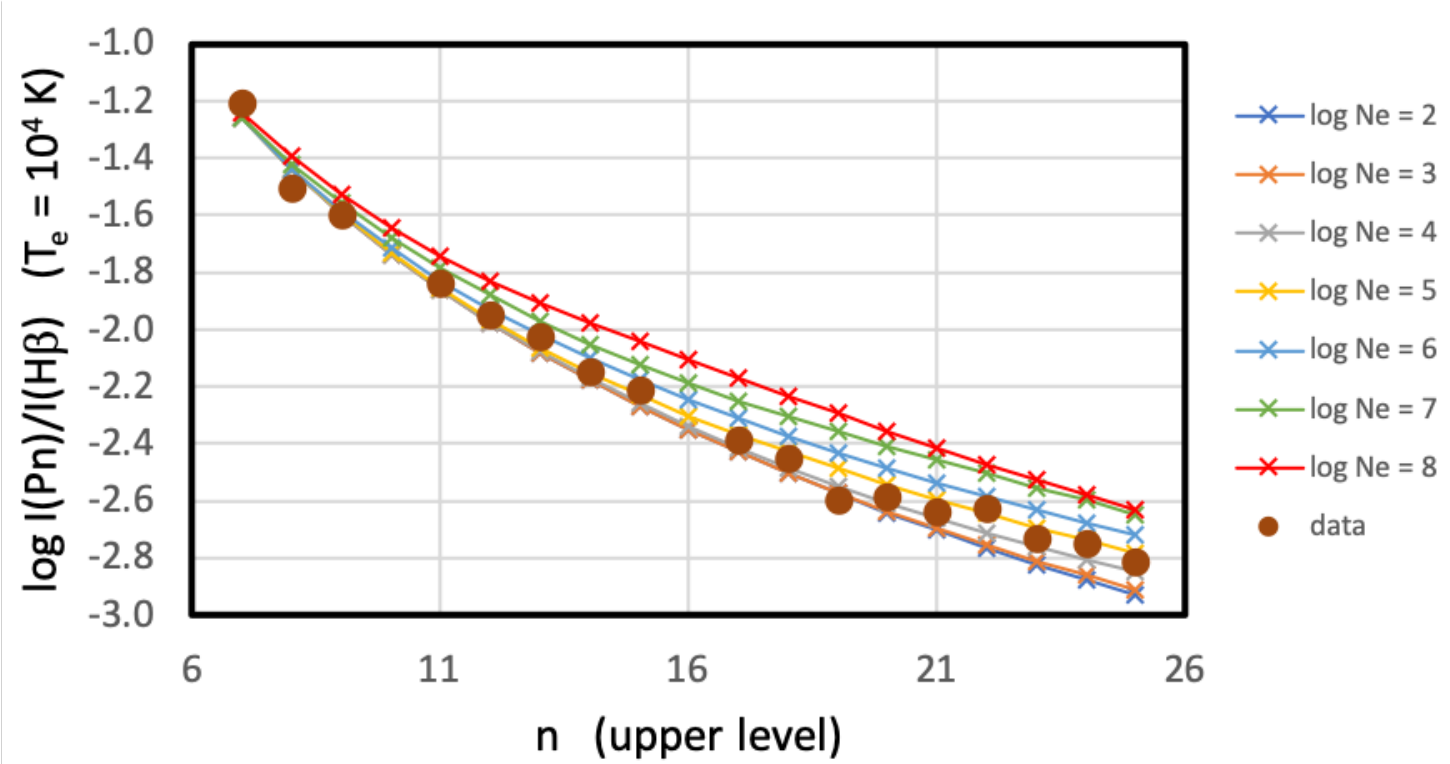}
\caption{These panels present the intensities of reddening-corrected, high-$n$ Paschen lines with respect to the intensity of H$\beta$ as a function of the upper level for assumed electron temperatures of 1,000\,K (left) and 10,000\,K (right).  The implied densities are of $\sim 1,000$\,cm$^{-3}$ at 1,000\,K and perhaps $\sim 10,000$\,cm$^{-3}$ at 10,000\,K.  
}
\label{fig_perm_NeH1_Paschen}
\end{figure*}

Turning to the electron density indicators, Figure \ref{fig_forb_NeS2_PV} presents the PV diagrams of the [\ion{S}{2}] $\lambda\lambda$6716,6731 lines and of the [\ion{S}{2}] electron density.  To compute the electron density, we assume an electron temperature of 10,000\,K (atomic data: Table \ref{tab_atomic_data}).  There is a clear gradient in the electron density implied by the [\ion{S}{2}] $\lambda\lambda$6716/6731 line ratio, with lower densities towards the periphery of the object, at either the velocities that differ most from the systemic velocity or at the greatest distance from the central star.  The variation in electron density appears to exceed a factor of 2, from approximately 2,000\,cm$^{-3}$ to $4,000-5,000$\,cm$^{-3}$.  The filament on the receding side of the main shell that is the brightest feature of the [\ion{S}{2}] $\lambda\lambda$6716,6731 PV diagrams is of low density, so it is presumably bright due to a change in the ionization stage for sulfur.  This feature is not evident in the H$\alpha$ line (Figure \ref{fig_pv_diagram}), so it is presumably of low mass.  
 
Figure \ref{fig_forb_NeCl3_PV} presents the PV diagrams of the [\ion{Cl}{3}] $\lambda\lambda$5517,5537 lines and 
the [\ion{Cl}{3}] density.  Again, 
we assume an electron temperature of 10,000\,K (atomic data: Table \ref{tab_atomic_data}).  The PV diagram of the [\ion{Cl}{3}] density is quite uniform over the entire area that includes [\ion{Cl}{3}] emission.  There is, however, a ``red border" around the outer edge of the area with [\ion{Cl}{3}] emission, as if the density drops at the outer edge of the Cl$^{2+}$ zone.  By and large, this is congruent with the variation in the density as traced by the [\ion{S}{2}] lines, since the [\ion{S}{2}] lines trace plasma with a lower degree of ionization.  The higher [\ion{Cl}{3}] density throughout most of the Cl$^{2+}$ zone agrees with the [\ion{S}{2}] electron density at the velocities and spatial coordinates in common.  However, one discrepancy is the higher density implied for the filament on the receding side of the main shell.  
 
Figure \ref{fig_forb_NeAr4_PV} presents the PV diagrams of the [\ion{Ar}{4}] $\lambda\lambda$4711,4740 lines and 
the [\ion{Ar}{4}] electron density (assuming 
an electron temperature of 10,000\,K; atomic data: Table \ref{tab_atomic_data}).  The PV diagram of the [\ion{Ar}{4}] electron density is very uniform.  However, the electron density implied by the [\ion{Ar}{4}] lines is substantially lower than that found from the [\ion{Cl}{3}] or [\ion{S}{2}] lines, even for the velocities and spatial coordinates in common, more like the low densities found at the periphery from [\ion{S}{2}].  Perhaps, a different atomic data set could resolve the issue.  It is unlikely due uncertainty in the line intensity ratio, since it implies a 20\% change, which we think unlikely for such strong lines.   

Based upon the [\ion{S}{2}] and [\ion{Cl}{3}] densities, the density throughout most of the main shell of NGC 6153 appears to be approximately $4,000-5,000$\,cm$^{-3}$ with a lower density region at the extreme velocities and distance from the central star (Figures \ref{fig_forb_NeS2_PV} and \ref{fig_forb_NeCl3_PV}).  The electron density implied by the [\ion{Ar}{4}] lines is about a factor of 2 lower throughout the nebular volume in common with the [\ion{S}{2}] and [\ion{Cl}{3}] emission.

\subsubsection{Permitted lines}\label{sec_physcond_perm}

\begin{figure*}
\includegraphics[width=0.33\linewidth]{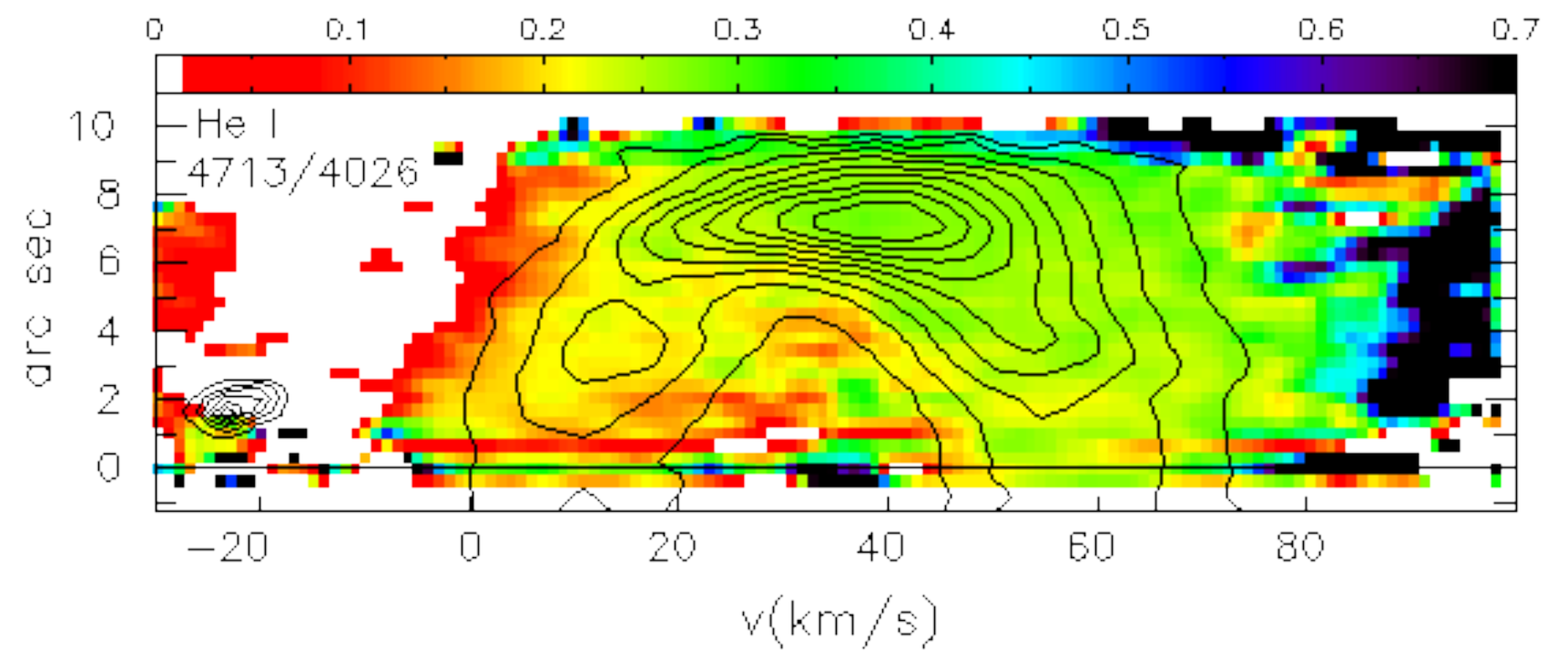}
\includegraphics[width=0.33\linewidth]{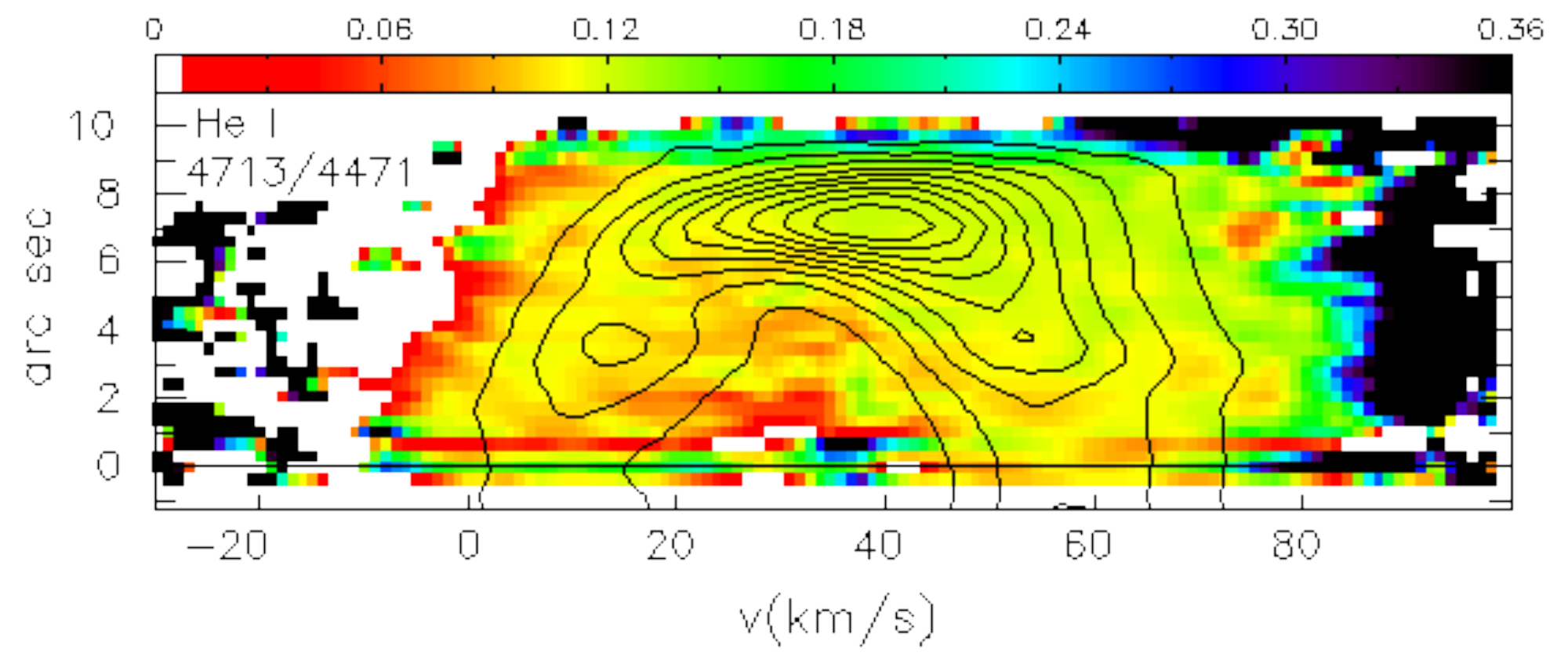}
\includegraphics[width=0.33\linewidth]{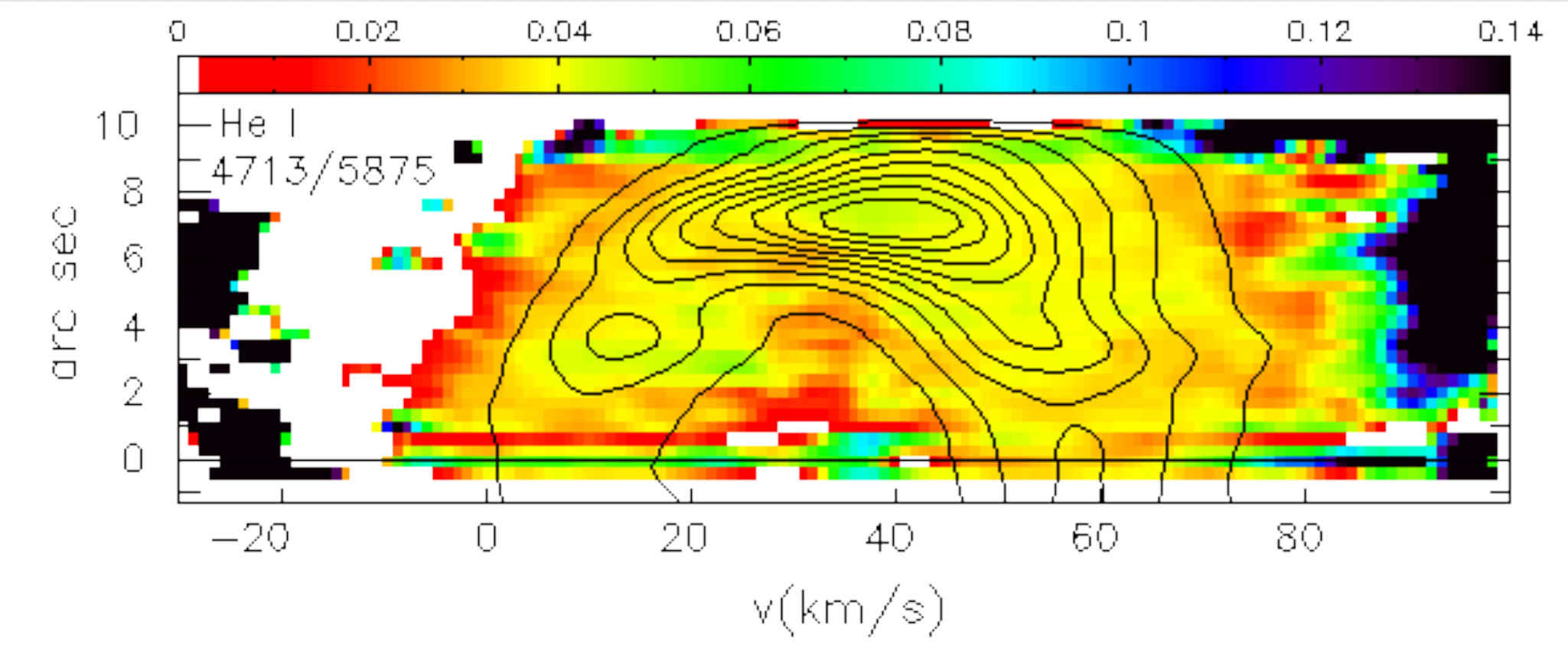}
\includegraphics[width=0.33\linewidth]{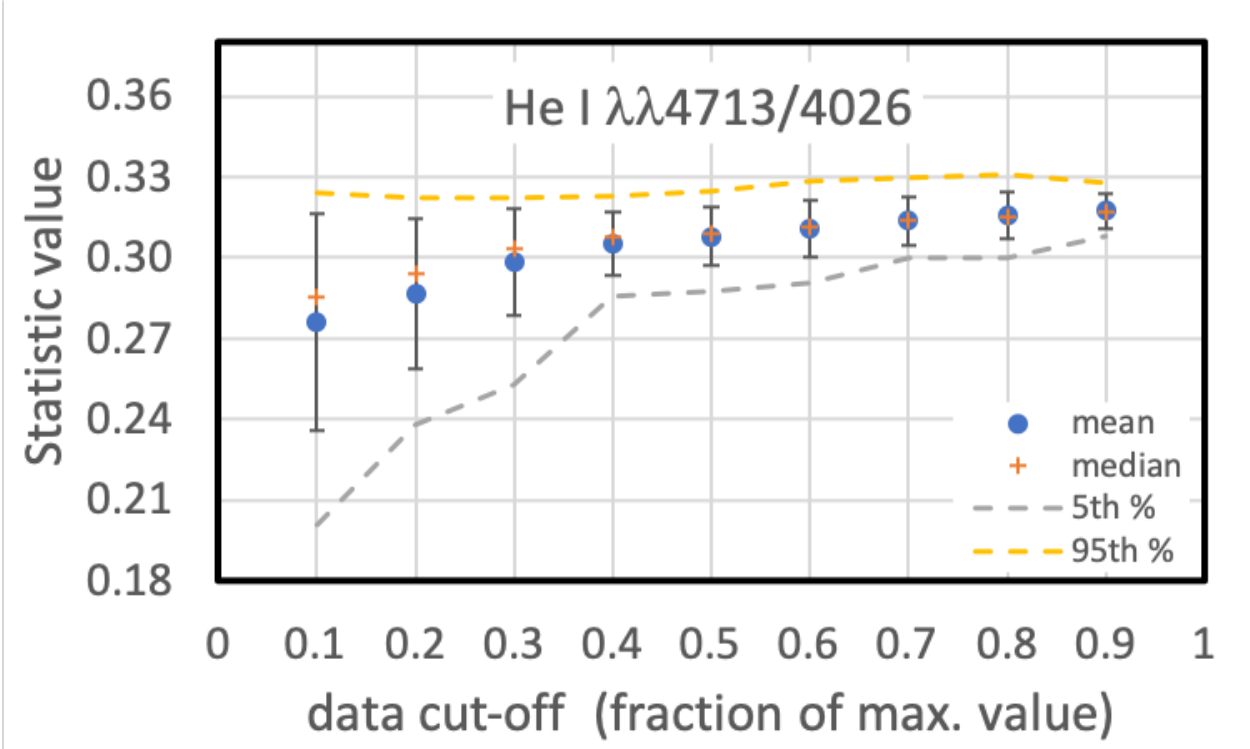}
\includegraphics[width=0.33\linewidth]{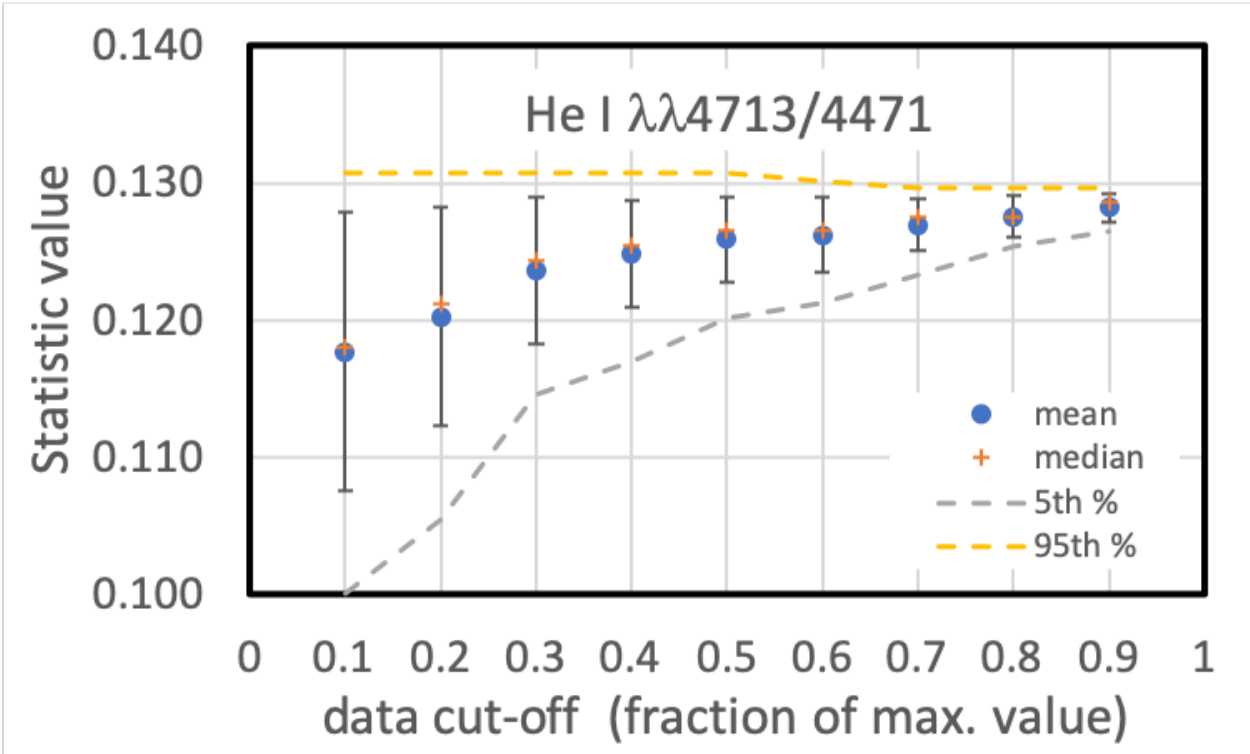}
\includegraphics[width=0.33\linewidth]{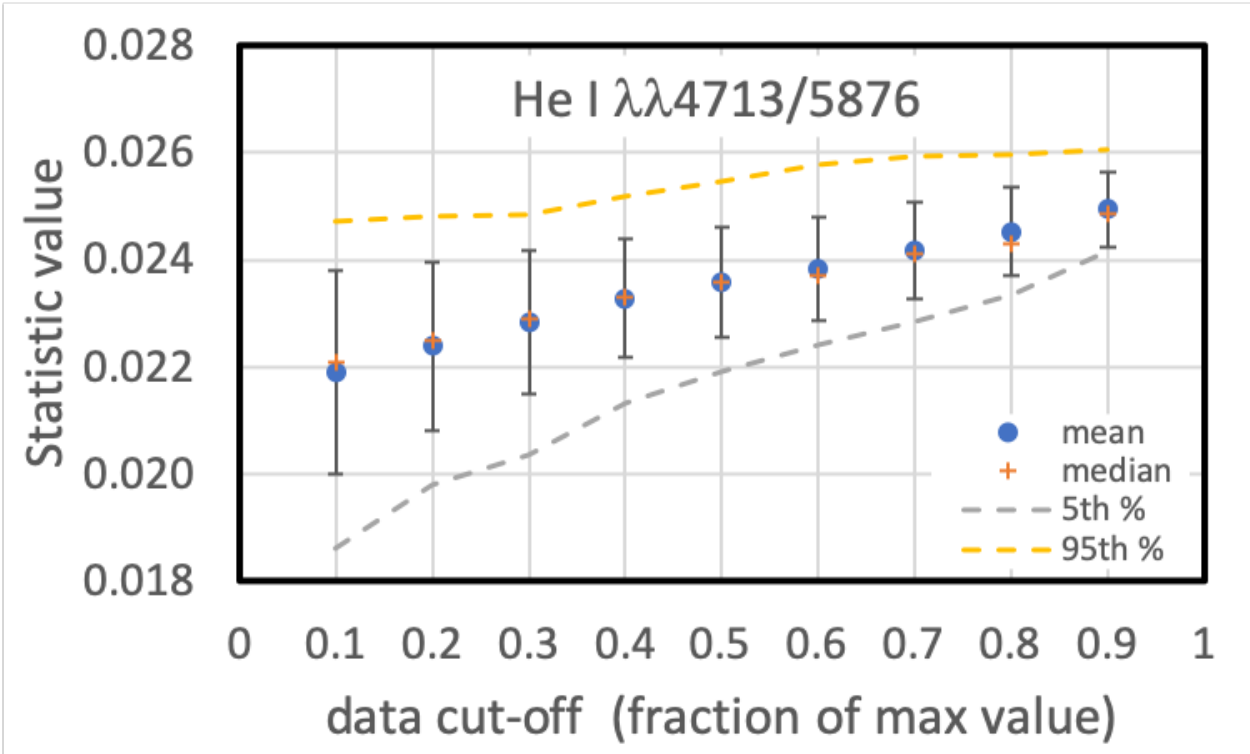}
\includegraphics[width=0.33\linewidth]{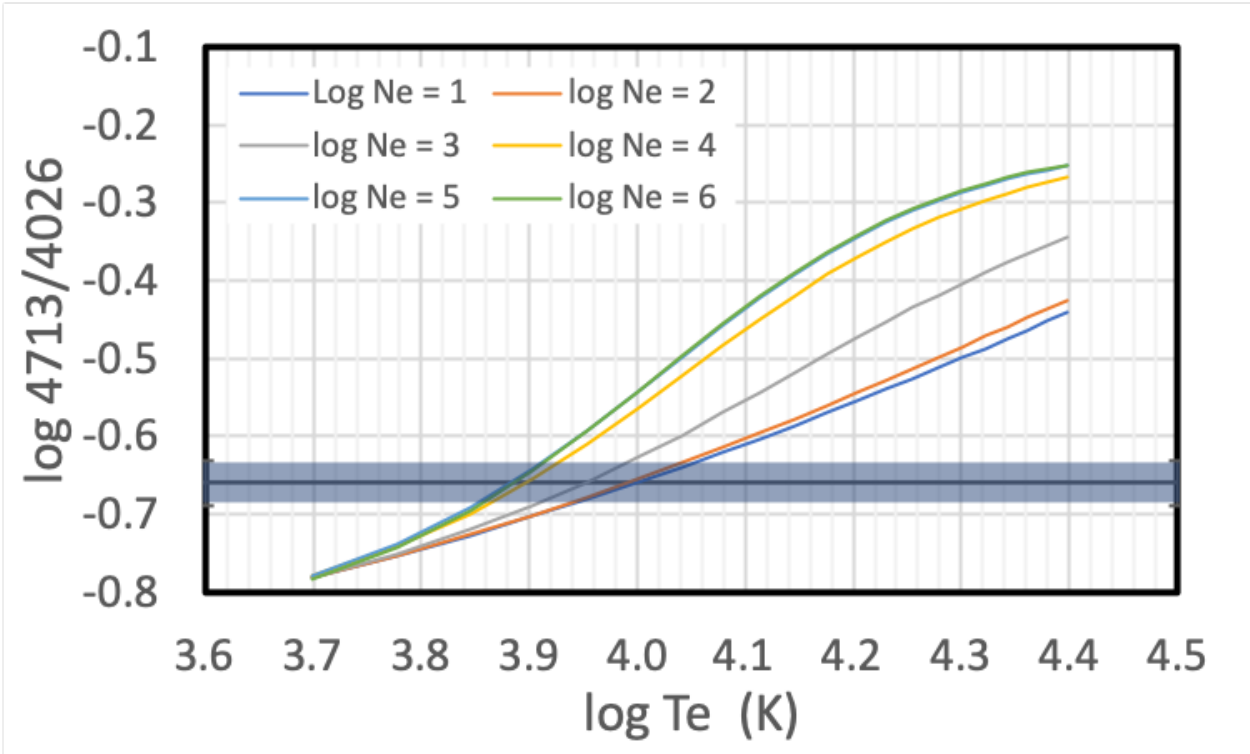}
\includegraphics[width=0.33\linewidth]{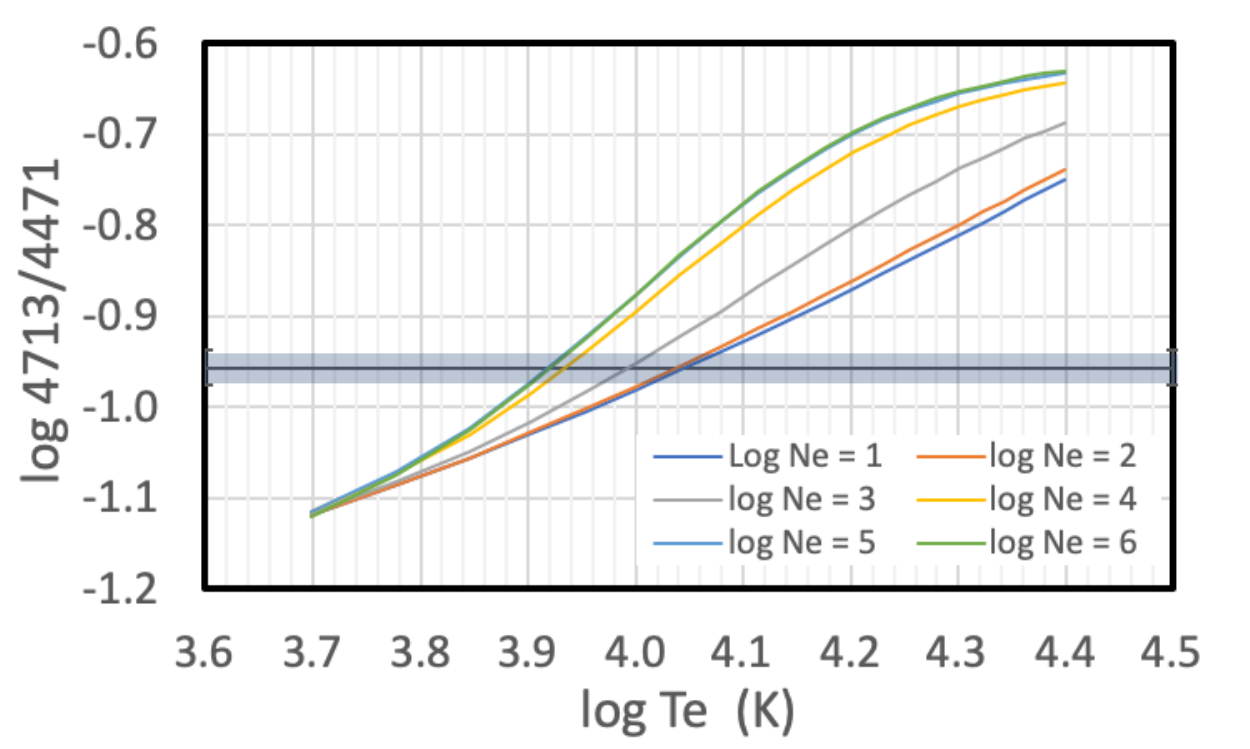}
\includegraphics[width=0.33\linewidth]{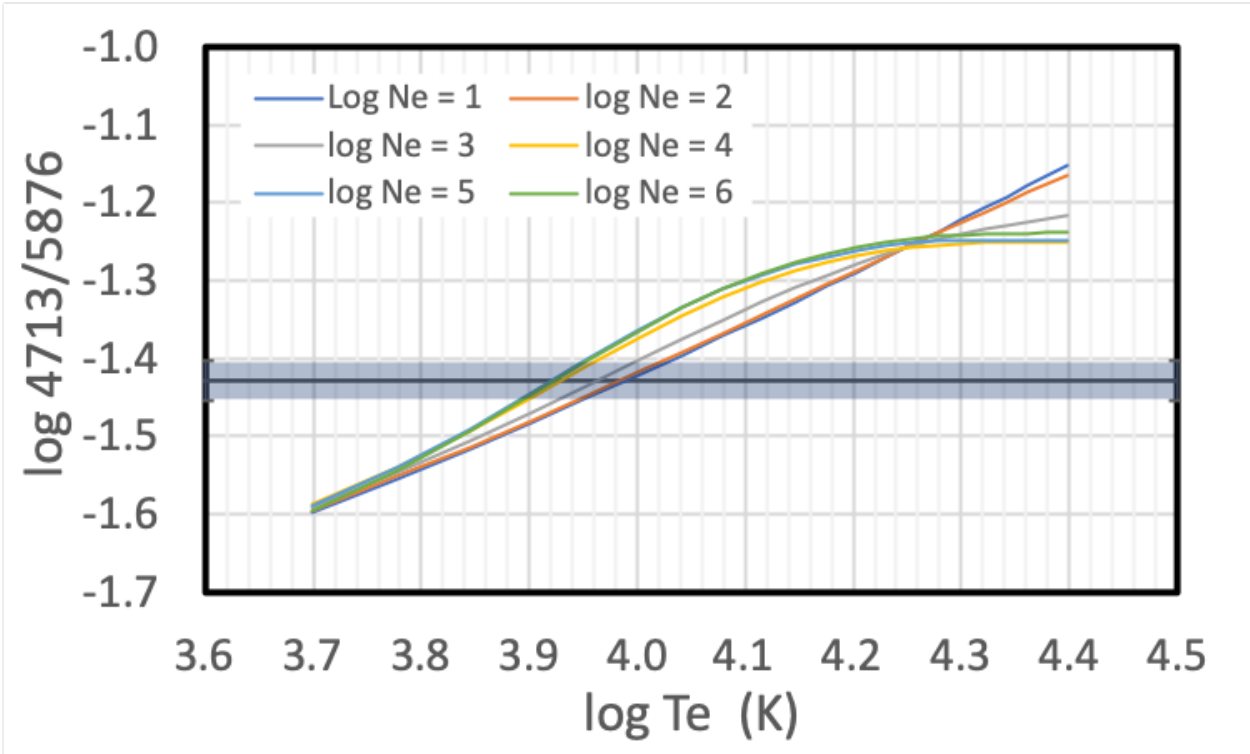}
\caption{Top row:  These panels present the PV diagrams of the line ratios \ion{He}{1} $\lambda\lambda$4713/4026 (left), \ion{He}{1} $\lambda\lambda$4713/4471 (middle), and \ion{He}{1} $\lambda\lambda$4713/5876 (right).  The contours show the intensity of the \ion{He}{1} $\lambda\lambda$4026, 4471, and 5876 lines (the denominator of the ratio).  Middle row:  These panels present various statistics for the ratios of the PV diagrams of \ion{He}{1} $\lambda\lambda$4713/4026 (left), \ion{He}{1} $\lambda\lambda$4713/4471 (middle), and \ion{He}{1} $\lambda\lambda$4713/5876 (right) as a function of the line intensities in the original PV diagrams of the \ion{He}{1} $\lambda\lambda$4026, 4471, and 5876 lines (denominator).  In the top and middle rows, the line ratios have been corrected for the flux scale factors (Table \ref{tab_flux_scale_factors}), but not for reddening.  Bottom row:  These panels compare the ratio of the line intensities of \ion{He}{1} $\lambda\lambda$4713/4026 (left), \ion{He}{1} $\lambda\lambda$4713/4471 (middle), and \ion{He}{1} $\lambda\lambda$4713/5876 (right), now corrected for reddening (Table \ref{tab_EBV_values}), with the theoretical line intensities from \citet{porteretal2013}.  We adopt the mean line ratio for all pixels exceeding 30\% of the maximum line intensity of the denominator and the uncertainty given by the standard deviation.  The \ion{He}{1} lines are compatible with temperatures $\log T_e = 3.91-4.07$ when considering densities up to $10^4\,\mathrm{cm}^{-3}$ (see Figure \ref{fig_perm_HeI_redd_dens}).  
}
\label{fig_perm_HeI_temperature}
\end{figure*}

The intensities of high-$n$ Balmer and Paschen lines may be used to infer the electron density, though this requires adopting an electron temperature.  We use the theoretical line emissivities for Case B (atomic data: Table \ref{tab_atomic_data}) and the line intensities measured from the 1-D spectrum (Table \ref{tab_HI_int_reddening}).  

Figure \ref{fig_perm_NeH1_Balmer} presents the intensities of the high-$n$ Balmer lines relative to the intensity of H$\beta$.  We consider temperatures of 1,000\,K (left panel) and 10,000\,K (right panel) as illustrative of the range of relevant values.  The H14 line (\ion{H}{1} $\lambda$3721) has an anomalously high intensity, presumably due to contamination ([\ion{S}{3}] $\lambda$3721).  As the temperature increases, so does the implied electron density, but the implied range of electron densities is modest, $<1,000$ to $<10,000$\,cm$^{-3}$ for this temperature range.  

Figure \ref{fig_perm_NeH1_Paschen} presents the intensities of the high-$n$ Paschen lines relative to the intensity of H$\beta$ (Table \ref{tab_HI_int_reddening}).  We consider the same representative electron temperatures.  In this case, an electron density of 1,000\,cm$^{-3}$ is found at a temperature of 1,000\,K, but the implied electron density is higher at 10,000\,K, with 10,000\,cm$^{-3}$ being perhaps the most representative value, except for the highest Paschen lines.  Given the difficulty of measuring the intensities of the highest Paschen lines accurately, due to both line crowding and telluric absorption, we shall give the electron density derived from them lower weight.  Generally, the \ion{H}{1} lines appear to indicate densities of $\lesssim 10,000\,\mathrm{cm}^{-3}$.  

\begin{figure*}
\begin{center}
\includegraphics[width=0.33\linewidth]{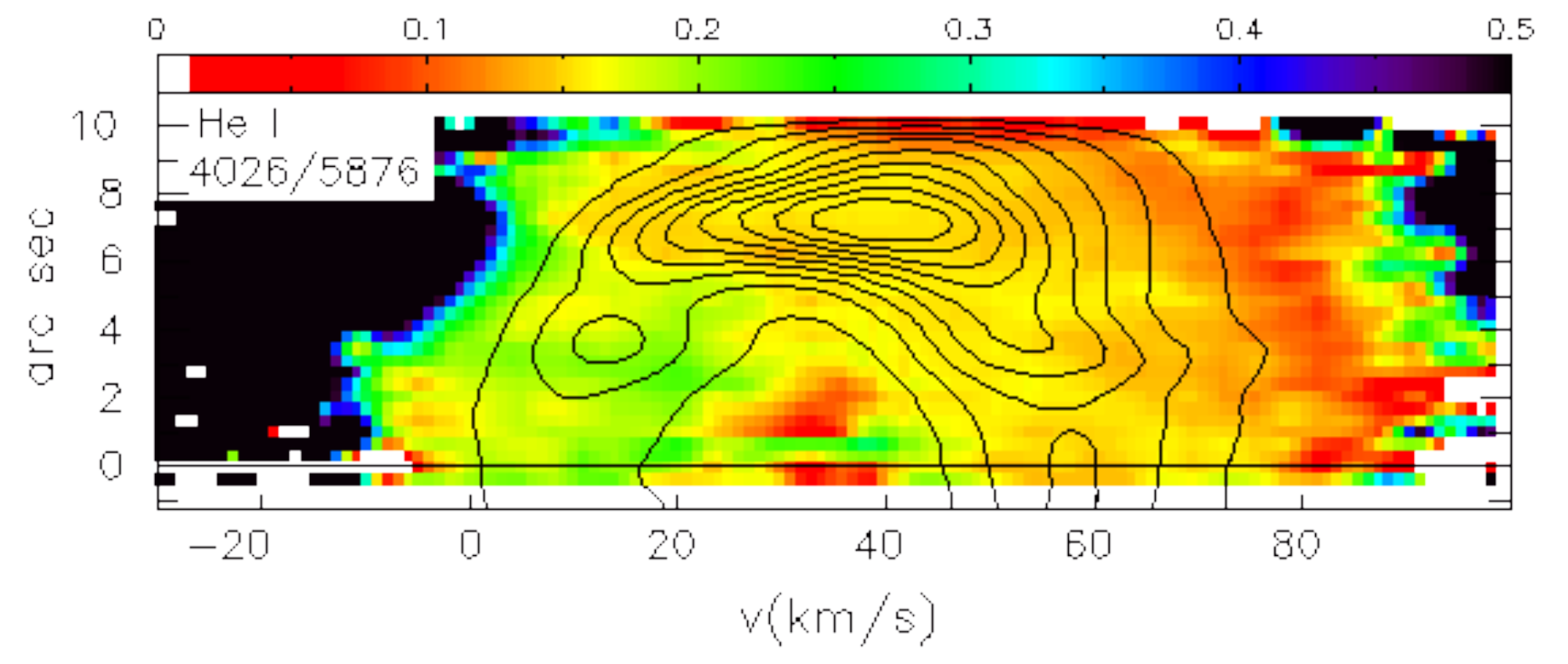}
\includegraphics[width=0.33\linewidth]{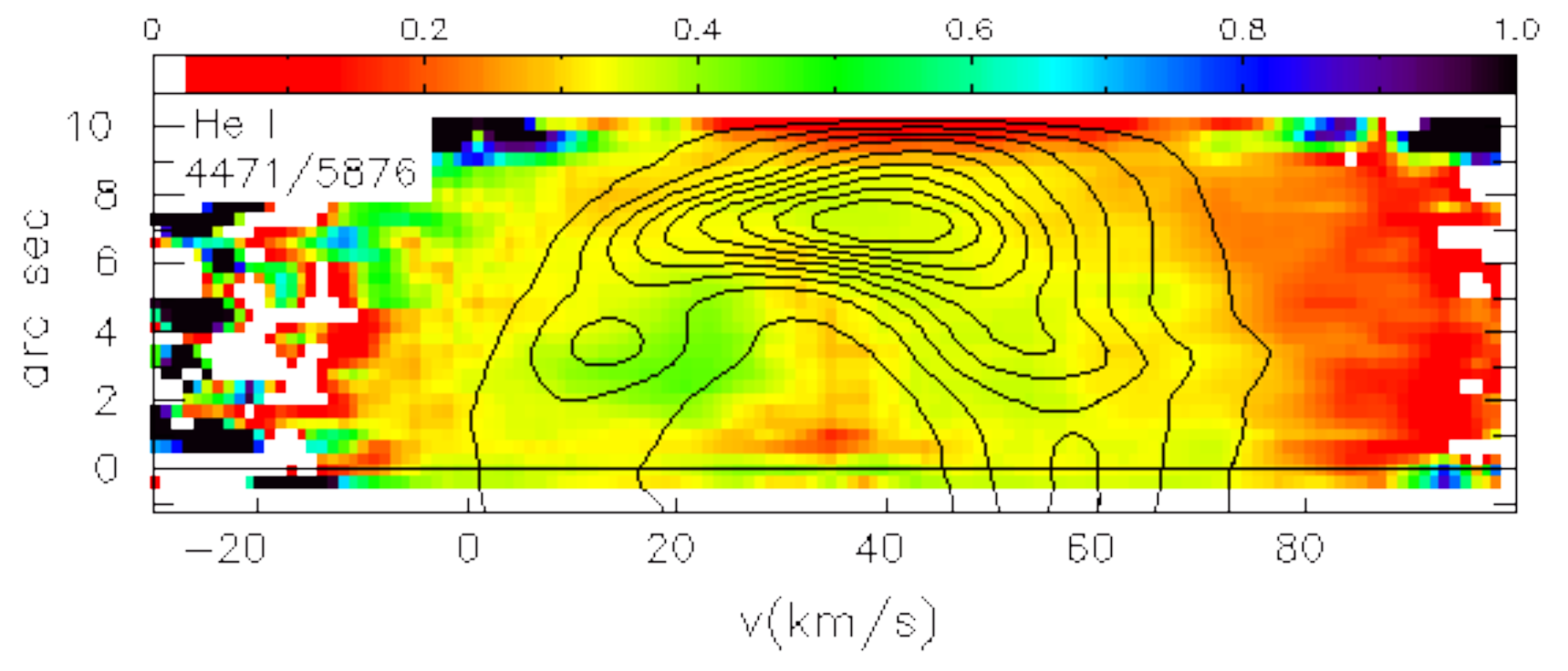}
\includegraphics[width=0.33\linewidth]{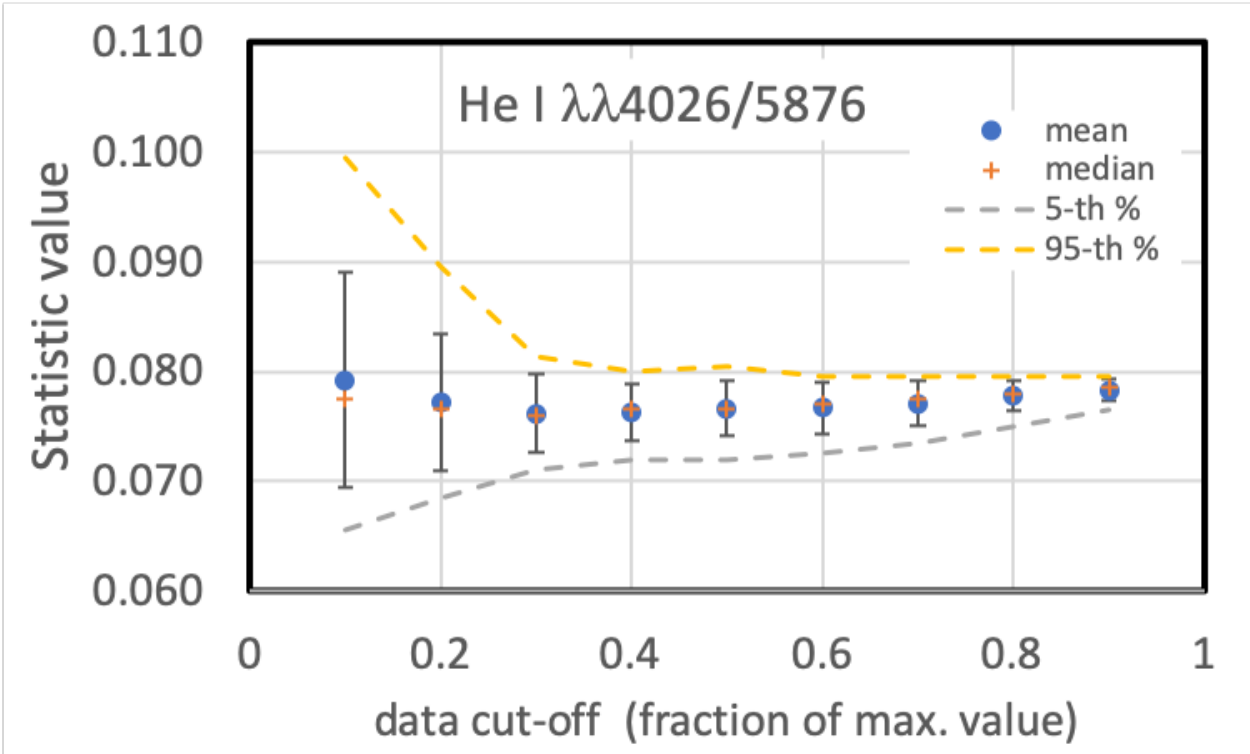}
\includegraphics[width=0.33\linewidth]{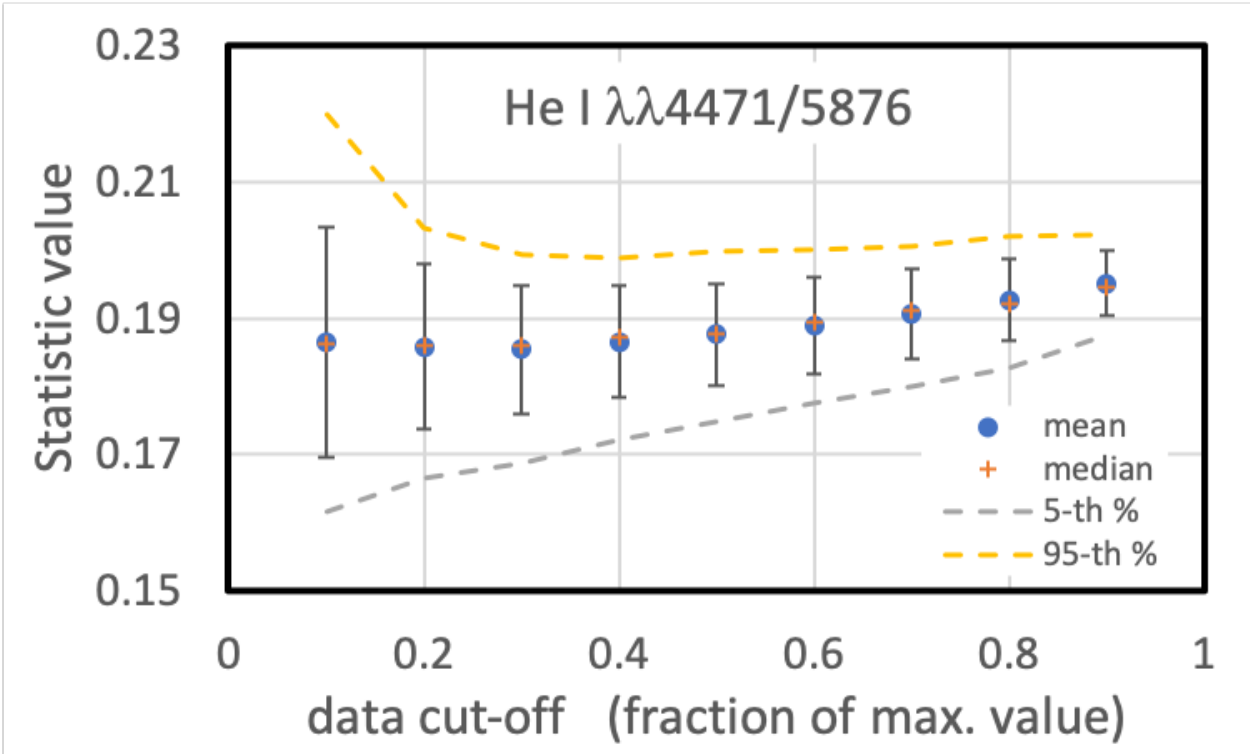}
\includegraphics[width=0.33\linewidth]{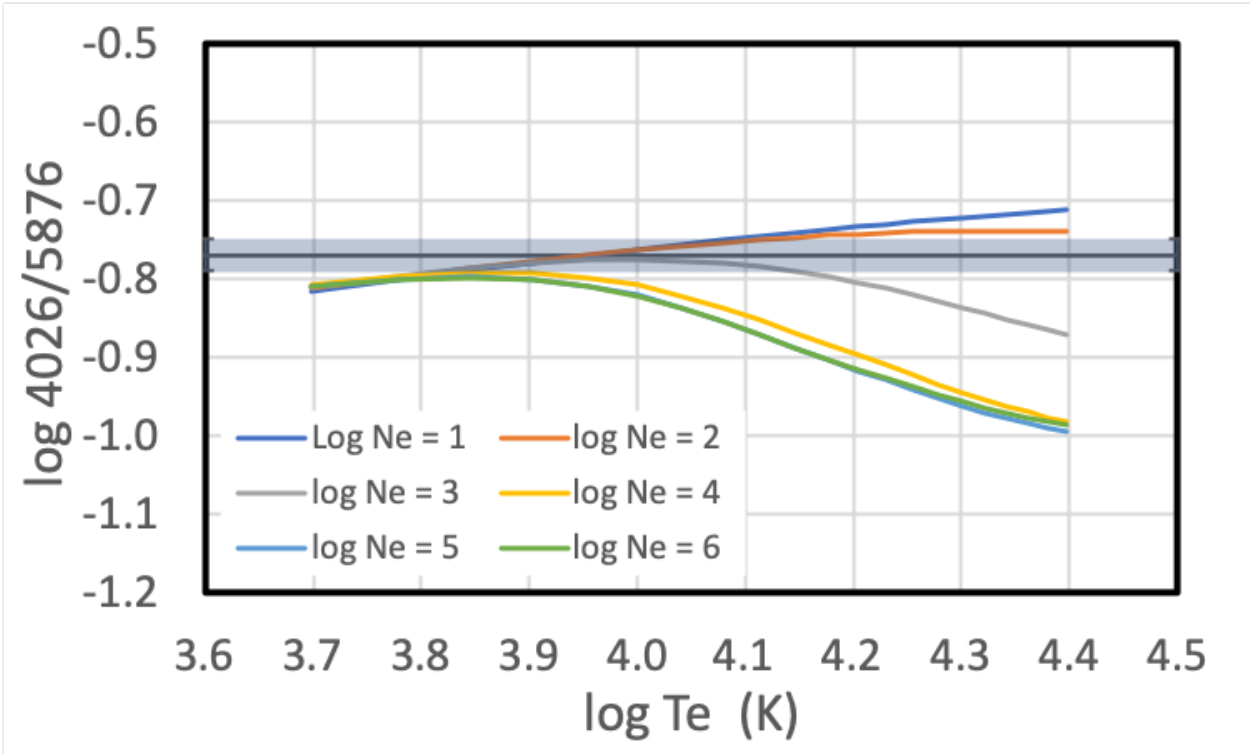}
\includegraphics[width=0.33\linewidth]{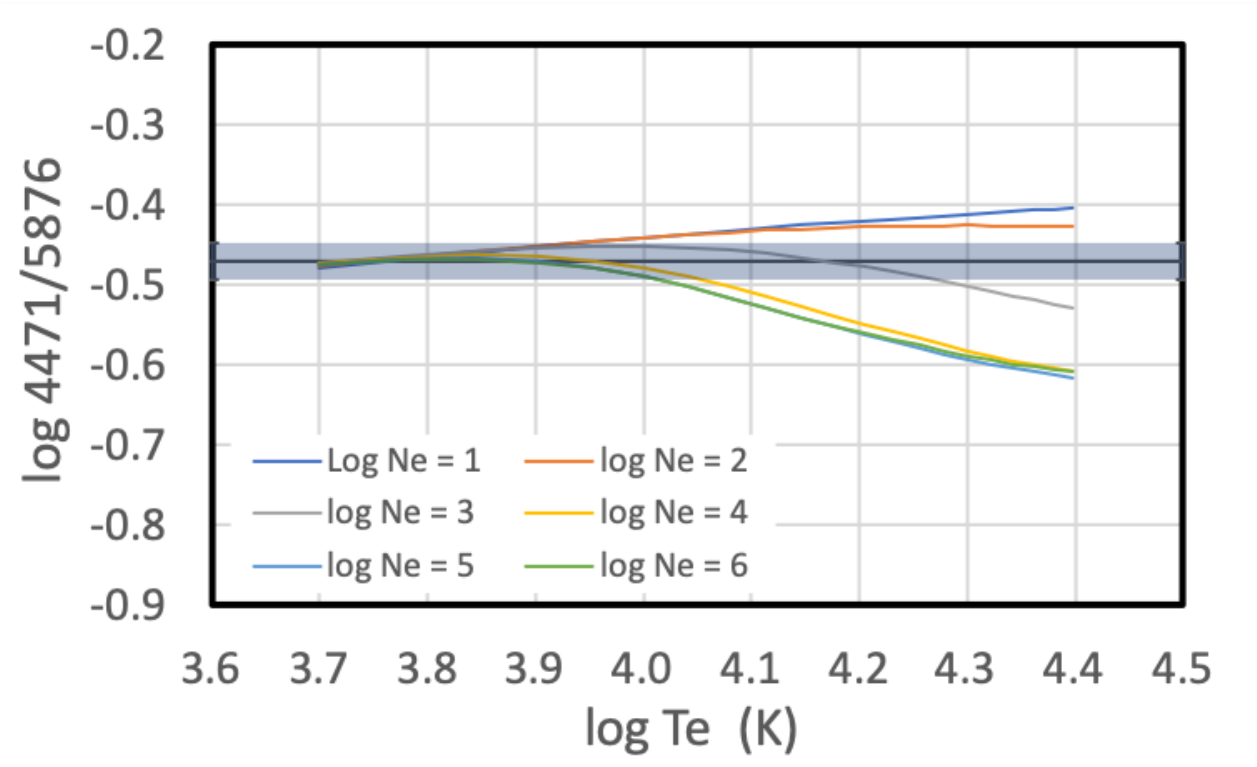}
\end{center}
\caption{Top row:  These panels present the PV diagrams of the line ratios \ion{He}{1} $\lambda\lambda$4026/5876 (left) and \ion{He}{1} $\lambda\lambda$4471/5876 (right).  The contours show the intensity of the \ion{He}{1} $\lambda$5876 line (the denominator).  Middle row:  These panels present various statistics for the line ratios of \ion{He}{1} $\lambda\lambda$4026/5876 (left) and \ion{He}{1} $\lambda\lambda$4471/5876 (right) as a function of the line intensities in the original PV diagram of the \ion{He}{1} $\lambda$5876 line (denominator).  These line ratios have been corrected for the flux scale factors (Table \ref{tab_flux_scale_factors}), but not reddening.  Bottom row:  These panels compare the ratio of the lines of \ion{He}{1} $\lambda\lambda$4026/5876 (left) and \ion{He}{1} $\lambda\lambda$4471/5876 (right), now corrected for reddening, with the theoretical line intensities from \citet{porteretal2013}.  These line ratios are insensitive to temperature, but have some sensitivity to density and are useful to check the reddening (Table \ref{tab_EBV_values}).  These \ion{He}{1} lines are compatible with  densities up to $10^4\,\mathrm{cm}^{-3}$.  
}
\label{fig_perm_HeI_redd_dens}
\end{figure*}

\begin{figure*}
\begin{center}\includegraphics[width=0.86\linewidth]{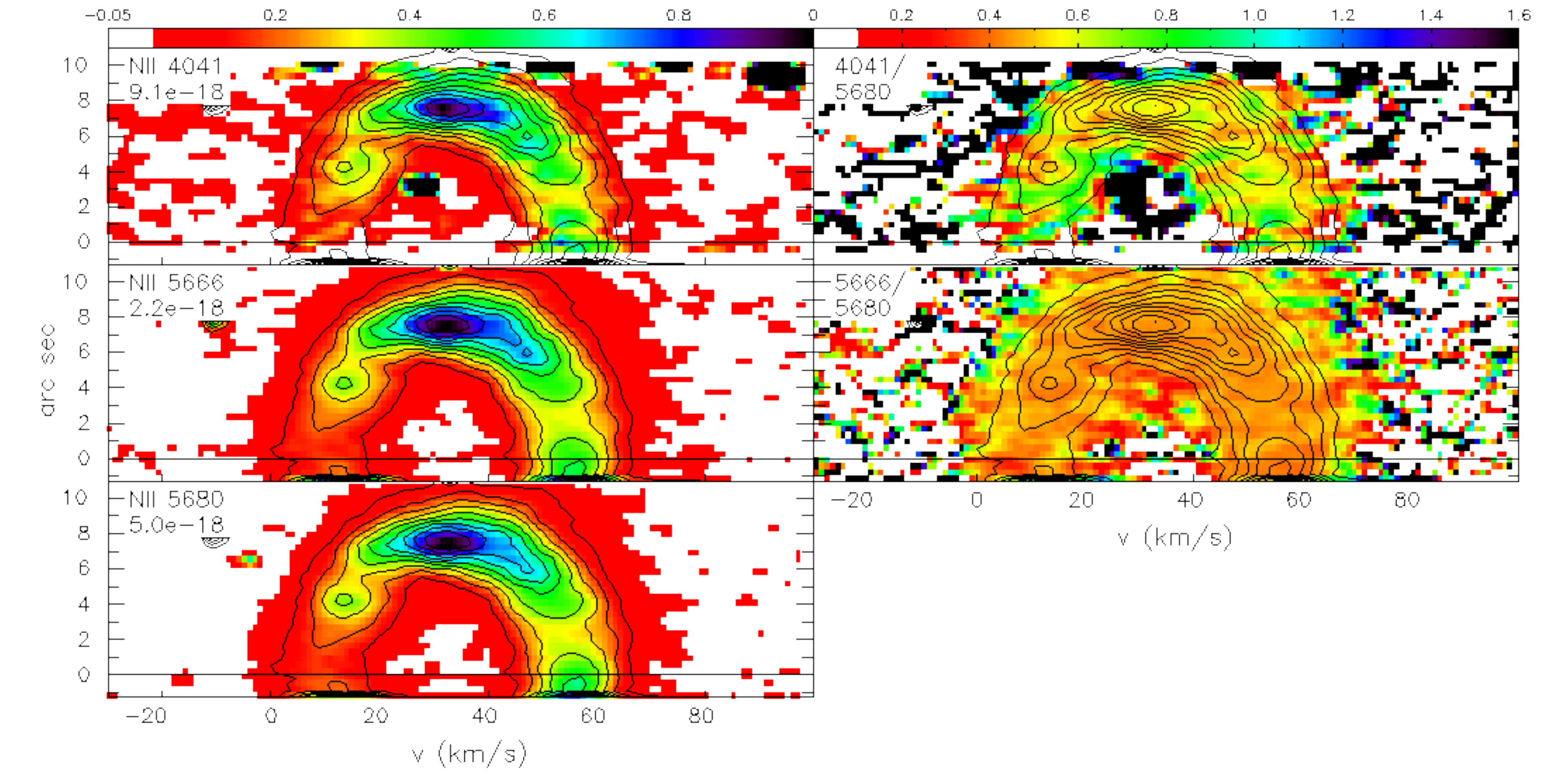}\end{center}
\caption{These panels present the PV diagrams for the lines (left) of \ion{N}{2} $\lambda$4041, \ion{N}{2} $\lambda$5666, and \ion{N}{2} $\lambda$5680, and for the line ratios (right) \ion{N}{2} $\lambda\lambda$4041/5680 and \ion{N}{2} $\lambda\lambda$5666/5680.  The contour lines are of the \ion{N}{2} $\lambda$5680 intensity.  The spatial coverage of the PV diagram of \ion{N}{2} $\lambda$4041 is less than for the others.  Both the temperature- and density-sensitive maps (4041/5680 and 5666/5680, respectively) are reasonably uniform.  
}
\label{fig_perm_N2_PV}
\end{figure*}

\begin{figure*}
\includegraphics[width=0.5\linewidth]{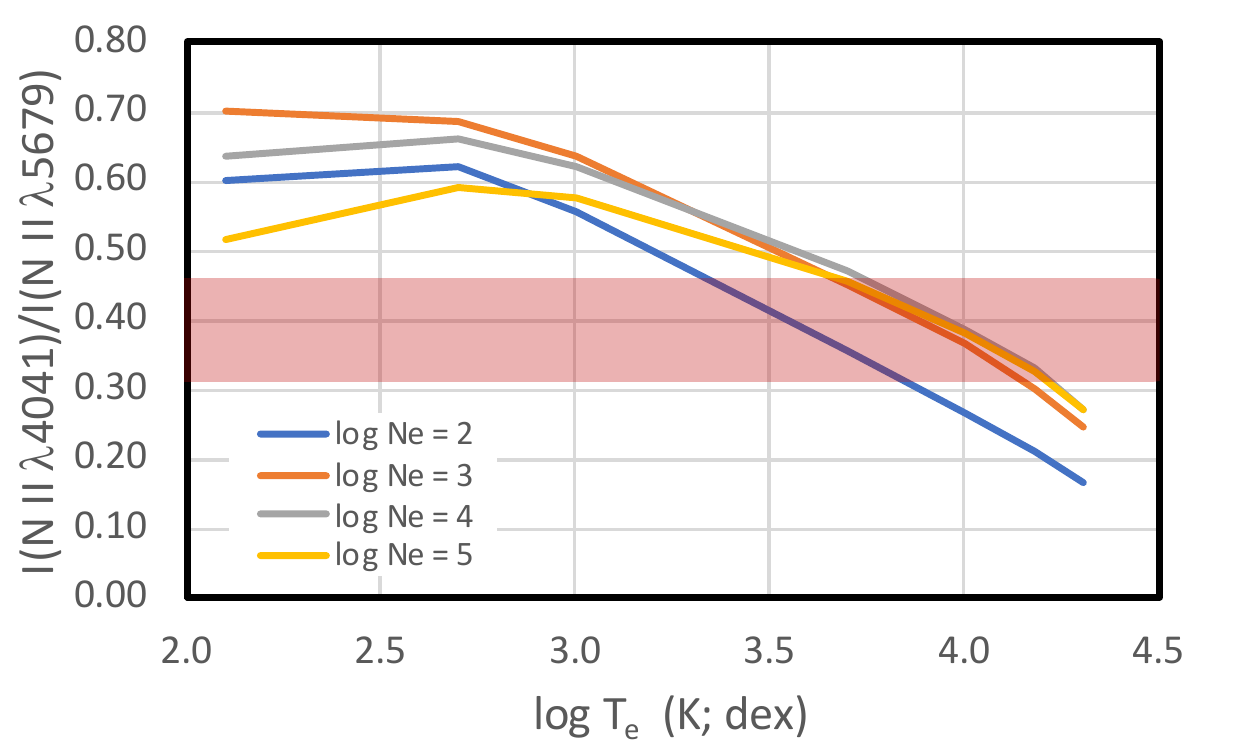}
\includegraphics[width=0.5\linewidth]{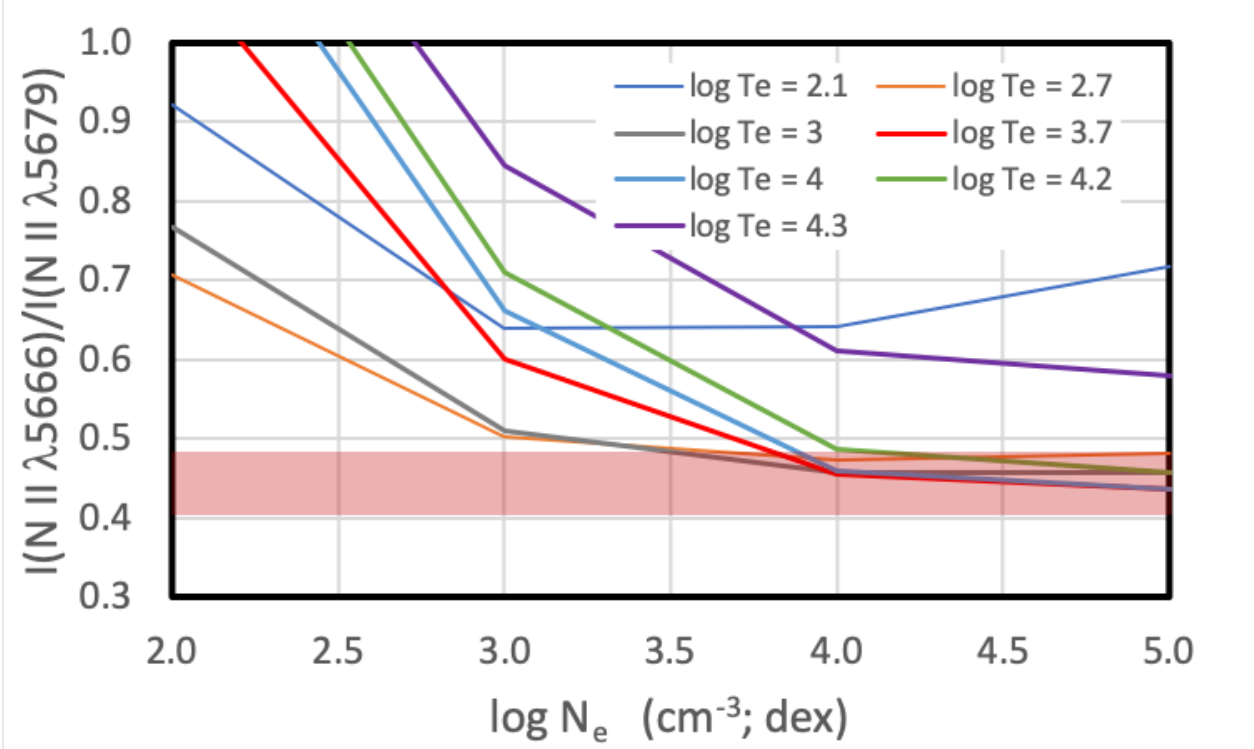}
\caption{Left:  We present the temperature-sensitive \ion{N}{2} $\lambda\lambda$4041/5680 ratio and its uncertainty (the pink region), based upon an intensity limit at 40\% of the maximum intensity and the 90\% width of the distribution of pixel values.  This ratio is corrected for both reddening and the scale factor between the CD2 and CD3b wavelength intervals.  $T_e > 2,000$\,K if the electron density is very low; $T_e > 5,000$\,K for higher densities.  Right:  We present the density-sensitive \ion{N}{2} $\lambda\lambda$5666/5680 ratio and its uncertainty (the pink region; same definitions).  The \ion{N}{2} $\lambda\lambda$5666/5680 ratio is compatible with high densities ($3,000-10^5$\,cm$^{-3}$) and a wide range of electron temperatures ($500-16,000$\,K).  
}
\label{fig_perm_N2_Te_Ne_diag}
\end{figure*}

The \ion{He}{1} lines also permit us to set limits on the electron temperature and density in NGC 6153 (atomic data: Table \ref{tab_atomic_data}).  
The ratio of $n\,\mathrm S-n^{\prime}\,\mathrm P$ transitions to $n\,\mathrm D-n^{\prime}\,\mathrm P$ transitions are sensitive to the electron temperature.  The ratios of like transitions, $n\,\mathrm S-n^{\prime}\,\mathrm P$ to $n^{\prime\prime}\,\mathrm S-n^{\prime}\,\mathrm P$ or $n\,\mathrm D-n^{\prime}\,\mathrm P$ to $n^{\prime\prime}\,\mathrm D-n^{\prime}\,\mathrm P$, have some limited sensitivity to density.  Here, we use the triplet lines \ion{He}{1} $\lambda$4713 ($4\,^3\mathrm S-2\,^3\mathrm P$) and \ion{He}{1} $\lambda\lambda$4026, 4471, 5876 ($n\,\mathrm D-2\,^3\mathrm P$; $n=5,4,3$, respectively) to illustrate the process.  (\ion{He}{1} $\lambda$4713 is the faintest of these lines.)  

In the top row of Figure \ref{fig_perm_HeI_temperature}, we present the ratios of the the PV diagrams of \ion{He}{1} $\lambda$4713 line to the \ion{He}{1} $\lambda\lambda$4026, 4471, 5876 lines to set limits upon the electron temperature.  As shown, these line ratios are not corrected for reddening.  The contours are of intensities of the \ion{He}{1} $\lambda\lambda$4026, 4471, 5876 lines (the denominator of the ratio).  
The line ratios are generally uniform, except perhaps the emission from the approaching side of the main shell in the \ion{He}{1} $\lambda\lambda$4713/4026 ratio.  Note that this part of the \ion{He}{1} $\lambda$4026 (4026.2\AA) profile suffers a slight contamination due to the \ion{He}{2} $\lambda$4025.6 line that contaminates the emission from the approaching side of the main shell.  

The three panels in the middle row of Figure \ref{fig_perm_HeI_temperature} present statistics for the \ion{He}{1} $\lambda\lambda$4713/4026, \ion{He}{1} $\lambda\lambda$4713/4471, and \ion{He}{1} $\lambda\lambda$4713/5876 ratios (left to right) for the pixels that exceed 10\%, 20\%, ..., 90\% of the intensity of the maximum intensity in the PV diagrams of the \ion{He}{1} $\lambda\lambda$4026, 4471, 5876 lines, respectively.  The error bars indicate the standard deviation of the distribution of the ratio values at each threshold level.  The crosses indicate the median values at each threshold level while the dashed lines indicate the values of the 5-th and 95-th percentiles of the distribution, again at each threshold level.  All statistics include only the $0.72\arcsec-9.00\arcsec$ spatial range to exclude the flat-fielding artefacts in the top few rows of PV diagrams (top row).  The mean and median values do not vary strongly with the threshold level, indicating that the temperature variation within the plasma emitting these \ion{He}{1} lines is modest.  The different threshold levels correspond to different volumes of the plasma that emit these lines.  At the 10\% threshold, plasma throughout the main shell contributes to the statistics of the line ratios in Figure \ref{fig_perm_HeI_temperature} while, at the 90\% threshold, the statistics reflect only the plasma along the line of sight to the end of the main shell, farthest from the central star.  

Given the limited variation in these line ratios (and temperatures) we adopt the mean value and its standard deviation for the threshold at 30\% of the maximum intensity as our temperature indicator and its uncertainty (this sample includes over 500 pixels).  The range of values spanned by the standard deviation at this threshold usually includes almost the entire range of variation in the mean or median at all threshold values.  In the bottom row of Figure \ref{fig_perm_HeI_temperature}, we compare the intensity ratio, now also corrected for reddening with the theoretical values (atomic data: Table \ref{tab_atomic_data}).  By this measure, the \ion{He}{1} lines imply an electron temperature of $\log T_e$ of $3.85-4.05$ (\ion{He}{1} $\lambda\lambda$4713/4026), $3.90-4.07$ (\ion{He}{1} $\lambda\lambda$4713/4471), and $3.89-4.03$ (\ion{He}{1} $\lambda\lambda$4713/5876) if electron densities range over $10^2-10^6\,\mathrm{cm}^{-3}$.  

In Figure \ref{fig_perm_HeI_redd_dens}, we present the \ion{He}{1} $\lambda\lambda$4026/5876 and \ion{He}{1} $\lambda\lambda$4471/5876 line ratios, which have some sensitivity to the electron density.  The three rows present the same information as in Figure \ref{fig_perm_HeI_temperature}.  Here, only the \ion{He}{1} $\lambda\lambda$4026/5876 ratio provides additional information, restricting the electron density to values of $10^4\,\mathrm{cm}^{-3}$ or lower.  Considering this restriction, the previous electron temperatures are restricted to the temperature ranges of $\log T_e$ of $3.91-4.05$ (\ion{He}{1} $\lambda\lambda$4713/4026), $3.97-4.07$ (\ion{He}{1} $\lambda\lambda$4713/4471), and $3.93-4.03$ (\ion{He}{1} $\lambda\lambda$4713/5876)

\begin{figure*}
\begin{center}\includegraphics[width=0.86\linewidth]{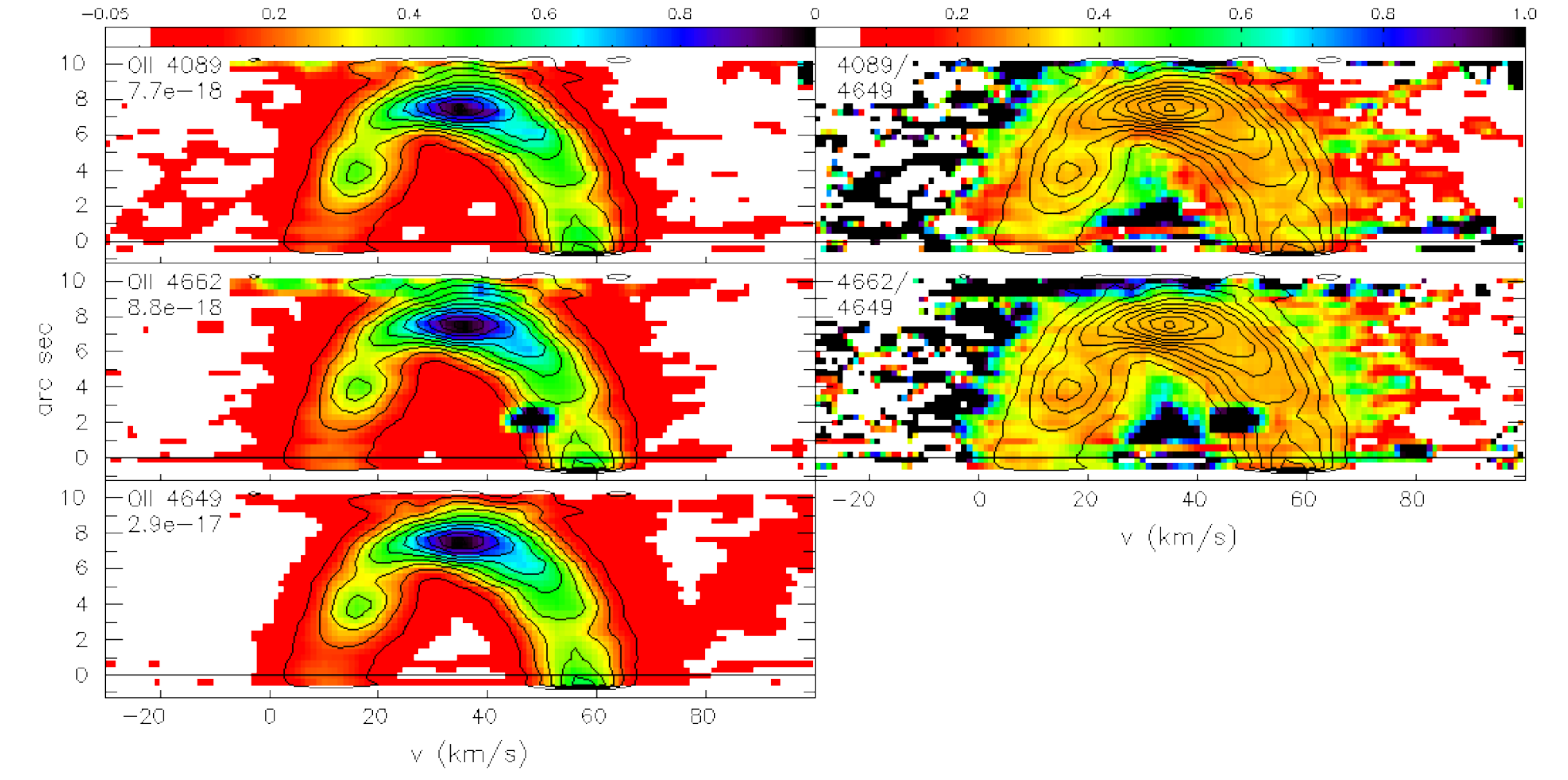}\end{center}
\caption{In the left column, these panels present the lines of \ion{O}{2} $\lambda$4089, \ion{O}{2} $\lambda$4662, and \ion{O}{2} $\lambda$4649.  In the right column, the panels present the PV diagrams of the line ratios \ion{O}{2} $\lambda\lambda$4089/4649 and \ion{O}{2} $\lambda\lambda$4662/4649.  The contours in the right panels are the intensity contours from the PV diagram of the \ion{O}{2} $\lambda$4649 line (bottom left).  The bright fringe at the top of the PV diagrams is due to imperfect subtraction of the nebular continuum.  A cosmic ray contaminates the PV diagram of the \ion{O}{2} $\lambda$4662 line.  
}
\label{fig_perm_O2_PV}
\end{figure*}

\begin{figure*}
\includegraphics[width=0.5\linewidth]{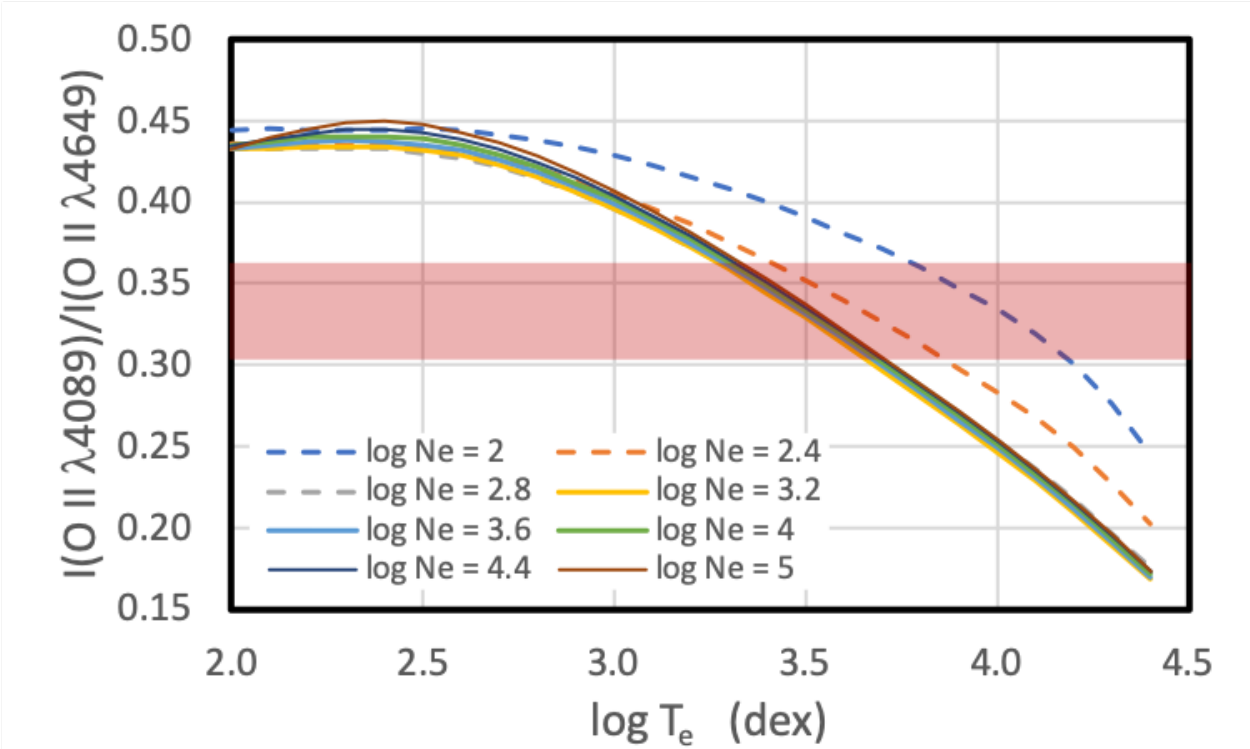}
\includegraphics[width=0.5\linewidth]{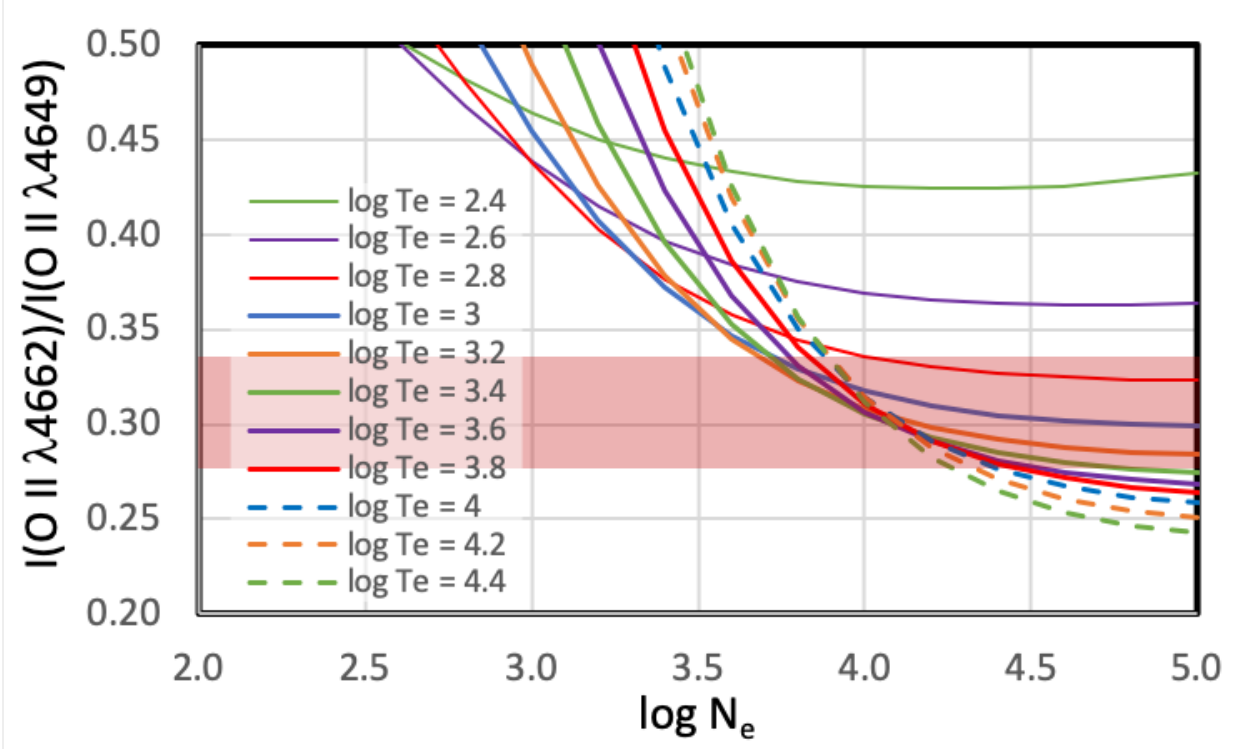}
\caption{Left:  We present the temperature-sensitive \ion{O}{2} $\lambda\lambda$4089/4649 ratio and its uncertainty as a function of the temperature (the pink band), based upon an intensity limit at 40\% of the maximum intensity and the 90\% width of its distribution of pixel values.  The temperature implied is rather low, $\log T_e = 3.25-3.7$\,dex, unless the density is very low.  Right:  We present the observed value of the density-sensitive ratio \ion{O}{2} $\lambda\lambda$4662/4649 and its uncertainty as a function of the electron density (the pink band; same definitions).  The observed \ion{O}{2} $\lambda\lambda$4662/4649 ratio implies high densities ($10^4-10^5$\,cm$^{-3}$) at lowest temperatures ($< 2,500$\,K) and densities of $5,000-30,000$\,cm$^{-3}$ at higher temperatures.  
}
\label{fig_perm_O2_Te_Ne_diag}
\end{figure*}

Finally, we can also use the \ion{N}{2} and \ion{O}{2} lines to compute the electron temperature and density.   
Figure \ref{fig_perm_N2_PV} presents the PV diagrams for the \ion{N}{2} $\lambda\lambda$4041, 5666, 5680 lines as well as the \ion{N}{2} $\lambda\lambda$4041/5680 (temperature) and \ion{N}{2} $\lambda\lambda$5666/5680 (density) line ratios.  The PV diagram for \ion{N}{2} $\lambda$4041 is the one with the worst S/N and is from a different wavelength interval.  Even so, the \ion{N}{2} $\lambda\lambda$4041/5680 ratio is decently uniform.  Certainly, there is no clear evidence of gradients or large-scale changes.  The \ion{N}{2} $\lambda\lambda$5666/5680 line ratio is very uniform, especially where the intensity of the lines is strong.  

In Figure \ref{fig_perm_N2_Te_Ne_diag}, we plot the \ion{N}{2} $\lambda\lambda$4041/5680 line ratio as a function of temperature (left panel) and the \ion{N}{2} $\lambda\lambda$5666/5680 line ratio as a function of density (right panel).  For both line ratios, we use all of the pixels that exceed 40\% of the maximum intensity in the PV diagram of the \ion{N}{2} $\lambda$5680 line.  The shaded region indicates the mean value and the range allowed by 90\% of the distribution (from the 5th to the 95th percentiles).  The theoretical ratio (atomic data: Table \ref{tab_atomic_data}), is shown for different values of the electron temperature (left panel) and density (right panel).  The \ion{N}{2} $\lambda\lambda$4041/5680 line ratio implies an electron temperature of $\log T_e = 3.3-4.2$\,dex.  The \ion{N}{2} $\lambda\lambda$5666/5680 line ratio implies an electron density of $\log N_e > 3.5$\,dex and is compatible with electron temperatures of $\log T_e = 2.7-4.2$\,dex.  Combining the restrictions from the two \ion{N}{2} line ratios, they are nominally compatible with $\log T_e = 3.7-4.2$\,dex and $\log N_e > 3.5$\,dex (and up to 5.0 dex given the atomic data available).  

Figure \ref{fig_perm_O2_PV} presents the PV diagrams of the \ion{O}{2} $\lambda\lambda$4089, 4649, 4662 lines (left column) and of the \ion{O}{2} $\lambda\lambda$4089/4649 (temperature) and \ion{O}{2} $\lambda\lambda$4662/4649 (density) line ratios (right column).  The PV diagrams for all of the lines have good S/N.  The temperature- (top right) and density-sensitive (middle right) line ratios are both quite uniform.  Hence, there is no evidence for any large scale variations in the temperature or density of the plasma emitting in the \ion{O}{2} lines.  

Figure \ref{fig_perm_O2_Te_Ne_diag} compares the \ion{O}{2} $\lambda\lambda$4089/4649 and \ion{O}{2} $\lambda\lambda$4662/4649 line ratios (left and right panels, respectively) with the theoretical values for different physical conditions (atomic data: Table \ref{tab_atomic_data}).  The pink shaded region indicates the observed value and the range containing 90\% of the distribution of values using all of the pixels that exceed 40\% of the maximum intensity in the \ion{O}{2} $\lambda$4649 PV diagram (denominator).  The temperature-sensitive \ion{O}{2} $\lambda\lambda$4089/4649 line ratio (Figure \ref{fig_perm_O2_Te_Ne_diag}; left panel) implies $\log T_e = 3.2-4.2$\,dex, considering all densities.  The density-sensitive \ion{O}{2} $\lambda\lambda$4662/4649 line ratio (Figure \ref{fig_perm_O2_Te_Ne_diag}; right panel) implies $\log N_e > 3.7$\,dex and an electron temperature of $\log T_e = 2.8-4.4$\,dex.  Combining the two \ion{O}{2} line ratios implies $\log T_e = 3.2-3.7$\,dex and $\log N_e > 3.7$\,dex (and up to 5.0 dex).  

For both the \ion{N}{2} and \ion{O}{2} lines, the density-sensitive indicator is compatible with temperatures substantially lower than the temperature-sensitive line ratio.  The density-sensitive line ratios are sufficiently close in wavelength that they are independent of the reddening correction and the scale factors between wavelength intervals, but one (\ion{O}{2}) or both (\ion{N}{2}) corrections are important for the temperature-sensitive ratios.  The density-sensitive line ratios also have higher S/N.  If the density-sensitive line ratios are considered alone, they 
are in excellent agreement, implying that the electron temperature may be as low as $\log T_e = 2.7-2.8$\,dex ($\sim 500$\,K).  
All things considered, we conclude that the \ion{N}{2} and \ion{O}{2} lines imply a ``low" electron temperature of $\log T_e \sim 3.2-3.7$\,dex and a ``high" electron density of $\log N_e > 3.7$\,dex (and up to 5.0 dex given the atomic data available).  If so, the \ion{N}{2} and \ion{O}{2} lines imply different physical conditions compared to the \ion{H}{1}, \ion{He}{1}, and forbidden lines.  Again, this argues for the presence of two plasma components.

\subsection{Contamination in forbidden lines}\label{sec_contamination}

\begin{figure*}
\begin{center}
\includegraphics[width=0.49\linewidth]{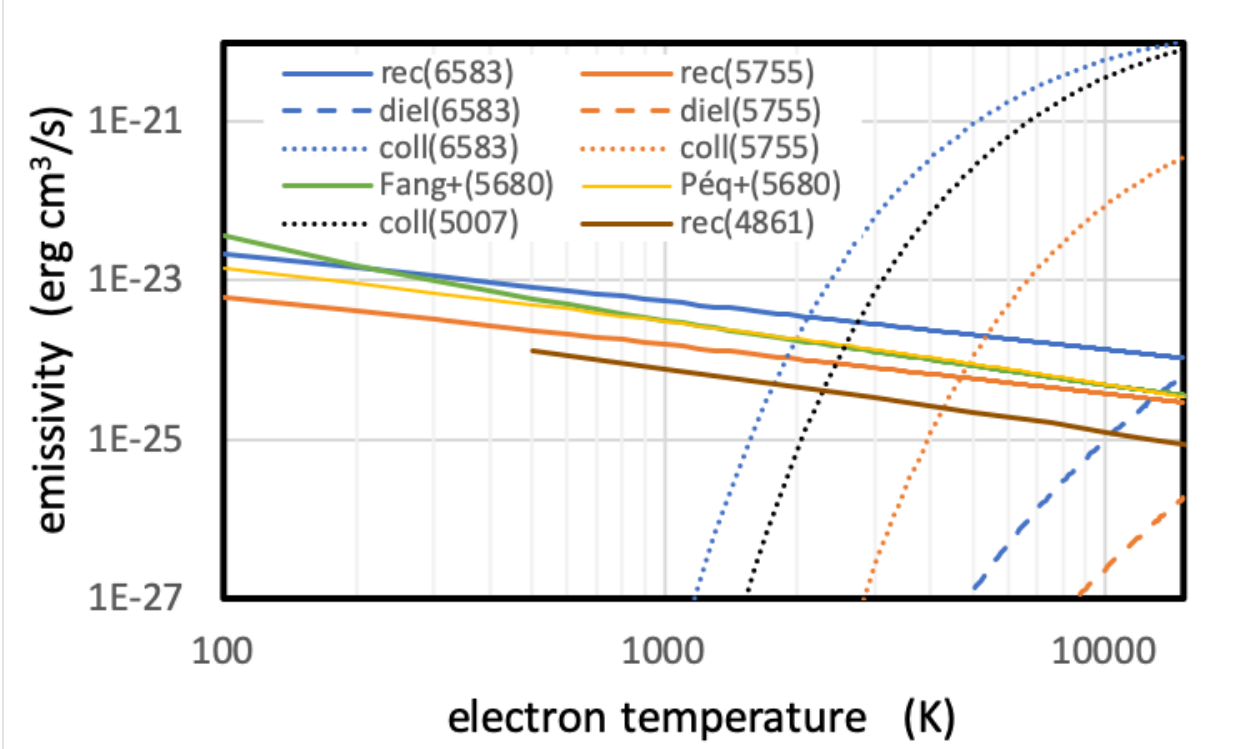}
\includegraphics[width=0.49\linewidth]{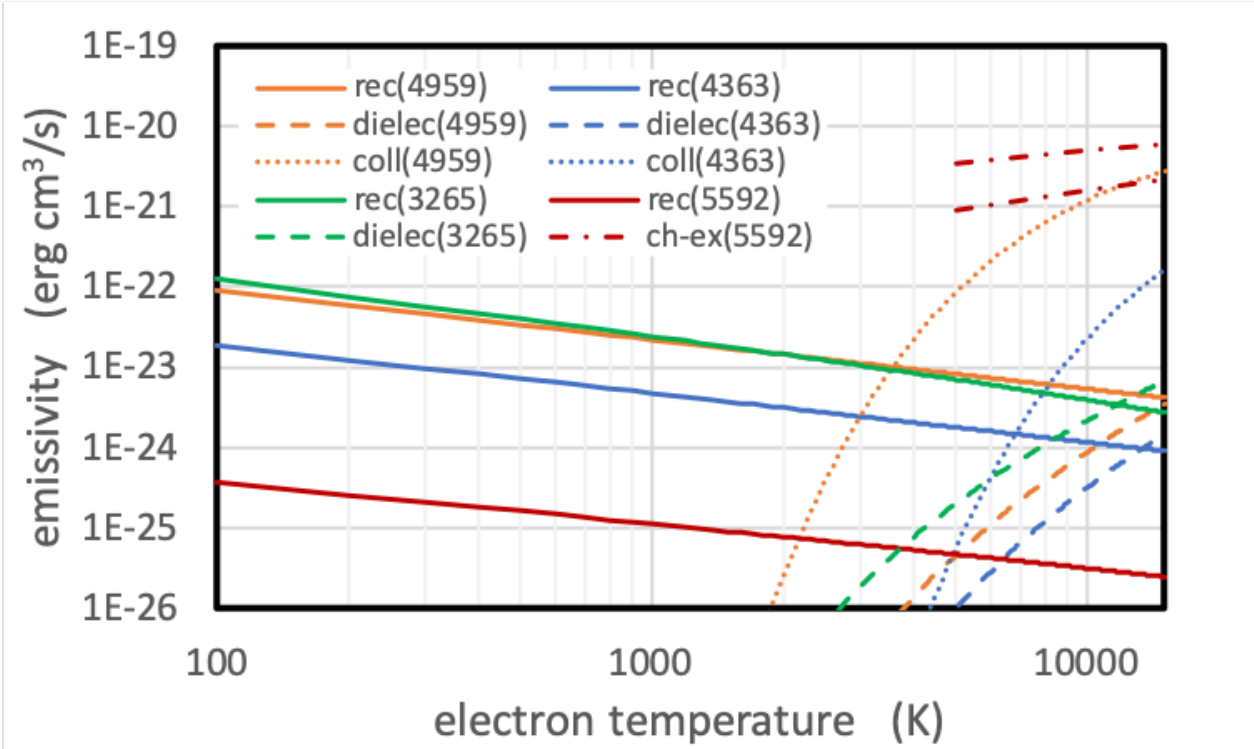}
\includegraphics[scale=0.7]{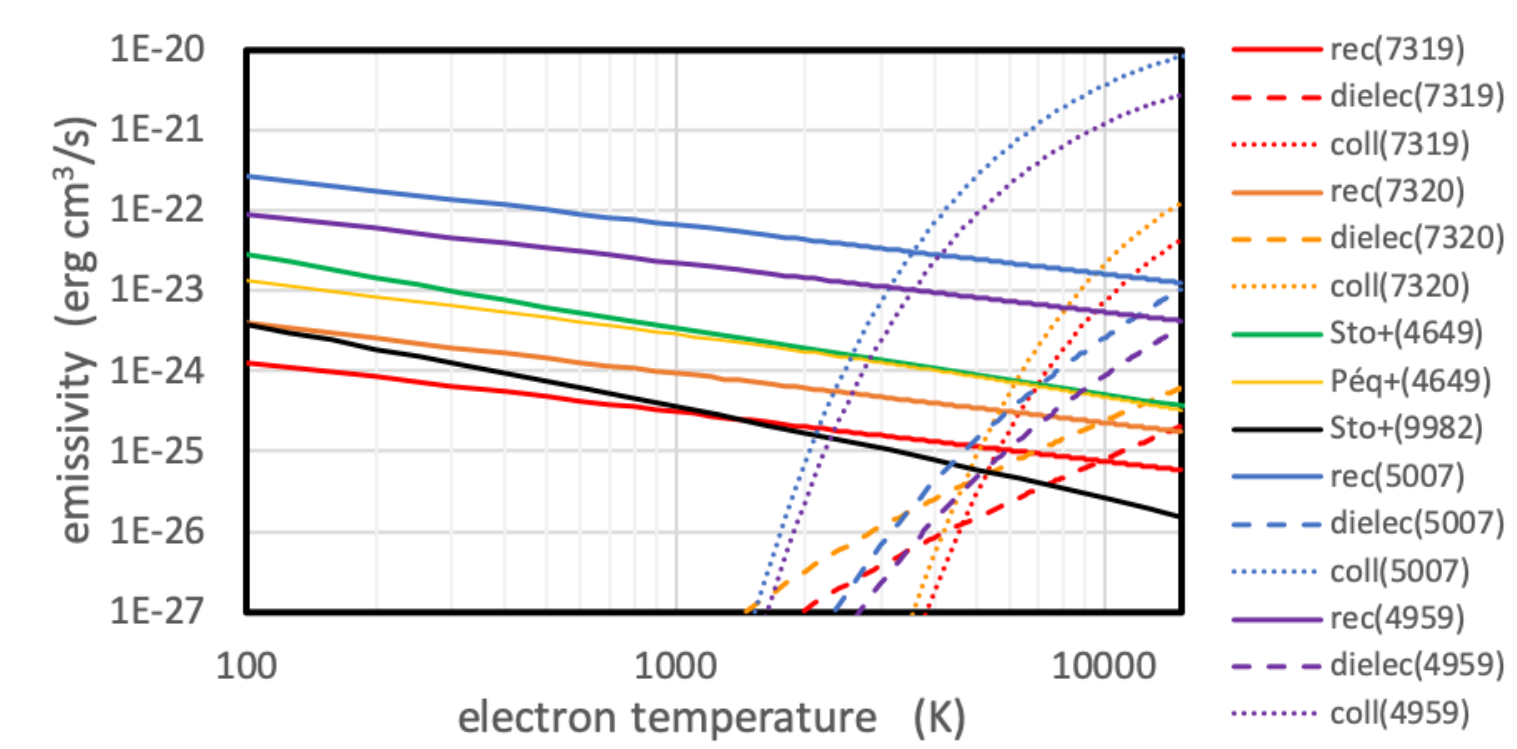}
\end{center}
\caption{We present the emissivities per ion and per electron for the lines required to decompose the PV diagram of [\ion{N}{2}] $\lambda$5755 (top left), [\ion{O}{3}] $\lambda$4363 (top right), and [\ion{O}{2}] $\lambda\lambda$7319,7320 (bottom).  The emissivity for charge exchange is per ion and per hydrogen atom.  The legends indicate the transition wavelength in brackets.  The line style indicates the physical process that excites the line:  solid for radiative recombination, dashed for dielectronic recombination, dotted for collisional excitation, and dot-dashed for charge exchange.  Atomic data:  See Table \ref{tab_atomic_data}.    
}
\label{fig_emissivities_N2_O2_O3}
\end{figure*}

Many emission lines may be excited by multiple processes.  Here, we consider the effects of excitation mechanisms other than collisional excitation on the [\ion{N}{2}] $\lambda$5755, [\ion{O}{2}] $\lambda\lambda$7319,7320, and [\ion{O}{3}] $\lambda$4363 lines.  
We shall call these additional excitation channels ``contamination", though Nature does not see it that way.  
The contamination is severe in the first two cases \citep{liuetal2000}, but it has a minimal effect on [\ion{O}{3}] $\lambda$4363 in NGC 6153.  We choose these examples deliberately because they are little affected by collisional de-excitation .  

\S\ref{sec_ionization_structure} and \S\ref{sec_physical_conditions} indicate that there are two sets of physical conditions that apply to two kinematical components.  One set of physical conditions applies to the \ion{H}{1}, \ion{He}{1}, and forbidden lines (the normal nebular plasma), where the electron temperature is $\sim 8,000$\,K and the electron density is $\sim 4,000-5,000$\,cm$^{-3}$.  For the normal nebular plasma, we adopt the physical conditions of $T_e=8,000$\,K and $N_e=5,000\,\mathrm{cm}^{-3}$.  The second set of physical conditions applies to the \ion{N}{2} and \ion{O}{2} lines (the additional plasma component), where the temperature is lower, $T_e\sim 1,600-5,000\,\mathrm K$, and the density higher, $N_e > 5,000\,\mathrm{cm}^{-3}$, possibly as high as $10^5\,\mathrm{cm}^{-3}$.  For the additional plasma component, we adopt the physical conditions of $T_e=2,000$\,K and $N_e=10,000\,\mathrm{cm}^{-3}$.   Thus, we shall decompose the PV diagrams for the [\ion{N}{2}] $\lambda$5755, [\ion{O}{2}] $\lambda\lambda$7319,7320, and [\ion{O}{3}] $\lambda$4363 lines, assuming that there are two plasma components in NGC 6153, each with distinct physical conditions.  

The goal of the decomposition is to obtain PV diagrams of the forbidden lines with \emph{only} the contribution due to collisional excitation, the only excitation process considered by the standard nebular analysis.  Thus, we need models or patterns of the emission due to recombination from the normal nebular plasma and the additional plasma component that we may subtract from the observed PV diagrams, leaving only the contribution due to collisional excitation.  To do this, we must scale the patterns of the PV emission using the emissivities of the lines involved for the physical conditions found in each of the plasma components.  The three panels of Figure \ref{fig_emissivities_N2_O2_O3} present the emissivities of the relevant lines due to H$^+$, N$^+$, O$^+$, and O$^{2+}$ as a function of the electron temperature (atomic data: Table \ref{tab_atomic_data}).  

\begin{figure}
\includegraphics[width=\linewidth]{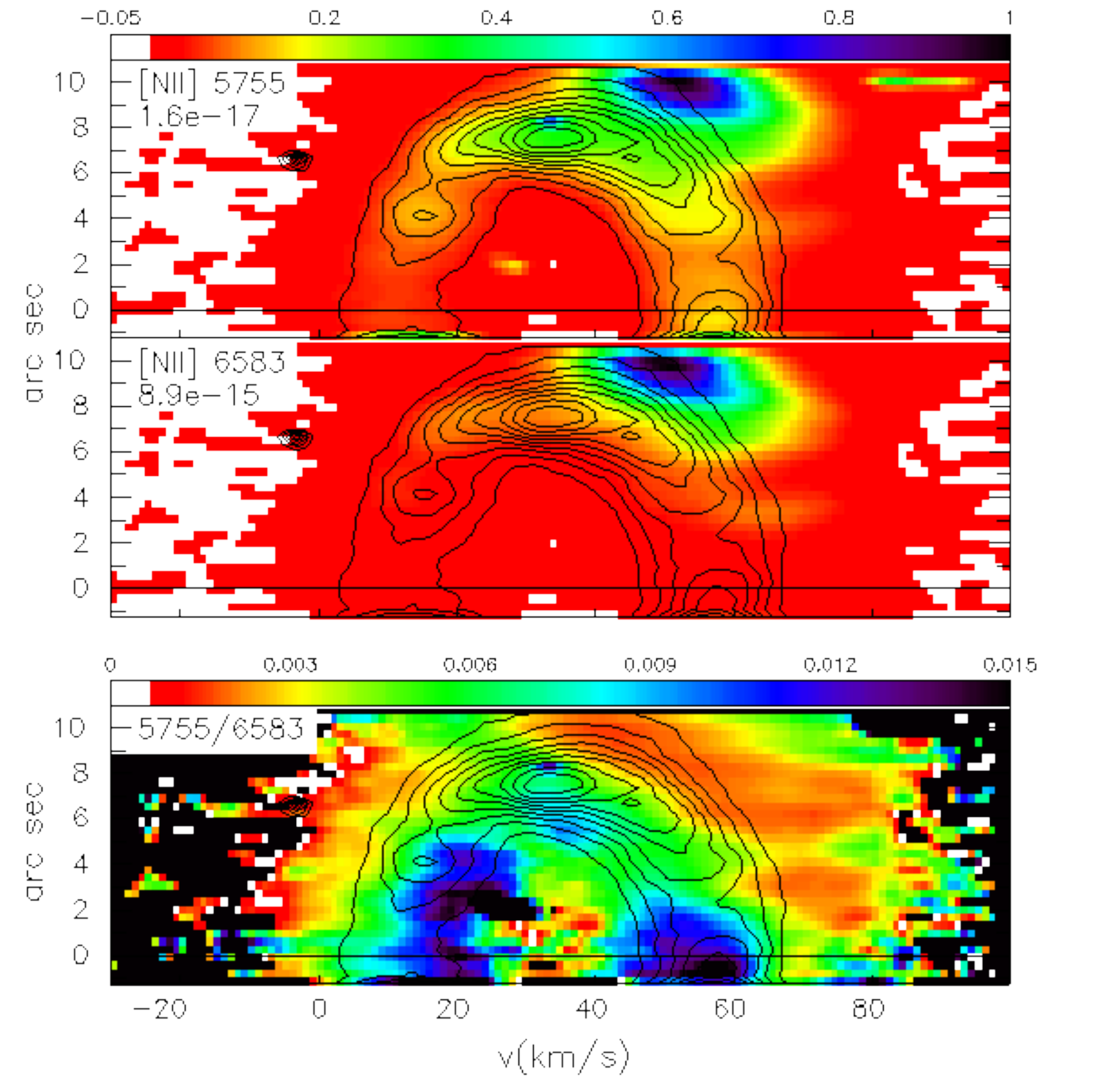}
\caption{We compare the PV diagrams of the forbidden [\ion{N}{2}] $\lambda\lambda$5755,6583 lines (top and middle, respectively) and of their ratio.   The contour lines are of the emission in the \ion{N}{2} $\lambda$5680 line (Figure \ref{fig_decompN2_PV_N2_perm}/Figure \ref{fig_PVC2}).  The main shell is notably brighter, compared to the filament on its receding side, in the [\ion{N}{2}] $\lambda$5755 line, as is the emission from the permitted \ion{N}{2} $\lambda$5680 line.  The ratio of the forbidden lines presents an excellent correspondence between the excess emission in the [\ion{N}{2}] $\lambda$5755 line and the emission from \ion{N}{2} $\lambda$5680.  
}
\label{fig_decompN2_PV_perm}
\end{figure}

\begin{figure}
\includegraphics[width=\linewidth]{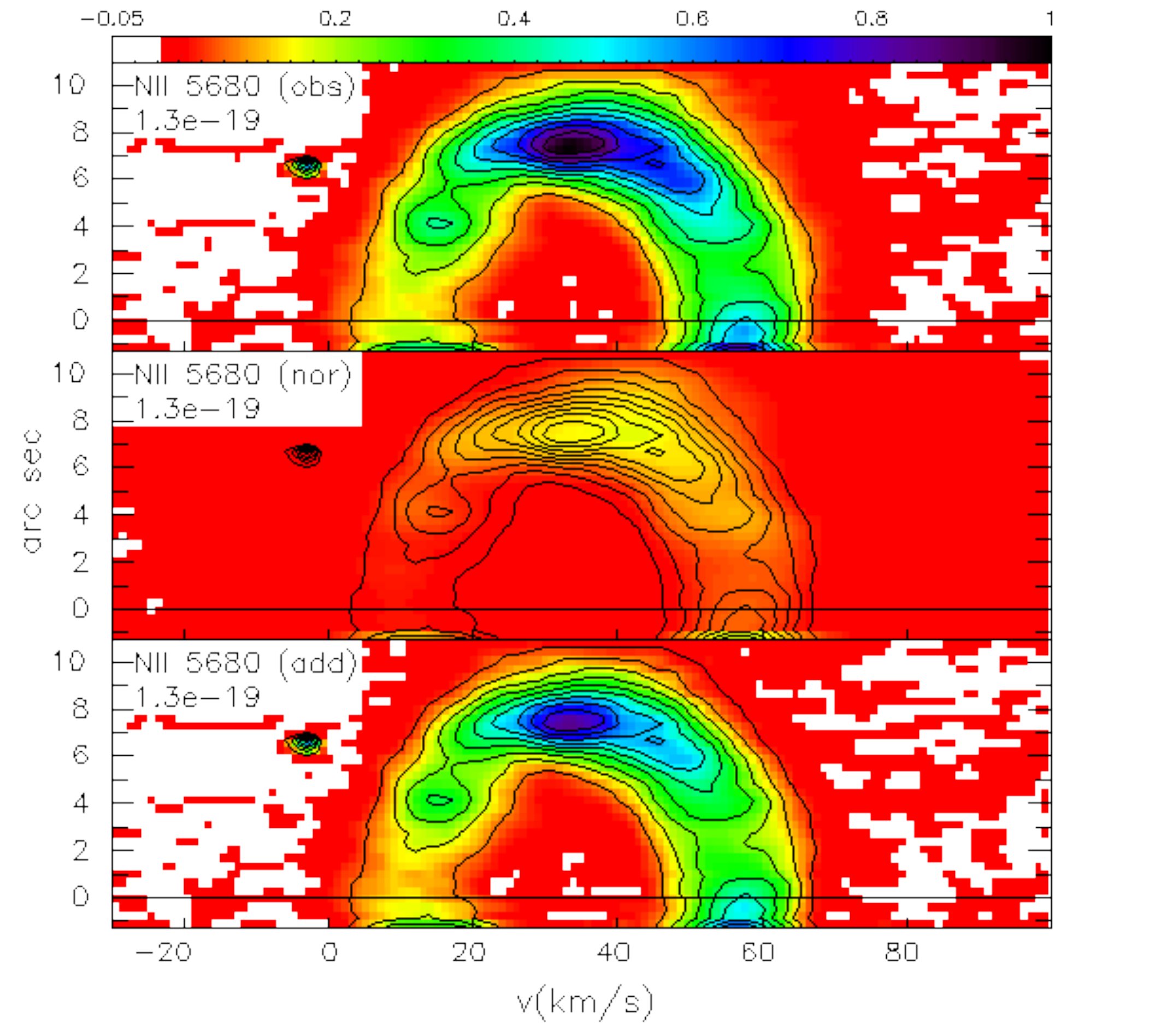}
\caption{These panels present the PV diagrams of the \ion{N}{2} $\lambda$5680 line as observed (top), the model of the recombination in this line from the normal nebular plasma (middle), and the difference between the previous two (bottom), which is interpreted as the recombination in this line due to the additional plasma component.  The contours in the three panels correspond to the contours for the \ion{N}{2} $\lambda$5680 line (top panel).  The three panels share the same intensity scale.  
}
\label{fig_decompN2_PV_N2_perm}
\end{figure}

\begin{figure*}
\begin{center}\includegraphics[width=0.86\linewidth]{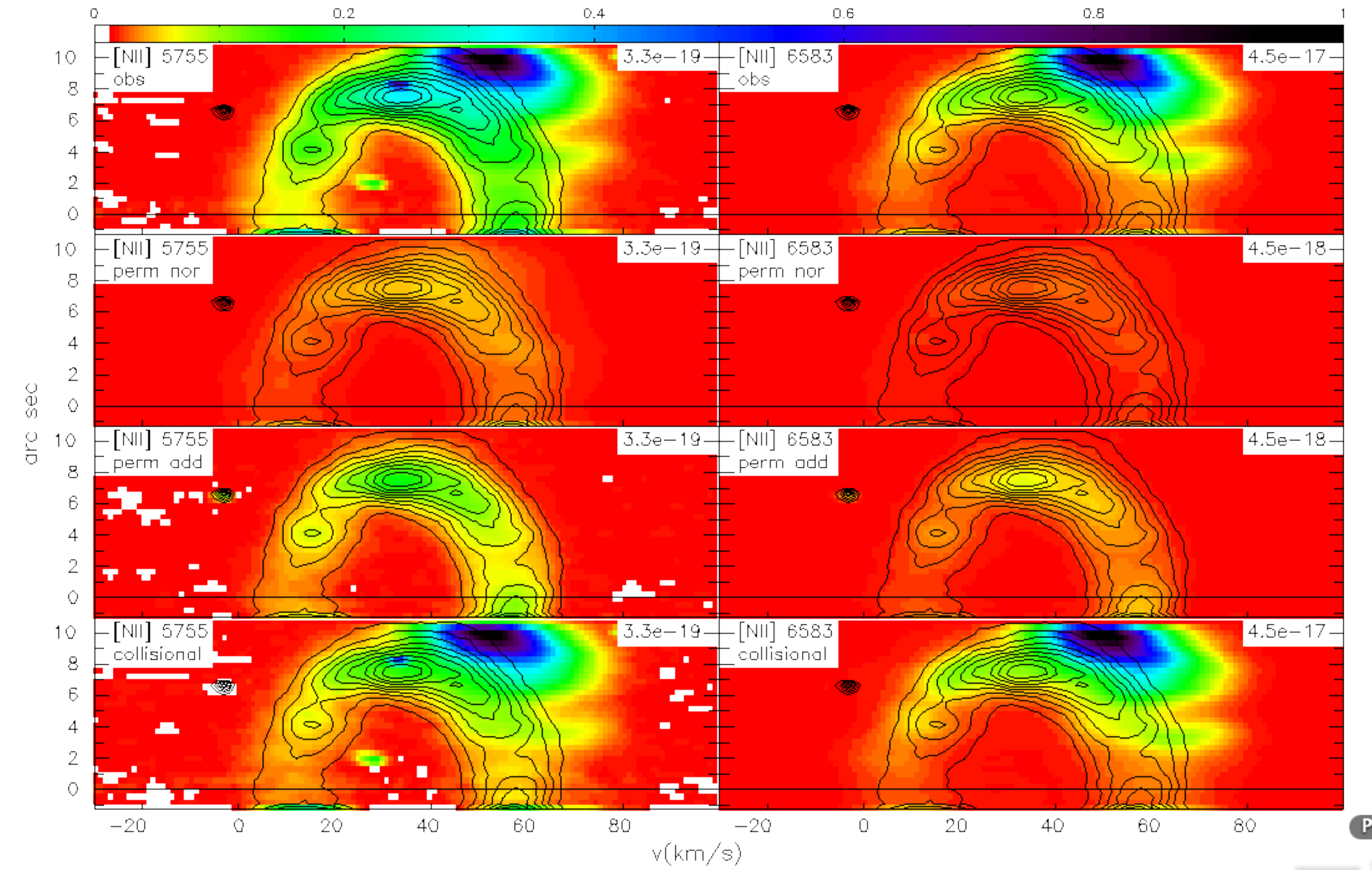}\end{center}
\caption{The left and right columns present the PV diagrams of [\ion{N}{2}] $\lambda$5755 and [\ion{N}{2}] $\lambda$6583, respectively.  In the top row, we present the observed PV diagrams of these lines.  In the second and third rows, we present the models of the permitted emission in these lines due to recombination in the normal nebular plasma and the additional plasma component, respectively.  In the bottom row, we present the difference between the observed PV diagram and the sum of the permitted emission, yielding what should be the emission due to collisional excitation only from the normal nebular plasma.  The maximum of the intensity/color scale is given at upper right in each panel (constant on the left; variable on the right).  The contour lines are of the intensity of the \ion{N}{2} $\lambda$5680 line (Figure \ref{fig_decompN2_PV_N2_perm}/Figure \ref{fig_PVC2}).  Clearly, the fractional contribution of recombination from both plasma components, but especially the additional plasma component, is much greater for [\ion{N}{2}] $\lambda$5755.  
}
\label{fig_decompN2_PV_N2_clean}
\end{figure*}

\begin{figure}
\includegraphics[width=\linewidth]{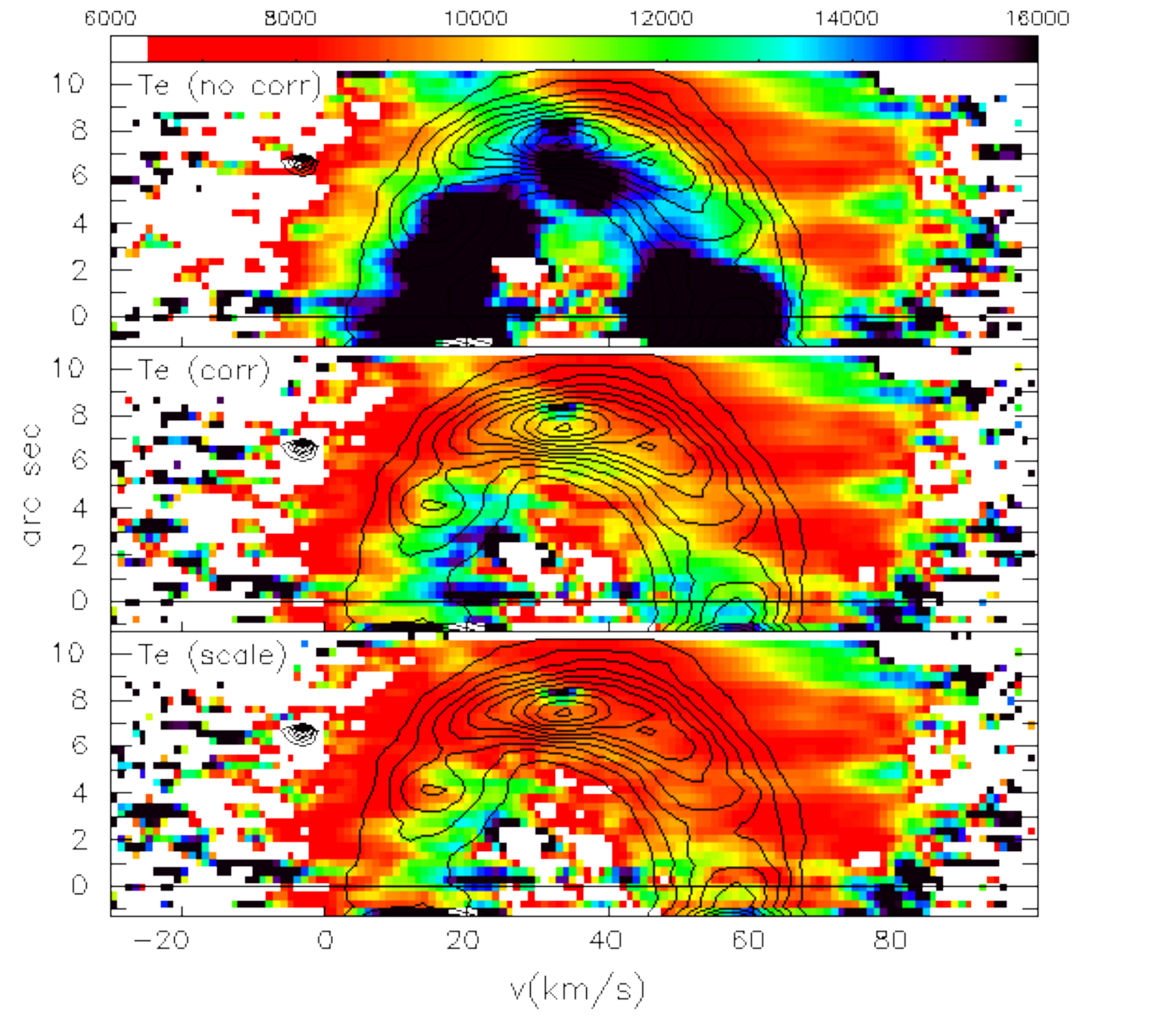}
\caption{These PV diagrams present the [\ion{N}{2}] temperature.  Top panel:  We use the observed PV diagrams of the [\ion{N}{2}] $\lambda\lambda$5755,6583 lines.  Middle panel:  We decompose the [\ion{N}{2}] lines using the atomic data as published (Figure \ref{fig_decompN2_PV_N2_clean}).  Bottom panel:  We scale the recombination coefficients (atomic data: Table \ref{tab_atomic_data}) by a factor of 1.2 when decomposing the [\ion{N}{2}] lines.  In all panels, the temperature scale spans the range $6,000-16,000$\,K and the contours are of the intensity of the \ion{N}{2} $\lambda$5680 line (Figure \ref{fig_decompN2_PV_N2_perm}).  After correcting for the recombination contribution to the [\ion{N}{2}] lines (middle), the [\ion{N}{2}] temperature is a much better approximation of the [\ion{O}{3}] and [\ion{Ar}{3}] temperatures, but may still be too high in the innermost regions.  The [\ion{N}{2}] temperature map in the bottom panel is nearly as uniform as the [\ion{O}{3}] (Figures \ref{fig_forb_TeO3}/\ref{fig_PV_decomp_O3_temp}) and [\ion{Ar}{3}] temperatures (Figure \ref{fig_forb_TeAr3}).  
}
\label{fig_decompN2_PV_N2_temp}
\end{figure}

\begin{figure*}
\begin{center}\includegraphics[width=0.86\linewidth]{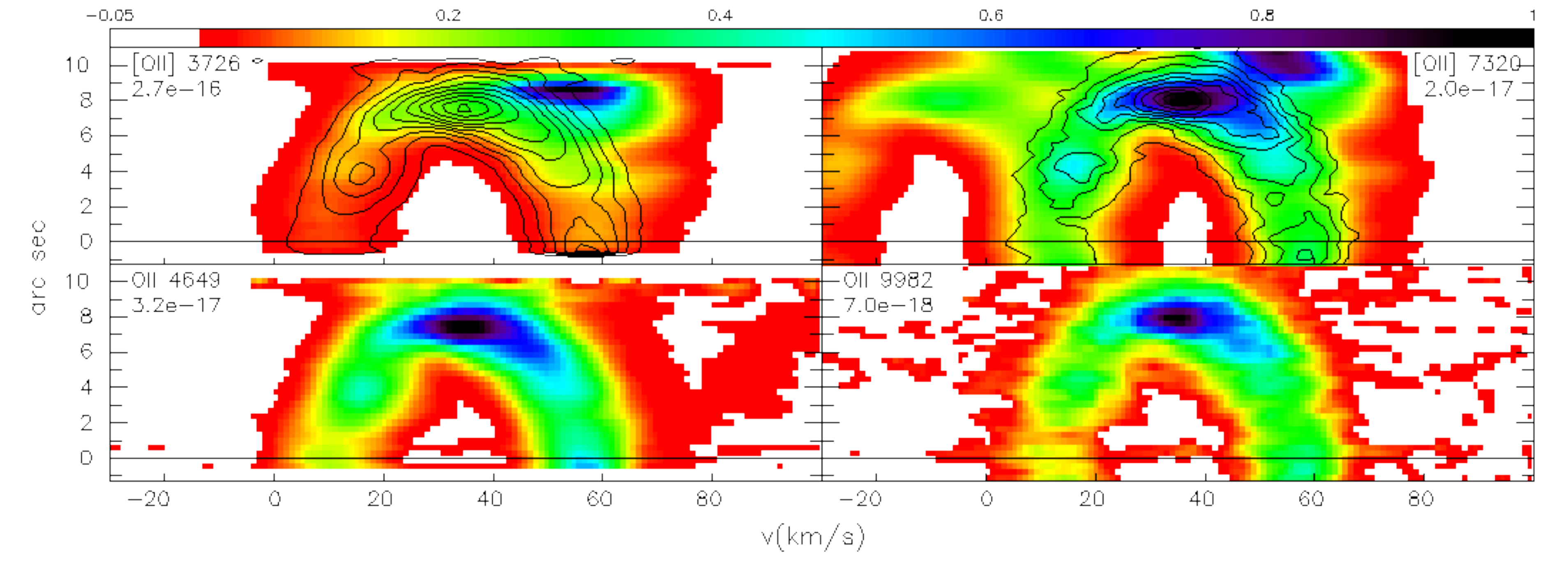}\end{center}
\caption{These PV diagrams present the [\ion{O}{2}] $\lambda$3726 and [\ion{O}{2}] $\lambda\lambda$7319,7320 lines (top row) and the \ion{O}{2} $\lambda$4649 and \ion{O}{2} $\lambda$9982 lines (bottom row).  The contours on the PV diagrams in the top row are of the intensity of the PV diagrams in the bottom row.  The morphology of the PV diagrams for the [\ion{O}{2}] lines are very different.  For the [\ion{O}{2}] $\lambda$7320 line, the main shell is the brightest feature and its emission matches very closely that from \ion{O}{2} $\lambda$9982.  Also, the velocity splitting for [\ion{O}{2}] $\lambda$3726 is greater than that for \ion{O}{2} $\lambda$4649 whereas the velocity splitting for [\ion{O}{2}] $\lambda$7320 is the same as for \ion{O}{2} $\lambda$9982.  
}
\label{fig_PV_O2_cont}
\end{figure*}

First, we consider [\ion{N}{2}] $\lambda$5755.  Figure \ref{fig_decompN2_PV_perm} compares the PV diagrams of the forbidden [\ion{N}{2}] $\lambda\lambda$5755,6583 lines.  
The bright filament is the brightest feature in both PV diagrams, but the contrast between it and the main shell is very different for the two lines, 
with the main shell being relatively much brighter in the [\ion{N}{2}] $\lambda$5755 line.  
The velocity splitting of [\ion{N}{2}] $\lambda$6583 is greater than in the [\ion{N}{2}] $\lambda$5755 line, 52.7\,km/s versus 46.2 km/s.  Finally, there is an excellent correspondence between the emission from the main shell in the [\ion{N}{2}] $\lambda$5755 line and the emission from the permitted \ion{N}{2} $\lambda$5680 line (contours).  All of these lines are from the CD3 wavelength intervals.  The bottom panel in Figure \ref{fig_decompN2_PV_perm} presents the [\ion{N}{2}] $\lambda\lambda$5755/6583 ratio and it is clear that the excess emission in the [\ion{N}{2}] $\lambda$5755 line correlates with the emission from the \ion{N}{2} $\lambda$5680 line (contours).  

Hence, in NGC 6153, the [\ion{N}{2}] $\lambda$5755 line appears to be contaminated due to an additional excitation mechanism whose PV distribution is similar to that of the \ion{N}{2} $\lambda$5680 line.  The \ion{N}{2} $\lambda$5680 line may be excited directly via recombination, but also indirectly by fluorescence in the \ion{He}{1} $\lambda$509 lines in the He$^+$ zone or by starlight in the N$^+$ zone \citep[][]{grandi1976, escalante2002, sharpeeetal2004}.  Were \ion{N}{2} $\lambda$5680 excited primarily by fluorescence, its PV diagram should be very similar to that of [\ion{N}{2}] $\lambda$6584 (Figure \ref{fig_PV_ionization}; N$^+$ zone) or \ion{He}{1} $\lambda$4471 (Figure \ref{fig_app_PVHbeta}; He$^+$ zone), as occurs for \ion{He}{2} $\lambda\lambda$3203,4686 and the Bowen fluorescence lines of \ion{O}{3} $\lambda$3444 and \ion{N}{3} $\lambda$4634 (Figure \ref{fig_app_PVHe2}).  So, in NGC 6153, fluorescence is unlikely to be the main excitation mechanism for the \ion{N}{2} $\lambda$5680 line.  The contamination pattern like \ion{N}{2} $\lambda$5680 in the [\ion{N}{2}] $\lambda$5755 line would thus appear to be due to recombination.  

Since the upper level of the [\ion{N}{2}] $\lambda$5755 line is a singlet state, fluorescence from the ground state (triplet) 
cannot excite it.  
Although the excitation of singlet states via recombination is disfavored relative to the triplet states, 
it is possible \citep{rubin1986, liuetal2000}.  The appearance of contamination with a PV pattern similar to that of a recombination line in [\ion{N}{2}] $\lambda$5755 is therefore not unreasonable.  If so, in the model of two plasma components previously outlined, both components will contribute to the observed PV diagrams of both [\ion{N}{2}] $\lambda\lambda$5755,6583 via recombination while the normal nebular plasma will contribute the collisional excitation, i.e., there are three contributions to the [\ion{N}{2}] $\lambda\lambda$5755,6583 lines.  

The key to decomposing the PV diagrams of [\ion{N}{2}] $\lambda\lambda$5755,6583 is to decompose the observed PV diagram of \ion{N}{2} $\lambda$5680 in terms of the recombination from each of the plasma components.  The parent ion that gives rise to the \ion{N}{2} lines is N$^{2+}$, which occupies a volume in the nebula very similar to O$^{2+}$ \citep[e.g., Figure 9 from][]{richeretal2013}.  Since [\ion{N}{2}] $\lambda$5755 is from the CD3b wavelength interval, we use the PV diagram of [\ion{O}{3}] $\lambda$5007 as a template for the PV pattern of the emissivity of the recombination contribution from the normal nebular plasma.  We estimate the \ion{N}{2} $\lambda$5680 emission due to the normal nebular plasma by scaling the PV diagram of [\ion{O}{3}] $\lambda$5007 by the abundance ratio \citep[we adopt $\mathrm N^{2+}/\mathrm O^{2+} = 0.5$;][]{liuetal2000, mcnabbetal2016}, and then by the ratio of the emissivities of the two lines at 8,000\,K (Figure \ref{fig_emissivities_N2_O2_O3}; adopting the atomic data for $N_e=10,000\,\mathrm{cm}^{-3}$ for \ion{N}{2} $\lambda$5680).  We subtract this model of the recombination contribution from the normal nebular plasma from the observed PV diagram of \ion{N}{2} $\lambda$5680 to obtain the PV diagram of the \ion{N}{2} $\lambda$5680 emission from the additional plasma component.  Mathematically, if,  $I(\lambda)_{x}$ indicates the PV diagram of the emission line at wavelength $\lambda$ due to recombination as observed ($x=o$), from the normal nebular plasma ($x=n$), or from the additional plasma component ($x=a$), the foregoing implies that the PV diagram of the \ion{N}{2} $\lambda$5680 due to the additional plasma component is
$$ I(5680)_{a} = I(5680)_{o} - I(5680)_{n} , $$  
\noindent where
$$ I(5680)_{n} = \frac{N(\mathrm N^{2+})}{N(\mathrm O^{2+})}\frac{\epsilon(5680)_n}{\epsilon(5007)_n} I(5007)_{o} $$
\noindent with $N(X)$ being the abundance of ion $X$ and $\epsilon(\lambda)_x$ the emissivity of line $\lambda$ at the temperatures previously indicated for each plasma component (subscript).  

Figure \ref{fig_decompN2_PV_N2_perm} demonstrates the decomposition of the PV diagram of the \ion{N}{2} $\lambda$5680 line.  Clearly, the contribution from the normal nebular plasma is a small fraction of the total observed emission.  However, its PV emission pattern differs from that of the additional plasma component.  

We use the two PV patterns of the recombination emission in the bottom two panels of Figure \ref{fig_decompN2_PV_N2_perm} to subtract these contributions from the PV diagrams of the [\ion{N}{2}] $\lambda\lambda$5755,6583 lines.  The scale factors are the ratios of the emissivities of the [\ion{N}{2}] $\lambda\lambda$5755,6583 lines with respect to that of the \ion{N}{2} $\lambda$5680 line at the temperatures of the two plasma components.  
For the [\ion{N}{2}] $\lambda$5755 line,  
$$ I(5755)_{a} = \frac{\epsilon(5755)_a}{\epsilon(5680)_a} I(5680)_{a} $$
$$ I(5755)_{n} = \frac{\epsilon(5755)_n}{\epsilon(5680)_n} I(5680)_{n} $$
$$ I(5755)_{c} = I(5755)_{o} - I(5755)_{a} - I(5680)_{n}$$
\noindent where $I(5755)_{c}$ indicates the PV diagram of [\ion{N}{2}] $\lambda$5755 due to collisional excitation only.  (The process for [\ion{N}{2}] $\lambda$6583 is equivalent.)  Figure \ref{fig_decompN2_PV_N2_clean} presents the result.  While the contamination due to recombination is completely negligible for [\ion{N}{2}] $\lambda$6583, this is not the case for [\ion{N}{2}] $\lambda$5755, since the additional plasma component makes a very significant contribution to its PV diagram.  Even the normal nebular plasma contributes noticeable emission due to recombination.  The morphologies of the PV diagrams for the [\ion{N}{2}] $\lambda\lambda$5755,6583 lines in the bottom row are now much more similar.   

Figure \ref{fig_decompN2_PV_N2_temp} presents the PV diagrams of the [\ion{N}{2}] temperature before and aftercorrecting for the recombination contributions to the [\ion{N}{2}] $\lambda\lambda$5755,6583 lines.  Clearly, the result after removing the recombination contributions is much more like the [\ion{O}{3}] and [\ion{Ar}{3}] temperature maps (Figures \ref{fig_forb_TeO3}-\ref{fig_forb_TeAr3}).  Even so, the [\ion{N}{2}] temperature still exceeds the [\ion{O}{3}] and [\ion{Ar}{3}] temperatures in the innermost regions.  The bottom panel in Figure \ref{fig_decompN2_PV_N2_temp} is an experiment to estimate the maximum contribution of recombination to the [\ion{N}{2}] $\lambda\lambda$5755,6583 lines in which the recombination contributions are arbitrarily increased by 20\% with respect to the values based upon the existing atomic data.  The resulting PV diagram of the [\ion{N}{2}] temperature is nearly as uniform as the maps of the [\ion{O}{3}] and [\ion{Ar}{3}] temperatures.  

Therefore, decontaminating the [\ion{N}{2}] lines using the nominal atomic data yields an electron temperature much more similar to those found using the [\ion{O}{3}] or [\ion{Ar}{3}] lines.  Though it might be argued that the nominal correction may underestimate the contamination due to recombination, the basic lesson is that using the raw PV diagrams (or line intensities) of the [\ion{N}{2}] lines leads to erroneous results in NGC 6153.  

The previous procedure differs from that employed by others \citep[e.g.,][]{liuetal2000, corradietal2015, ruizescobedopena2022, garciarojasetal2022}.  Those analyses suppose that the additional plasma component emits all of the permitted emission whereas we find that the additional plasma component contributes 76\% of the total \ion{N}{2} $\lambda$5680 emission (and 84\% of the total \ion{O}{2} $\lambda$4649 emission), which is not a great difference.  Our method may be applied when two-dimensional data are available, whether in PV or spatial coordinates.  

Turning now to [\ion{O}{2}], Figure \ref{fig_PV_O2_cont} compares the PV diagrams of [\ion{O}{2}] $\lambda$3726, [\ion{O}{2}] $\lambda\lambda$7319,7320, and \ion{O}{2} $\lambda\lambda$4649,9982.  The morphologies of the PV diagrams of the [\ion{O}{2}] lines are very different (considering only one line of the [\ion{O}{2}] $\lambda$7319,7320 doublet).  For [\ion{O}{2}] $\lambda$3726, the filament on the receding side of the main shell is by far the brightest feature, but, in the auroral lines, the brightest feature is the brightest part of the main shell!  The contour lines superposed on both panels are those of the \ion{O}{2} lines with the same spatial coverage and they are an excellent approximation to the emissivity of the main shell in [\ion{O}{2}] $\lambda$7320.  This suggests that, like [\ion{N}{2}] $\lambda$5755, the [\ion{O}{2}] $\lambda\lambda$7319,7320 lines suffer very significant contamination due to recombination.  

\begin{figure*}
\begin{center}\includegraphics[width=0.86\linewidth]{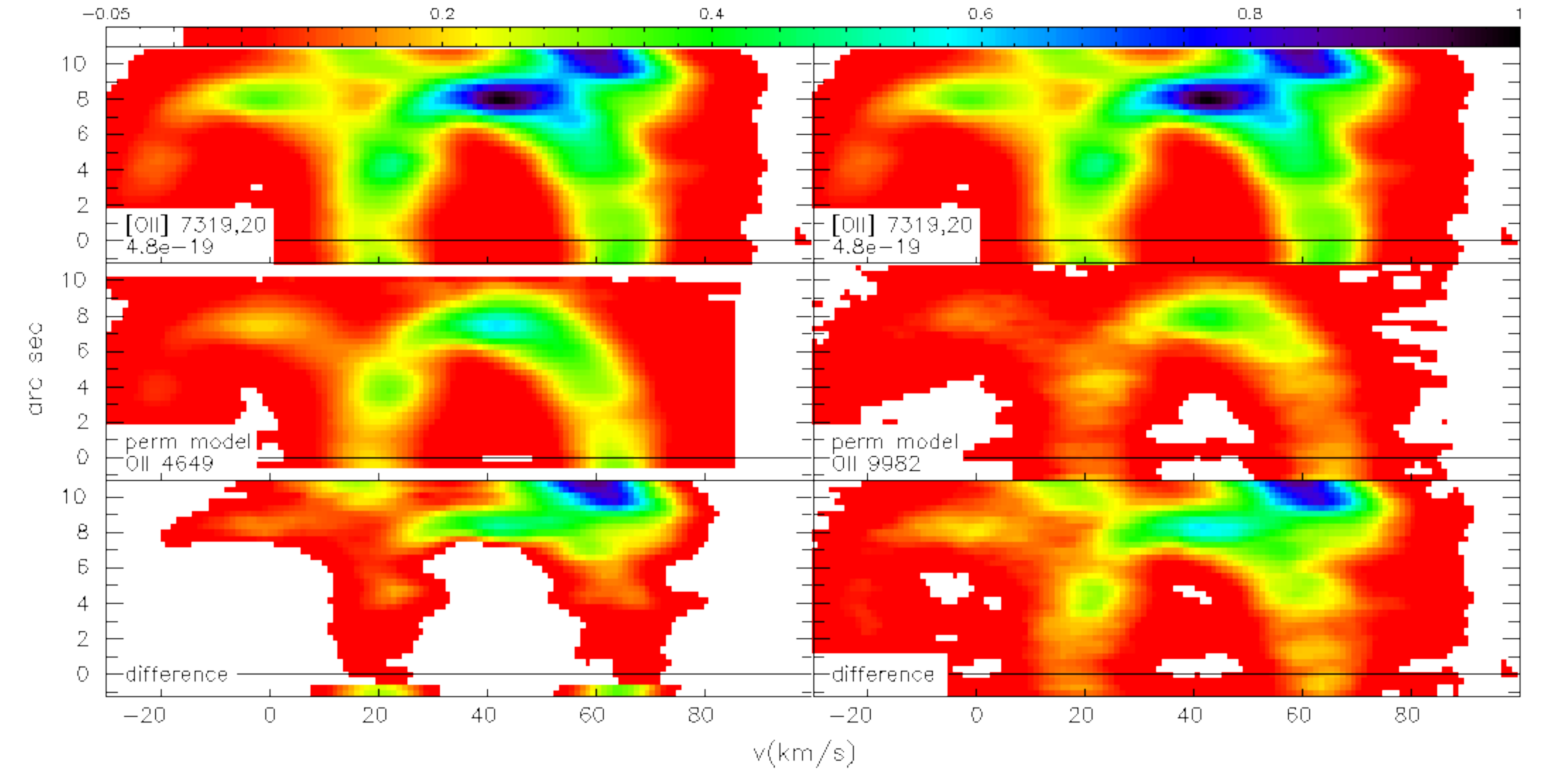}\end{center}
\caption{These panels present the decomposition of the [\ion{O}{2}] $\lambda\lambda$7319,7320 line profiles (top row, both columns).  The middle row presents the total permitted emission in both lines using \ion{O}{2} $\lambda$4649 (left) and \ion{O}{2} $\lambda$9982 (right) as models.  The bottom row is the difference between the observations (top) the models (middle).  All panels share a common intensity scale.  The model based upon the \ion{O}{2} $\lambda$4649 line accounts for \emph{almost all} of the emission from the main shell in the [\ion{O}{2}] $\lambda\lambda$7319,7320 lines.  
}
\label{fig_decompO2_PV_O2_forb}
\end{figure*}

We model the [\ion{O}{2}] $\lambda\lambda$7319,7320 lines since their critical densities for collisional de-excitation are well above the densities found for the normal nebular plasma in NGC 6153.  (This is not the case for [\ion{O}{2}] $\lambda\lambda$3726,3729.)  We begin by decomposing the \ion{O}{2} $\lambda\lambda$4649,9982 lines to obtain the recombination contributions from the two plasma components, as we did for [\ion{N}{2}] $\lambda$5680.  We use the [\ion{O}{3}] $\lambda\lambda$4959,5007 lines to model the recombination contribution from the normal nebular plasma (emissivities in Figure \ref{fig_emissivities_N2_O2_O3}).  We use [\ion{O}{3}] $\lambda$5007 to match the spatial extent of the \ion{O}{2} $\lambda$9982 line, though it was not obtained simultaneously with the other lines and may suffer from a slightly different slit placement.
As for \ion{N}{2} $\lambda$5680, the additional plasma component contributes the majority of the emission due to recombination.  
We construct models of the recombination contribution to each of the [\ion{O}{2}] $\lambda\lambda$7319,7320 lines based upon each of the \ion{O}{2} $\lambda\lambda$4649,9982 lines.  
We subtract the sum of the models from the observed PV diagram of the [\ion{O}{2}] $\lambda\lambda$7319,7320 lines to obtain the PV diagram of the emission due to collisional excitation only.  

Figure \ref{fig_decompO2_PV_O2_forb} presents the result of this decomposition for the [\ion{O}{2}] $\lambda\lambda$7319,7320 lines.  The observed PV diagram is in the top row (both columns).  The middle row presents the model of the permitted emission, based upon the \ion{O}{2} $\lambda$4649 (left) and \ion{O}{2} $\lambda$9982 lines (right).  The two models differ, with that based upon \ion{O}{2} $\lambda$9982 having approximately two thirds of the flux of the model based upon \ion{O}{2} $\lambda$4649.  The models also differ because the shapes of the \ion{O}{2} $\lambda$4649 and \ion{O}{2} $\lambda$9982 PV diagrams differ slightly, with the latter being somewhat ``taller", likely due to the longer slit used in the red arm of the spectrograph.  The taller \ion{O}{2} $\lambda$9982 line profile is a better match to the shape of the [\ion{O}{2}] $\lambda$7319,7320 PV diagram.  

The bottom row in Figure \ref{fig_decompO2_PV_O2_forb} presents the difference between the observed emission and the modeled permitted emission, again using the models based upon the \ion{O}{2} $\lambda$4649 and \ion{O}{2} $\lambda$9982 lines.  The PV diagrams in the bottom row in Figure \ref{fig_decompO2_PV_O2_forb} should include only the emission due to collisional excitation.  Evidently, collisional excitation dominates the [\ion{O}{2}] $\lambda\lambda$7319,7320 emission from the filament, but recombination provides the overwhelming majority of the [\ion{O}{2}] $\lambda\lambda$7319,7320 emission from the main shell itself, as expected given the comparison of the PV diagrams for [\ion{O}{2}] $\lambda$3726 and [\ion{O}{2}] $\lambda\lambda$7319,7320 in Figure \ref{fig_PV_O2_cont}.  The PV diagram of the [\ion{O}{2}] $\lambda\lambda$7319,7320 at bottom left is very similar to the PV diagrams of the [\ion{N}{2}] $\lambda$6583, [\ion{S}{2}] $\lambda$6716 or [\ion{Cl}{2}] $\lambda\lambda$8578,9123 lines (Figure \ref{fig_app_PVO2}).  

On the other hand, the PV diagram of the [\ion{O}{2}] $\lambda\lambda$7319,7320 lines after subtracting the model based upon \ion{O}{2} $\lambda$9982 (Figure \ref{fig_decompO2_PV_O2_forb}, bottom right) still has significantly more emission from the main shell than those of the [\ion{N}{2}] $\lambda$6583, [\ion{S}{2}] $\lambda$6716 or [\ion{Cl}{2}] $\lambda\lambda$8578,9123 lines (Figure \ref{fig_app_PVO2}).  (The emission in the filament is unaffected.)  In order to obtain a PV diagram for [\ion{O}{2}] $\lambda\lambda$7319,7320 similar to those using the model of the permitted emission based upon \ion{O}{2} $\lambda$9982, the emissivity for this line from \citet{storeyetal2017} should be multiplied by a factor of about 1.5.  \citet{storeyetal2017} note the limitations of their calculations for the $2\mathrm s^22\mathrm p^2\,(^3\mathrm P)4f$ and $5f$ configurations.  Since the upper level of the \ion{O}{2} $\lambda$9982 transition, $(^3\mathrm P_2)\,5g[6]_{13/2,11/2}$ \citep{wenaker1990}, is even higher in energy, within 2.2\,eV of the ionization limit (O$^{2+}\ ^3\mathrm P_2$), a discrepancy of $\sim 50\%$ in its emissivity is perhaps not unexpected. 

We suppose that the decontamination of the [\ion{O}{2}] $\lambda\lambda$7319,7320 lines using the model based upon \ion{O}{2} $\lambda$4649 (Figure \ref{fig_decompO2_PV_O2_forb}) is the more reliable result.  However, the most important lesson is that recombination contributes the majority of the emission in the [\ion{O}{2}] $\lambda\lambda$7319,7320 lines for the main shell of NGC 6153 regardless of which model we use.  Hence, were the [\ion{O}{2}] $\lambda\lambda$7319,7320 emission to be analyzed as if due to collisional excitation, the result would be erroneous for the main shell (but not the filament).  Although Figure \ref{fig_decompO2_PV_O2_forb} concerns the [\ion{O}{2}] $\lambda\lambda$7319,7320 lines, recombination undoubtedly contributes to the emission in the [\ion{O}{2}] $\lambda\lambda$3726,3729 lines as well.  All of the recombination contribution to the [\ion{O}{2}] $\lambda\lambda$7319, 7320, 7330, 7331 lines will contribute to the excitation of the [\ion{O}{2}] $\lambda\lambda$3726,3729 lines as will direct recombination to the upper levels of the [\ion{O}{2}] $\lambda\lambda$3726,3729 lines, as has been argued elsewhere \citep[e.g.,][]{liuetal2000, wessonetal2018}.  

\begin{figure}[]
\includegraphics[width=\linewidth]{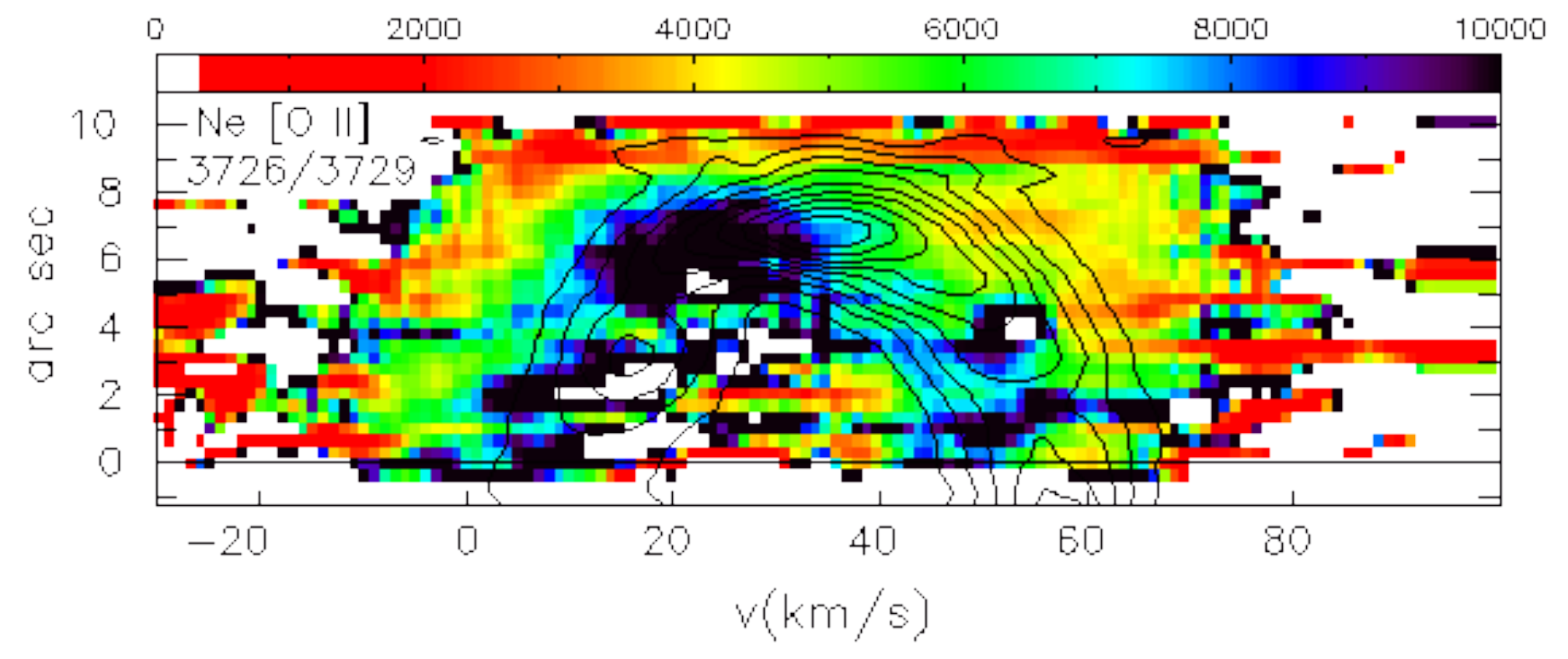}
\caption{We present the PV diagram of the density derived from the [\ion{O}{2}] $\lambda\lambda$3726,3729 lines.  The contours are those of the intensity of the emission in the \ion{O}{2} $\lambda$4649 line (see Figure \ref{fig_PVC2}).  Overall, the density implied by the [\ion{O}{2}] lines is higher than implied by the [\ion{S}{2}], [\ion{Cl}{3}], or [\ion{Ar}{4}] lines.  In addition, there is a similar structure of high densities where the [\ion{N}{2}] $\lambda\lambda$5755/6583 ratio is high (Figure \ref{fig_decompN2_PV_perm}).   
}
\label{fig_PV_Ne_O2forb}
\end{figure}

\begin{figure*}
\begin{center}\includegraphics[width=0.86\linewidth]{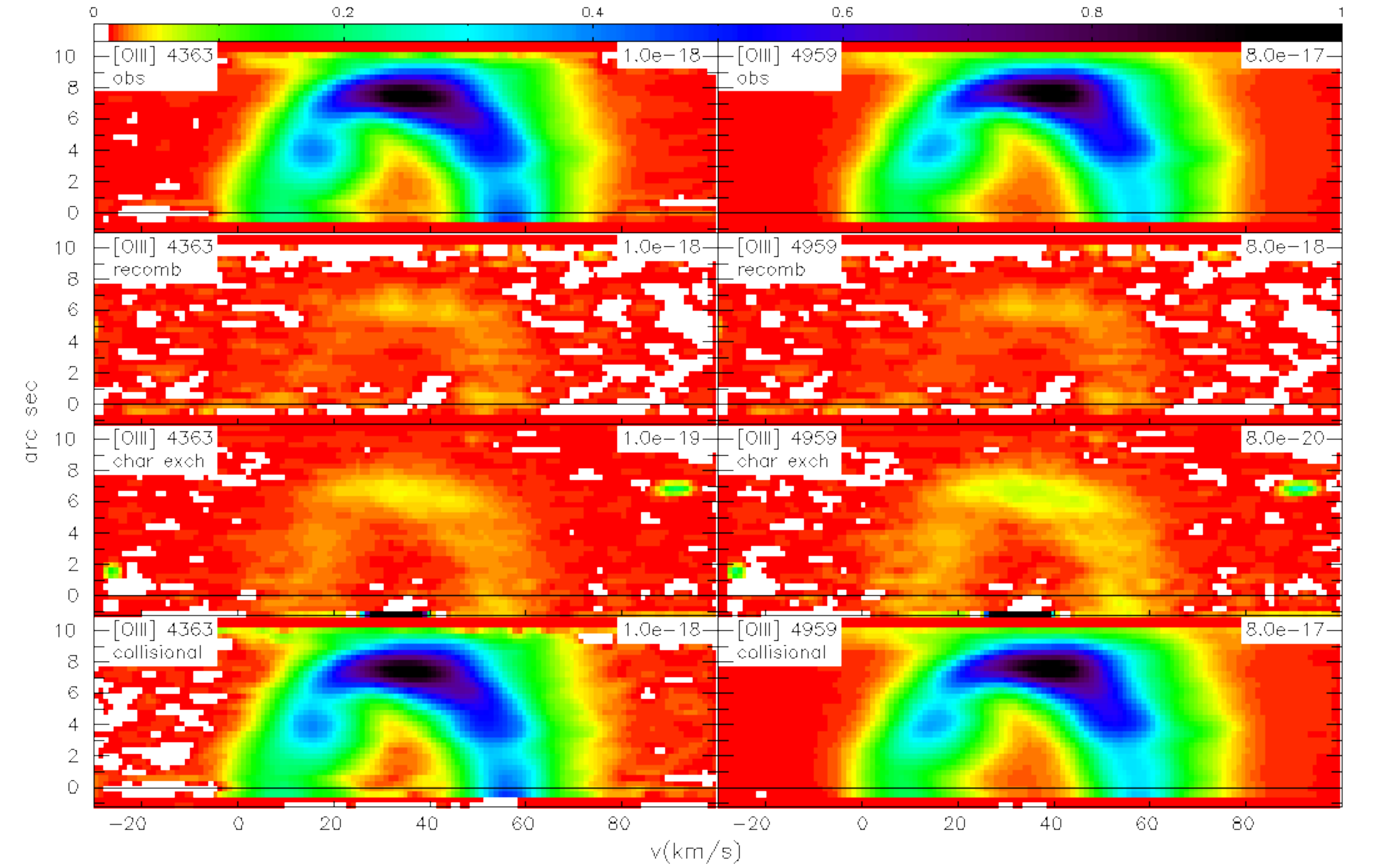}\end{center}
\caption{These panels present PV diagrams for [\ion{O}{3}] $\lambda$4363 (left) and [\ion{O}{3}] $\lambda$4959 (right) of the observed emission (top), the contamination due to recombination (second), the contamination due to charge exchange (third), and the emission due to collisional excitation (bottom).  The PV diagrams in the bottom row are the result of subtracting those in the second and third rows from the PV diagrams in the top row.  Since the emission due to recombination and charge exchange is so faint, the normalization value for the intensity/color scale varies in the middle two rows by factors of $10-1,000$ with respect to the top and bottom rows.  While recombination makes a small contribution to the [\ion{O}{3}] $\lambda\lambda$4363,4959 PV diagrams, charge exchange is much less relevant.  
}
\label{fig_PV_decomp_O3}
\end{figure*}

\begin{figure}
\includegraphics[width=\linewidth]{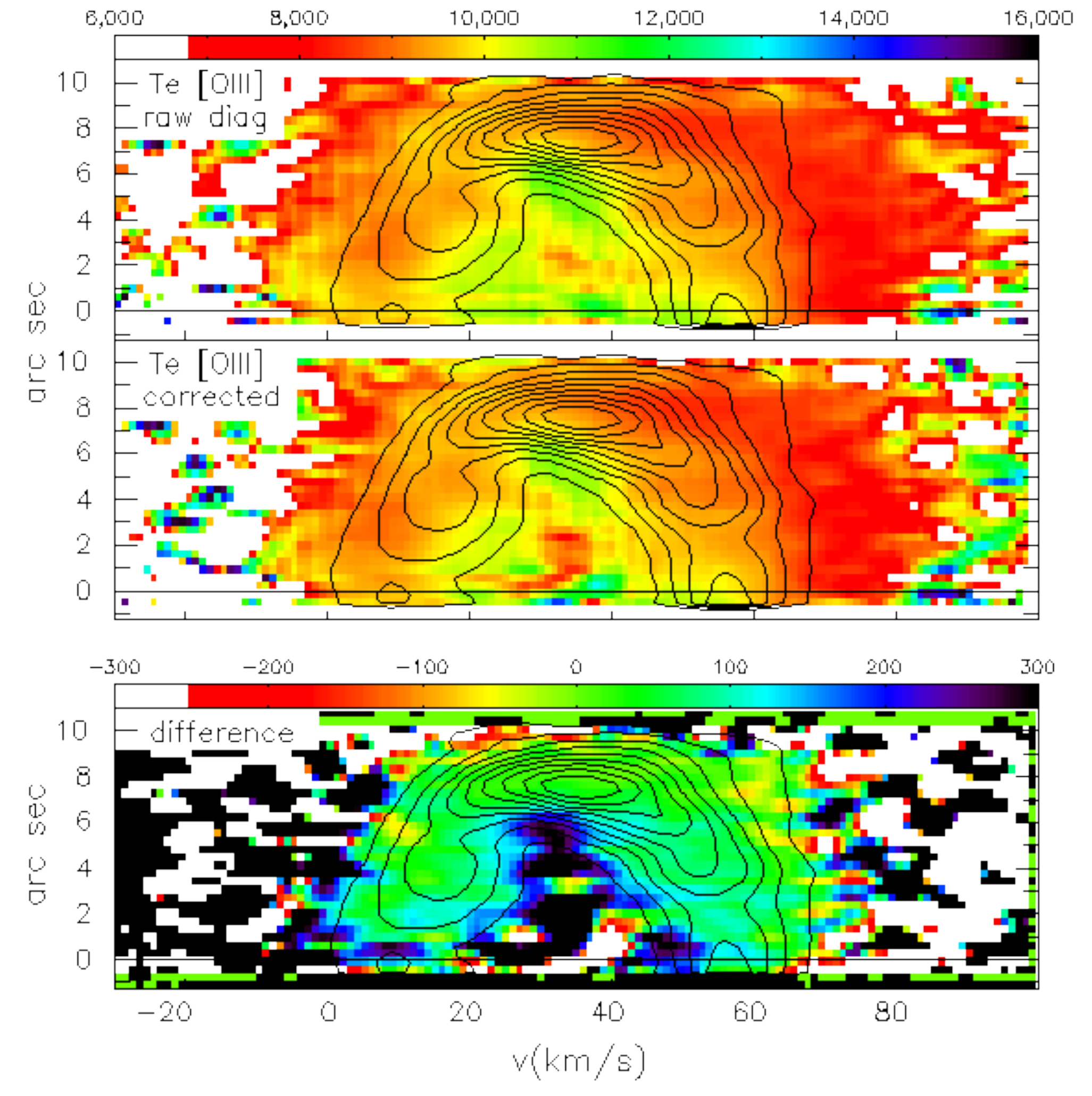}
\includegraphics[width=\linewidth]{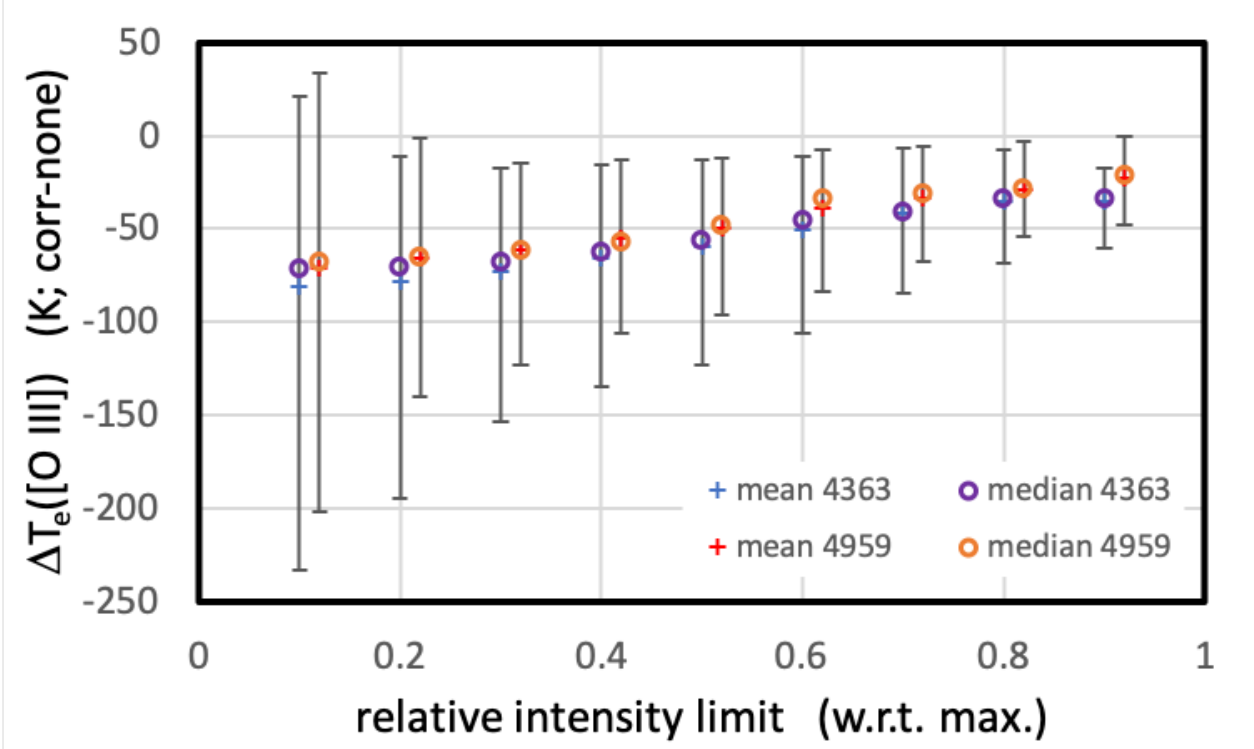}
\caption{The top two panels present the [\ion{O}{3}] electron temperature using the raw PV diagrams and those corrected for contamination due to recombination and charge exchange.  The third panel presents the PV diagram of the difference map between the corrected and raw temperature maps.  In all panels, the contours are of the [\ion{O}{3}] $\lambda$4363 intensity.  At bottom, we plot the mean and median for the difference map as a function of the relative intensity in the [\ion{O}{3}] $\lambda$4363 line.  The error bars span from the 5-th to 95-th percentiles.
}
\label{fig_PV_decomp_O3_temp}
\end{figure}

Even though the [\ion{O}{2}] $\lambda\lambda$3726,3729 lines are contaminated by recombination excitation, they remain a useful density indicator.  However, the [\ion{O}{2}] $\lambda\lambda$3726/3729 ratio will sense the density in the two plasma components.  
Given the low critical densities for the [\ion{O}{2}] $\lambda\lambda$3726,3729 lines, $\sim 1,000$ and 4,000\,cm$^{-3}$ at 9,000\,K, respectively (atomic data: Table \ref{tab_atomic_data}), the [\ion{O}{2}] $\lambda\lambda$3726,3729 lines will suffer more 
collisional de-excitation in the additional nebular plasma.  Where a given plasma component dominates the emission in these lines in the PV diagram, the [\ion{O}{2}] $\lambda\lambda$3726/3729 ratio will reflect the density in that plasma component.  Necessarily, there will be PV coordinates where the two plasma components will contribute and the density will not represent either component.  Figure \ref{fig_PV_Ne_O2forb} presents the density computed from the ratio of these lines, assuming an electron temperature of 10,000\,K.  Its structure is more complicated than the PV diagrams for the densities computed from the [\ion{S}{2}], [\ion{Cl}{3}], [\ion{Ar}{4}], \ion{N}{2}, and \ion{O}{2} lines.  However, it is possible to recognize that, at PV coordinates corresponding to the outer part of the main shell, the density is near 5,000\,cm$^{-3}$, as for the other forbidden lines, and, for the inner part of the main shell, the densities reach 10,000\,cm$^{-3}$, compatible with the results from the \ion{N}{2} and \ion{O}{2} lines.  

Finally, we consider the PV diagrams of [\ion{O}{3}] $\lambda\lambda$4363,4959.  These emission lines are the result of three physical processes.  The dominant process is collisional excitation, but recombination and charge exchange also contribute.  So, we must model and subtract the latter two contributions from the observed PV diagrams to yield PV diagrams for the [\ion{O}{3}] $\lambda\lambda$4363,4959 lines due to collisional excitation only.  We attribute the contributions from both recombination and charge exchange to the normal nebular plasma, since we assume that the additional plasma component does not emit in \ion{O}{3} lines (\S\ref{sec_ionization_structure}, Appendix \ref{app_ionization_structure}).  

We model the charge exchange contribution using the \ion{O}{3} $\lambda$5592 line.  The upper level of the \ion{O}{3} $\lambda$5592 line will be dominantly excited by charge exchange, though recombination also contributes a minority ($< 10$\%; atomic data: Table \ref{tab_atomic_data}).  The lower level of the \ion{O}{3} $\lambda$5592 line decays to the upper levels of the [\ion{O}{3}] $\lambda$4363 or [\ion{O}{3}] $\lambda\lambda$4959,5007 lines \citep[for a useful Grotrian diagram, see][]{dalgarnosternberg1989}.  The upper level may also decay directly to the upper levels of the [\ion{O}{3}] $\lambda$4363 or [\ion{O}{3}] $\lambda\lambda$4959,5007 lines and it may also decay indirectly to these levels via other intermediate levels.  

Since we use the observed PV diagram for \ion{O}{3} $\lambda$5592, the scaling to account for the contribution of charge exchange to [\ion{O}{3}] $\lambda$4363 requires only the branching ratio from the upper and lower levels of the \ion{O}{3} $\lambda$5592 line (the Einstein A-values) for populating the upper level of the [\ion{O}{3}] $\lambda\lambda$4363,4959 lines, i.e., the physical conditions are not involved (atomic data: Table \ref{tab_atomic_data}).  We find that the excitation to the upper levels of both lines is at most 48\% of the intensity of the \ion{O}{3} $\lambda$5592 line.  

We use the PV diagram of the \ion{O}{3} $\lambda\lambda$3260,3265 lines as the emission pattern for the recombination contribution to the [\ion{O}{3}] $\lambda\lambda$4959,5007 lines (atomic data: Table \ref{tab_atomic_data}), as all of the bright \ion{O}{3} lines are enhanced by Bowen fluorescence (often dominated by it) and so cannot be used.  However, the \ion{O}{3} $\lambda\lambda$3260,3265 lines are just to the blue of our flux calibration limit at 3300\,\AA.  Since the \ion{He}{2} $\lambda$3203 line over-estimates the $\mathrm{He}^{2+}/\mathrm H^+$ ionic abundance (Table \ref{tab_HeII_int_abun}), the \ion{O}{3} $\lambda\lambda$3260,3265 lines may over-estimate the recombination contribution to the [\ion{O}{3}] lines.  

Figure \ref{fig_PV_decomp_O3} presents the results for [\ion{O}{3}] $\lambda\lambda$4363,4959.  Clearly, the contamination due to recombination and charge exchange is minor.  Only for [\ion{O}{3}] $\lambda$4363 is contamination noticeable, and only in the case of recombination.  Although the effect of this contamination upon the total emission from [\ion{O}{3}] $\lambda$4363 is minor, it affects the ``inside" of the line profile where the flux is lower.  

Figure \ref{fig_PV_decomp_O3_temp} presents the [\ion{O}{3}] electron temperature before and after correcting for the contamination due to recombination and charge exchange.  Figure \ref{fig_PV_decomp_O3_temp} also presents the change in temperature and, as expected, it is largest on the ``inside" of the line profile, near the systemic velocity and at spatial positions near the central star.  Even so, the change is small, amounting to a decrease in the electron temperature of up to 200\,K.  The final panel in Figure \ref{fig_PV_decomp_O3_temp} presents the change in temperature over the area covered by the contours at 10\%, 20\%, ..., 90\% of the maximum intensity of the [\ion{O}{3}] $\lambda$4363 line.  The biggest change occurs for the areas enclosed by the lower limiting fluxes.  These areas also show the widest range of change.  This occurs because these areas of the PV diagram correspond to larger nebular volumes within which there is a greater range of electron temperature, 
including both the hotter interior of the main shell and the lower temperatures in the diffuse emission at the most redshifted velocities.  

In NGC 6153, the effect of recombination and charge exchange upon the electron temperature deduced from the [\ion{O}{3}] $\lambda\lambda$4363,4959 lines is small to negligible.  
In particular, the central volume of the nebula (near the systemic velocity and near the central star) appears to be hotter than the rest.  This is very likely due to the extra heating as a result of the presence of He$^{2+}$ ions.  Since the appearance of He$^{2+}$ is accompanied by the disappearance of O$^{2+}$, the most important source of nebular cooling, there is good reason to accept that the electron temperature really does increase in the central volume of the nebula.

\section{Discussion:  $T_e$ and consequences} \label{sec_discussion}

\begin{deluxetable*}{lcll}
\tablecaption{Summary of main results\label{tab_main_results}}
\tablewidth{0pt}
\tablehead{
\colhead{Parameter} & \colhead{Details} & \colhead{Indicator} & \colhead{Result} 
}
\decimalcolnumbers
\startdata
$ E(B-V)$ & \S\ref{sec_reddening} & \ion{H}{1} and \ion{He}{1} lines & $0.535\pm 0.053$\,mag \\
temperature & \S\ref{sec_kinematic_temperature} & kinematics & $\sim 8000$\,K \\
            & \S\ref{sec_contamination} & [\ion{N}{2}] $\lambda\lambda$5755/6583 & $\sim 9,000$\,K; contaminated by recombination \\
            & \S\ref{sec_physcond_forb} & [\ion{O}{3}] $\lambda\lambda$4363/4959 & $\sim 9,000$\,K, higher in inner main shell \\
            & \S\ref{sec_physcond_forb} & [\ion{S}{3}] $\lambda\lambda$6312/9069 & unusable due to telluric absorption \\
            & \S\ref{sec_physcond_forb} & [\ion{Ar}{3}] $\lambda\lambda$5191/7751 & $\sim 9,000$\,K, higher in inner main shell \\
            & \S\ref{sec_physcond_perm} & \ion{He}{1} lines & $8,000-12,000$\,K, rather uniform \\
            & \S\ref{sec_physcond_perm} & \ion{O}{2} $\lambda\lambda$4089/4649 & $1,800-5,000$\,K, little variation  \\
            & \S\ref{sec_physcond_perm} & \ion{N}{2} $\lambda\lambda$4041/5680 & $> 2,500$\,K \\
            & \S\ref{sec_temp_fluc} & Peimbert (1967) $t^2$ & $T_0\sim 8000$\,K and $t^2\sim 0.03$ in large volume \\
density & \S\ref{sec_contamination} & [\ion{O}{2}] $\lambda\lambda$3726/3729 & $\sim 5,000-10,000$\,cm$^{-3}$; contaminated by recombination \\
        & \S\ref{sec_physcond_forb} & [\ion{S}{2}] $\lambda\lambda$6716/6731 & $\sim 2,000$\,cm$^{3}$ at edges, $5,000-6,000$\,cm$^{-3}$ closest to centre\\
        & \S\ref{sec_physcond_forb} & [\ion{Cl}{3}] $\lambda\lambda$5517/5537 & $5,000-6,000$\,cm$^{-3}$, rather uniform \\
        & \S\ref{sec_physcond_forb} & [\ion{Ar}{4}] $\lambda\lambda$4711/4740 & $3,000-4,000$\,cm$^{-3}$ \\
        & \S\ref{sec_physcond_perm} & high Balmer lines & $< 10,000$\,cm$^{-3}$ \\
        & \S\ref{sec_physcond_perm} & \ion{He}{1} lines & $< 10,000$\,cm$^{-3}$ \\
        & \S\ref{sec_physcond_perm} & \ion{O}{2} $\lambda\lambda$4662/4649 & $> 5,000$\,cm$^{-3}$, rather uniform \\
        & \S\ref{sec_physcond_perm} & \ion{N}{2} $\lambda\lambda$5666/5680 & $> 10,000$\,cm$^{-3}$, rather uniform \\
ADF(O$^{2+}$) & \S\ref{sec_PV_ADF} & ORL/CEL & $\mathrm{ADF}(\mathrm O^{2+})$ up to 12 in main shell, $\sim 1.2$ in diffuse emission \\
N$^{++}$ mass fraction & \S\ref{sec_rel_mass_add_comp} & additional component & $23-53$\%\,@\,10,000\,cm$^{-3}$; $3-10$\%\,@\,100,000\,cm$^{-3}$ \\                                 
O$^{++}$ mass fraction & \S\ref{sec_rel_mass_add_comp} & additional component & $30-62$\%\,@\,10,000\,cm$^{-3}$; $4-13$\%\,@\,100,000\,cm$^{-3}$ \\                                 
H$^+$ mass fraction    & \S\ref{sec_H_mass_additional} & additional component & $3-5$\% \\
\enddata
\end{deluxetable*}

\begin{figure*}
\includegraphics[width=0.49\linewidth]{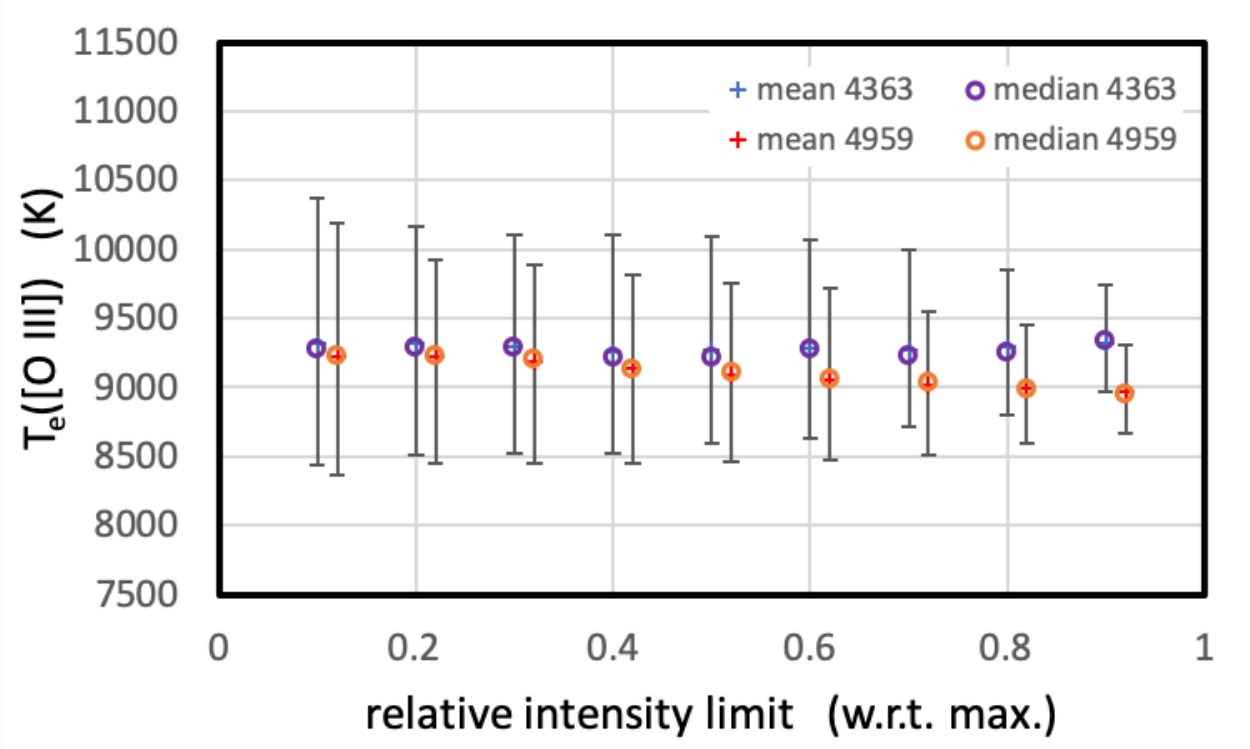}
\includegraphics[width=0.49\linewidth]{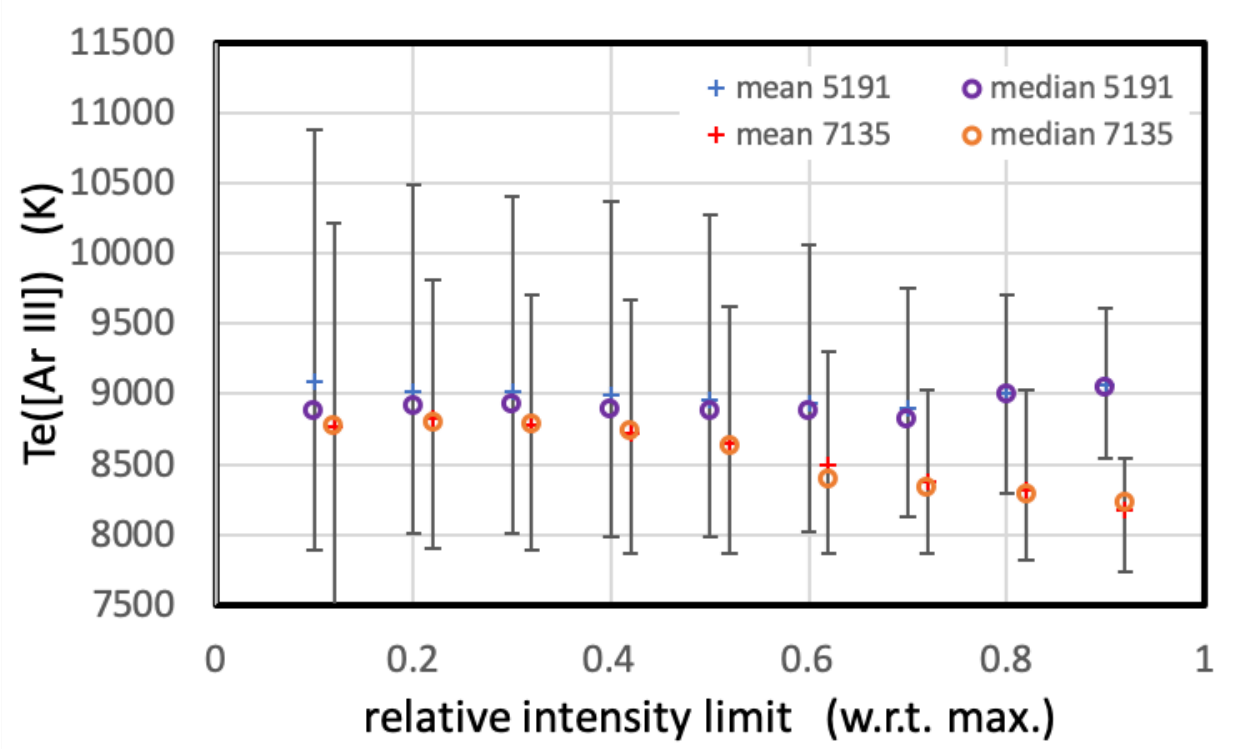}
\caption{The left panel presents mean and median values of the [\ion{O}{3}] temperature as a function of the limiting intensities of [\ion{O}{3}] $\lambda$4363 or [\ion{O}{3}] $\lambda$4959 with respect to the maximum value for each line.  The error bars represent the extent of the distributions from the 5th to 95th percentiles.  The right panels presents the same information for the [\ion{Ar}{3}] temperatures as a function of the line intensity limit for the [\ion{Ar}{3}] $\lambda\lambda$5191,7135 lines.  In both cases, 
there is a systematic difference depending upon whether they are weighted by the intensity of the nebular or auroral line, with the values weighted using the auroral line being systematically higher.  
}
\label{fig_forb_TeO3_TeAr3_limit}
\end{figure*}
mgro.astroph

Although it is not a direct result, our most notable finding is that the plasma in NGC 6153 is complex.  Whether we consider the kinematics or the physical conditions in the nebular shell, we find complexity.  Usually, there are complementary and congruent lines of evidence for this complexity.  For instance, both the kinematics and the physical conditions imply the existence of two plasma components.  
In spite of the complexity, we also find very strong support for the conventional astrophysics of nebular plasmas.  The physical conditions implied by the \ion{H}{1}, \ion{He}{1}, and forbidden lines are generally congruent, as are the kinematics found for the \ion{H}{1}, \ion{He}{1}, and [\ion{O}{3}] $\lambda$4959 lines.  The kinematics of the great majority of the emission lines, both permitted and forbidden, conform to the canonical results of \citet{wilson1950}.  

Table \ref{tab_main_results} collects the physical conditions computed in the previous section for easier comparison.  This table also includes results from the subsections that follow.  As will become clear, the theme of complexity becomes more and more apparent.

\subsection{Variations in the electron temperatures}\label{sec_temp_fluc}

The two panels in Figure \ref{fig_forb_TeO3_TeAr3_limit} quantify the variation of the [\ion{O}{3}] and [\ion{Ar}{3}] temperatures (Figures \ref{fig_forb_TeO3} and \ref{fig_forb_TeAr3}, respectively).  The left panel presents the mean and median [\ion{O}{3}] temperature computed based upon different limiting intensities of the [\ion{O}{3}] $\lambda\lambda$4363,4959 lines.  To compute these values, only the temperatures from the pixels whose intensities exceed the indicated intensity relative to the maximum intensity are used.  For instance, for a relative intensity limit of 0.1, only the non-zero temperature values from the pixels whose line intensities exceed 10\% of the maximum intensity for the indicated line are used ([\ion{O}{3}] $\lambda\lambda$4363,4959 in the left panel).  The mean and median values of the temperatures are similar, indicating that the temperature distributions are approximately symmetric.  
For a given line, either [\ion{O}{3}] $\lambda$4363 or [\ion{O}{3}] $\lambda$4959, the variation of the mean and median temperatures as a function of the limiting line intensity is modest. 
The right panel in Figure \ref{fig_forb_TeO3_TeAr3_limit} presents the same information for the [\ion{Ar}{3}] $\lambda\lambda$5191,7135 lines.  (The analogous figure for the temperature based upon the [\ion{Ar}{3}] $\lambda\lambda$5191,7751 lines is very similar.)  

The mean or median temperatures at different limiting line intensities measure the temperature over very different volumes of the nebula.  At the 10\% limiting intensity, plasma throughout the entire volume that emits the [\ion{O}{3}]/[\ion{Ar}{3}] lines is included, but, at the 90\% limiting intensity, only part of the plasma seen along the line of sight towards the edge of the nebula's main shell is included.  Thus, the wider distribution of temperature values at lower limiting intensities is not due primarily to signal-to-noise, but to the wider range of physical conditions that occur in the larger volumes included by the lower limiting intensity limits.  

In both panels of Figure \ref{fig_forb_TeO3_TeAr3_limit}, there is a systematic variation in the difference between the mean/median temperature weighted by the auroral and nebular lines, increasing as the limiting intensity increases.  The mean/median values based upon the intensity of the auroral lines ([\ion{O}{3}] $\lambda$4363, [\ion{Ar}{3}] $\lambda$5191) are systematically higher than those based upon the intensity of the nebular lines ([\ion{O}{3}] $\lambda$4959, [\ion{Ar}{3}] $\lambda$7135).  Since the excitation energy for the auroral lines is more than double that for the nebular lines, it makes sense that the temperatures weighted by the auroral lines are the larger ones since they are excited more easily where the temperature is higher and so will be biased to regions of higher temperature.  

\begin{figure}
\includegraphics[width=\linewidth]{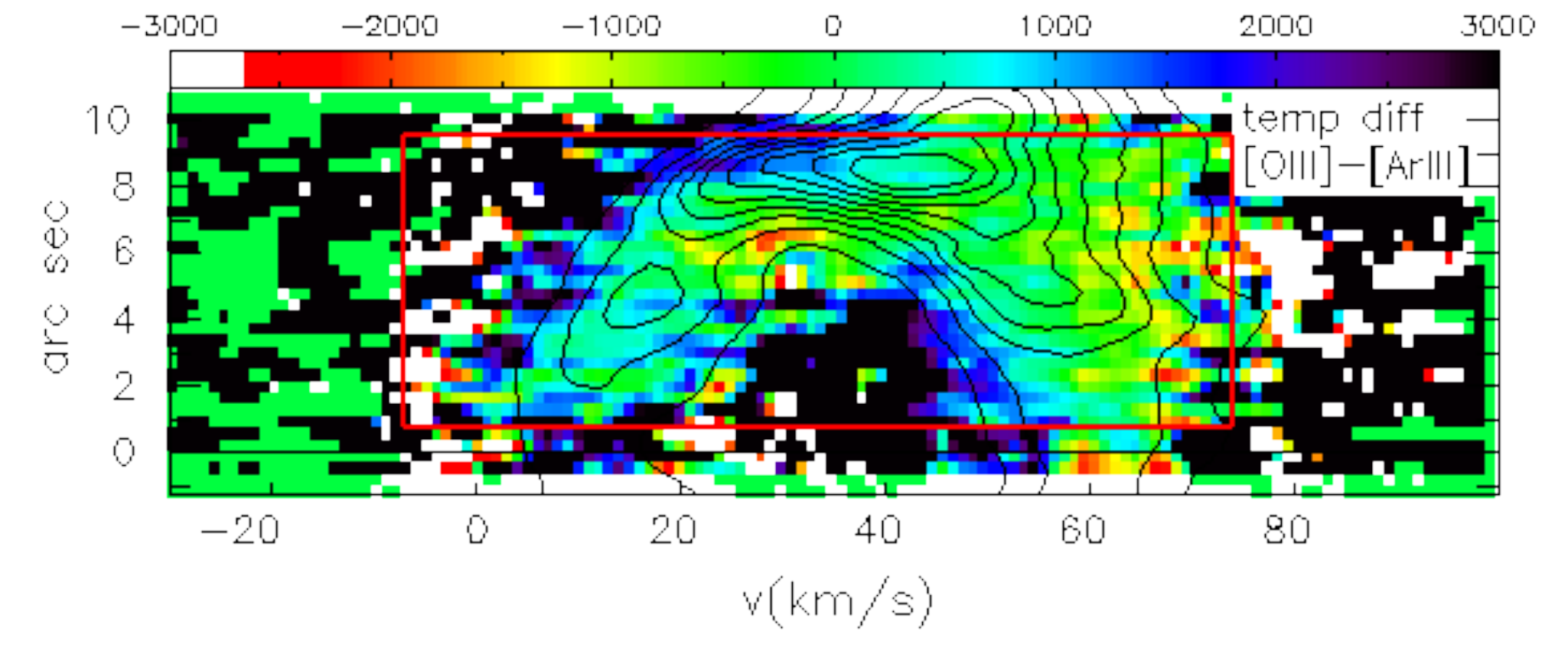}
\includegraphics[width=\linewidth]{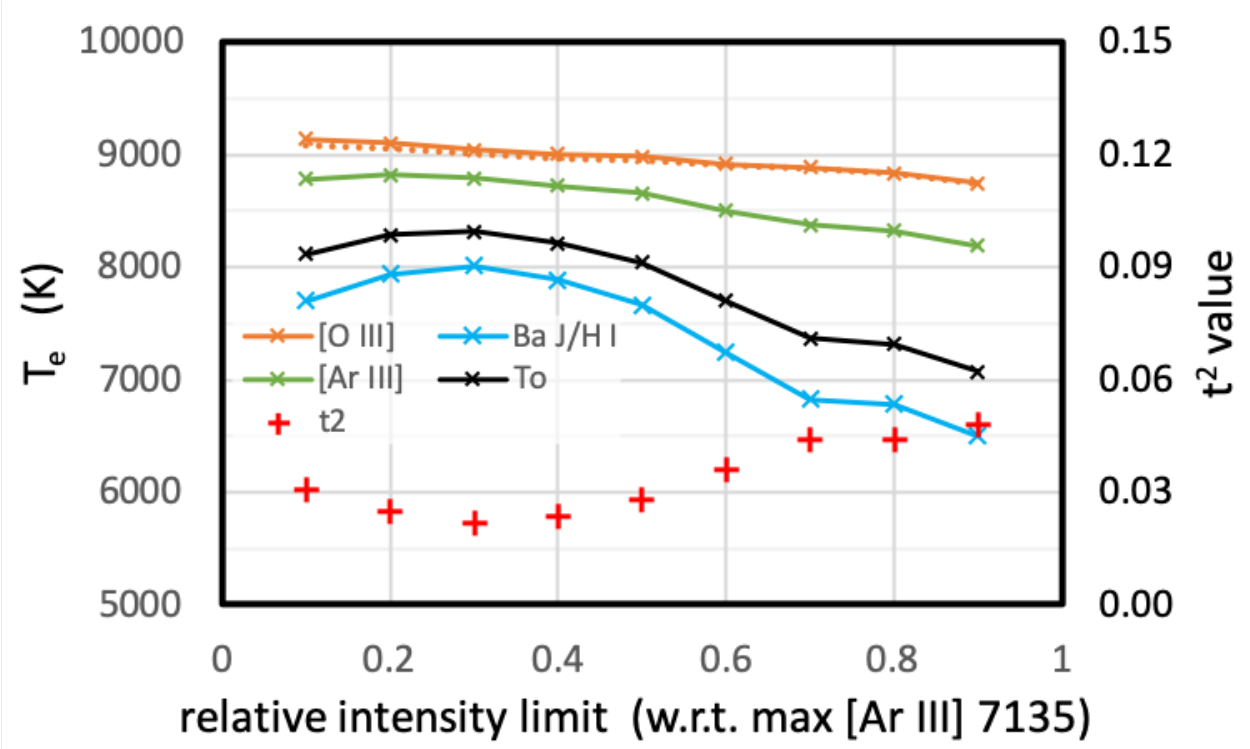}
\caption{The top panel presents the PV diagram of the difference between the [\ion{O}{3}] and [\ion{Ar}{3}] temperatures.  The contours are of the intensity of [\ion{Ar}{3}] $\lambda$7135.  The rectangle indicates the columns and rows considered for the statistics.  In the bottom panel, we plot the mean [\ion{O}{3}] and [\ion{Ar}{3}] temperatures as a function of the limiting relative intensity of the [\ion{Ar}{3}] $\lambda$7135 line, i.e., both temperatures are averaged over the same PV coordinates.  (The dotted line shows the effect of correcting for the contamination in the [\ion{O}{3}] $\lambda$4363 line.)  As in Figure \ref{fig_forb_TeO3_TeAr3_limit}, the [\ion{O}{3}] temperature is systematically higher.  We also plot the $T_0$ and $t^2$ parameters for the temperature fluctuation formalism of \citet{peimbert1967} for the same limiting relative intensities of the [\ion{Ar}{3}] $\lambda$7135 line, as well as the predicted temperature for the Balmer jump following \citet{peimbert2019}.  When large volumes are considered, $T_0$ is similar to the kinematic temperature (\S\ref{sec_kinematic_temperature}).  
}
\label{fig_temp_fluc}
\end{figure}

Comparing the two panels in Figure \ref{fig_forb_TeO3_TeAr3_limit}, there is a systematic difference between the [\ion{O}{3}] and [\ion{Ar}{3}] temperatures, with the latter being systematically lower.  While the volumes occupied by the O$^{2+}$ and Ar$^{2+}$ ions largely coincide, the Ar$^{2+}$ volume is biased towards the outer part of the O$^{2+}$ volume,  
so it may be cooler.  However, exciting the [\ion{Ar}{3}] lines requires less energy, so they will also be more easily excited 
where the temperature is cooler.  Figure \ref{fig_temp_fluc} plots the mean [\ion{O}{3}] and [\ion{Ar}{3}] temperatures weighted using the [\ion{Ar}{3}] $\lambda$7135 line, i.e., \emph{averaging over the same volume of the nebula}.  Again, the [\ion{O}{3}] temperature is systematically higher than the [\ion{Ar}{3}] temperature.  
This would appear to be a clear example of the temperature sensitivity of the excitation of forbidden lines affecting the temperature derived from them \citep{peimbert1967}.  An alternative explanation is that, for some reason, one or both of the [\ion{Ar}{3}] or [\ion{O}{3}] temperatures are systematically wrong.  

Both of the [\ion{O}{3}] and [\ion{Ar}{3}] temperatures (Figures \ref{fig_forb_TeO3} and \ref{fig_forb_TeAr3}) are greater than the $\sim 8,000$\,K temperature that matches the thermal line widths of the \ion{H}{1}, \ion{He}{1}, and [\ion{O}{3}] $\lambda$4959 lines (Table \ref{tab_main_results}).  The lower end of the range allowed by the \ion{He}{1} lines also matches the kinematic temperature, though temperatures derived from recombination lines should actually under-estimate the true mean temperature \citep{peimbert1967}.  

The difference between the temperatures found when weighting with the auroral and nebular lines in Figure \ref{fig_forb_TeO3_TeAr3_limit} or the difference between the [\ion{O}{3}] or [\ion{Ar}{3}] temperatures in Figure \ref{fig_temp_fluc} nicely illustrates the central concern discussed by \citet{peimbert1967}:  The temperature sensitivity of the emission mechanism influences the temperature obtained if temperature fluctuations are present.  In the temperature fluctuation formalism of \citet{peimbert1967}, 
normally a pair of temperatures is used, one sensitive to cooler regions and the other sensitive to hotter ones, but, in principle, any pair with a different temperature sensitivity should work.  So, we use the [\ion{O}{3}] and [\ion{Ar}{3}] temperatures.  
These temperatures are related to $T_0$ and $t^2$, the mean square temperature fluctuations (atomic data: Table \ref{tab_atomic_data}), by
$$ T_e([\mathrm{O}\,\mathrm{III}])\approx T_0\left(1+\left(\frac{91183}{T_0}-3\right)t^2\right) $$
$$ T_e([\mathrm{Ar}\,\mathrm{III}])\approx T_0\left(1+\left(\frac{676981}{T_0}-3\right)t^2\right)\,. $$
\noindent  Ideally, we would study the temperature fluctuations on a pixel-by-pixel basis in the PV diagram, but the PV diagram of the [\ion{Ar}{3}] temperature has insufficient S/N for this, so instead we consider the same limiting intensities of the [\ion{Ar}{3}] $\lambda$7135 line as in Figures \ref{fig_forb_TeO3_TeAr3_limit} and \ref{fig_temp_fluc}.  We weight both temperatures using this line intensity so as to compare exactly the same volume of the nebula.  

In Figure \ref{fig_temp_fluc}, we present the values of true mean temperature and the $t^2$ value at each flux limit of the [\ion{Ar}{3}] $\lambda$7135 line.  Evidently, the true mean temperature is substantially lower than the [\ion{Ar}{3}] or [\ion{O}{3}] temperatures, by $1,000-2,000$\,K.  For the low flux limits, whose volume is similar to that sensed by the kinematic temperature, $T_0$ is similar to the kinematic temperature.  For the high flux limits, corresponding to the small volume towards the rim of the main shell, the temperature is lower.  (The kinematic temperature is not sensitive to small volumes because of the thermal line width of the \ion{H}{1} lines.)  The large decrease in $T_0$ for the smallest volumes is mostly driven by the similar, but smaller, change in the [\ion{Ar}{3}] temperature.  Given the difference in the true mean temperature for small and large volumes, we would not be surprised if deeper data found large scale variations in $T_0$, as we find for the [\ion{Ar}{3}] and [\ion{O}{3}] temperatures.   

In Figure \ref{fig_temp_fluc}, like $T_0$, the $t^2$ value varies substantially, but in the opposite direction.  Given the definition of $t^2$ \citep{peimbert1967}, the relative amplitude of the temperature fluctuations varies from 15\% in the largest volumes to 22\% in the smallest.  These are large temperature fluctuations.  The absolute value of these fluctuations mirrors the variation in $t^2$, spanning $1250-1550$\,K.  Hence, in NGC 6153, we find substantial temperature fluctuations using only forbidden line temperatures.  Since the forbidden lines arise in the normal nebular plasma, these temperature fluctuations pertain to that plasma component alone.  

Figure \ref{fig_temp_fluc} illustrates the risk of using two forbidden lines to determine $t^2$ and $T_0$.  The value of $t^2$ is driven primarily by the difference between the [\ion{O}{3}] and [\ion{Ar}{3}] temperatures, increasing as the difference increases.  Meanwhile, the value of $T_0$ is driven by the value of the lower of the two temperatures, also increasing as the lower temperature increases.  Provided the [\ion{Ar}{3}] temperature is lower than the [\ion{O}{3}] temperature, a solution can always be found.  $T_0$ can easily differ very substantially from the forbidden line temperatures, so it is important that they be well-constrained.  In the usual scenario, a temperature based upon permitted emission is included, usually the ratio of an \ion{H}{1} line to the Balmer jump, which is lower than the true mean temperature, $T_0$, as shown in Figure \ref{fig_temp_fluc}.  

The above supposes that the difference in the [\ion{Ar}{3}] and [\ion{O}{3}] temperatures is due to temperature fluctuations.  An alternative explanation is that one or both of these temperatures could be systematically in error.  Our [\ion{O}{3}] temperatures are similar, though up to $\sim 200$\,K higher than, those of \citet[][; 8940\,K]{kingsburghbarlow1994} and \citet[][; 9030\,K and 9070\,K, adopting the same atomic data; Table \ref{tab_atomic_data} in both cases]{liuetal2000}, so our [\ion{O}{3}] temperature appears reasonable and secure considering that our observations do not coincide spatially with theirs.  

As for the [\ion{Ar}{3}] temperature, based upon the atomic data we use (Table \ref{tab_atomic_data}), we find values of 10,125\,K and 9,900\,K for the minor axis and whole nebula spectra from \citet{liuetal2000}, $\sim 1,000$\,K higher than we find, so the agreement is not good, but perhaps not unreasonable.  \citet{mcnabbetal2016} report an [\ion{Ar}{3}] temperature of 9,350\,K, similar to our result, but we are unable to convert this to the same atomic data that we use since we derive a much higher temperatures based upon the line intensities in their Table 2.  Objectively, we have no strong cross-check on the flux calibration for the blue end of the CD4b wavelength interval (\S\ref{sec_reddening}), so it is possible that the flux calibration for the [\ion{Ar}{3}] $\lambda\lambda$7135,7751 lines is incorrect.  If these line intensities were lower by 20\% or more, the [\ion{O}{3}] and [\ion{Ar}{3}] temperatures would be very similar and they could not be used to derive the amplitude of temperature fluctuations in Figure \ref{fig_temp_fluc}.  However, even such an error would not eliminate the evidence for temperature fluctuations from the [\ion{O}{3}] and [\ion{Ar}{3}] temperature maps (Figure \ref{fig_forb_TeO3_TeAr3_limit}).  

Considering the independent evidence for temperature fluctuations from the [\ion{O}{3}] and [\ion{Ar}{3}] temperature maps and the lower kinematic temperature (\S\ref{sec_kinematic_temperature}), we do not doubt the presence of temperature fluctuations in the normal nebular plasma in NGC 6153.  

For the diffuse emission beyond the receding side of the main shell, the [\ion{O}{3}] temperature falls to values near 8,000\,K (Figure \ref{fig_forb_TeO3}).  This temperature agrees with the kinematic temperature, which appears to apply to the PV coordinates of this emission since Figures \ref{fig_tkin_HbHgO3}, \ref{fig_tkin_HeIHbHg}, and \ref{fig_tkin_HeIO3} are all relatively uniform at the PV coordinates of this emission.  As we show below (\S\ref{sec_PV_ADF}), the ADF for this volume of the normal nebular plasma is near 1.0.  So, the temperature fluctuations in this part of the normal nebular plasma are apparently small.  Unfortunately, we cannot check this directly since we do not detect the [\ion{Ar}{3}] $\lambda$5191 line in this part of the PV diagram.  Therefore, the amplitude of the temperature fluctuations in the normal nebular plasma appears to vary, from very small in the diffuse emission beyond the receding side of the main shell to very substantial within the main shell itself.  

In summary, not only does the normal nebular plasma in NGC 6153 contain large scale temperature gradients (Figures \ref{fig_forb_TeO3} and \ref{fig_forb_TeAr3}), but it also contains small scale temperature fluctuations, all based only upon forbidden lines (Figures \ref{fig_forb_TeO3_TeAr3_limit} and \ref{fig_temp_fluc}).  The amplitudes of both the large- and the small-scale effects are similar.  Hence, even after decomposing the permitted \ion{O}{2} emission as in \S\ref{sec_contamination}, were we to use the \ion{O}{2} emission from the normal nebular plasma to compute the O$^{2+}$ abundance, it would not coincide with that computed from the [\ion{O}{3}] lines and the [\ion{O}{3}] temperature because of temperature fluctuations.  Our use of the kinematic temperature for the normal nebular plasma in previous sections anticipates this result.

\subsection{The Balmer jump temperature}\label{sec_balmer_jump}

\citet{liuetal2000} find a Balmer jump temperature that is spatially uniform across NGC 6153's minor axis with values scattering about a mean value of 6,080\,K.  From a spatially-integrated spectrum, \citet{zhangetal2004} report a value of $6,000 \pm 400$\,K.  \citet{mcnabbetal2016} find a Balmer jump temperature of $6250^{+150}_{-100}$\,K.  Nominally, these temperatures do not agree with those we find here by other means based upon the kinematics or the \ion{He}{1} and forbidden lines.  

However, all of these Balmer jump temperatures were computed supposing a single plasma component.  If we assume, as we illustrate below (\S\ref{sec_H_mass_additional}), that approximately 5\% of the mass of H$^+$ is at a temperature of 2,000\,K and the rest at 8,000\,K, using the definition of \citet{liuetal2000}, we find a ratio of Balmer jump to H11 of 0.161\,\AA$^{-1}$ using the atomic data of \citet{storeyhummer1995} and \citet{ercolanostorey2006}, which implies an electron temperature of 6,000\,K when we assume a single plasma component.  Therefore, interpreted in the framework of two plasma components, the existing measurements of the Balmer jump temperatures are indeed compatible with the temperatures found here for both plasma components.  

The complication in the foregoing is that the spatial extent of the two plasma components in NGC 6153 is not the same.  The additional plasma component is more centrally concentrated \citep[][here, e.g., Figure \ref{fig_tkin_O2O3}]{liuetal2000, tsamisetal2008}.  In principle, given the difference in the spatial distributions of the plasma components, \citet{liuetal2000} should have found a lower Balmer jump temperature in the central part of NGC 1653, where the additional plasma component is found and where it contributes to increase the Balmer jump (w.r.t. the normal nebular plasma).  Their Figure 14 shows no clear difference between the interior and exterior of the nebula, but it could be a S/N issue.  

\citet{garciarojasetal2022} find similar results for the Paschen jump temperature in NGC 6881, Hf 2-2, and M 1-42.

\subsection{The PV variation of the ADF}\label{sec_PV_ADF}

\begin{figure}
\includegraphics[width=\linewidth]{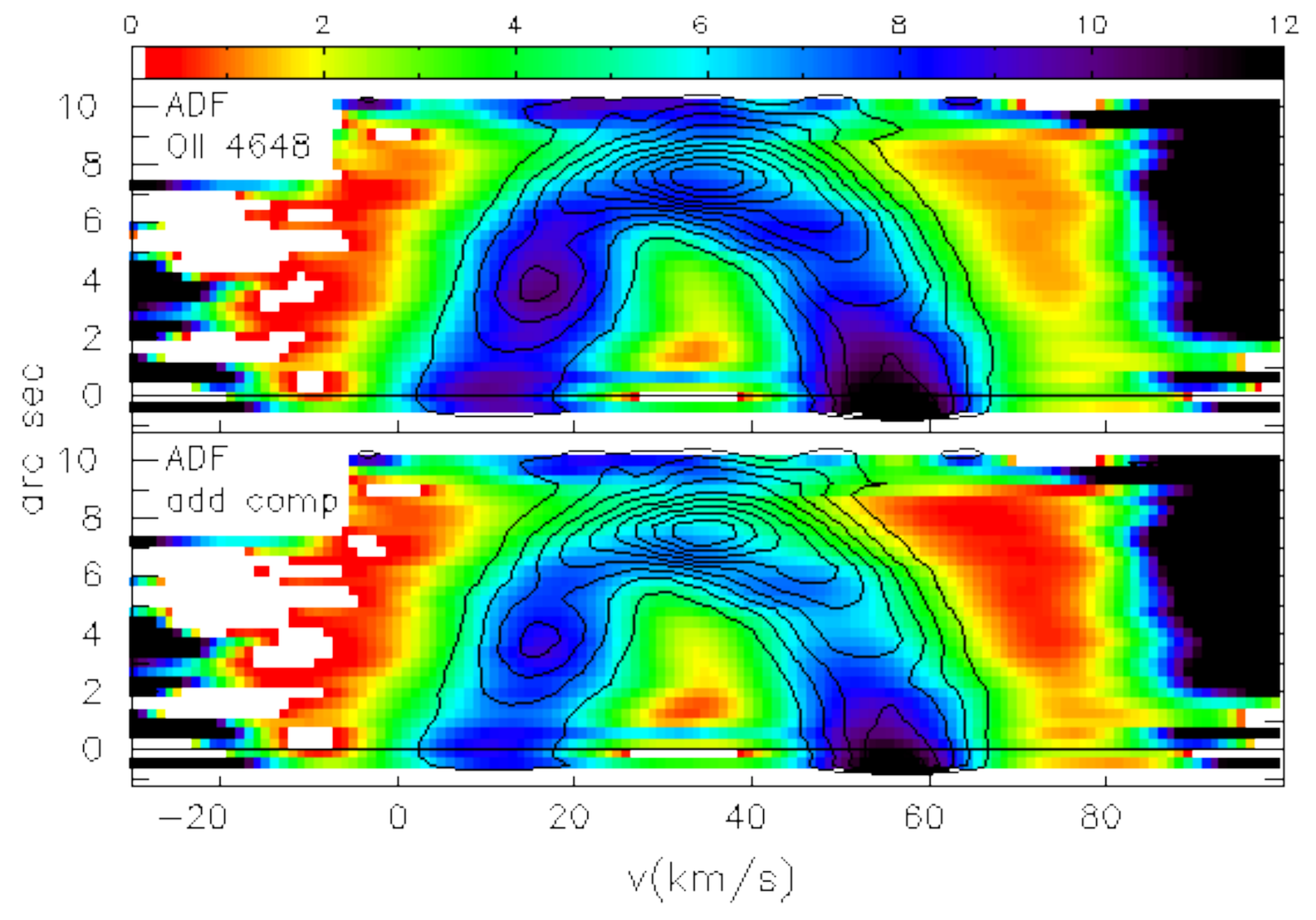}
\caption{These PV diagrams present the variation of the ADF for O$^{2+}$ as a function of position and velocity within our slit.  The top panel presents the ADF computed using all of the permitted \ion{O}{2} $\lambda$4649 emission while the bottom panel uses only the emission due to the additional plasma component.  The scale is common to the two panels, but the contours are not.  At top, the contours are of the intensity of the observed \ion{O}{2} $\lambda$4649 intensity.  At bottom, they are of the intensity of this line due to the additional plasma component only.  In the top panel, the ADF for the diffuse emission beyond the receding side of the nebula is near 1.0, implying that this emission is due to the normal nebular plasma, i.e., there is no abundance discrepancy when considering this volume of the normal nebular plasma.  In both panels, the apparently high values at the largest velocities are due to the \ion{C}{3} $\lambda$4650 line in the PV diagram of \ion{O}{2} $\lambda$4649.
}
\label{fig_PV_ADF_OII}
\end{figure}

We may use the decomposition of the \ion{O}{2} $\lambda$4649 line (\S\ref{sec_contamination}) to construct the PV diagram of the ADF for O$^{2+}$.  We compute the O$^{2+}$ ionic abundance using the \ion{O}{2} $\lambda$4649/H$\beta$ and the [\ion{O}{3}] $\lambda$4959/H$\beta$ ratios using the emissivities from Figure \ref{fig_emissivities_N2_O2_O3}.  We adopt our usual choices for the physical conditions in the two plasma components, $T_e = 8,000$\,K and $N_e = 5,000$\,cm$^{-3}$ for the normal nebular plasma and $T_e = 2,000$\,K and $N_e = 10,000$\,cm$^{-3}$ for the additional plasma component.  Before computing either line ratio, we must broaden the PV diagrams of the oxygen lines to the width of the hydrogen lines at the adopted temperatures, which is different for the \ion{O}{2} $\lambda$4649 and [\ion{O}{3}] $\lambda$4959 lines.  Otherwise, we would introduce spurious structure in the PV diagram of the ADF.  

Figure \ref{fig_PV_ADF_OII} presents the resulting PV diagram of the ADF.  We compute the ADF in two ways.  First, following common practice, we compute the \ion{O}{2} $\lambda$4649/H$\beta$ ratio using the observed PV diagram for \ion{O}{2} $\lambda$4649,   
broadened to match the thermal width expected for the H$\beta$ line.  This is shown in the top panel of Figure \ref{fig_PV_ADF_OII}.  The ADF for O$^{2+}$ is highest where the \ion{O}{2} $\lambda$4649 emission is most intense.  Given the kinematics of the \ion{O}{2} $\lambda$4649 line, we expect the ADF to be highest at velocities closest to the systemic velocities and positions close to the central star, as observed.   

\begin{figure*}
\includegraphics[width=0.5\linewidth]{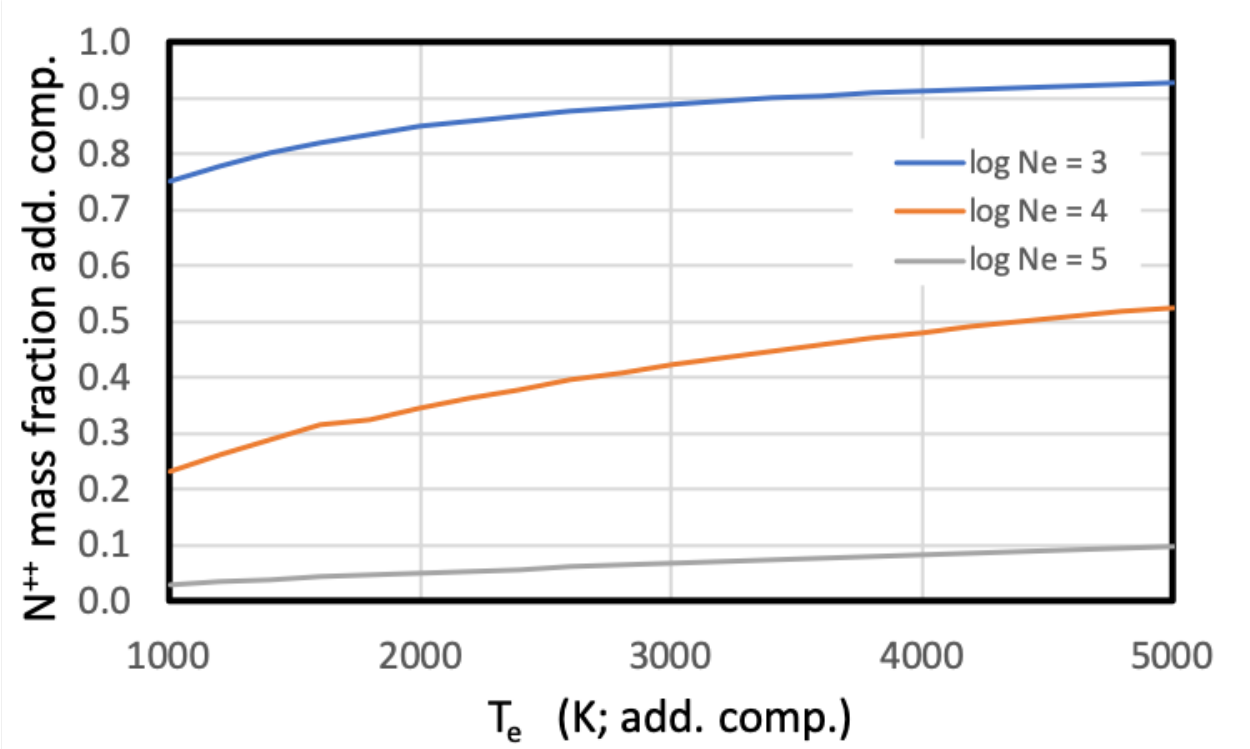}
\includegraphics[width=0.5\linewidth]{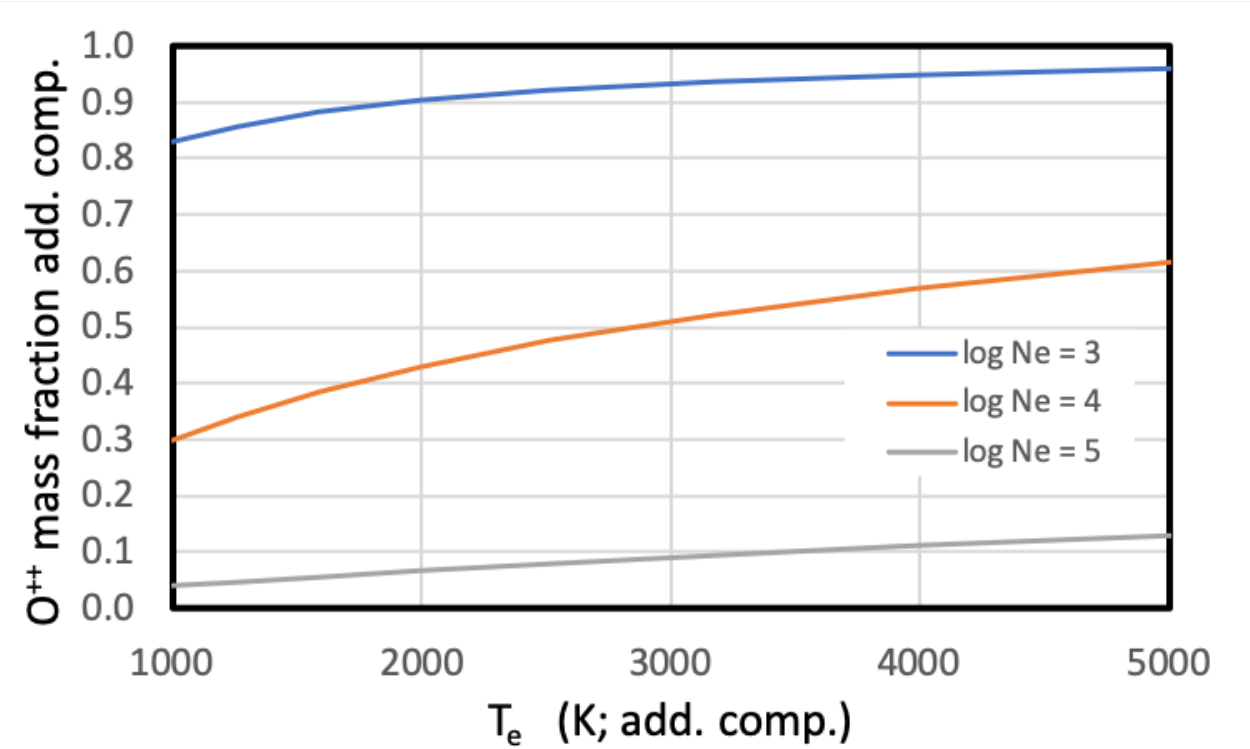}
\caption{These panels present the mass fraction of N$^{2+}$ (left) and O$^{2+}$ (right) contributed by the additional plasma component for different densities and temperatures.  The temperature range is that indicated by the \ion{N}{2} $\lambda\lambda$4041/5680 and \ion{O}{2} $\lambda\lambda$4089/4649 line ratios (\S\ref{sec_physcond_perm}).  The \ion{N}{2} $\lambda\lambda$5666/5680 and \ion{O}{2} $\lambda\lambda$4662/4649 line ratios imply electron densities $>5,000-100,000$\,cm$^{-3}$.  If so, the additional plasma component accounts for up to about half of the N$^{2+}$ and O$^{2+}$ mass fractions.  
}
\label{fig_mass_frac}
\end{figure*}

The other important result from Figure \ref{fig_PV_ADF_OII} is that the ADF in the diffuse emission beyond the receding side of the main shell has a value near 1 (the actual value is $\sim 1.1-1.2$).  An ADF near unity is congruent with the results that the temperatures determined using the [\ion{O}{3}] lines, the kinematics, and the $t^2$ formalism are similar there (\S\ref{sec_kinematic_temperature}, \S\ref{sec_physcond_forb}, \S\ref{sec_temp_fluc}), so we expect that the diffuse emission beyond the main shell is due to the normal nebular plasma alone and that temperature fluctuations are small in that volume of plasma.  

Second, we compute the \ion{O}{2} $\lambda$4649/H$\beta$ ratio using only the \ion{O}{2} $\lambda$4649 emission from the additional plasma component.  
The bottom panel of Figure \ref{fig_PV_ADF_OII} presents this result.  When two plasma components are present, this method is the more reasonable, since it discounts the permitted emission due to the normal nebular plasma.  Given that the additional plasma component emits most of the \ion{O}{2} $\lambda$4649 emission (\S\ref{sec_rel_mass_add_comp}), the general pattern is similar to the top panel.  The main difference is that the ADF for the diffuse emission beyond the receding side of the main shell is now near zero since the additional plasma component does not emit at those velocities.  Regardless of the details, the basic result is that the ADF for O$^{2+}$ is highest in the inner part of the main shell, where the additional plasma component has most of its mass in O$^{2+}$ (\S\ref{sec_rel_mass_add_comp}).

\subsection{The relative masses of N$^{2+}$ and O$^{2+}$ in the two plasma components}\label{sec_rel_mass_add_comp}

\begin{figure*}
\begin{center}\includegraphics[width=0.86\linewidth]{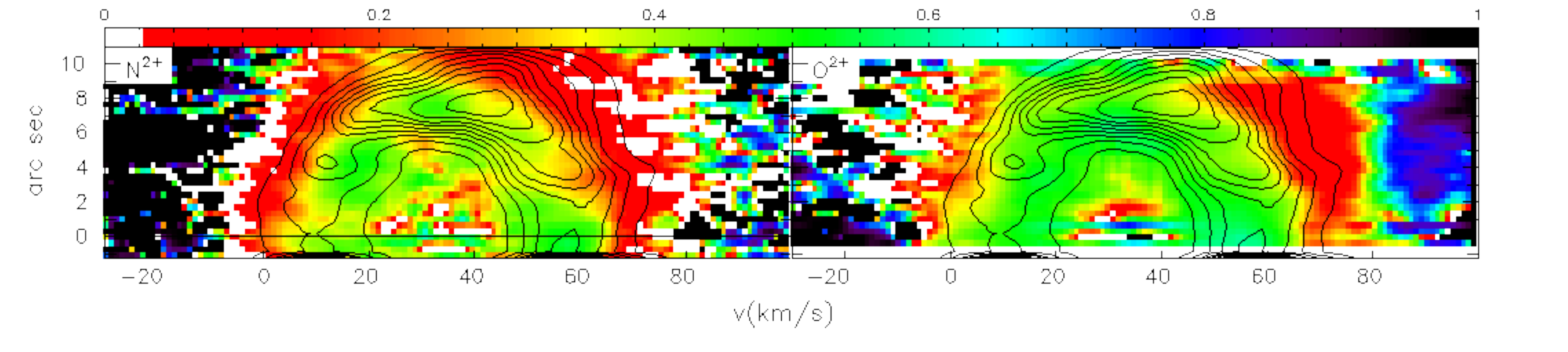}\end{center}
\caption{These PV diagrams present the fraction of the mass in the additional plasma component for N$^{2+}$ (left) and O$^{2+}$ (right) computed assuming $(T_e,\,N_e)=(2,000\,\mathrm K,\,10,000\,\mathrm{cm}^{-3})$ for the additional plasma component and $(T_e,\,N_e)=(8,000\,\mathrm K,\,5,000\,\mathrm{cm}^{-3})$ for the normal nebular plasma.  The contours are of the intensity of the [\ion{O}{3}] $\lambda$4959 line.  Locally, the additional plasma component represents up to 50-60\% of the total mass of N$^{2+}$ and O$^{2+}$ and is concentrated in the inner part of the main shell.  The region of apparently high fractional mass at the right side of the PV diagram for O$^{2+}$ is due to contamination from the \ion{C}{3} $\lambda$4650 line adjacent to the \ion{O}{2} $\lambda$4649 line.
}
\label{fig_mass_frac_PV}
\end{figure*}

We may use the decompositions of the \ion{N}{2} $\lambda$5680 and \ion{O}{2} $\lambda$4649 lines into components due to the normal nebular plasma and the additional plasma component to estimate the relative masses of these two plasma components.  We adopt our usual physical conditions (\S\ref{sec_contamination}):  $T_e = 8,000$\,K and $N_e = 5,000$\,cm$^{-3}$ for the normal nebular plasma and $T_e = 2,000$\,K and $N_e = 10,000$\,cm$^{-3}$ for the additional plasma component (emissivities: Figure \ref{fig_emissivities_N2_O2_O3}).  
The mass of ion $X$ in the additional plasma component with respect to that in the normal nebular plasma is 
$$\frac{M(X)_a}{M(X)_n} = \frac{I(\lambda)_a}{I(\lambda)_n}\frac{N(X)_n}{N(X)_a}\frac{\epsilon(\lambda)_n}{\epsilon(\lambda)_a}$$
where $I(\lambda)$ indicates the intensity, $N(X)$ the density, and $\epsilon(\lambda)_x$ the emissivities at the appropriate temperatures for the two plasma components.  So the mass fraction in the additional component is 
$$\mathrm{mass\ fraction} = 1-\left(1+\frac{M(X)_a}{M(X)_n}\right)^{-1}\ . $$

For the \ion{N}{2} $\lambda$5680 line, we integrate the emission from the middle and bottom panels from Figure \ref{fig_decompN2_PV_N2_perm} to obtain the total emission (within our slit) due to the two plasma components.  We find that the additional plasma component emits 76\% of the total emission in the \ion{N}{2} $\lambda$5680 line.  However, given the adopted temperatures and densities for the two plasma components, the additional plasma component accounts for 35\% of the mass of N$^{2+}$ for the volume of the nebula intercepted by our spectrograph slit.  If the density is as high as $10^5$\,cm$^{-3}$ (\S\ref{sec_physcond_perm}), the mass fraction drops to 5\%.  The left panel of Figure \ref{fig_mass_frac} presents the variation of the mass fraction of the additional plasma component for densities spanning $10^3$\,cm$^{-3} - 10^5$\,cm$^{-3}$ and temperatures of $1,000-5,000$\,K.  Unless the density is lower than indicated by the \ion{N}{2} and \ion{O}{2} lines, the additional plasma component accounts for at most half of the mass of N$^{2+}$ in NGC 6153.  

For \ion{O}{2} $\lambda$4649, we proceed in a similar manner.  We find that the additional plasma component emits 84\% of the \ion{O}{2} $\lambda$4649 emission for the volume of the nebula seen by the spectrograph slit, but accounts for only 43\% of the O$^{2+}$ mass at the adopted electron densities and temperatures for the two plasma components.  This mass fraction falls to 7\% if the density is as high as $10^5$\,cm$^{-3}$.  The right panel in Figure \ref{fig_mass_frac} presents the fraction of the mass accounted for by the additional plasma component for a range of relevant temperatures and densities.  Again, we find that, unless the density for this plasma component has been significantly under-estimated, the additional plasma component accounts for about half of the mass of O$^{2+}$ in NGC 6153 seen by the spectrograph slit.  This agrees with the results of \cite[][theoretical models]{gomezllanosmorisset2020} and \citet[][observations of planetary nebulae with high ADFs]{garciarojasetal2022}.  

Figure \ref{fig_mass_frac_PV} presents the fractional mass of the additional plasma component in the PV diagram based upon the decomposition of the \ion{N}{2} $\lambda$5680 and \ion{O}{2} $\lambda$4649 lines using the nominal physical conditions above.  Locally, the additional plasma component accounts for up to 50-60\% of the total mass of N$^{2+}$ or O$^{2+}$ ions.  As expected from the kinematics of the \ion{C}{2}, \ion{N}{2}, \ion{O}{2}, and \ion{Ne}{2} lines, the maximum of the mass in the additional plasma component is concentrated to the inner part of the main shell, both in velocity and spatial coordinates, in agreement with the spatial distribution of these emission lines \citep{liuetal2000, tsamisetal2008}.  

\subsection{The H mass in the additional plasma component}\label{sec_H_mass_additional}

\begin{figure}
\includegraphics[width=\linewidth]{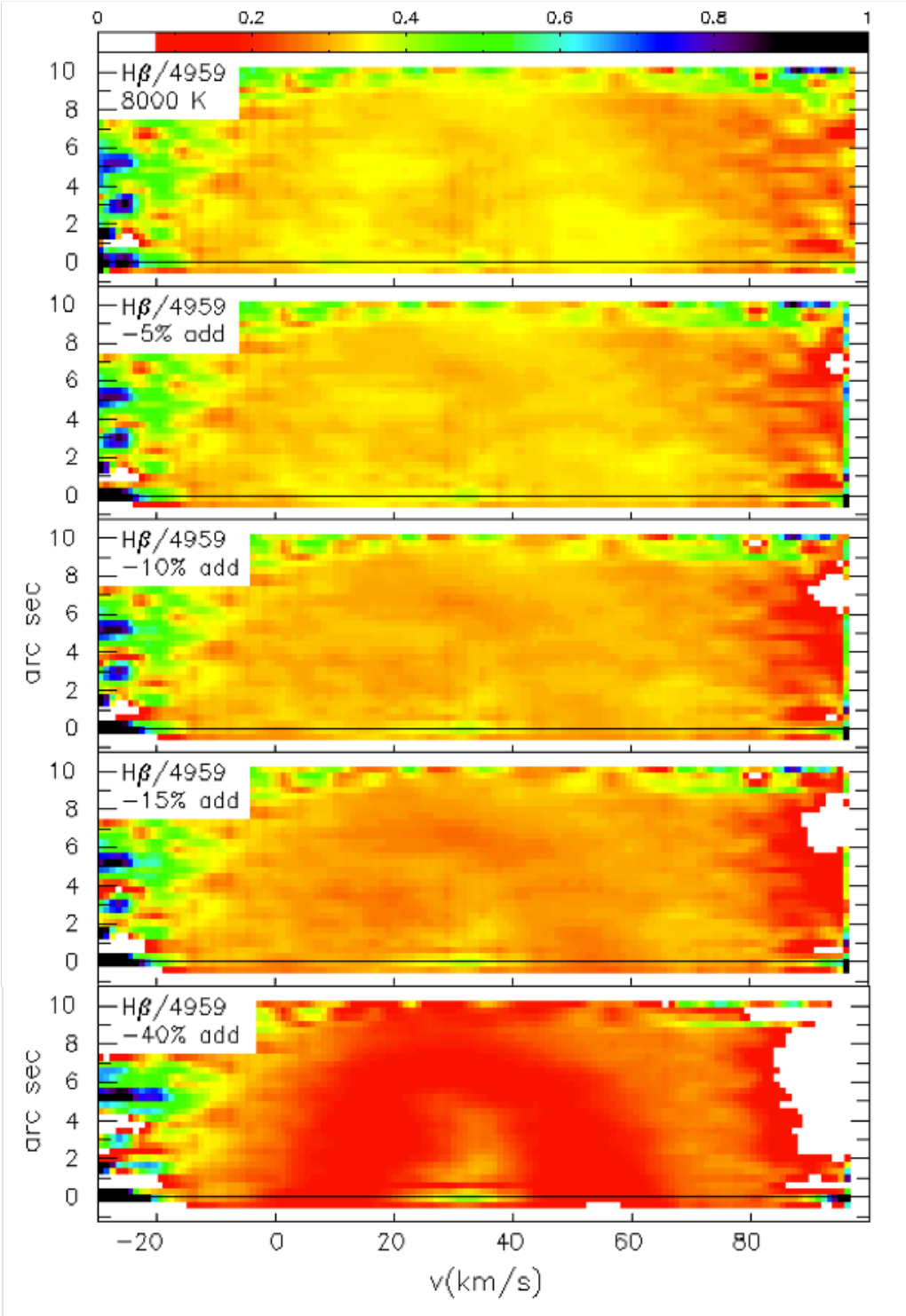}
\caption{We present the ratio of the PV diagram of the H$\beta$ line and the model of this line based upon the [\ion{O}{3}] $\lambda$4959 line broadened for a temperature of 8,000\,K.  The top panel considers the observed PV diagram of H$\beta$ (Figure \ref{fig_tkin_HbHgO3}).  The next four panels consider the observed PV diagram of H$\beta$ after subtracting a contribution due to the additional plasma component, which we model using the \ion{O}{2} $\lambda$4649 line broadened for a temperature of 2,000\,K, equivalent to 5\%, 10\%, 15\%, and 40\% of the total H$\beta$ emission.  Contributions below 5\% are not noticeable.  For contributions of 5\% and 10\% (second and third panels), the result is a flatter ratio, presumably indicating that removing the emission due to the additional plasma component improves the agreement for the normal nebular plasma.  For contributions of 15\% or more (fourth and fifth panels), there is a deficit of H$\beta$ emission at the positions and velocities expected for the additional plasma component.  
}
\label{fig_add_H_mass_v1}
\end{figure}

Having determined the fraction of the mass of O$^{2+}$ contained in the additional plasma component, it is of interest to investigate the fraction of the mass of hydrogen it may contain.  As \citet{gomezllanosmorisset2020} illustrate convincingly, it may be a small minority, making it difficult to determine.  We can think of no direct way of probing the H mass fraction, but here present three estimates.  The two of the three estimates indicate that the additional plasma component contributes of order $10-15$\% of the emission in H$\beta$, and so of order $3-5$\% of the mass of hydrogen within the area intercepted by our slit in NGC 6153.  

The simplest means to estimate the mass of hydrogen in the additional plasma component is to consider its effect upon the kinematic temperature (\S\ref{sec_kinematic_temperature}).  Clearly, the additional plasma component must contribute a minority of the H$\beta$ emission; otherwise it would distort the comparison between the H$\beta$, H$\gamma$ and [\ion{O}{3}] $\lambda$4959 lines (Figure \ref{fig_tkin_HbHgO3}), since we assume that [\ion{O}{3}] $\lambda$4959 arises only in the normal nebular plasma.  

To estimate the H$\beta$ emission from the additional plasma component, we assume that its emission will follow that from the \ion{O}{2} lines and that its temperature is 2,000\,K, as found from the \ion{N}{2} and \ion{O}{2} lines (\S\ref{sec_physcond_perm}).  We construct a model of H$\beta$ line based upon the \ion{O}{2} $\lambda$4649 line and broadened assuming a temperature of 2,000\,K (as described in \S\ref{sec_kinematic_temperature}, subscript $th$ below).  We then scale this model to different fractions of the total H$\beta$ emission, $f$, subtract the model from the observed PV diagram for H$\beta$, $I(\mathrm H\beta)_o$, and compare the remaining H$\beta$ emission with that from the [\ion{O}{3}] $\lambda$4959 line (the same as used in \S\ref{sec_kinematic_temperature}).  Arithmetically, 
$$ I(\mathrm H\beta/4959)_n = \frac{I(\mathrm H\beta)_o - f\times I(4649)_{o,th}}{I(4959)_{o,th}}\ . $$
\noindent 
Above some scale factor, subtracting the contribution of the H$\beta$ emission due to the additional plasma component will cause a deficit of H$\beta$ emission compared to [\ion{O}{3}] $\lambda$4959, similar to that observed in the top left panel of Figure \ref{fig_tkin_HbHgO3}.  Once that occurs, we assume that the contribution from the additional plasma component is too great.  

\begin{figure}
\includegraphics[width=\linewidth]{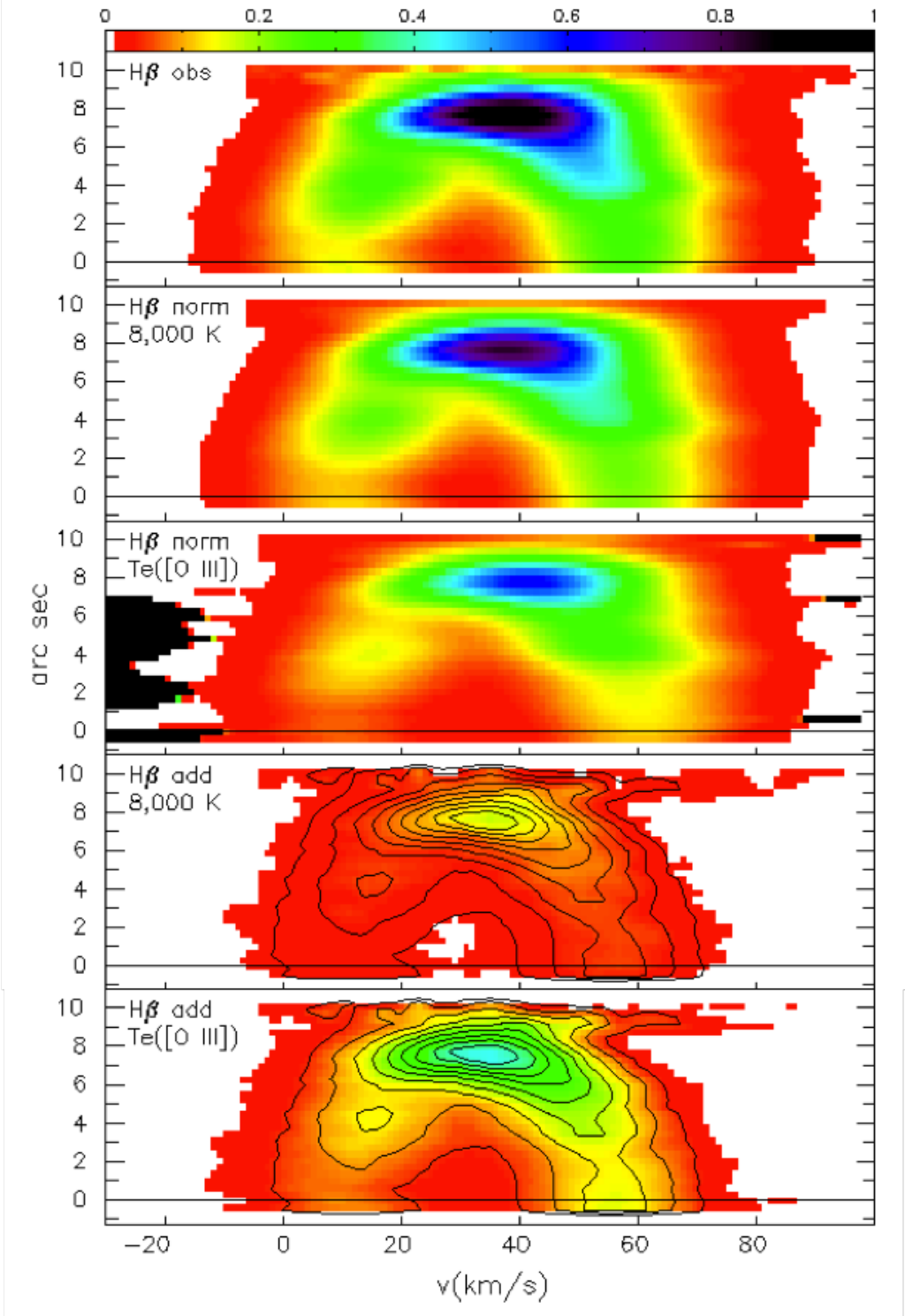}
\caption{We present the PV diagrams of the observed H$\beta$ line (top panel), predictions of the H$\beta$ emission from the normal nebular plasma based upon the [\ion{O}{3}] $\lambda$4959 line assuming (a) the kinematic temperature of 8,000\,K (second panel) and (b) the [\ion{O}{3}] temperature map (third panel), and the residual H$\beta$ emission attributable to the additional plasma component assuming a temperature of 8,000\,K (fourth panel) and the [\ion{O}{3}] temperature map (bottom panel).  The contours in the bottom two panels are of the model of the H$\beta$ emission based upon the \ion{O}{2} $\lambda$4649 line broadened for a temperature of 2,000\,K (used in Figure \ref{fig_add_H_mass_v1}).  All panels share a common color scale.  The residual emission amounts to 15.3\% (fourth panel) and 41.7\% (bottom panel) of the total H$\beta$ emission, i.e., the bottom two panels correspond closely to the contributions supposed in the bottom two panels of Fig. \ref{fig_add_H_mass_v1}.  
}
\label{fig_add_H_mass_v2}
\end{figure}

Figure \ref{fig_add_H_mass_v1} presents the results.  We consider contributions from the additional plasma component equivalent to $1\%-40\%$ of the total H$\beta$ emission.  Contributions below 5\% do not cause noticeable changes compared to the observed PV diagram of H$\beta$ (top panel).  A 5\% contribution (second panel) begins to be noticeable, but in a positive way, making the ratio map even flatter than before.  At a 10\% contribution (third panel), the ratio is flattest.  Beginning at contributions of 15\% (fourth panel), there is a deficit of H$\beta$ emission that becomes very notable at 40\% (bottom panel).  Assuming that the flattest ratio map indicates the most likely estimate of the contribution of the additional plasma component to the H$\beta$ emission, we attribute 10\% of the total H$\beta$ emission to this plasma component, though we probably cannot exclude the range from 5\% to 15\%.  Given the change in emissivity for H$\beta$ between the temperatures of 8,000\,K and 2,000\,K \citep{storeyhummer1995}, a 10\% contribution for the emission from the additional plasma component represents 3-4\% of the total mass of H.  

Instead of estimating the H$\beta$ emission of the additional plasma component, we can also estimate the H$\beta$ emission from the normal nebular plasma.  We assume that the emission in the [\ion{O}{3}] $\lambda$4959 line is entirely due to the normal nebular plasma.  By adopting an electron temperature, we may predict the H$\beta$/[\ion{O}{3}] $\lambda$4959 ratio via Figure \ref{fig_emissivities_N2_O2_O3}.  The product of the PV diagram of the [\ion{O}{3}] $\lambda$4959 line (broadened for a temperature of 8,000\,K) and the H$\beta$/[\ion{O}{3}] $\lambda$4959 ratio gives the predicted H$\beta$ emission per O$^{2+}$ ion.  We then scale the result by the $\mathrm H^+/\mathrm O^{2+}$ ionic ratio and subtract the result from the observed PV diagram of H$\beta$ to obtain the residual H$\beta$ emission that we attribute to the additional plasma component,  $I(\mathrm H\beta)_a$.  When scaling, we fit the predicted H$\beta$ emission to that observed in the diffuse plasma beyond the main shell where the ADF is near unity (\S\ref{sec_PV_ADF}, Figure \ref{fig_PV_ADF_OII}).  This second estimate of the H$\beta$ emission from the additional plasma component is 
$$ I(\mathrm H\beta)_a = I(\mathrm H\beta)_o - I(4959)_{o,th} \times \frac{\epsilon(\mathrm H\beta, T_e)}{\epsilon(4959, T_e)}  \times \frac{N(\mathrm H^+)}{N(\mathrm O^{2+})} \ . $$
As regards the electron temperature, we proceed in two ways.  First, we assume that the electron temperature is the kinematic temperature of $T_e = 8,000$\,K (\S\ref{sec_kinematic_temperature}).  The $\mathrm H^+/\mathrm O^{2+}$ ionic ratio required in this case is $1200\pm 100$.  Second, we assume that the temperature varies as shown in the PV diagram of the [\ion{O}{3}] temperature, $T_e = T_e([\mathrm{O~III}])$ computed after broadening the [\ion{O}{3}] $\lambda\lambda$4363,4959 lines to the width of the H$\beta$ line for a temperature of 8,000\,K.  The $\mathrm H^+/\mathrm O^{2+}$ ratio required in this case is $1400\pm 100$.  The $\mathrm H^+/\mathrm O^{2+}$ ionic abundances for the two temperatures are similar since the [\ion{O}{3}] temperature in the diffuse emission beyond the main shell is near 8,300\,K.  

Figure \ref{fig_add_H_mass_v2} illustrates both versions of this second method.  The top panel presents the observed H$\beta$ emission.  The second panel presents the predicted H$\beta$ emission for the normal nebular plasma supposing an electron temperature of 8,000\,K.  The third panel presents the predicted H$\beta$ emission supposing the electron temperature given by the [\ion{O}{3}] temperature map.  The predicted H$\beta$ emission from the normal nebular plasma is less in this second case because the adopted electron temperature is higher in the main shell, $9,000-10,000$\,K.  The fourth panel presents difference between the observed H$\beta$ emission (top panel) and that predicted for the normal nebular plasma for an electron temperature of 8,000\,K (second panel).  The PV diagram of the residual H$\beta$ emission has a morphology similar to that of the PV diagrams of the \ion{C}{2}, \ion{N}{2}, \ion{O}{2}, and \ion{Ne}{2} lines (Figure \ref{fig_PVC2}).  This PV diagram is strictly positive, as it must be if the predicted H$\beta$ emission in the normal nebular plasma is reasonable.  The residual emission in the fourth panel amounts to 15.3\% of the total H$\beta$ emission.  Hence, this estimate implies that the additional plasma component contains about 5\% of the total H mass.  The bottom panel presents difference between the observed H$\beta$ emission (top panel) and that predicted for the normal nebular plasma for an electron temperature according to the [\ion{O}{3}] temperature map (third panel).  The residual emission is greater in this case, 41.7\% of the total H$\beta$ emission (approximately 14\% of the H mass), but still has a morphology similar to that expected for the additional plasma component.  

The estimate of the H mass in the additional plasma component found by subtracting a hypothetical contribution from the observed PV diagram of H$\beta$ agrees with the estimate found by adopting the kinematic temperature as the true electron temperature in the normal nebular plasma.  These estimates imply that the H mass of the additional plasma component is $3-5$\% of the total H mass.  The estimate of the H mass in the additional plasma component found when adopting the temperature given by the PV diagram of the [\ion{O}{3}] temperature is very different.  Indeed, Figure \ref{fig_add_H_mass_v1} indicates that it is not feasible that the additional plasma component contain so much H if we insist that the H$\beta$ emission from the normal nebular plasma follows that from [\ion{O}{3}] $\lambda$4959.  We interpret this discrepancy as the result of adopting an incorrect estimate of the electron temperature.

Finally, if the two plasma components contain similar masses of O$^{2+}$ (\S\ref{sec_rel_mass_add_comp}), but the additional plasma component contains only $3-5$\% of the mass of H, then the $\mathrm O^{2+}/\mathrm H^+$ ionic abundance ratios differ by a factor of $20-30$ between the two plasma components.  This result confirms the finding from \citet{gomezllanosmorisset2020} that the ADF is not a particularly good estimate of the difference in chemical composition between the two plasma components.  If the $\mathrm O^{2+}/\mathrm H^+$ ionic ratio is $1/1200$ or $1/1400$ in the normal nebular plasma, $12+\log(\mathrm O^{2+}/\mathrm H^+)=8.85-8.92$\,dex, including the additional plasma component implies an overall abundance ratio of approximately $1/600$ or $1/700$, $12+\log(\mathrm O^{2+}/\mathrm H^+)=9.15-9.22$\,dex.  (This neglects a -0.013 dex correction for the H in the additional plasma component.)

\subsection{On the abundance discrepancy in NGC 6153}

The classical understanding of the abundance discrepancy is that permitted and forbidden lines emitted by a given ion indicate different chemical abundances.  The temperatures used to compute these abundances are critical, as \citet{peimbert1967} first pointed out.  With few exceptions, the analyses prior to that of \citet{liuetal2000} typically supposed a single plasma component and often a single electron temperature for the permitted and forbidden lines.  

Our study of NGC 6153 finds conditions that differ markedly from that framework.  The kinematics, physical conditions, and relative masses of N$^{2+}$ and O$^{2+}$ yield congruent results regarding the two plasma components that apparently coexist in the nebular shell of NGC 6153 (\S\ref{sec_ionization_structure}, \S\ref{sec_physical_conditions}, \S\ref{sec_rel_mass_add_comp}).  Considering two plasma components with very different physical conditions and chemical compositions allows explaining the anomalous electron temperature and density derived from the [\ion{N}{2}] $\lambda\lambda$5755,6583 and [\ion{O}{2}] $\lambda\lambda$3726,3729 lines, respectively, as well as the Balmer jump temperature determined by others (\S\ref{sec_contamination} and \S\ref{sec_balmer_jump}).  Even considering the normal nebular plasma alone, there are large-scale temperature gradients and small-scale temperature fluctuations (\S\ref{sec_physical_conditions} and \S\ref{sec_temp_fluc}), both of which are substantial, but the temperature fluctuations also apparently differ in different volumes of the plasma.  All of these are confounding factors that, if not taken into account, will inevitably lead to the conclusion that there is an abundance discrepancy in NGC 6153.

However, if we consider the diffuse emission beyond the main shell in the normal nebular plasma, there is no abundance discrepancy in NGC 6153.  There, the ADF is near unity and the [\ion{O}{3}] and kinematic temperatures are similar, near 8,000\,K.  In this volume of the normal nebular plasma, it is particularly simple to determine the O$^{2+}$ ionic abundance (\S\ref{sec_H_mass_additional}).  
Assuming that this ionic abundance holds throughout the normal nebular plasma and that there are similar masses of N$^{2+}$ and O$^{2+}$ ions in the two plasma components, we can estimate the total abundance of O$^{2+}$ ions in the plasma within our spectrograph slit (\S\ref{sec_H_mass_additional}).  
That we obtain similar estimates of the H$^+$ mass in the additional plasma component assuming (1) an emission pattern like the \ion{O}{2} emission from the additional plasma component and its nominal physical conditions and (2) an emission pattern like the [\ion{O}{3}] emission from the normal nebular plasma and the kinematic temperature (equivalently, $T_0$; \S\ref{sec_H_mass_additional}), 
indeed implies that it is reasonable to suppose a constant chemical composition within the normal nebular plasma.

To determine the O$^{2+}$ ionic abundance in other volumes of the normal nebular plasma is more difficult, because choices are required.  In the main shell, the [\ion{O}{3}] and [\ion{Ar}{3}] temperatures are substantially higher than the kinematic temperature, the Balmer jump temperature, or $T_0$, forcing a choice as to which to use.  Using the kinematic temperature or $T_0$ and the forbidden line intensity would yield the same O$^{2+}$ abundance.  However, the ADF is substantial in the main shell, so, to compute the abundance from the permitted lines requires subtracting the majority of the emission that arises in the additional plasma component.  Both of these options are possible only because of the quantity of information we have for NGC 6153.  This illustrates very clearly the central issue concerning the abundance discrepancy in NGC 6153:  It is an information and analysis issue.  

Normally, tradition favors using the forbidden line temperatures, because they are much more commonly available and have higher S/N.  Also, separating the emission from the two plasma components is usually impossible.  So, it is not difficult to imagine an apparent abundance discrepancy arising if the conditions in NGC 6153 are typical.  Even if only one plasma component is present, the fundamental lesson from \citet{peimbert1967} is that all temperature indicators are biased to some extent (Figure \ref{fig_temp_fluc}), so their interpretation is key.

\subsection{Beyond NGC 6153}

Our findings in NGC 6153 emphasize the complexity that may exist in nebular plasmas.  These findings pertain to only the small part of NGC 6153 observed by our spectrograph slit, so there is probably even more complexity lurking within.  Although none of the sources of this complexity are new, the combination of multiple plasma components \citep{liuetal2000} and variable temperature fluctuations \citep{peimbert1967} implies a degree of complexity not considered previously.  The plasma in other objects may be less, or more, complex.  When possible, this complexity should be taken into account when computing the elemental/ionic abundances in planetary nebulae, at least.  

The plasma's complexity manifests itself directly in the line intensities and so the interpretation of the line intensities requires care.  In particular, it is necessary to correct the intensities of the auroral lines of [\ion{N}{2}] and [\ion{O}{2}] for contributions due to recombination if the volume of O$^{2+}$ observed is large.  Otherwise, the derived electron temperatures will be too large, biasing the abundance calculations, even if only one plasma component is present.  The density determined from the [\ion{O}{2}] nebular lines may be significantly affected if the volume observed includes multiple plasma components, as this density will reflect the density of different plasma components in different parts of the nebula.  There is considerable documentation of the contamination of the [\ion{N}{2}] $\lambda$5755 and [\ion{O}{2}] lines in the literature \citep[e.g.,][]{liuetal2000, tsamisetal2003, wessonetal2005, liuetal2006, corradietal2015, ruizescobedopena2022}, and we advocate that the contamination of these lines be taken into account in those cases where it is reasonable to expect it to be important.  

If the volume of He$^{2+}$ included in the observations is large, the intensity of the [\ion{O}{3}] $\lambda$4363 line may also need correction.  As \citet{gomezllanosetal2020} point out, the excitation of the [\ion{O}{3}] $\lambda$4363 due to recombination is less sensitive to temperature than many lines, varying much less than the usual $\sim T^{-1}$ assumption, so the recombination contribution may be more important from the normal nebular plasma than from the colder additional plasma component.  In this case, we caution against using the \ion{O}{3} lines involved in the Bowen fluorescence cascade to deduce the corrections necessary since they will not reflect the true abundance of O$^{3+}$ \citep[e.g.,][]{liudanziger1993, kastnerbhatia1996, selvellietal2007}.  

The [\ion{O}{3}] temperature is sensitive to the presence of the He$^{2+}$ zone, since this zone may include a considerable O$^{2+}$ content as a result of the efficiency of charge exchange with O$^{3+}$.  Normally, the temperature is higher in the He$^{2+}$ zone due to the extra heating that occurs there \citep[e.g.,][]{gomezllanosmorisset2020}, which will affect the intensity of [\ion{O}{3}] $\lambda$4363 preferentially.  Hence, the often-adopted tactic of correcting for the recombination contribution to the [\ion{O}{3}] $\lambda$4363 line assuming a flat spatial distribution of the [\ion{O}{3}] temperature through the central zone of the nebula (where the He$^{2+}$ is found) is likely to over-correct for the contribution of recombination to the excitation of this line, i.e., one problem substitutes for another.  

If the physical conditions indicated by the \ion{N}{2} and \ion{O}{2} lines are representative (\S\ref{sec_physcond_perm}), the additional plasma component is colder by a factor of $\sim 3$ and denser by a factor of $\gtrsim 2$ compared to the normal nebular plasma in NGC 6153.  If so, the two plasma components appear to be close to being in pressure equilibrium.  Using similar means, \citet{richeretal2019} found that the two plasma components may be in pressure equilibrium in NGC 7009.  \citet{peimbertetal2014} found a similar result for their sample of planetary nebulae, with the pressures computed using forbidden and permitted lines being within a factor of a few of being in equilibrium.  

Although there is no commonly-accepted explanation of how they are maintained, temperature fluctuations appear to be important in the normal nebular plasma in NGC 6153, so they should be taken into account.  While it is tempting to argue that any abundance calculation that does not consider temperature fluctuations should be considered as only a lower limit to the true value, both this study and that of \citet{richeretal2019} of NGC 7009 demonstrate that there may be regions within any given object where temperature fluctuations are negligible.  So, while it is not simple to decide how to proceed, having multiple reliable temperature indicators is the best guide as to which temperature to adopt when calculating chemical abundances.  

NGC 6153 joins a growing list of objects whose permitted lines of \ion{C}{2}, \ion{N}{2}, \ion{O}{2}, and \ion{Ne}{2} have kinematics that are anomalous with respect to the ionization structure defined by other lines \citep{sharpeeetal2004, barlowetal2006, otsukaetal2010, richeretal2013, richeretal2017, penaetal2017, ruizescobedopena2022}.  Unfortunately, multiple plasma components can only be recognized via high resolution spectroscopy or if the spatial profile of the emission lines is available \citep[e.g.,][]{liuetal2000, garnettdinerstein2001, tsamisetal2008, monrealiberowalsh2020, garciarojasetal2022}.  

In most previous studies of this issue, the additional plasma component is represented only through the emission of the \ion{C}{2}, \ion{N}{2}, \ion{O}{2}, and \ion{Ne}{2} lines.  However, \citet{corradietal2015} and \citet{garciarojasetal2022} find that the \ion{O}{1} $\lambda\lambda$7771,7774,7775 lines imply an O$^+$ abundance discrepancy in Ou 5, NGC 6778, M 1-42, and Hf 2-2.  Here, we also find that the emission from the permitted quintuplet lines of \ion{O}{1}, which cannot be excited by fluorescence from the ground state, have kinematics that are compatible with belonging to the additional plasma component.  Thus, \ion{O}{1} represents a second ionization stage of oxygen identified in the additional plasma component.  This is perhaps the most significant evidence that the additional plasma component is a distinct plasma from the normal nebular plasma, since the presence of \ion{O}{1} emission indicates that there are two ionization structures.  

Here, we also find similar kinematics for the additional plasma component using lines of \ion{O}{2} that span a wide range of energy levels (Table \ref{tab_line_splitting}), e.g., the upper level of \ion{O}{2} $\lambda$6501.4 is $2\mathrm p^2\,6\mathrm g$ within 1.5\,eV of the ionization limit while the lower level of the V1 multiplet, $2\mathrm p^2\,3\mathrm s$, is the first excited state above the ground level (for the $2\mathrm p^2\,(^3\mathrm P)$ core configuration).  This demonstrates that fluorescence, including indirect fluorescence, is not a viable explanation of the kinematics of the additional plasma component in NGC 6153, or of the abundance discrepancy problem more generally.  

Both \citet{liuetal2000} and \citet{tsamisetal2008} found that the $\mathrm O^{2+}/\mathrm H$ ratio from permitted lines was more centrally concentrated in NGC 6153 than the same ratio computed from forbidden lines.  Figures \ref{fig_PV_ADF_OII} and \ref{fig_mass_frac_PV} indicate the same result here.  This agrees with essentially all studies that can address the issue, the spatial distribution of the emission from the permitted lines of \ion{C}{2}, \ion{N}{2}, \ion{O}{2}, and \ion{Ne}{2} is more compact than that of the forbidden lines from the same ions \citep[e.g.,][]{barker1982, barker1991, garnettdinerstein2001, luoliu2003, garciarojasetal2022}.  Thus, while the kinematics of the \ion{C}{2}, \ion{N}{2}, \ion{O}{2}, and \ion{Ne}{2} lines can address the spatial location of the plasma only indirectly, via the hypothesis that the multiple plasmas follow a typical relation between velocity and ionization energy \citep{wilson1950}, the spatial distribution confirms that both are found in the inner part of the nebula, close to the central star, supporting a common \citet{wilson1950}-like relation for all plasma components.  So far, though, no study is able to spatially isolate a ``clump" of the additional plasma component \citep{garciarojasetal2022}, so its topology within the nebular shell remains a mystery.  

As \citet{gomezllanosmorisset2020} point out, the abundance discrepancy factor is a poor descriptor to relate the chemical compositions of the two plasmas because of the difficulty of apportioning the emission from hydrogen between the two plasma components.  Fortunately, they find that the normal nebular plasma emits the great majority of the hydrogen emission.  While it would be useful to have a more general study of this effect, it seems likely that the basic result will remain, as we find for NGC 6153.  Under these circumstances, it is probably more instructive, as \citet{liuetal2006} first pointed out, to compare the masses of O$^{2+}$ (or CNONe elements) in the two plasma components.  That way, the ``usual" techniques to determine the chemical composition of the normal nebular plasma may be leveraged to estimate the overall chemical composition considering both plasma components.  Here, it appears that the overall abundance of N$^{2+}$ and O$^{2+}$ is no more than double the abundance of these ions in the normal nebular plasma in NGC 6153.   Similar results have been found for NGC 7009, NGC 6778, M 1-42, and Hf 2-2 \citep{liuetal2006, richeretal2019, garciarojasetal2022}.  

An important observation is that abundance ratios, such as $\mathrm C/\mathrm O$, are similar in both plasma components, as was first noted by \citet[][]{liuetal2000}.  When two plasma components are present with arbitrary absolute abundances of C, N, and O, it is not evident why the relative abundances in the two components need be maintained.  Hence, it is unlikely that the additional plasma component has an origin in an event such as a very late thermal pulse, since that would dramatically change the relative abundances \citep[e.g., as observed in A 38, A 58, and A 78;][]{jacobyford1983,lauetal2011}.  However, as \citet{garciarojasetal2022} point out, common relative abundances could arise naturally if the additional plasma component is ejected later from a more internal layer of the progenitor star.  The question would then be why that would happen or be evident only in some objects, such as NGC 6153, and not in all.  

The work of \citet{bautistaahmed2018}, who model a nebula whose central star is periodically eclipsed, may offer an alternative to the two plasma components we advocate here.  The particular nebula they model is not a good approximation to NGC 6153, but, if the amplitude of their resonant temperature fluctuations varies sufficiently rapidly with depth, it might mimic the behavior of two plasma components within a single plasma and binary stars appear to be very common among planetary nebulae with large abundance discrepancies \citep[e.g.,][]{wessonetal2018}.  While the return to a single plasma is a major simplification, the model's time-variable ionizing flux and the plasma's departure from a steady state equilibrium are very significant complications.  We encourage further work on this front as it is congruent with our argument that it is necessary to better consider the complexity of the nebular plasma.

\section{Conclusions}\label{sec_conclusions}

We present a detailed study of the kinematics of the emission lines in NGC 6153 based upon echelle spectra obtained with the UVES spectrograph at the ESO VLT.  Our first conclusion is that the plasma in NGC 6153's nebular shell is complex.  We find that there are two plasma components.  What we call the normal nebular plasma emits in all lines and behaves generally as described in textbooks.  What we call the additional plasma component emits in the lines of \ion{O}{1}, \ion{C}{2}, \ion{N}{2}, \ion{O}{2}, and \ion{Ne}{2}.  The additional plasma component contributes a small fraction of the \ion{H}{1} emission.  Within both plasma components, the usual physics of nebular plasmas appears to hold and we find consistent results for the physical conditions from multiple lines of evidence.  

The kinematics of the two plasma components are distinct.  Except for the permitted lines of \ion{O}{1} (quintuplet states), \ion{C}{2}, \ion{N}{2}, \ion{O}{2}, and \ion{Ne}{2}, all lines follow a single relation between velocity splitting and ionization energy \citep{wilson1950}, and is the means by which we define the normal nebular plasma.  These permitted and forbidden lines arise from a variety of physical processes (recombination, collisional excitation, Bowen fluorescence, and charge exchange).  The permitted lines of \ion{O}{1} (quintuplets), \ion{C}{2}, \ion{N}{2}, \ion{O}{2}, and \ion{Ne}{2} all have the same velocity splitting even though the energies needed to create the parent ions differ widely, defining the additional plasma component.  Thus, we confirm a second ionization state, O$^+$, for the additional plasma component \citep[see also][]{corradietal2015, garciarojasetal2022}.  In the case of the \ion{O}{2} lines, lines arising from energy levels spanning from just below the ionization threshold to the first excited state above ground all present similar kinematics.  Multiple ionization stages of oxygen and the similar kinematics from a wide variety of lines of \ion{O}{2} argue that two plasma components coexist in the nebular shell of NGC 6153.  As has been found in studies of the spatial distribution of \ion{C}{2}, \ion{N}{2}, \ion{O}{2}, and \ion{Ne}{2} lines, we also find that these lines are concentrated towards the inner part of NGC 6153's main shell \citep[e.g.,][]{barker1982, barker1991, liuetal2000, garnettdinerstein2001, tsamisetal2008, garciarojasetal2016, garciarojasetal2022}.  

We determine the physical conditions in both plasma components.  For the normal nebular plasma, the kinematics, the \ion{He}{1} lines and the forbidden lines of [\ion{N}{2}] (once corrected), [\ion{O}{3}], and [\ion{Ar}{3}] all imply temperatures of $8,000-10,000$\,K while the \ion{H}{1} lines and the forbidden lines of [\ion{S}{2}], [\ion{Cl}{3}], and [\ion{Ar}{4}] imply electron densities in the neighborhood of 5,000\,cm$^{-3}$.  For the additional plasma component, we determine the electron temperature and density from the \ion{N}{2} and \ion{O}{2} lines, finding $T_e\sim 2,000$\,K and $N_e>5,000$\,cm$^{-3}$ (and possibly as high as $10^5$\,cm$^{-3}$).  The [\ion{O}{2}] $\lambda\lambda$3726,3729 lines imply a density of $\sim 10,000$\,cm$^{-3}$ where their emission may be dominated by recombination excitation in the additional plasma component.  Given these physical conditions, the additional plasma component contains a mass of N$^{2+}$ and O$^{2+}$ similar to that found in the normal nebular plasma (for $N_e=10,000$\,cm$^{-3}$; a smaller fraction if the density is higher).  When we compute the ADF for O$^{2+}$ resolved in velocity and position, we find that it peaks in the inner part of NGC 6153's main shell, but that the ADF falls to unity in the diffuse emission beyond the receding side of the main shell, i.e., there is no abundance discrepancy in that volume of the plasma.  

Supposing the existence of two plasma components, we find that recombination excitation from the additional plasma component dominates the emission in the [\ion{N}{2}] $\lambda$5755 and [\ion{O}{2}] $\lambda$7320,7330 auroral lines in some volumes of the nebular shell \citep[e.g.,][]{liuetal2000, tsamisetal2003, corradietal2015, ruizescobedopena2022}.  However, there are also volumes of the plasma where the collisional excitation dominates the emission and the usual interpretation of these lines applies.  We recommend caution in using the auroral transitions of the [\ion{N}{2}] and [\ion{O}{2}] lines in any circumstance when the observation includes a large volume of the O$^{2+}$ zone.  A similar exercise shows that the emission in the [\ion{O}{3}] $\lambda$4363 line is dominated by the normal nebular plasma and its use as a temperature indicator is unaffected in NGC 6153.  However, as for the auroral lines of [\ion{N}{2}] and [\ion{O}{2}], the [\ion{O}{3}] $\lambda$4363 line may be suspect when the He$^{2+}$ zone dominates the volume included in the observation.   

There are clear temperature variations within the normal nebular plasma.  There are large-scale variations, for instance a higher temperature where helium is doubly ionized, though even the He$^{2+}$ zone is not near isothermal.  Although the [\ion{Ar}{3}] and [\ion{O}{3}] temperatures are very similar, differing by only $\sim 300-600$\,K, they both exceed the temperature found from the kinematics of the H$\beta$, H$\gamma$, \ion{He}{1} $\lambda\lambda$4471,4922, and [\ion{O}{3}] $\lambda$4959 lines, which should be sensitive to the motions of the ions involved.  The average Balmer jump temperature in NGC 6153 found by others is consistent with the temperatures we find for the two plasma components \citep{liuetal2000, zhangetal2004, mcnabbetal2016}.  However, we cannot explain why it is constant over the minor axis \citep{liuetal2000} since the concentration of the additional plasma component to the central part of the nebula should lower it there, as found by \cite{garciarojasetal2022}.  

When two plasma components are present, with distinct physical conditions, the relevance of an abundance discrepancy becomes less obvious.  Furthermore, the ADF may be a poor description of the abundances in the additional plasma component \citep{gomezllanosmorisset2020, garciarojasetal2022}, so instead we advocate following \citet{liuetal2006} in computing the fraction of O$^{2+}$ (or CNONe ions) contributed by the additional plasma component as a clearer indicator of the relevance of the additional plasma component.  The observation that the relative abundances of the CNONe elements appear to be similar in both plasma components indicates that the two plasmas are not completely independent \citep{liuetal2000}.  

\begin{acknowledgments}

We gratefully acknowledge financial support from UNAM DGAPA grant PAPIIT IN103519, CONACyT grant A1-S-15140, and the Departamento de Investigaci\'on of the Universidad Iberoamericana.  We thank the anonymous referee for helping to improve this work very substantially.

\end{acknowledgments}

\software{IDL, IRAF \citep{tody1986, tody1993}, Microsoft Office \citep{microsoftcorporation}, Python \citep{vanrossumdrake2009}, PyNeb \citep{luridianaetal2015}, SAOImageDS9 \citep{joyemandel2003}}

\bibliography{}{}
\bibliographystyle{aasjournal}



\appendix

\section{Morphology of the PV diagram}\label{app_ionization_structure}

Here, we compare the morphology observed in the PV diagrams for emission lines emitted by many ions.  Our primary intention is to demonstrate the robustness of the results presented in \S\ref{sec_ionization_structure}, since PV diagrams provide more information than may be represented in the Wilson diagram (Figure \ref{fig_Wilson_diagram}).

\begin{figure*}
\begin{center}\includegraphics[width=0.86\linewidth]{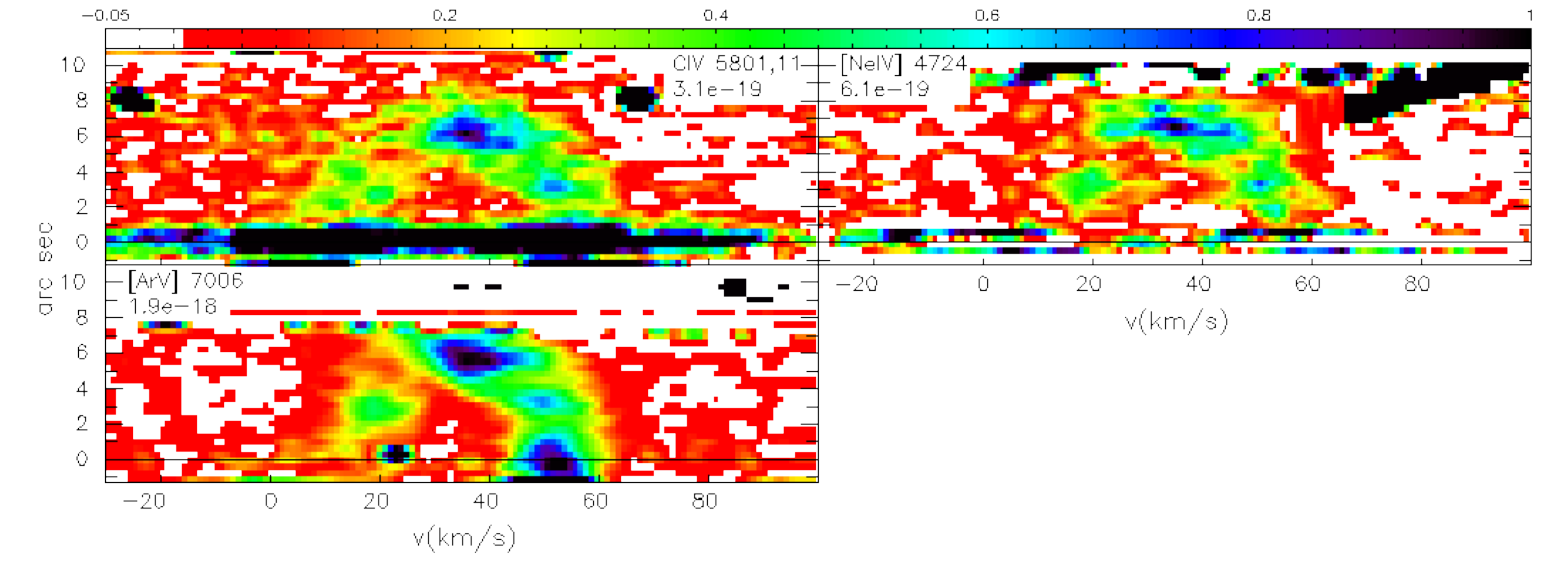}\end{center}
\caption{These panels present the PV diagrams of the lines of highest ionization: C IV $\lambda\lambda$5801,5811, [Ne IV] $\lambda$4724, and [Ar V] $\lambda$7006.  Note that there is also C IV emission from the central star.  In general, the receding side of the main shell and its outer edge (at top) are the brightest features.  The approaching side of the main shell is fainter, but a feature half-way to the outer edge is visible in all three PV diagrams.  Although it is discontinuous, the velocity ellipse has a semicircular shape in these PV diagrams.  
}
\label{fig_app_PVC4}
\end{figure*}

\begin{figure*}
\begin{center}\includegraphics[width=0.86\linewidth]{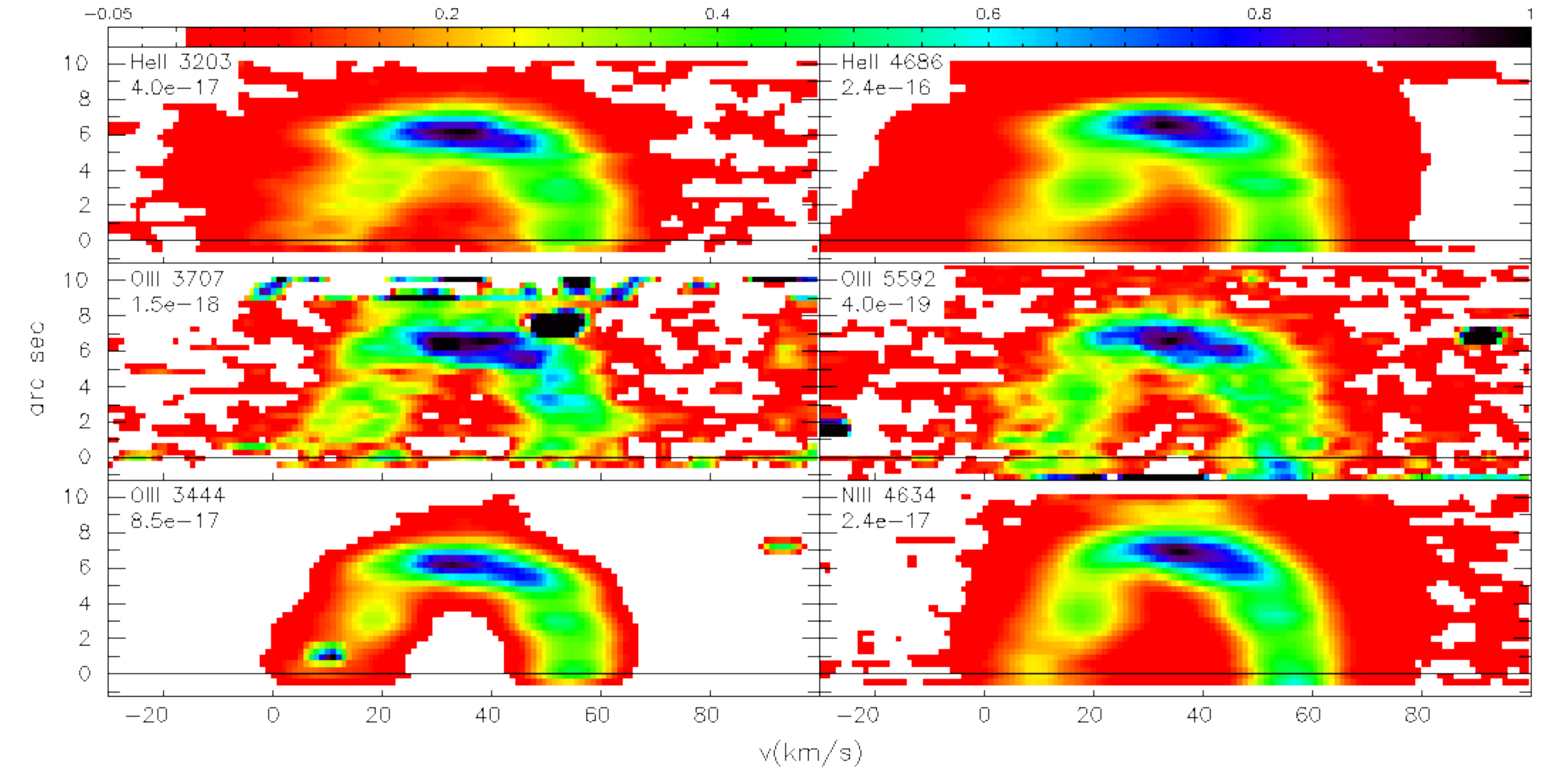}\end{center}
\caption{These panels present lines of high ionization:  \ion{He}{2} $\lambda$3203, \ion{He}{2} $\lambda$4686, \ion{O}{3} $\lambda$3707 (recombination), \ion{O}{3} $\lambda$5592 (charge exchange), \ion{O}{3} $\lambda$3444 (Bowen fluorescence), and \ion{N}{3} $\lambda$4634 (Bowen fluorescence).  In terms of shape, these PV diagrams are more square than those in Figure \ref{fig_app_PVC4} and extend to a larger distance from the central star.  
}
\label{fig_app_PVHe2}
\end{figure*}

\begin{figure*}
\begin{center}\includegraphics[width=0.86\linewidth]{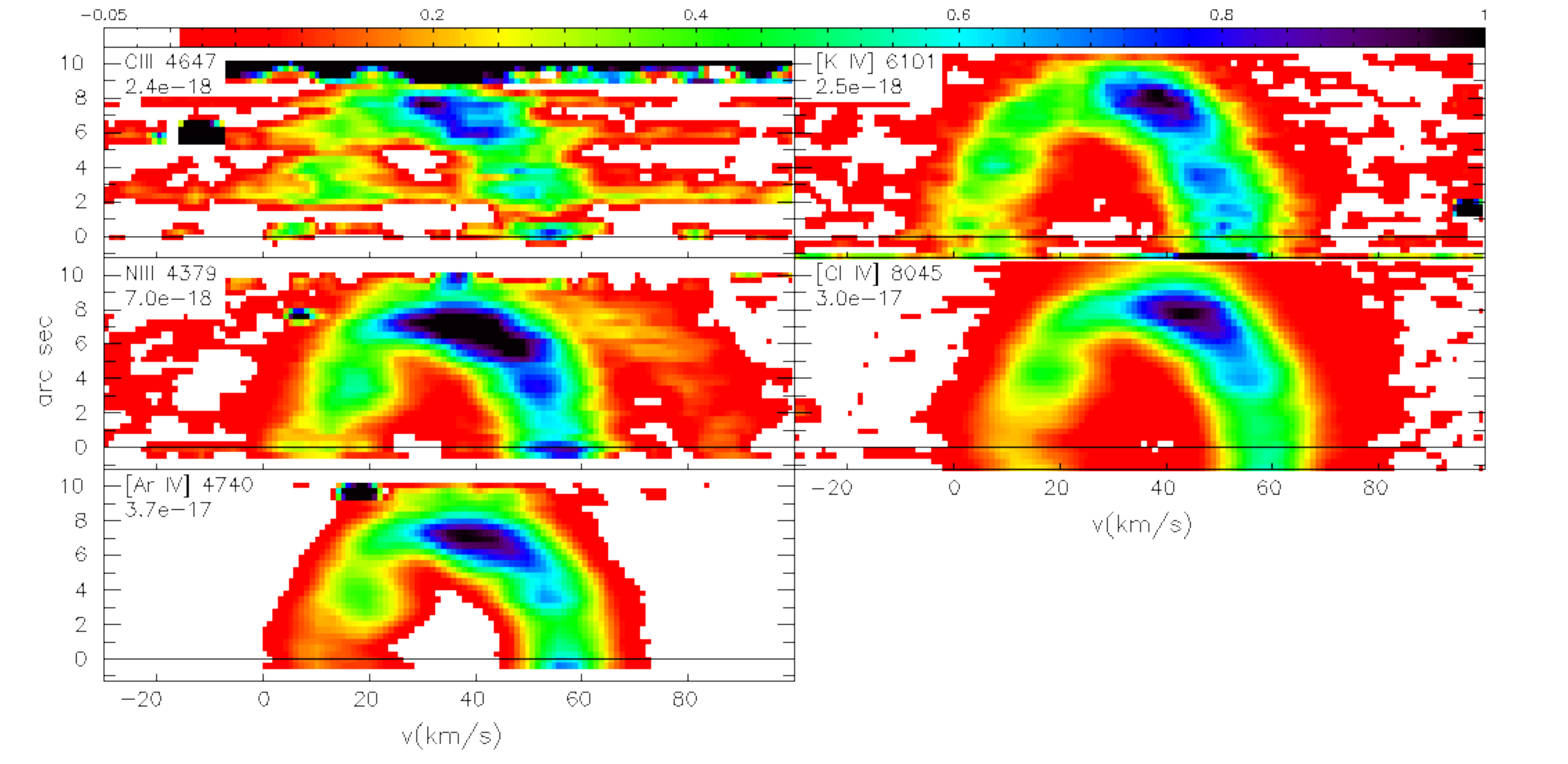}\end{center}
\caption{These panels present the PV diagrams for the lines of \ion{C}{3} $\lambda$4647, [\ion{K}{4}] $\lambda$6101, \ion{N}{3} $\lambda$4379 (\ion{O}{3} $\lambda$4379.6 is the second/redder line), [\ion{Cl}{4}] $\lambda$8045, and [\ion{Ar}{4}] $\lambda$4740.  The velocity ellipse in these PV diagrams has returned to a rounder shape than in Figure \ref{fig_app_PVHe2}, largely because of the greater velocity splitting.  
}
\label{fig_app_PVC3}
\end{figure*}

\begin{figure*}
\begin{center}\includegraphics[width=0.86\linewidth]{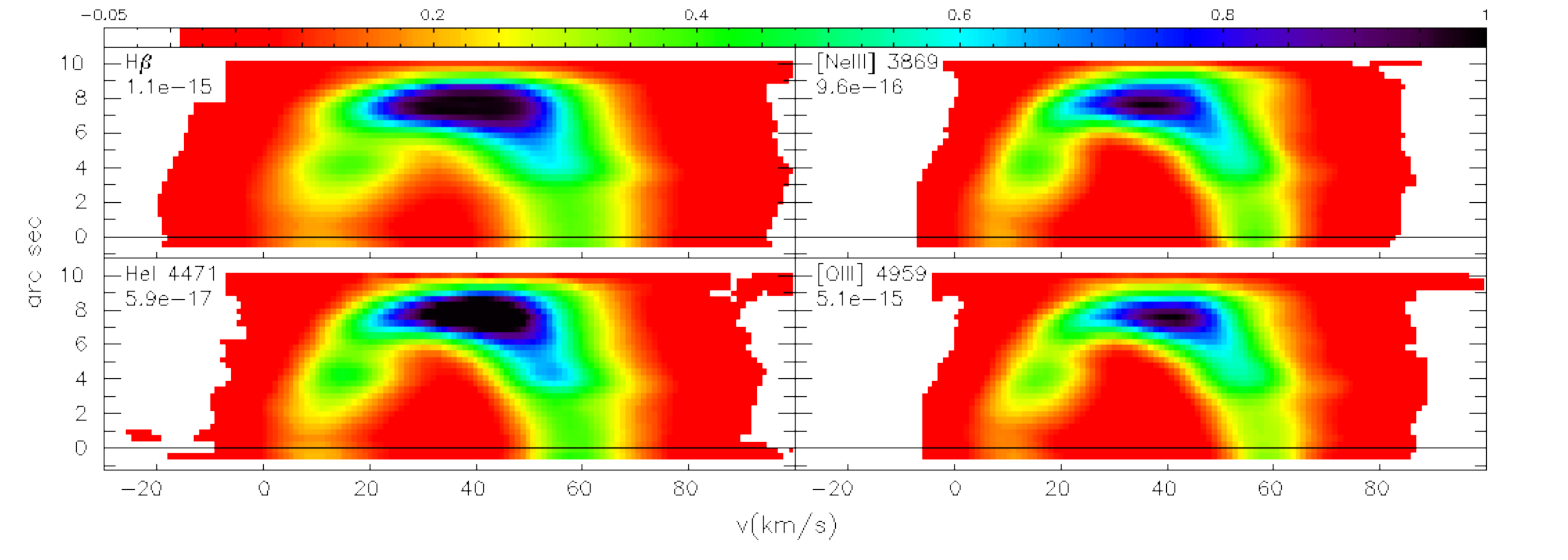}\end{center}
\caption{These panels present the PV diagrams in the lines of \ion{H}{1} $\beta$, [\ion{Ne}{3}] $\lambda$3869, \ion{He}{1} $\lambda$4471, and [\ion{O}{3}] $\lambda$4959, lines that sample the majority of the mass of the nebula (all of it for \ion{H}{1}), all from the CD2 wavelength interval.  Compared to the lines in Figure \ref{fig_app_PVC3}, there is clearly less emission from the receding side of the main shell for lines of sight close to the central star.  
}
\label{fig_app_PVHbeta}
\end{figure*}

\begin{figure*}
\begin{center}\includegraphics[width=0.86\linewidth]{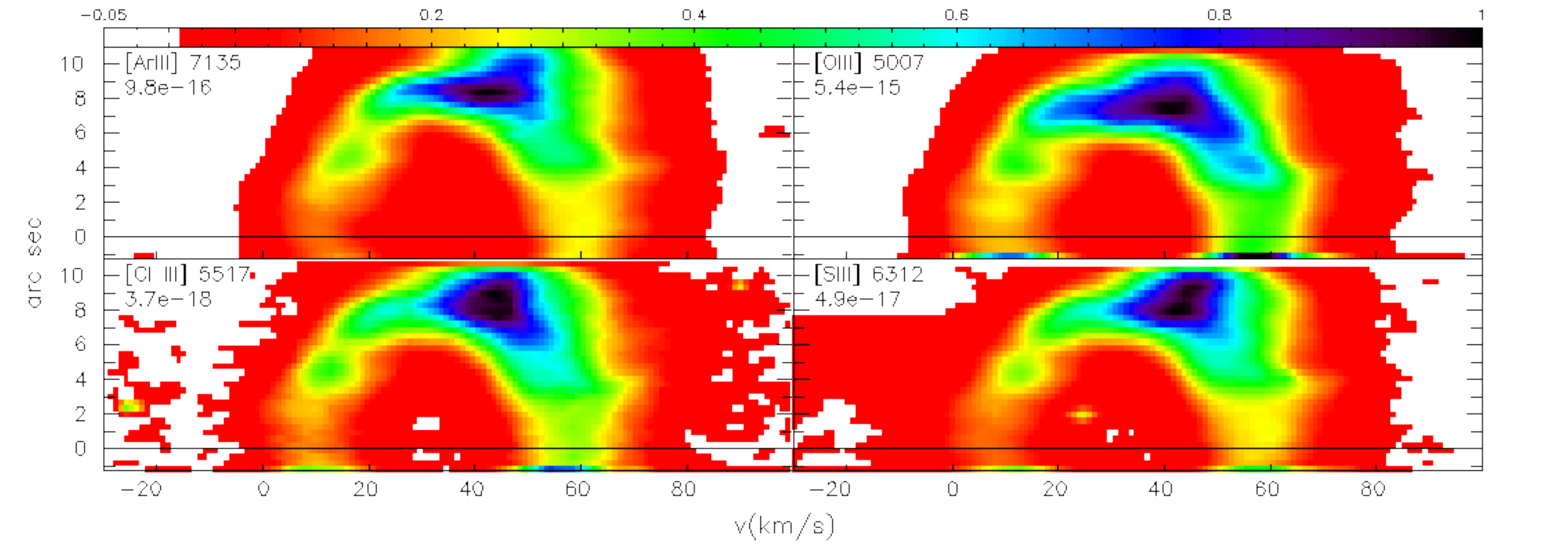}\end{center}
\caption{These panels present the PV diagrams for [\ion{Ar}{3}] $\lambda$7135, [\ion{O}{3}] $\lambda$5007, [\ion{Cl}{3}] $\lambda$5517, and [\ion{S}{3}] $\lambda$6312.  Compared to [\ion{O}{3}] $\lambda$5007, the other three lines have even less emission from the receding side of the main shell along lines of sight near the central star, but more emission from the filament beyond the receding side of the main shell (Figure \ref{fig_PV_ionization}).  The difference between [\ion{O}{3}] $\lambda$5007 and [\ion{O}{3}] $\lambda$4959 (Figure \ref{fig_app_PVHbeta}) is due to the longer slit for lines to the red of 5000\,\AA\ and a slight shift in the position of the slit perpendicular to its length.  
}
\label{fig_app_PVAr3}
\end{figure*}

\begin{figure*}
\begin{center}\includegraphics[width=0.86\linewidth]{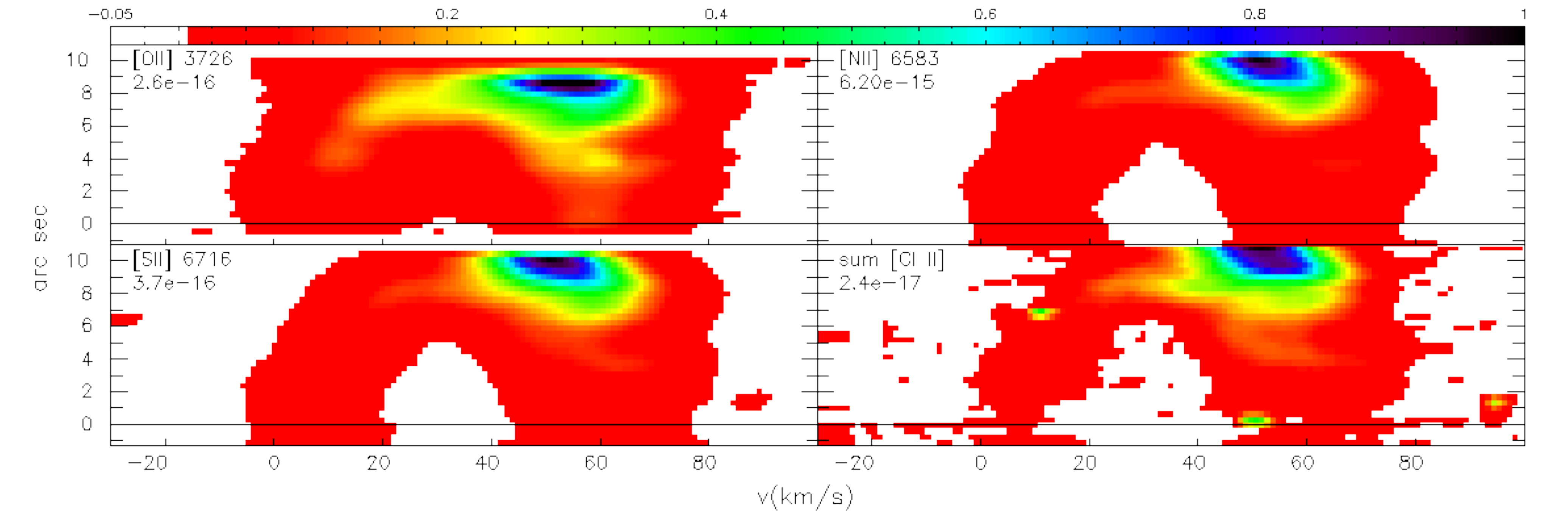}\end{center}
\caption{These panels present the PV diagrams for the lines of [\ion{O}{2}] $\lambda$3726, [\ion{N}{2}] $\lambda$6583, [\ion{S}{2}] $\lambda$6716, and the sum of [\ion{Cl}{2}] $\lambda\lambda$8578,9123.  In all of these lines, the emission from the main shell is much weaker than in the previous figures while the only bright emission comes from the filament outside the main shell.  Even so, the main shell is brighter in the [\ion{O}{2}] $\lambda$3726 line than in the other lines, the result of collisional excitation (\S\ref{sec_contamination}).  
}
\label{fig_app_PVO2}
\end{figure*}

\begin{figure*}
\begin{center}\includegraphics[width=0.86\linewidth]{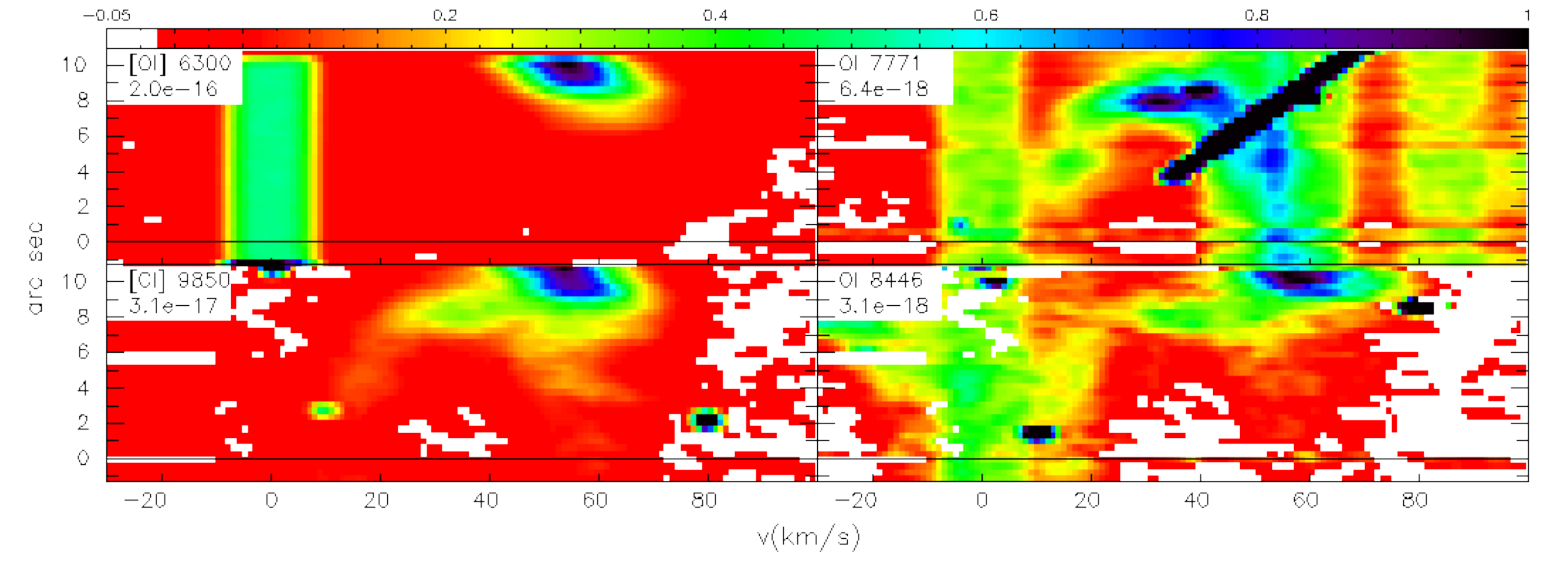}\end{center}
\caption{The panels in the left column present the PV diagrams in the lines of lowest ionization potential, [\ion{O}{1}] $\lambda$6300 and [\ion{C}{1}] $\lambda$9850.  In [\ion{O}{1}] $\lambda$6300, emission from the main shell is now nearly absent.  In the right column, the panels contrast the PV diagrams of the \ion{O}{1} $\lambda$7771 and \ion{O}{1} $\lambda$8446 lines.  The different morphologies likely arise due to different physical processes contributing to the two lines.  The black streak is a cosmic ray.  
}
\label{fig_app_PVC1}
\end{figure*}

Figures \ref{fig_app_PVC4}-\ref{fig_app_PVC1} present a selection of the many PV diagrams from which we derive the Wilson diagram (Figure \ref{fig_Wilson_diagram}).  The order of these figures is from the lines arising from the ions with the highest degree of ionization (e.g., C$^{4+}$; Figure \ref{fig_app_PVC4}) through the lines arising from the ion with the lowest degree of ionization (e.g., O$^0$; Figure \ref{fig_app_PVC1}).  The morphologies of these PV diagrams present coherent trends and so define a kinematic structure that varies continuously as a function of ionization.  We recall that the slit length is shorter for wavelengths below 5000\AA\ than for longer wavelengths.  

Considering Figures \ref{fig_app_PVC4}-\ref{fig_app_PVO2} individually, the component PV diagrams in each figure present similar morphologies.  That is, in Figure \ref{fig_app_PVC4}, for example, although the S/N varies (and is never high), the PV diagrams for \ion{C}{4} $\lambda\lambda$5801,5811, [\ion{Ne}{4}] $\lambda$4724, and [\ion{Ar}{5}] $\lambda$7006 show similar structures.  In all lines, the emission from the main shell is broken into 3-4 components that define a velocity ellipse.  The fourth component, due to the receding side of the main shell along the line of sight to the central star, is only visible in the [\ion{Ar}{5}] $\lambda$7006 line due to residuals in the subtraction of the emission from the central star.  (In the case of \ion{C}{4} $\lambda\lambda$5801,5811, the central star has emission lines that are not removed by the continuum subtraction.)  In all lines, the top of the main shell is the brightest feature while the approaching and receding sides of the main shell are fainter.  Also, the spatial extent of the main shell in all of these highly ionized ions is similar, i.e., the top of the ellipse defined by the main shell is at the same distance from the central star in all of the PV diagrams.  Although the details vary, in each of Figures \ref{fig_app_PVC4}-\ref{fig_app_PVO2}, the morphologies of the component PV diagrams are similar, indicating that the lines presented in each of these figures arise from volumes of plasma of with similar kinematics and similar spatial distributions, i.e., they are largely the same volume of plasma in each figure.  

There are also systematic changes when proceeding from figure to figure in the sequence from Figure \ref{fig_app_PVC4} to Figure \ref{fig_app_PVC1}.  For instance, the  velocity ellipse in the three PV diagrams in Figure \ref{fig_app_PVC4} has a shape that is approximately a semicircle while it is somewhat more square in Figure \ref{fig_app_PVHe2}.  Also, the edge of the main shell is farther from the position of the central star in Figure \ref{fig_app_PVHe2} than in Figure \ref{fig_app_PVC4}.  The similar morphologies of the PV diagrams in Figure \ref{fig_app_PVHe2} indicates that they arise in similar volumes of the nebular plasma, in spite of the variety of physical processes involved.  Comparing Figure \ref{fig_app_PVHe2} ($\mathrm{He}^{++}/\mathrm O^{3+}$ zone and $\mathrm O^{3+}\rightarrow \mathrm O^{++}$ transition zone) and Figure \ref{fig_app_PVC3} (the zone containing C$^{3+}$, N$^{3+}$, Cl$^{3+}$, Ar$^{3+}$, and K$^{3+}$), the emission from the receding side of the main shell becomes more spatially continuous, i.e., a full arc is now visible from the receding side of the main shell.  These changes reflect a change in the volume of the plasma that emits in the lines presented in the two figures.  Proceeding from Figure \ref{fig_app_PVC3} to Figure \ref{fig_app_PVHbeta}, the main change is a decrease in the brightness of the receding side of the main shell along lines of sight near the central star in Figure \ref{fig_app_PVHbeta}, which presents the PV diagrams for the lines that sample most (or all) of the nebular mass.  This effect continues in Figure \ref{fig_app_PVAr3}, where the emission is quite faint from the receding side of the nebular shell along lines of sight near the central star (except for [\ion{O}{3}] $\lambda$5007, which repeats from the previous figure).  However, in Figure \ref{fig_app_PVAr3}, the emission from the filament on the approaching side of the nebula is now fainter and the emission from the filament, beyond the main shell on the receding side of the nebula (Figure \ref{fig_pv_diagram}), is now clearly present.  From Figure \ref{fig_app_PVAr3} to Figure \ref{fig_app_PVO2}, the emission from the entire main shell decreases significantly while the brightness of the filament increases substantially.  Finally, considering the PV diagram for [\ion{O}{1}] $\lambda$6300 in Figure \ref{fig_app_PVC1}, the emission from the main shell is nearly absent while that from the filament is now very bright. 

Hence, the sequence from Figure \ref{fig_app_PVC4} to Figure \ref{fig_app_PVO2} (and [\ion{O}{1}] $\lambda$6300 in Figure \ref{fig_app_PVC1}) shows that lines with similar ionization potentials, expected to be found together in the nebular plasma, present PV diagrams with similar morphologies.  Since the morphology of the PV diagram is a result of the kinematics and spatial distribution of the volume of plasma that emits a given line, similar morphologies mean that the lines with similar ionization potentials are emitted together.  Also, progressing from along the sequence, the changes are gradual, and continuous.  Based upon the PV diagrams in Figure \ref{fig_app_PVC4} to Figure \ref{fig_app_PVO2}, it is clear that the approaching and receding sides of the main shell are more highly ionized than the end of the main shell or the material outside the main shell on the receding side of the nebula.  This structure results naturally if the sides of the main shell are closer to the central star than is the end of the main shell or if the end of the main shell has a higher density or a greater depth.  

We now consider the puzzles presented by several lines, starting with the \ion{O}{1} $\lambda\lambda$7771,8846 lines in Figure \ref{fig_app_PVC1}.  The first is a quintuplet line that cannot be excited by fluorescence from the atom's ground state (triplet), but the \ion{O}{1} $\lambda$8446 triplet line is part of a decay cascade that can be excited by the \ion{H}{1} Ly$\beta$ line.  The emission from the \ion{O}{1} $\lambda$7771 line traces only the main shell, but the \ion{O}{1} $\lambda$8446 line also has emission from the filament outside the main shell.  The velocity splitting of the \ion{O}{1} $\lambda$7771 line and the morphology of its PV diagram are very similar to those for the \ion{C}{2}, \ion{N}{2}, \ion{O}{2}, and \ion{Ne}{2} lines.  (Two sky lines perturb the shape of the PV diagram of \ion{O}{1} $\lambda$7771 and the S/N is only modest.)  As for \ion{O}{1} $\lambda$8446, its velocity splitting may be compatible with that of the \ion{C}{2}, \ion{N}{2}, \ion{O}{2}, and \ion{Ne}{2} lines, but the details of its PV diagram are even less clear since it is fainter and the blue side of the line profile is blended with a cluster of 3 sky lines.  It is conceivable that the emission from the main shell in the \ion{O}{1} $\lambda$8446 line is due to the plasma that emits in the \ion{C}{2}, \ion{N}{2}, \ion{O}{2}, and \ion{Ne}{2} lines, in which case the emission from the main shell would be due to recombination (the O$^+$ ion).  It is likely that the emission in \ion{O}{1} $\lambda$8446 from the filament is due to fluorescence (and so the O$^0$ ion).  

The [\ion{C}{1}] $\lambda$9850 line in Figure \ref{fig_app_PVC1} may, like \ion{O}{1} $\lambda$8446, also represent a mixture of C ions and physical processes.  The velocity splitting we measure is intermediate between those of the additional plasma component and of [\ion{O}{1}] $\lambda\lambda$6300,6364 (Table \ref{tab_line_splitting}).  Emission from the main shell is clearly present.  However, the brightest feature in the PV diagram is the filament on the receding side of the main shell.  In this case, the emission from the main shell may be a combination of recombination (the C$^+$ ion) and collisional excitation (the C$^0$ ion), but the latter would dominate in the filament (the C$^0$ ion).  If so, the [\ion{C}{1}] $\lambda$9850 line may represent a second ionization stage of C from the additional plasma component.  

\begin{figure}
\begin{center}\includegraphics[width=0.46\linewidth]{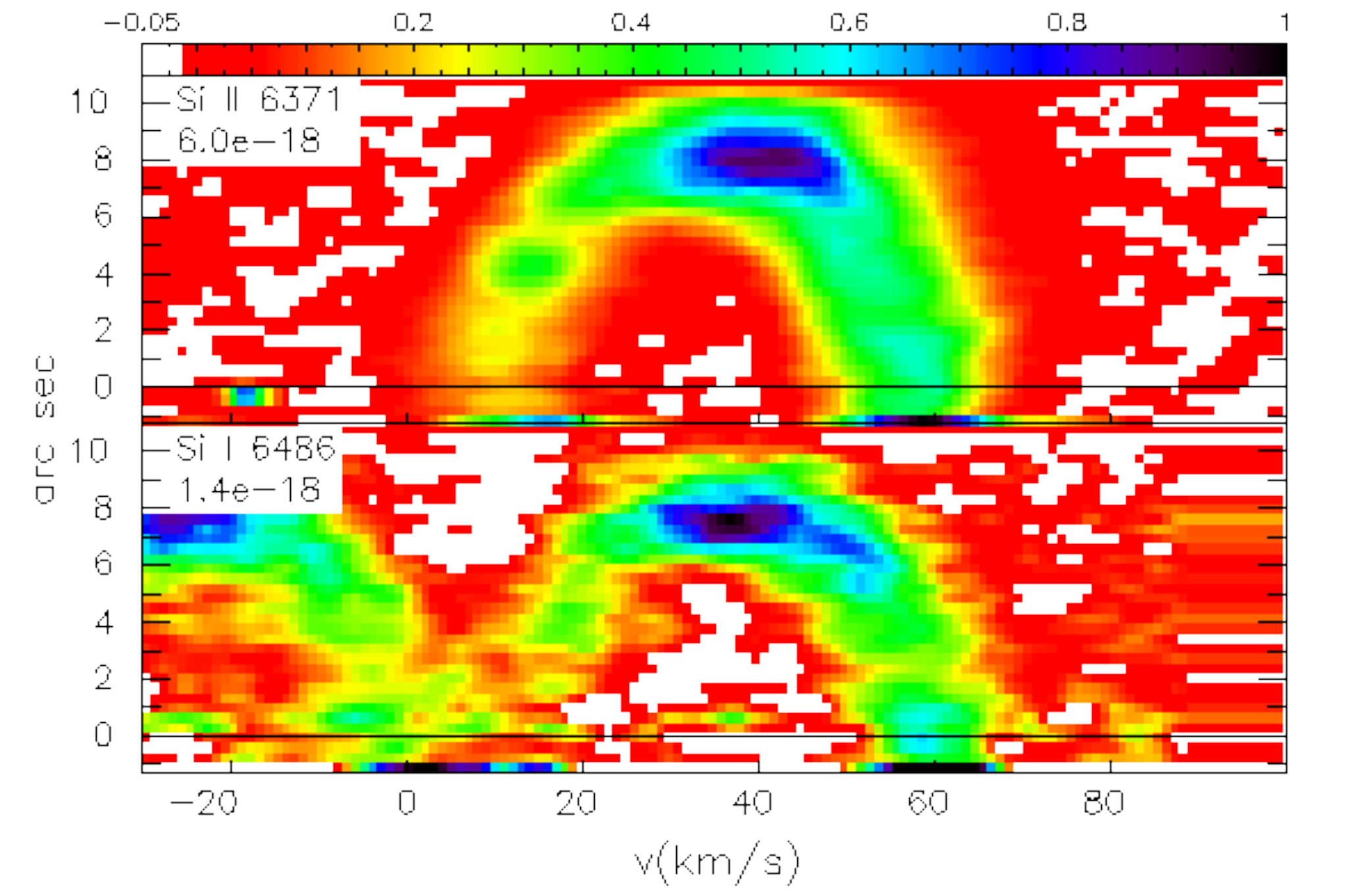}
\includegraphics[width=0.46\linewidth]{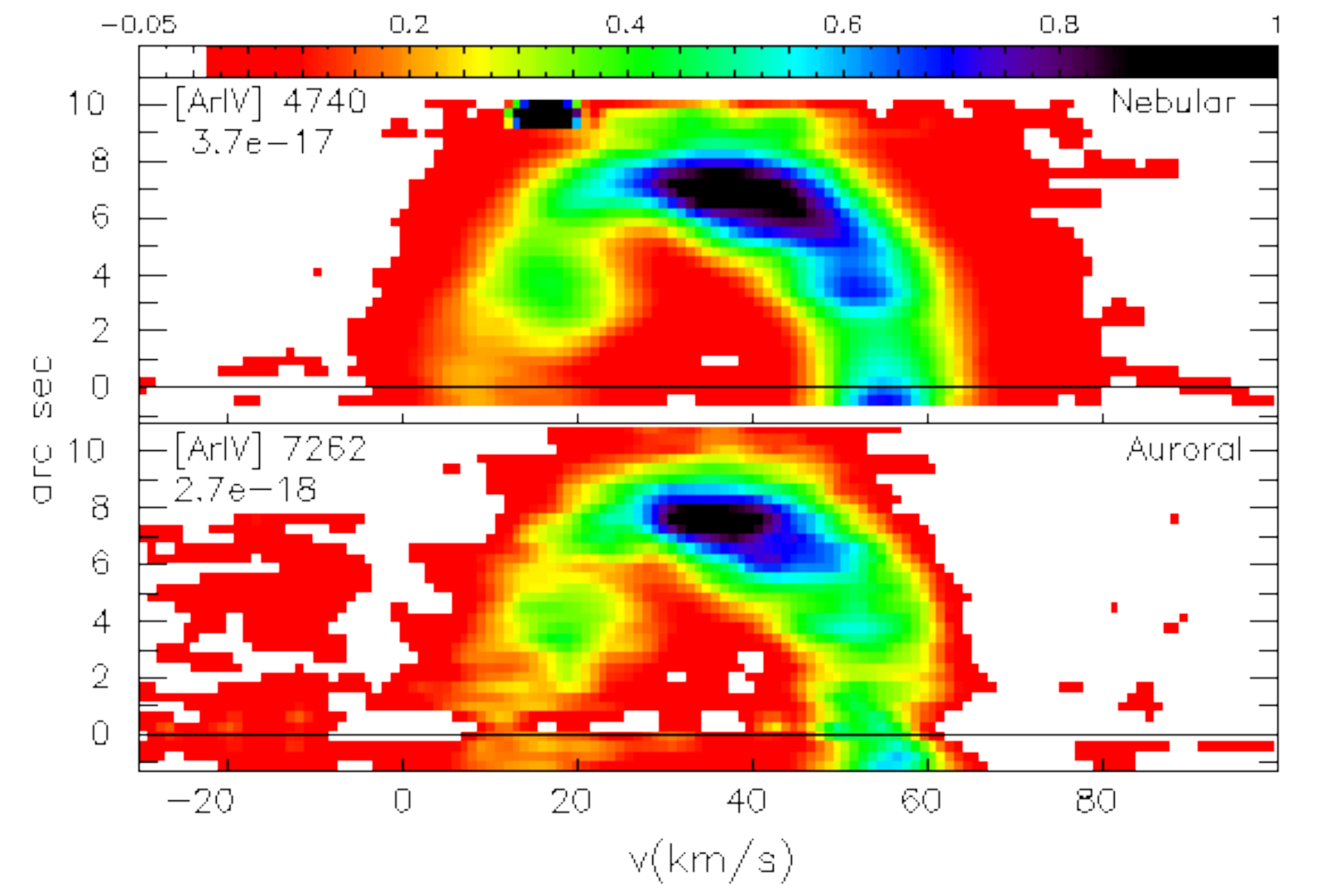}\end{center}
\caption{Left column:  These panels present the PV diagrams in the lines of \ion{Si}{2} $\lambda$6371 and \ion{Si}{1} $\lambda$6486.  Right column:  These panels present the PV diagrams in the lines of [\ion{Ar}{4}] $\lambda$4740 (nebular line) and [\ion{Ar}{4}] $\lambda$7262 (auroral line).  The vertical stripes of low counts (white) to the blue and red of the [\ion{Ar}{4}] $\lambda$7262 line profile are due to atmospheric absorption, which affects the measurement of the velocity splitting (Table \ref{tab_line_splitting}; Figure \ref{fig_Wilson_diagram}).    
}
\label{fig_app_PVSi2}
\end{figure}

The morphology of the PV diagram of the \ion{Si}{1} $\lambda$6486 line (Figure \ref{fig_app_PVSi2}; left column) is similar to that of the \ion{C}{2}, \ion{N}{2}, \ion{O}{2}, and \ion{Ne}{2} lines (Figure \ref{fig_PVC2}).  However, its velocity splitting is ``normal" for its ionization energy (Table \ref{tab_line_splitting}, Figure \ref{fig_pv_diagram}), though it is uncertain since the blue side of the line profile has low S/N and is blended with the adjacent lines to the blue.  

In the PV diagram of the \ion{Si}{2} $\lambda$6371 line (Figure \ref{fig_app_PVSi2}; left column), like those of the \ion{Si}{2} $\lambda\lambda$5041,6347 lines, the shape of the receding side of the main shell is reminiscent of that in Figures \ref{fig_app_PVC3} and \ref{fig_app_PVHbeta}, which represent more highly ionized plasma.  The velocity splitting of $46.7\pm 0.6$\,km/s is lower than expected ($\sim 50$\,km/s) for Si$^+$'s ionization energy, assuming the lines arise from recombination in the normal nebular plasma.  \ion{Si}{2} $\lambda$5041 is from multiplet V5 while \ion{Si}{2} $\lambda\lambda$6347,6371 are from multiplet V2.  The lower level of V5 is the upper level of V2.  The upper level of V5 can be radiatively excited from the Si$^+$ ground state \citep[989.87\AA][]{morton2003}.  Given the line splitting and the morphology of the PV diagram, there may be a fluorescence contribution in these lines in addition to recombination.  A potential fluorescence source is emission in the \ion{Ne}{3} $\lambda\lambda$989.90,989.91 lines from the He$^{2+}$ zone \citep[][\ion{Fe}{5} $\lambda\lambda$989.93,989.96 lines is a less likely possibility]{vanhoof2018}.  The population of Si$^+$ ions in the He$^{2+}$ zone will be minuscule, so the fluorescing ions will be from zones of lower ionization, though preferentially near the He$^{2+}$ zone since the efficiency of fluorescence drops with distance due to dilution.  If so, the kinematics inferred from these \ion{Si}{2} lines in NGC 6153 may mimic the kinematics of zones of higher ionization, and not necessarily the zone where S$^+$ is the dominant ionization stage. 

Finally, in Figure \ref{fig_app_PVSi2} (right column), the [\ion{Ar}{4}] $\lambda$7262 auroral line profile has a smaller velocity splitting than its nebular counterpart (Table \ref{tab_line_splitting}).  In this case, the difference appears to be due to telluric absorption affecting both the bluest emission from the approaching side of main shell and the reddest emission from the receding side, both of which bias the velocity splitting to lower values.  Unfortunately, we cannot easily check this conjecture since the other auroral lines of [\ion{Ar}{4}] are all contaminated by stronger lines ([\ion{Fe}{4}] $\lambda$7171, \ion{C}{2} $\lambda$7236, [\ion{O}{2}] $\lambda\lambda$7330,7331).

\end{document}